\documentclass[11pt]{article}
\pdfoutput=1
\usepackage{jheppub}
\usepackage[dvipsnames]{xcolor}
\usepackage{graphicx}
\usepackage{placeins}
\usepackage{wrapfig}
\usepackage{amsmath}
\usepackage{amsfonts}
\usepackage{amssymb}
\usepackage{multirow}
\usepackage{slashed}
\usepackage{physics}
\usepackage{dsfont}
\usepackage{soul}
\usepackage{ulem}
\usepackage{verbatim}
\usepackage{import}

\usepackage{cleveref}

\usepackage[utf8]{inputenc}
\usepackage{dcolumn} 
\usepackage{graphicx}
\usepackage{slashed,color}
\usepackage{amsmath,amsfonts,bm}
\usepackage{amssymb}

\usepackage{placeins}
\usepackage{calligra}
\usepackage[T1]{fontenc}

\def\OMIT#1{{}}

\title{
   Gaussian Processes for Inferring Parton Distributions
}
\newcommand*{\WM}{Physics Department, William \& Mary, Williamsburg, VA 23187, USA}
\newcommand*{\JLAB}{Thomas Jefferson National Accelerator Facility, Newport News, VA 23606, USA}
\newcommand*{\CNRS}{Aix Marseille Univ, Universit\'e de Toulon, CNRS, CPT, Marseille, France.}

\author[a] {Yamil Cahuana Medrano,}
\author[b,a]{Herv\'e Dutrieux,}
\author[c]{ Joseph Karpie,}
\author[a]{ Kostas Orginos,}
\author[b]{ Savvas Zafeiropoulos}
\author{\\(on behalf of the HadStruc Collaboration)}
\affiliation[a]{\WM}
\affiliation[b]{\CNRS}
\affiliation[c]{\JLAB}
\emailAdd{yacahuanamedra@wm.edu}
\emailAdd{herve.dutrieux@cpt.univ-mrs.fr}
\emailAdd{jkarpie@jlab.org}
\emailAdd{kostas@wm.edu}
\emailAdd{savvas.zafeiropoulos@cpt.univ-mrs.fr}
\newcommand{\ignore}[1]{}
\abstract{The extraction of parton distribution functions (PDFs) from experimental or lattice QCD data is an ill-posed inverse problem, where regularization strongly impacts both systematic uncertainties and the reliability of the results. We study a framework based on Gaussian Process Regression (GPR) to reconstruct PDFs from lattice QCD matrix elements. Within a Bayesian framework, Gaussian processes serve as flexible priors that encode uncertainties, correlations, and constraints without imposing rigid functional forms. We investigate a wide range of kernel choices, mean functions, and hyperparameter treatments. We quantify information gained from the data using the Kullback–Leibler divergence. Synthetic data tests demonstrate the consistency and robustness of the method. Our study establishes GPR as a systematic and non-parametric approach to PDF reconstruction, offering controlled uncertainty estimates and reduced model bias in lattice QCD analyses.}

\begin{document}

\date{\today}
\preprint{JLAB-THY-25-4579}
\maketitle

\section{Introduction}
Many scientific problems face an inverse problem in the form of functional inference, i.e., determining an underlying function from a finite set of observations. Such problems are often ill-posed, as infinitely many functions can reproduce the same finite dataset, and additional regularization is required to obtain stable and physically meaningful solutions. These ``observations'' can either be experimental or the result of ab initio computations such as those performed with lattice Quantum Chromodynamics (LQCD), which will be the focus of this work. 

While LQCD is a powerful ab initio method to study strong interaction physics non-perturbatively, many important physical problems cannot be directly addressed through the calculation of Euclidean correlation functions. These include, among others, the transport coefficients of the strongly coupled quark-gluon plasma (QGP)~\cite{Meyer:2007ic}, inclusive decay rates~\cite{ExtendedTwistedMass:2024myu}, the R-ratio~\cite{ExtendedTwistedMassCollaborationETMC:2022sta,Fowlie:2023cta}, scattering amplitudes~\cite{Salg:2025now} and parton distribution functions~\cite{Liu:1993cv, Aglietti:1998ur, Detmold:2005gg, Braun:2007wv, Ji:2013dva,Monahan:2015lha,Monahan:2016bvm,Radyushkin:2017cyf, Chambers:2017dov,Karpie:2018zaz, Ma:2017pxb,Gao:2023lny,Shindler:2023xpd} and their evolution~\cite{Dutrieux:2023zpy}. In all the aforementioned cases, the difficulty translates mathematically to an ill-posed inverse problem. 
In the case of inferring functions, unless the unknown function's form is precisely determined by a finite number of parameters, the ill-posed nature is clear due to the lack of uniqueness, since the finite number of data points could be produced by an infinite number of functions. 

The ill-posed functional inference problem can be worsened by the fact that the data points have statistical noise or that the kernel relating them to the function has exponentially decaying eigenvalues (ill conditioning)~\cite{Karpie:2019eiq,Xiong:2025obq}. To resolve this issue, some adequate regularization must be adopted. 

In hadronic physics, we wish to learn parton distributions which describe the universal pattern in which quarks and gluons arrange themselves within a hadron. Global analysis of experimental cross sections has been performed for decades to infer the parton distribution functions (PDFs) which describe hadron structure in terms of the parton's fraction of the hadron's longitudinal momentum given by $x$; see Refs.~\cite{Bailey:2020ooq,PhysRevD.104.016015,refId0,Chiefa:2025loi} for some recent results on the nucleon PDFs. Matrix elements of fields with space-like separation can be calculated in lattice QCD and are related to PDFs by integral relations similar to experimental cross-sections. The quasi-PDF~\cite{Ji:2013dva} and pseudo-PDF~\cite{Radyushkin:2017cyf} are two such objects determined from a finite number of numerically calculated hadronic matrix elements. Although the focus of this work 
is on pseudo-PDFs, we will briefly refer to quasi-PDFs in Sec.~\ref{sec:fitting_param_funcs} 
to illustrate common strategies used to regularize Fourier transforms of the same lattice data.
Other lattice PDF approaches can calculate different matrix elements, which can be represented by their own integrals of the PDF~\cite{Liu:1993cv, Aglietti:1998ur, Detmold:2005gg, Braun:2007wv, Monahan:2015lha, Monahan:2016bvm, Chambers:2017dov,Karpie:2018zaz, Ma:2017pxb,Gao:2023lny,Shindler:2023xpd}. In all those cases, obtaining the PDF as a function of its momentum fraction $x$ from any of these datasets requires resolving some type of integral inverse problem. For reviews of lattice QCD studies on matrix elements of the nonlocal Wilson-line quark bilinears relevant to quasi and pseudo-PDF calculations, see Refs.~\cite{Lin:2017snn,Constantinou:2020hdm,Constantinou:2020pek,Lin:2023kxn}.
 In this manuscript, we study how to regulate the inverse problem which arises in pseudo-PDF computations of PDFs using the Gaussian process regression (GPR)~\cite{RasmussenW06,gramacy2020surrogates}, 
 which has previously been used in lattice QCD for PDF extractions~\cite{Alexandrou:2020tqq,Candido:2024hjt,Dutrieux:2024rem,Dutrieux:2025jed}, for spectral function reconstructions~\cite{Horak:2021syv,Pawlowski:2022zhh,Horak:2023xfb}, and in broader applications of hadron structure~\cite{DelDebbio:2024lwm,Jay:2025dzl,Salg:2025now}.

In lattice QCD computations of pseudo-PDFs, matrix elements with an operator with space-like separated quark fields connected by a Wilson line are calculated 
\begin{equation}
    M^\mu(p,z) = \langle p | \bar{\psi}_q(z) \gamma^\mu W(z;0) \psi_q(0) | p\rangle = p^\mu \mathcal{M}(\nu, z^2) + z^\mu \mathcal{N}(\nu,z^2)\,,\label{eq:matelem_definition}
\end{equation}
where the Ioffe time~\cite{Braun:1994jq} is given by $\nu=p\cdot z$, a Lorentz invariant generalization of the term's original meaning. 
The pseudo-PDF is defined by taking the inverse Fourier transform of $\mathcal{M}$ with respect to $\nu$ while fixing the Lorentz invariant scale $z^2$ to some non-zero space-like value $z^2<0$. 
Given an appropriate short distance $z^2$, the pseudo-PDF has an integral relationship to the PDF with a perturbatively calculable kernel, just like the experimental cross sections have. On the other hand, to accurately obtain the pseudo-PDF from lattice QCD data through the Fourier transform requires a large $p_z$ so that a sufficiently large range of Ioffe time is obtained. Because large momentum on the lattice introduces significant statistical and systematic uncertainty, data points at large Ioffe time come with large uncertainty.  

While performing the inverse Fourier transform of the matrix element is possible analytically, in practice, one does not have access to the infinite, continuous Ioffe time range required to perform the integral. Numerical approximations for integrating discrete and finite range data can be recast as just another set of assumptions on the form of the unknown (pseudo-)PDF that are regulating the inverse problem, though in some cases the problem will remain quite singular and ill-posed~\cite{Karpie:2019eiq,Xiong:2025obq}. Since lattice QCD can only provide a discrete set of data, it is more appropriate to write the functions as the forward integral. For the pseudo-PDF, the integral is
\begin{equation}
   \mathcal{M}(\nu, z^2) = \int_{-1}^1 dx \, e^{ix\nu} P(x, z^2)\label{eq:pseudo_def}
\end{equation}
where the bounds of the pseudo-PDF are $x\in [-1,1]$, as shown in~\cite{Radyushkin:2017cyf}. 
This support for $x$ follows from Lorentz invariance and therefore holds for fixed spacelike separations as well as at the light cone limit where it is often derived with arguments based on large-$p^+$~\cite{Radyushkin:2019mye}.

Before continuing, it is useful to break the problem of the inverse Fourier transform into its real and imaginary components as inverse cosine and sine transforms. In fact, this makes physical sense, since the real/imaginary component comes from the CP even/odd components of the operator in Eq.~\eqref{eq:matelem_definition} respectively. In the case of the light cone PDF, if the Fourier transform of the light-cone matrix element $\mathcal{M}(\nu,z^2=0)$ is $f_q(x)$ which has the domain $x\in[-1,1]$, then the object can be interpreted as the quark PDF $q$ for $x>0$ and the antiquark PDF $\bar{q}$ for $x<0$ with the definitions 
\begin{equation}
q(x) = f_q(x) \quad;\quad \bar{q}(x) = -f_q(-x) 
\end{equation}  where $q$ and $\bar{q}$ have the domain $x\in[0,1]$. The sum and difference of $q$ and $\bar{q}$ will be related to the sine and cosine transform of $\mathcal{M}(\nu,z^2=0)$ respectively,
\begin{eqnarray}
    {\rm Re}\, \mathcal{M}(\nu) = \int_{0}^1 dx\,\cos(\nu x) \left[q(x) - \bar{q}(x)\right], \nonumber\\
    {\rm Im}\, \mathcal{M}(\nu) = \int_{0}^1 dx\,\sin(\nu x) \left[q(x) + \bar{q}(x)\right],
\end{eqnarray}
leading to the CP projected PDFs $q_\pm(x)=q(x) \pm \bar{q}(x)$.
Similarly, the real and imaginary components of the off lightcone $\mathcal{M}(\nu,z^2)$ can be identified in terms of the CP projected pseudo-PDFs. In the case of the isovector flavor $q=u-d$, the nucleon matrix element has a known value $\mathcal{M}(\nu=0,z^2)=1$ which effectively normalizes the PDF or pseudo-PDF results and can be added as a constraint. On the other hand, the normalization of $q_+$ is not constrained. When analyzing the PDF and pseudo-PDF, a fixed point constraint is added such that they vanish at $x=1$.

Many solutions to the functional inference of parton distributions have been studied for
both experimental and lattice QCD datasets. One class of approaches constrains the unknown functions to a low-dimensional parametric form, typically involving a few parameters for each PDF~\cite{Sato:2016tuz,Martin:2009iq,Dulat:2015mca}. 
Neural-network parameterizations~\cite{Forte:2002fg,Kades:2019wtd,Cichy:2019ebf,Karpie:2019eiq,DelDebbio:2020rgv,Dutrieux:2021wll,Khan:2022vot,Chowdhury:2024ymm,Chu:2025jsi}, as well as expansions in orthogonal~\cite{Parisi:1978jv,Barker:1980wu,Barker:1982rv,Chyla:1986eb,Krivokhizhin:1987rz,Krivokhizhin:1990ct,AtashbarTehrani:2007odq,Khorramian:2009asi,Taghavi-Shahri:2016idw,Karpie:2021pap,Kotz:2023pbu} polynomials fall into this category although the number of parameters might be larger. Beyond these, for analyzing lattice QCD data, non-parametric approaches have been studied~\cite{Karpie:2019eiq, Liang:2019frk, Alexandrou:2020tqq,Candido:2024hjt, Dutrieux:2024rem, Dutrieux:2025jed, Xiong:2025obq}. Many of these approaches, including the parametric ones, rely on Bayes' theorem to define a probability distribution for the PDF given the data and the prior assumptions. The prior information on the function regulates the ill-posed nature of the inverse problem. In Gaussian Process Regression, the prior distribution is given by a Gaussian Process (GP) defined by its central value and covariance. Variations of this approach were previously studied in Ref.~\cite{Alexandrou:2020tqq,Candido:2024hjt, Dutrieux:2024rem, Dutrieux:2025jed, Xiong:2025obq} specifically for lattice calculations studied here. GP regression is a highly flexible and non-parametric approach. Its aim is not only to have minimal biases but also to allow for a reliable assessment of uncertainty for the extraction of PDFs from lattice QCD data.

In this work, we attempt to study the effect of varying many of the possible choices within a GPR-based analysis. In the following, we
\begin{itemize}
    \item propose several forms of the covariances which could be used for the Gaussian prior distribution as well as prescriptions for building others, 
    \item demonstrate the consistency of results with different prior distributions and treatment of hyperparameters,
    \item provide a statistical measure to identify which regions of $x$ were impacted by the data,
    \item study model averaging of those results using weights from various information criteria.
\end{itemize}
In the following closure tests, a significant number of the very different GPR implementations provide results that reproduce the true central values with variances that are either consistent or larger than the true variance, demonstrating the robustness of the method. The information criteria provided allow for a rigorous framework of weighting the relative quality of non-pathological choices, and model averaging becomes a systematically improvable method for controlling model dependence. This approach can be applied to other PDF extractions, both from the lattice and from phenomenology, to better quantify uncertainty with significant control over scientist-made choices.

This work is organized as follows. Sec.~\ref{sec:inverse_overview} overviews the integral inverse problem as a Bayesian statistics problem with specific integrals to evaluate for expectation value and covariance. Sec.~\ref{sec:gp_prior} reviews the concept of Gaussian process regression and describes the solutions of the functional inference at hand. In Sec.~\ref{sec:data_tests}, we study the implementation of GP priors on synthetic data for a closure test of the method. In Sec.~\ref{sec:realdata}, we apply our technique on real lattice data of the proton isovector PDF from Ref~\cite{Egerer:2021ymv}. Finally, in Sec.~\ref{sec:conclusion} we conclude and discuss future prospects.

\section{Probabilistic interpretation of the inverse problem} \label{sec:inverse_overview}
The integral inverse problem of functional inference can be cast as a Bayesian statistics question. Given what one already knows about the unknown function and given the newly available data, what results can be expected for the unknown function(s) and what are the (co)variances on those expectations at different values of $x$?

In the following, for simplicity, we will restrict ourselves to one dataset $M_\nu$, which is the integral of a desired unknown 1-dimensional function $q(x)$ and a known convolution kernel $B(\nu,x)$. The approach can be extended to higher dimensions in a similar manner or to include multiple datasets and unknown functions for the systematic errors as proposed in Ref~\cite{Karpie:2021pap}. It can also be extended to have unknown parameters in the convolution kernel, which could be treated like other hyperparameters described later. Up to statistical and systematic errors, the theoretical model of the data $M^{th}(\nu)$ is related to the unknown function through an integral equation of the form 
\begin{equation}
M^{th}(\nu) = \int\, dx B(\nu,x) q(x)\,,\label{eq:linear_inverse_integral}
\end{equation}
no matter what the dimension of the data or model spaces $\nu$ and $x$ are. When $x$ integration finally becomes necessary, the specific numerical details can be chosen based upon the dimensionality and domain of the integral. For the applications to parton structure from lattice QCD, we could be doing a simple Fourier transform for the pseudo-PDF or we could have in mind a Fourier transform with QCD's short distance corrections to obtain the PDF in the inverse, i.e., $B(\nu,x) = \cos(\nu x) + O(\alpha_s)+\dots$. In the following numerical studies, we will focus solely on the Fourier transform, which dominates $B$, but we derive our results in generality.

In the Bayesian approach, our ultimate goal is to determine the unknown function's expectation value and covariance from the functional integrals
\begin{equation}
\langle q(x) \rangle = \int Dq\, q(x) P[q | M, I],
\end{equation}
and
\begin{equation}
\langle {\rm Cov}[q(x),q(x')] \rangle = \int Dq\, \bigg(q(x) - \langle q(x)\rangle\bigg)\bigg(q(x') - \langle q(x')\rangle\bigg) P[q | M, I],
\end{equation}
where $q(x)$ is the unknown function at a given $x$, and $P[q | M, I]$ is the posterior distribution of the unknown function given the data $M$ and any prior information $I$. Then $\int Dq$ is a functional integral over all possible unknown functions. The posterior distribution is given by Bayes' theorem
\begin{equation}
P[q | M, I] = \frac{P[M | q] P[q | I] }{ P[M | I] }\,.
\end{equation}

We use the Gaussian approximation for the data likelihood probability
\begin{equation}
    P[M|q] \propto \exp\left[ - \chi^2/2\right] \quad ;\quad \chi^2= r^T C^{-1}r \,,
\end{equation} 
where $\chi^2$ is the correlated square of the residual difference between data $M_\nu$ and theoretical model $M^{th}(\nu)$
\begin{equation} r = M^{th}(\nu) -M_\nu
\end{equation} 
and $C^{-1}$ is the inverse of the data covariance. The evidence probability 
 is given by 
 \begin{equation}
     P[M|I]=\int Dq \, P[M | q] P[q | I] \,.
 \end{equation} With this form, the evidence normalizes the posterior distribution to integrate to 1. Finally, $P[q|I]$ is the prior information on the unknown function. 

It is the choice of the prior that defines how the fit will proceed. Some choices of priors will appear too restrictive and bias the result artificially~\cite{Dutrieux:2025jed,Dutrieux:2025axb}. Other choices of priors will be too forgiving and allow the result to vary in an unphysical manner based on the fluctuations and truncations of the data. The goal is to find a sufficiently stringent prior to regulate the ill-posed nature of the inverse, but loose enough to accurately represent our actual knowledge, or more typically our ignorance, of the unknown functions. This work focuses on using GPs to define a prior distribution and studies the choices made. For instance, in Ref.~\cite{Dutrieux:2024rem}, the prior was chosen based on a general expectation for a reasonable extrapolation uncertainty and a loose positivity of the resulting PDF. In Ref.~\cite{Candido:2024hjt}, the prior distribution's variance was designed to inflate in the region where the function was supposed to be less known. For the remainder of this section, the Bayesian analysis for fitting parameterized functions is highlighted to show the connection where traditional fits are ultimately a specifically constraining limit of GPR, where the prior is a $\delta$ function in the space of unknown functions $q$.

\subsection{Fitting a parameterized functional form}\label{sec:fitting_param_funcs}
One of the most common approaches to the inverse problem is to restrict the unknown $q(x)$ to be in a family of functions defined by a few parameters. This approach is adopted in phenomenological global analyses~\cite{Sato:2016tuz} of PDFs. For simplicity, let us use the classic parameterization of the PDF with parameters $N$, $\alpha$, and $\beta$ given by
\begin{equation}
Q(x;N, \alpha,\beta) = N x^\alpha (1-x)^\beta / B(\alpha+1, \beta+1),\label{eq:most_loved_model}
\end{equation}
where $B$ denotes the Beta function ensuring normalization of the PDF to $N$. $P[N, \alpha,\beta | I]$ is the prior distribution of the model parameters $N$, $\alpha$ and $\beta$. 
This parametric fit is equivalent to a GPR with the following prior distribution
\begin{equation}
P[q|I] = \int dN d\alpha d\beta\, \delta\big(q - Q(\cdot;N,\alpha, \beta)\big) P[N,\alpha,\beta | I] ,\label{eq:prior_for_model_fits}
\end{equation}
where the $\delta$-function constrains the unknown function to lie within the parametric family. This $\delta$-function is meant to act on the full functional space such that
\begin{equation}
    \int Dq\, q(x) \delta(q - Q) = Q(x)\,.
\end{equation}
 Given this prior, the $Dq$ integral can be performed explicitly in the definition of the expectation value and covariance of the unknown function to obtain the following simpler, finite-dimensional integrals,
\begin{equation}
\langle q(x) \rangle = \int dN d\alpha d\beta\, Q(x;N,\alpha, \beta) P[N,\alpha,\beta | M, I ]\, , 
\end{equation}
and
\begin{equation}
\langle {\rm Cov}[q(x),q(x')]  \rangle =  \int dN d\alpha d\beta\, {\rm Cov}[Q(x;N, \alpha, \beta),Q(x';N, \alpha, \beta)] P[N, \alpha,\beta | M, I ] \,.
\end{equation}
 The new posterior of the model parameters $N,\alpha,\beta$ is given by
\begin{equation}
 P[N,\alpha,\beta | M, I ] = \frac{1}{P[M|I]} \exp\big[ -\chi^2(M ; N,\alpha,\beta)/2\big] P[N,\alpha,\beta | I]\, ,
\end{equation}
where the residuals in $\chi^2$ are now evaluated with the model function $q(x)=Q(x;N, \alpha, \beta)$ and the evidence $P[M|I]$ collects all the normalizations and other factors independent of $N$, $\alpha$ and $\beta$. This choice of prior distribution regulates the inverse problem by taking the infinite-dimensional space of $q$ and restricting it to the low-dimensional space of functions parametrized by $(N,\alpha,\beta)$. The above description also holds with a neural network parameterization~\cite{Forte:2002fg}.
In Sec.~\ref{sec:gp_prior} we show that, unlike parametric fits which explicitly restrict the functional space to a small family, GPR retains the full space of functions 
but assigns them different probabilities according to the specific Gaussian process prior.
 This weighting favors functions consistent with smoothness or other prior assumptions, thereby reducing the risk of overconstraining the solution and introducing model bias.
 
Even a Discrete Fourier Transform (DFT) can be described in this manner, as a model with exactly as many parameters as data points in the DFT. The model is given by
\begin{equation} 
Q^{DFT}(x;m_i) = \sum_i e^{i\nu_i x} m_i \,.
\end{equation}
The $\chi^2$ will be minimized by setting the parameters $m_i$ to be equal to their corresponding data point $M_{\nu_i}$ at $\nu_i$. This exactly constrained system has a solution, but is subject to more severe overfitting compared to models with far fewer parameters than data points.

 The DFT approach is also used in the quasi-PDF framework~\cite{Ji:2013dva,Lin:2017snn,Constantinou:2020hdm}, where the limited range of lattice 
data in coordinate space requires additional regularization of the Fourier transform. Several methods have been proposed, including most successfully discrete Fourier transforms supplemented with large-$|z|$ 
extrapolations~\cite{Gao:2021dbh,Chen:2025cxr}, where the associated model is given by 
\begin{equation}
    Q(x;m_i, c_j) = Q^{DFT}(x;m_i) + Q^{{\rm large\,} z}(x;c_j)\,.
\end{equation}
The data likelihood is split into low and large $z$ regions, and they are constructed with the corresponding part of the model. Just as traditional parametric fits, this is formally equivalent to a GPR with a particular prior distribution. 

 As will be common with all the methods described, the objective is to handle the selection of a $q(x)$ from the space of all continuous functions that are consistent with the data. This is done by assigning a prior probability distribution to that space, disfavoring unphysical results. The above do so in a very restrictive manner with an explicit $\delta$-function. In the non-parametric approaches in the next sections, the priors will be introduced with less restrictive forms, with the hope of lessening model biases.

\subsection{Diagonal Non-parametric Bayesian methods}\label{sec:diag_priors}
In this section, the function $q(x)$ will be discretized into a finite number of points on a grid. The value of the function on this grid becomes the degrees of freedom, and in the continuum limit, it represents a universal function approximation. The numerical implementation of the Gaussian Process Prior in Sec.~\ref{sec:gp_prior} will include a similar discretization via finite elements. Provided that all the integrals can be evaluated, a complete derivation of the GP regression with a continuous function $q(x)$ can proceed until one is left with an integral over several hyperparameters, similar to the above analysis of a nonlinear model. 

Besides parameterized functional forms, another class of prior information is methods such as the Maximum Entropy Method (MEM)~\cite{MEM,MEM2}, Bayesian reconstruction (BR)~\cite{Burnier:2013nla}, and the Quadratic Difference Ratio (QDR), studied for this specific Fourier inverse problem in Ref.~\cite{Karpie:2019eiq}. The ``model parameters'' $q_i$ are the values of the unknown function at some grid in $x$ which spans the domain of $q(x)$. After picking a set of numerical integration rules, the integrals can then be performed to obtain models for $M$. The function $Q(x;{q_i})$ is an interpolation between these points as defined by that integration rule. The choice of the prior's form for these model parameters $q_i$ is what distinguishes these different Bayesian techniques and generates their model dependence. For example, in the QDR method, the prior is given by
\begin{equation}
 P_{QDR}[q_i| I ] \propto  \delta\big(q - Q(\cdot;q_i)\big) \exp\left[ -\frac{\alpha}2 \int dx \, \frac{\big( Q(x; q_i) - g(x;c)  \big)^2 }{ \sigma^2(x) } \right]\, ,
\end{equation}
where $\alpha$, $g(x;c)$ and $\sigma^2(x)$ are hyperparameters of the prior. Similarly, MEM has a prior form with the exponential of an integral based upon the Shannon-Jaynes entropy. Note that in the limit $\alpha \to \infty$ or $\sigma^2\to0$, it reduces to a $\delta$ function fixing the unknown to the model function $g(x;c)$, which could depend on other hyperparameters $c$. In this limit, the QDR reduces to a parameterized fit. Note that the Tikhonov regulator suggested in Ref~\cite{Xiong:2025obq} is a special case of QDR, which is itself a special case of GPR outlined in the following section.

\section{Gaussian Process as a prior distribution}\label{sec:gp_prior}

A GP can be used to define a prior distribution that correlates the unknown function at different values of $x$, unlike the QDR. The prior distributions will take the form~\footnote{Note that since we will be dealing primarily with Gaussian distributions, probability distributions will be given by -2 times their logarithm. This exposes the useful quadratic form in the variables being handled.}
\begin{eqnarray}
&-2\log P_{GP}[q | \theta, I] =  \int dx \int dx' \big(q(x) - g(x;\theta)\big)
K^{-1}(x,x';\theta)\big(q(x') - g(x';\theta)\big) \nonumber\\
&\hspace{150pt}+ \log \det (2\pi K) \,, \label{eq:gp_prior_def}
\end{eqnarray}
where $K(x,x';\theta)$ is the GP covariance kernel and $g(x;\theta)$ is the prior's mean, both of which depend on the hyperparameters $\theta$. The inverse kernel $K^{-1}$ in Eq.~\eqref{eq:gp_prior_def}, is defined by 
\begin{equation}
    \delta(x-x') = \int dy K(x, y) K^{-1}(y,x')\,.
\end{equation}
Given a kernel $K(x,x')$, there may not be a simple form for $K^{-1}(x,x')$. In generality, the hyperparameters may also have a prior distribution $P[\theta|I]$ which multiplies the GP prior distribution. This term will be discussed in Sec.~\ref{sec:vary_hyp} where it will first be required. The kernel $K$ plays the role of a covariance in the prior distribution and can be used to impose desired levels of correlation and uncertainty on the unknown function. Through the particular form of $K$ and how it depends on hyperparameters, the value of $q$ becomes correlated at different positions in $x$. In regions where data are no longer constraining, the prior will send $q(x)\to g(x;\theta)$ with a variance of $K(x,x)$. In the limit of $K$ vanishing for all $x,x'$, this Gaussian prior distribution becomes the $\delta$ function prior in Sec.~\ref{sec:fitting_param_funcs} and the problem is reduced to fitting the parameterized model $g$. The QDR method in the previous section is a GPR with a diagonal prior covariance.

This prior can also be recast into $\nu$ space if the inverse convolution kernel $B^{-1}$ is known. Substituting $q(x) = \int d\nu B^{-1}(\nu, x) M^{th}(\nu)$ and $g(x) = \int d\nu B^{-1}(\nu, x) G(\nu)$, then the GP prior becomes 
\begin{eqnarray}
&-2\log P_{GP}[M^{th} | \theta, I] =  \int d\nu \int d\nu' \left(M^{th}(\nu) - G(\nu;\theta)\right)
k^{-1}(\nu,\nu';\theta)\left(M^{th}(\nu') - G(\nu';\theta)\right) \nonumber \\ &\hspace{150pt} + \log \det (2\pi k)  \,, \label{eq:gp_prior_def_ioffe}
\end{eqnarray}
where 
\begin{equation}
    k^{-1}(\nu,\nu';\theta) = \int dx dx' B^{-1}(\nu, x) K^{-1}(x,x';\theta) B^{-1}(\nu',x')\,. \label{eq:prior_in_nu}
\end{equation}
The GP prior in $x$ is formally related to a GP prior in $\nu$, and it is the domain of these integrals that changes the practicality of Bayesian analysis in the two spaces. When using CP projection to study cosine/sine transforms, Ioffe time scans an infinite range $\nu\in [0,\infty]$ while for (pseudo-)PDFs the momentum fraction is bounded with $x\in [0,1]$. In the following, we choose to handle $q$ as our unknown.

\subsection{Examples of kernels}\label{sec:examples_of_kernels}

\subsubsection{Radial Basis Functions and similar}
 One of the most popular kernels, the square exponential Radial Basis Function (RBF), given by
\begin{equation}
K_{\rm RBF}(x,x'; \theta=\{\sigma, l^2\}) = \sigma \exp\left(-\frac{||x-x'||^2}{2 l^2}\right)\,,
\end{equation}
allows for exponentially decaying correlations as $x$ and $x'$ separate, based upon the correlation length $l$ and prior strength $\sigma$. As an RBF kernel, it depends on $||x-x'||$. A modification of this kernel, proposed in~\cite{gibbs1998bayesian}, allows for the correlation length to vary with $x$, as $x$ and $x'$ approach 0
\begin{equation}
K_{{\rm Gibbs}}(x,x'; \theta=\{\sigma, l_0\}) = \sigma \sqrt{\frac{2l(x) l(x')}{l(x)^2 + l(x')^2}} \exp\left(-\frac{||x-x'||^2}{2 (l(x)^2 + l(x')^2)}\right)\,.
\end{equation}
In what will be referred to as $K_{CDGP}$, Ref~\cite{Candido:2024hjt} proposed $l(x) = l_0 *(x+\epsilon)$ where small $\epsilon$ regulates a divergence at $x=0$ as well as to multiply the Gibbs kernel by $(xx')^\alpha$ for $\alpha\in[-1,0)$ to inflate the variance at low $x$ to represent the lack of information in that regime. The goal of this kernel and choice of $l(x)$ is to begin to decrease the correlation length in the small $x$ regime so it can be modeled more independently. In the specific inverse problems for PDFs, which exist on a bounded domain $x\in[0,1]$, the regions of $x$ very close to 0 and 1 lack constraint from most lattice and experimental results. The decorrelation of the low $x$ regime is useful to increase the uncertainty in the regime where the data is lacking. 

Another generalization of the RBF kernel can allow for a function to redefine the distance in $x$ used in the correlations as
\begin{equation}
K_{f-{\rm RBF}} (x,x'; \theta=\{\sigma, l^2\}) = \sigma \exp\left(-\frac{||f(x)-f(x')||^2}{2 l^2}\right) \,,\label{eq:exp-rbf}
\end{equation}
where $f$ is some non-pathological choice of a function. For use in PDF analyses, where exponentially small $x$ should be relatively decoupled from larger $x\gtrsim 0.1$, in Ref~\cite{Dutrieux:2024rem,Dutrieux:2025jed} $f:=\log$ was used in the numerical implementation with some regulation at $x=0$ as needed. Naturally, this can again be extended with $l\to l(x)$ if more generality is desired.

All stationary kernels, defined as a function of $\tau=x-x'$ and including the RBF kernels which are functions of $|\tau|$, can be described by their Fourier transform or spectral density $S(s)$ via Bochner's theorem~\cite{bochner,stein1999interpolation,gihmanskorohod1974}
\begin{equation}
    K(\tau = x-x') = \frac1{2\pi}\int ds \,e^{i \tau s} S(s)\,.
\end{equation} For the specific case of the square exponential RBF function, with an unbounded domain for the unknown function $q$, the spectral density is a Gaussian centered at $\nu=0$. In Ref.~\cite{pmlr-v28-wilson13}, the square exponential RBF kernel was extended to give greater expressivity by attributing more weight to suppressed regions of the spectrum, producing the Spectral Mixture (SM) kernel. The new spectral density is given by a series of pairs of Gaussians with non-zero means, placed symmetrically about 0, that is, each pair comes with all parameters the same except the means, which have opposite signs. This even symmetry in $s$ forces the stationary kernel $K(\tau=x-x')$ to be a real function. Given this spectral density with $Q$ pairs of Gaussians with means and widths given by $\mu_q$ and $1/l^2_q$, up to rescaling by powers of $2\pi$ that are irrelevant for our purpose, the kernel is given by
\begin{equation}
    K_{\rm SM} (x,x') = \sum_{q=1}^Q \sigma_q \exp\left[-\frac{(x-x')^2}{2l_q^2}\right] \cos[(x-x') \mu_q]\,,\label{eq:sm_def}
\end{equation}
 where $\sigma_q$ are weights to adjust the heights of the pairs of Gaussians. In Ref.~\cite{pmlr-v28-wilson13}, extrapolations from data with oscillating patterns on top of long-term trends were significantly better represented by results from the SM than from other kernel choices, including the square exponential RBF. While the system we are studying is a Fourier inverse, it is not expected that an actual periodic signal will appear in the data. Since many spectral densities can be reproduced by a series of Gaussians, in Ref.~\cite{pmlr-v28-wilson13}, it was shown that this form could be used to approximate other stationary kernels quite well. The limited bounds of the PDF's integration with $x\in [-1,1]$ mean that the $S(s)$ for these kernels will not exactly be these Gaussians, but that does not invalidate the use of Eq.~\eqref{eq:sm_def} as a kernel. Unfortunately, as shown in App.~\ref{app:poorkernels}, as demonstrated, this implementation of an SM kernel results in biases from poor extrapolations or overestimation of error in the region where data exists. Still better kernels designed by their spectral density may have useful applications with relevant Fourier inverse problems.

\subsubsection{Kernel Trick}\label{sec:ktrick}
Another class of kernels can be generated, whose effect is similar to fitting a (possibly infinite) series of model functions $\phi_n(x)$ with linear coefficients $c_n$, so that
\begin{equation}
    Q(x;c_n, \alpha, \beta) = \sum_{n=0}^\infty c_n \phi_n(x;\alpha,\beta). \label{eq:poly_model}
\end{equation}
In what's called the ``kernel trick''~\cite{Rasmussen2006Gaussian}, given the possibly infinite basis of model functions $\phi_n(x;\theta)$ and the prior on the linear model parameters $c_n$ with a covariance $\Sigma_{n n'}(\theta)$, then the result of a least square minimization of $c_n$ is equivalent to the result of a Gaussian Process using the kernel 
\begin{equation}
    K(x,x';\theta) = \sum_{nn'} \phi_n(x;\theta) \Sigma_{nn'}(\theta) \phi_{n'}(x';\theta)\,.\label{eq:kernel_trick}
\end{equation}
 For generality, we allow the model functions $\phi_n$ and the covariance $\Sigma$ to both depend on some set of hyperparameters $\theta$. In fact, the result of any GP can be described by a fit of the eigenfunctions of the specific kernel~\cite{Rasmussen2006Gaussian}. For the square exponential RBF kernel, the eigenfunctions are a Gaussian multiplying Hermite polynomials~\cite{Zhu1998}. By analyzing this GP, one could create estimates of $q(x)$ or even $c_n$ from a possibly infinite number of terms without necessarily requiring a truncation of the sum in the model, as long as the infinite sums can be evaluated. 

In order to obtain kernels of relevance for the phenomenology of PDFs, it is important to encode in the family of model functions $\phi_n$ and the prior covariance $\Sigma_{nn'}$ the relevant physics we want to enforce. For instance, to guarantee a posterior which vanishes at $x = 1$, it is useful to choose $\phi_n$ such that $\phi_n(x = 1) = 0$. To model the divergent behavior of the PDF at small $x$ and offer it an enhanced freedom in that region, one could target $K(x, x) \rightarrow \infty$ when $x \rightarrow 0$.

A very simple kernel that satisfies those constraints is obtained by considering
\begin{equation}
    \phi_n(x) = (1-x)^n\,,\ \  n \geq 1 \quad;\qquad  \Sigma_{nn'} = \sigma \delta^{nn'}\,.
\end{equation}
This leads straightforwardly to the kernel
\begin{equation}
    K_{plog-1}(x, x') = \sigma \sum_{n = 1}^\infty (1-x)^n(1-x')^n = \sigma \frac{(1-x)(1-x')}{1-(1-x)(1-x')}\,.
\end{equation}
By using a simple $\sigma \delta^{nn'}$ as prior covariance of the $c_n$ coefficients, we are assuming uncorrelated and equal-sized contributions of all fitted functions $(1-x)^n$ in the prior. As $n$ increases, the $(1-x)^n$ functions become increasingly peaked at $x = 0$, and that singular point is not constrained by the dataset in Ioffe time. Therefore, at small $x$, the variance in the posterior obtained with this kernel will typically diverge as $\sum_{n = 1}^\infty \sqrt{\sigma} (1-x)^n \sim \sqrt{\sigma} /x$.

This may be an excessive rate of divergence of uncertainty at small $x$ for a non-singlet PDF. A simple way to moderate the rate of divergence is to decrease the size of the prior contained in $\Sigma_{nn'}$ as $n$ increases. For instance,
\begin{equation}
    \Sigma_{nn'} = \sigma \frac{\delta^{nn'}}{n}\,,
\end{equation}
which leads to
\begin{equation}
    K_{plog-2}(x, x') = \sigma \sum_{n = 1}^\infty \frac{(1-x)^n(1-x')^n}{n} = -\sigma \log(1-(1-x)(1-x'))\,.
\end{equation}

More generally, the strength of the divergence at small $x$ can be parameterized by $\theta$ using
\begin{equation}
    \Sigma_{nn'} = \sigma \frac{\delta^{nn'}}{n^\theta}\,.
\end{equation}
which leads to the kernel
\begin{equation}
    K_{plog-\theta}(x, x') = \sigma \sum_{n = 1}^\infty \frac{(1-x)^n(1-x')^n}{n^\theta} = \sigma \textrm{Li}_\theta((1-x)(1-x'))\,,
\end{equation}
where $\textrm{Li}$ denotes the polylogarithm function. $\theta = 0$ and 1 correspond to the cases explored before.

This method is powerful for designing kernels with sophisticated properties. If used improperly, it can also lead to faulty kernels. For instance, using $\Sigma_{nn'} = t^{n+n'}$ where $t$ is a parameter between 0 and 1 could seem advantageous, as it allows a reduction in size of the prior as $n$ increases and correlates neighboring coefficients $c_n$. But the covariance $\Sigma_{nn'}$ is then of rank 1. A rank 1 correlation matrix (excluding variables with no variance) is exclusively made of 1 and -1, so it offers no nuanced correlations. Similarly, had we switched $\phi_n(x) = (1-x)^n$ for $x^n$, for instance, our kernel would not offer enhanced modelling freedom in the small $x$ region, and the result would be very rigid and unadapted. Examples of kernels with poor performance are given in Appendix~\ref{app:poorkernels}.

Finite-size models, even with a large number of linear parameters or a more general $\Sigma$, could also be utilized in the same form without requiring a generating function or some other identity to handle infinite sums. The B\'ezier curve series of Bernstein polynomials in Ref.~\cite{Kotz:2023pbu} could be handled in a similar way.

\subsubsection{Changepoints}
Considering the different physics and data constraints at low $x$ and high $x$, one may wish to give different kernels for different regions of the space. Applying the concept of ``Changepoints''~\cite{duvenaud_2014}, we can incorporate this structure into a combined kernel in a smooth manner.
For example, by using the sigmoid function \(\sigma(x;s,x_0) = \frac{1}{1 + e^{-s(x-x_0)}}\), we can smoothly transition between different kernels in different regions of $x,x'$ as
\begin{gather}\label{eq:combined}
  K_{\text{combined}}\bigg(x,x';\theta=\{\theta_1, \theta_2, s,x_0\}\bigg) = \sigma(x;s,x_0)K_1\bigg(x,x';\theta_1\bigg)\sigma(x';s,x_0) + \nonumber\\  (1-\sigma(x;s,x_0))K_{2}\bigg(x,x';\theta_2 \bigg)(1-\sigma(x';s,x_0))
\end{gather}
where the hyper-parameter $s$ controls how abrupt the change of kernels will be at the $x_0$ changepoint. With this kernel, for example, we could combine the Gibbs kernel in equation \eqref{eq:exp-rbf} for low $x$ and the RBF kernel for a large $x$ region. Another potential application is to provide different kernels for the DGLAP and ERBL regions of a GPD via a changepoint at $x=\xi$. Finally, to specify which pair of kernels is combined in subsequent sections, the notation $K_{\text{kernel}_1}^{\text{kernel}_2}$ will be adopted.

\subsubsection{Constraints}
Finally, one's prior information on unknown functions is not always limited to simple smoothness or other types of correlations in the function. For example, one may wish to enforce that the (pseudo-)PDFs vanish at $x=1$. Some PDFs have normalizations and other integrals constrained by sum rules. Frequently, explicit information in the form of known integral normalizations $\int dx\, q(x)=N$ or fixed points $q(x=1)=0$ is a powerful constraint to help reduce the severity of the ill-posedness. When enforced by the prior distributions, the effect of some kernels to decorrelate specific regions will remain true only up to the need to enforce the constraint. Constraints can be enforced exactly or approximately as the situation demands. 

Consider the extended prior distribution
\begin{eqnarray}
    -2\log \widetilde{P}_{GP}[q | \theta, I] =-2\log P_{GP}[q | \theta, I]
 + \frac{1}{\lambda_{N}} \left|N- \int dx q(x)\right|^2  \nonumber\\+  \frac{1}{\lambda_{FP}} \left|\int dx \delta(1-x)q(x) \right|^2 +  C_n\,, \label{eq:constraints}
\end{eqnarray}
where the constraining power of the normalization and fixed point are governed by the size of $\lambda_N$ and $\lambda_{FP}$ respectively. The constants $C_n$ represent the normalizing factors from the constraints. In the limit that these $\lambda$ parameters go to 0, their respective pieces of prior information are constrained exactly. By design, all these constraints come in the form of quadratic functions of $q(x)$, so that the total sum can be rewritten as a single quadratic with a modified kernel $\tilde{K}$. 

At this point, it is convenient to define a convolution notation. One of the convolutions in the Gaussian process can be written as 
\begin{equation}
     K^{-1} \circ q = \int dx' K^{-1}(x,x') q(x'), \label{eq:notation}
\end{equation}
and our constraints from the integrals
\begin{eqnarray}
    n\circ q = \int dx\, n(x) q(x) \, : \quad n(x)=1, \nonumber \\
    d\circ q = \int dx\, d(x) q(x) \, : \quad d(x)=\delta(1-x) \,. \label{eq:constraint_convols}
\end{eqnarray}

The new kernel $\widetilde{K}$ can be given by the Woodbury identity
\begin{equation}
    \widetilde{K} = (K^{-1} + U  \Lambda^{-1} U^T)^{-1} = K - K\circ U (\Lambda + U^T\circ  K \circ U)^{-1} U^T\circ  K 
\end{equation}
where 
\begin{equation}
    U= \begin{pmatrix}
        n & d
    \end{pmatrix} \qquad \Lambda = {\rm diag}[\lambda_N , \lambda_{FP}] \,.
\end{equation}
In this notation, $\Lambda$ is a diagonal matrix of weights, $U$ integrates each convolution kernel with the object on its left to create a row vector, and $U^T$ convolutes with the object on its right to make a column vector. Implicit in the derivation of a unique modified kernel $\widetilde{K}$ is an assumption that the prior's mean $g(x;\theta)$ satisfies the constraints being applied to the unknown function $q(x)$, which is a logical choice. Ultimately, using a prior's mean which satisfies exactly the constraints is, however, not necessary. Then one will not enjoy the advantage of a single modified kernel, but one can still continue forward with the prior in Eq.~\eqref{eq:constraints}. We will argue later that, to avoid the many convolutions in the Woodbury formula, in the numerical implementation, the constraints are better placed elsewhere than literally in the prior covariance. Instead, the additional terms in Eq.~\eqref{eq:constraints} can be recast as additional ``data points'' whose central value sets the value of the normalization or the value of the unknown at a fixed point. The potentially vanishing error $\lambda$, uncorrelated to any other actual data points, enforces the strength of the constraint to the desired precision. The $B$ transformation used for the actual data will be substituted by $n$ and $d$. This can be implemented such that the posterior is identical to when using the modified $\widetilde{K}$. 

Another type of constraint that can be applied is on the asymptotic behavior of the function. For use in determining the quasi-PDF, some propose to build up an extrapolation model using physical constraints~\cite{Ji:2020brr}, such as the exponential decay of the data. The practical implementation of these constraints leaves room for different choices, such as order of truncation, some of which may imprint an excessive model-dependence on the result~\cite{Dutrieux:2025jed, Chen:2025cxr, Dutrieux:2025axb}. In Ref.~\cite{Dutrieux:2025jed} it was shown how the exponential decay could be imposed by a prior in $\nu$ space with zero central value and covariance that decays with $\nu$. It was also shown that the exact nature of the exponential is less relevant than other choices being made, which were more akin to having different choices of the function multiplying the exponential.

\subsection{Posterior distribution of $q$}
It is standard to consider the log of the data likelihood as a quadratic form $\chi^2$. 
In the case of integral inverse problems, the $\chi^2$ requires mapping from the space of the unknown function to the space of the (finite number of) data points. Again, the transformation of the PDF model $q$ in $x$ space into the Ioffe time distribution (ITD) model $M^{th}$ in $\nu$ space is given by 
\begin{equation}
M^{th}(\nu) = B_\nu \circ q =  \int dx B(\nu,x) q(x)\,.
\end{equation}
The $\chi^2$ is given by the inner product of the vector of residuals $r_\nu=M^{th}(\nu)-M_\nu$ and the inverse covariance
\begin{equation}
    \chi^2=\left( B \circ q - M \right)^T_{\nu} C^{-1}_{\nu\nu'}\left( B \circ q - M \right)_{\nu'} \,.
\end{equation}
The normalized data likelihood function is given by
\begin{eqnarray}
    -2\log P[M|q] = \chi^2 + \log \det [2\pi C]. 
\end{eqnarray}

Combining the data likelihood with the GP prior distributions, we obtain the posterior, up to the $1/P[M|I]$ factor
\begin{eqnarray}
L_q^2=& -2\log P[M|q] P[q | \theta, I ]P[\theta|I]= \left( B \circ q  - M \right)^T_\nu C^{-1}_{\nu\nu'}\left( B \circ q  - M \right)_{\nu'} \nonumber\\ &+ (q-g) \circ K^{-1}\circ (q-g) + \log \det[2\pi C] + \log \det[2\pi K] \,.
\end{eqnarray}
This complicated expression that combines ``products'' in both $\nu$ and $x$ spaces deserves some explanation. The $x$ space functions, $q$, $g$ and $K^{-1}$, are being convoluted together and the $\nu$ space vectors and matrices, $M$ and $C^{-1}$, are being matrix multiplied. To transition between spaces, $B$ is a vector-valued function of $x$. Its $x$ dependence can be convoluted with $q$ to give a standard vector in $\nu$ space, which can multiply the central values and covariance of the data. Similarly, in $B^T C^{-1}B$, the $\nu$ space vector indices of $B$ are contracted with those of the inverse covariance to form a kernel in $x,x'$ function space.

By completing the square, any quadratic function can be rearranged to the form
\begin{equation}
    L_q^2 = (q-\bar{q})\circ H^{-1}\circ (q-\bar{q}) + L_0^2.
\end{equation}
In this case, the minimum of $L_q^2$ is given by
\begin{equation}
    L_0^2 = -\bar{q} \circ H^{-1} \circ \bar{q} +  M^T C^{-1} M + g \circ K^{-1} \circ g + \log \det [2\pi C] + \log \det [2\pi K]   \,.
\end{equation}
The posterior covariance $H$ is given by
\begin{equation}
H = [K^{-1} + B^T C^{-1} B]^{-1} \,.\label{eq:h_inverse_bad}
\end{equation}
Finally, the mean of the distribution is given by
\begin{align}
    \bar{q} &= H \circ \left( B^T C^{-1} M + K^{-1} \circ g \right)\nonumber\\
    &= g + K \circ B^{T} \left[ C + B \circ K \circ B^T \right]^{-1} \left[ M - B \circ g \right]\,. \label{eq:mean_woodbury}
\end{align}
The Woodbury identity~\cite{woodbury}
\begin{equation}
H =  K - K \circ  B^T \left[ C + B \circ K \circ B^T \right]^{-1} B \circ K \label{eq:woodbury}
\end{equation}
is used to modify the expression for $\bar{q}$ to avoid requiring the inverse kernel $K^{-1}$ in Eq.~\eqref{eq:h_inverse_bad}
and is not easily determined more directly. Instead, a matrix inverse in the space of data $\nu$ is required.

The mean and covariance define a Gaussian distribution from which we can numerically sample the posterior of $q$. To do this sampling, integrals of the kernel must be evaluated. One way to evaluate the integrals is with finite elements similar to the discretization of the integrals in Sec.~\ref{sec:diag_priors}. The function is approximated by its value on a grid of points $x_i$ with grid spacings $\Delta x_i$. An integration rule defines a finite element $e_i(x)$ which gives how much the PDF on $i^{\rm th}$ grid point, given by $q_i$, contributes to the integrand at point $x$. In this way the convolutions become a matrix relationship $B\circ q = B_{\nu i} q_i$ between the $\nu$ and $x$ spaces with the matrix 
\begin{equation}
    B_{\nu i} = \int dx B(\nu, x) e_i(x)\,.
\end{equation} Similarly $K$ and $H$ can be replaced by matrices between points in the $x$ space such as
\begin{equation}
    K^{-1}_{ij} = \int dx dx' e_i(x) K^{-1}(x,x') e_j(x') \,. \label{eq:inverse_kernel_fe}
\end{equation}

 When using the finite element approach, the unknowns in the Bayesian problem are restricted from the continuous function $q(x)$ to $q_i$, the function's values on grid points $x_i$. It is reasonable to add prior information only on the points on the grid that are being sampled, since those are the actual degrees of freedom. In the same way we could define the matrix $K_{ij} = K(x_i, x_j)$ which is used in the prior distribution
\begin{equation}
    -2\log {P}_{\rm FE}[q|\theta, I] = \bigg(q_i - g(x_i)\bigg) \big[ K^{-1} \big]_{ij} \bigg(q_j - g(x_j)\bigg)+ \log\det(2\pi K)\,,
\end{equation} 
where $[K^{-1}]_{ij}$ is the typical matrix inverse of $K_{ij}$. In practice, we will intend to use, at minimum, roughly an order of magnitude more finite elements than data points, so using the Woodbury formula in Eq.~\eqref{eq:woodbury} can give a large gain in numerical efficiency by replacing an inverse of a matrix whose size is the number of grid points to an inverse of a matrix whose size is the number of data points.

This finite element approach is a choice we make here on how to evaluate these integrals in the data studies of later sections. In the continuum limit of the grid, this prior distribution will correspond to the one with a continuous function kernel. Instead, a completely continuous form of all these expressions can be evaluated, given the integrals and determinants of the kernels can be computed.

\subsection{KL divergence and information gained}

To quantitatively estimate the information obtained from the data in the GPR, we use the Kullback–Leibler (KL) divergence to measure the difference between the prior and the posterior. Given that we have access to the analytical expression, in the context of GPR, we obtain

\begin{gather}
    D_{KL}(P[q|M,\theta,I]||P[q|\theta,I])= \int Dq \,\, P[q|M,\theta,I] \log \left( \frac{P[q|M,\theta,I]}{P[q|\theta,I]}\right)\nonumber\\
    =\frac{1}{2} \left( {\rm Tr}\left[K^{-1}H - \mathbf{1}\right] - \log{ \frac{\det(H)}{\det (K)}} + (\bar{q}-g)^{T}K^{-1}(\bar{q}-g)  \right)\,.\label{eq:full_KL}
\end{gather}

It should be noted that we are comparing probability distributions associated to the full function $q(x)$ across its full range. However, the comparison can be made at each of the individual points of the grid in $x$ as suggested in Ref.~\cite{ValentineSambridgeGP} for diagonal covariance matrices. The derivation for the non-diagonal case is provided in App.~\ref{app:kl} with the definition

\begin{equation}
     D_{KL}(x_i)\equiv D_{KL}(P[q_i|M,\theta,I]||P[q_i|\theta,I])=\int dq_i \, P[q_i|M,\theta,I] \log \left( \frac{P[q_i|M,\theta,I]}{P[q_i|\theta,I]}\right).
\end{equation}
This will measure the difference between the posterior and prior in one point $x_i$ of our grids, given by the full posterior and prior distribution, and integrating the dynamical variables $q_j$ for all $j\neq i$. This point-by-point version of the KL divergence is
\begin{eqnarray}
     D_{KL}(x_i)=\frac 12 \sqrt{\frac{\det\hat{H}^{-1}}{\tilde{H}_{ii}\det H}}  \Bigg(\frac{\tilde{K}_{ii}}{\tilde{H}_{ii}} -1 -\log \frac{\det H \det \hat{K}^{-1}}{\det \hat{H}^{-1} \det K}   + (\bar{q}(x_i)- g(x_i))^2\tilde{K}_{ii}\Bigg),\label{eq:pointKLx}
\end{eqnarray}
where $\hat{H}$ is $H^{-1}$ with the $i$-th column and row removed and $\tilde{H}_{ii}=H^{-1}_{ii}-\sum_{jj'\neq i}H^{-1}_{ij}\hat{H}^{-1}_{jj'}H^{-1}_{j'i}$ is the Schur complement. There are similar definitions for $\hat{K}$ and $\tilde{K}$. In the limited case where $K$ and $H$ are diagonal matrices, then Eq.~\eqref{eq:pointKLx} reduces to Eq 23 of Ref.~\cite{ValentineSambridgeGP}. 
Analogously, we can compare the probability distributions in the $\nu$-space, such that
\begin{eqnarray}
     D_{KL}(\nu_i)=\frac 12 \sqrt{\frac{\det\hat{h}^{-1}}{\tilde{h}_{ii}\det h}}  \Bigg(\frac{\tilde{k}_{ii}}{\tilde{h}_{ii}}-1 -\log \frac{\det h \det \hat{k}^{-1}}{\det\hat{h}^{-1} \det k}  + (M^{th}(\nu_i)- G(\nu_i))^2\tilde{k}_{ii}\Bigg),\label{eq:pointKLnu}
\end{eqnarray}
where the lower case $h$ and $k$ represent the kernels transformed to Ioffe time space as in Eq. \eqref{eq:prior_in_nu}. The point-by-point KL divergences provide a quantitative metric to study where in $x$ or $\nu$ space is being more impacted by the data than other regions. A comparison of analyses with varying data ranges allows one to assess how different intervals of $\nu$ contribute information about distinct regions of $x$

\subsection{Varying hyperparameters}\label{sec:vary_hyp}

Given a fixed set of hyperparameters $\theta$, the quantities $K$ and $g$—and therefore $H$ and $\bar{q}$—are determined. In this case, the analytic forms of $\bar{q}$ and $H$ provide $\langle q(x) \rangle$ and its covariance. Such a solution represents the most likely $q$ under the chosen prior $\theta$, but leaves open the question of what $\theta$ to use. Fixing $\theta$ explicitly is equivalent to taking the hyperparameter prior $P[\theta|I]$ as a $\delta$-function. 

One option is to choose $\theta_0$ that minimizes the functional $E^2(\theta)$, called the effective evidence, given by
\begin{eqnarray}
    E^2(\theta) \equiv& -2\log P[\theta|M,I] \nonumber\\
    =& -2\log P[M|\theta,I]P[\theta|I] = -2\log \int Dq \exp[ - L^2_q/2] P[\theta|I]\nonumber\\=& L_0^2 +  \log \det(2\pi H ) -2 \log P[\theta|I]\,.\label{eq:eff_evi}
\end{eqnarray}
The minimum $\theta_0$ is the mode of the posterior in $\theta$ of the probability distribution $P[\theta|M,I]$ or the Maximum a posteriori (MAP). By integrating away the $q(x)$ degrees of freedom, one is left with purely a function of $\theta$ to handle. The log-posterior for $\theta$, $E^2(\theta)$, will be called in the following the effective evidence, since it is proportional to $P[M|\theta,I]$, the evidence in the definition of the posterior for $q$ given a specific $\theta$. It can be rewritten as
\begin{equation}
    E^2 = (M - B\circ g)^T [ C+ B\circ K\circ B^T  ]^{-1} (M - B\circ g) + \log \det  [2\pi(C+ B\circ K\circ B^T)  ] - 2\log P[\theta | I]  \label{eq:evidence_def} \,.
\end{equation}
Note that $ [ C+ B\circ K\circ B^T  ]$ is a finite matrix in the space of $\nu$ and it does not require evaluation of the inverse kernels. See that in this form, there is no need to explicitly use $K^{-1}$ if we wish to work with continuous functions instead of finite element matrices.

The goal now is to evaluate the expectation value integral for $\langle q\rangle$, which, given the GP prior, is given by
\begin{eqnarray}
    \langle q(x)\rangle =&  \int d\theta \int Dq \,q(x) \exp[-L^2_q(\theta)/2]/P[M|I]\nonumber\\
    =& \int d\theta\, \bar{q}(x;\theta) \exp [-E^2(\theta)/2]/ P[M | I]\,, \label{eq:evidence_integral}
\end{eqnarray}
where the full evidence is the integral over hyperparameters of the effective evidence $P[M|I] = \int d\theta \exp[-E^2(\theta)/2]$. For numerical integration of the $\theta$ integral, the method of importance sampling is to focus on the regions where $E^2$ is smallest while ignoring the regions where $E^2$ is so large that the integrand is negligible.
In the extreme case, one may approximate the integral by evaluating $\bar{q}$ at the most probable $\theta$, assuming it is the dominant contribution.
From that optimal $\theta_0$ value which maximizes the distribution, the mean and covariance of the posterior can be taken as the expectation values of the PDF and its covariance.

When integrating over the full space of $\theta$, one must be careful to take care of all effects in the estimate of the covariance. For brevity, we explicitly derive the variance
\begin{eqnarray}
   \langle \Big(q(x) - \langle q(x)\rangle \Big)^2\rangle  =& \int d\theta \int Dq P[q | M, \theta, I]  \left(q^2(x)- 2 q(x) \langle q(x) \rangle + \langle q(x)\rangle^2\right)  \nonumber \\
    =& \int d \theta P[M|\theta, I]  \left( H_{xx}(\theta) + \big(\bar{q}(x;\theta) -\langle q(x)\rangle\big)^2\right)/ P[M | I]\label{eq:weighted_errors}\,.
\end{eqnarray}
To derive the second line, we used the Gaussian integrals 
in Eq.~\ref{eq:evidence_integral} and 
\begin{equation}
\int Dq \, q^2 \exp[-L_q^2/2]/ P[M | I]  = \left(H + \bar{q}^2\right) \exp[-E^2/2]/ P[M | I]\,.
\end{equation}
Similarly derived, the covariance is given by the off-diagonal term $H_{x x'}$ and the product $(\bar{q}(x;\theta) - \langle q(x)\rangle)(\bar{q}(x';\theta) - \langle q(x')\rangle)$. Note that the covariance can be evaluated on the same Monte Carlo samples as the central value since they both can use the same probability distribution $\exp[-E^2/2]/P[M|I]$. A similar expression appears when performing a posterior weighted average over different model parameterizations~\cite{leamer1978specification, kass1995, Jay:2020jkz} such as those discussed in Sec.~\ref{sec:fitting_param_funcs}. In those cases, one is handling sums of discrete choices of priors given by the different models in the prior's $\delta$-function, such as that in Eq.~\eqref{eq:prior_for_model_fits}. Here, with Gaussian processes, we are integrating over a continuous choice of priors differing by the values of $\theta$. The logic of the derivation and the resulting covariance are essentially the same.

The total covariance of $q$ about $\langle q \rangle$ has two contributions. The first reflects intrinsic fluctuations at fixed $\theta$, described by $H(\theta)$. The second arises from the variation of $\bar q (\theta)$ across hyperparameter space. Together, these contributions add in quadrature to give the total variance. In the approach of extremizing $E^2$ to find an optimal $\theta_0$, the second term gives exactly zero since by assumption $\langle q(x) \rangle = \bar{q}(x;\theta_0)$ and is frequently not written. Instead, when integrating over the space of $\theta$ by some sampling approach, these contributions can be significant and should be included. 

In Refs.~\cite{ValentineSambridgeGP, Candido:2024hjt}, the error from $H$ was broken into 2 pieces to observe the effect of resolution of the inverse problem separated from the propagation of statistical errors. One can define a resolution matrix, which describes the uncertainty in the infinite statistics limit,
\begin{equation} 
R= K\circ B^T [B\circ K\circ B^T]^{-1}B\,.
\end{equation}
It describes how the limited information in $\nu$ space restricts the information in $x$ space. Even in the case of zero statistical error and fixed hyperparameters, the reconstructed PDF $\bar{q}$ would differ from the true PDF $q^{\rm true}$ as
\begin{equation}
    \bar{q} - g = R \circ(q^{\rm true} - g)
\end{equation}
This represents a finite resolution of the unknown whose error will not be removed even with infinite precision on the data points. 

For the case of PDF calculations from lattice QCD, low $x$ behavior is governed by large $\nu$, which requires systematically noisy large momentum results. Realistic limitations on the $\nu$ range of available data will generate a fundamental lack of information to constrain the low $x$ regime of the parton distributions. This was demonstrated in Ref.~\cite{Candido:2024hjt} where this ``reconstruction error'' dominates $x<10^{-2}$. The other piece represents the residual propagation of statistical error of the data through the GP result. Both of these effects are included within the $H$ term in Eq.~\eqref{eq:weighted_errors}. 

The latter $(\bar{q} - \langle q\rangle )^2$ term in Eq.~\eqref{eq:weighted_errors} was not included in Ref.~\cite{ValentineSambridgeGP}, which was assuming optimized, yet fixed, hyperparameters in the assessment of error from only $H$. In Ref.~\cite{Candido:2024hjt} where hyperparameters are sampled, the posterior $P[q, \theta | M,I]$ is evaluated by first performing a MCMC trajectory of $P[\theta| M,I]\propto \exp[-E^2/2]$ then subsequent sampling from the Gaussian $P[q|\theta, M,I]\propto \exp[-L_q^2/2]$ for that $\theta$ sample. The covariance of the resulting $q$ samples will produce an estimate for $\langle \textrm{Cov}[q(x),q(x')]\rangle$, but using a single sample for each $\theta$ may require more Monte Carlo samples for accurate covariance estimation compared to analytically including the part of the error from $H$. With so few parameters being sampled, there is possibly only a small difference in cost.

\subsection{Weighted model averaging}\label{sec:model_average}
Following the same principles used to average over discrete model choices~\cite{leamer1978specification, kass1995, Jay:2020jkz}, we combine results obtained from different prior distributions. For the $i$-th prior, defined by kernel $K_i$, mean function $g_i$, and hyperparameter prior $P[\theta|I]$, one obtains the posterior mean  $\langle q(x)\rangle_i$, covariance $ \langle \Big(q(x) - \langle q(x)\rangle_i \Big)^2\rangle_i$, log-evidence $\langle E^2\rangle_i$ and mode of the evidence $\theta_0^i$. The expected value and covariance across all priors would come from the weighted sums
\begin{equation}
    \langle q(x) \rangle = \sum_i w_i \langle q(x) \rangle_i
\end{equation}
and
\begin{equation}
    \langle  \Big(q(x) - \langle q(x)\rangle \Big)^2 \rangle = \sum_i  w_i \left [ \langle \Big(q(x) - \langle q(x)\rangle_i \Big)^2 \rangle_i  + \Big(\langle q(x)\rangle_i - \langle q(x)\rangle\Big)^2\right ] \,,
\end{equation}
with weights $w_i=\exp[- IC_i /2 ] / \sum_j \exp[- IC_j /2 ]$ where $IC_i$ is the estimate of an information criterion (IC). In the above equation, the first term reflects variance within each prior, and the second captures the spread of posterior means across priors.
Although framed as model averaging, the weighting can, in practice, reduce to effective model selection if nearly all weights vanish except one.

Information criteria aim to approximate the KL divergence between the estimated posterior $E^2$ from a finite number of samples for the data and the (unknown) true posterior distribution at infinite statistics. Lower divergence corresponds to higher weight. The posterior log-likelihood from finite data is a biased estimator, so one derives an additive correction. The different approximations used in deriving that bias give the various information criteria.
We compare results from 3 choices. 
The Posterior Averaged Information Criterion (PAIC)~\cite{zhou2009} given by
\begin{equation}
    PAIC_i = \langle E^2 \rangle_i + 2\tr[J^{-1}(\theta_0^i)I(\theta_0^i)]
\end{equation}
where the Fisher information $I$ and the Hessian $J$ are given by
\begin{equation}
    I_{ij} = \frac{\partial E^2}{\partial \theta_i}\frac{\partial E^2}{\partial \theta_j} \qquad J_{ij}=\frac{\partial^2 E^2}{\partial \theta_i\partial \theta_j}\,.
\end{equation}
Compared to the other criteria studied in Ref~\cite{zhou2009}, the PAIC most accurately reproduced the true bias in the Gaussian and non-Gaussian distributions tested. To correctly estimate the relative weights, one must be very careful to include all normalization constants in the definitions of $\langle E^2\rangle _i$, such as for the hyperparameter prior, which may differ across models. The posterior averaged log-evidence could be evaluated on the same MC samples as will be used for $\bar{q}$ and $H$, so there is no significant additional cost to use the PAIC.  

An alternative information criterion, the Bayesian Takeuchi Information Criterion (BTIC)~\cite{takeuchiinfo,Neil:2022joj}, approximates $\langle E^2\rangle_i \to E^2(\theta^i_0)$ to avoid integration over the space of (hyper-)parameters. 
\begin{equation}
    BTIC_i =  E^2(\theta_0^i) +2 \tr[J^{-1}(\theta_0^i)I(\theta_0^i)]\,.
\end{equation}
This form is appropriate when the MAP approximation is used.
In the following data analyses, the results from PAIC and BTIC averaging are very similar.

Finally the Bayesian Akaike Information Criterion (BAIC)~\cite{akaikeinfo,Jay:2020jkz,Neil:2022joj} approximates $J^{-1}(\theta_0^i)=I^{-1}(\theta_0^i)$ to give
\begin{equation}
    BAIC_i =  E^2(\theta_0^i) +2 k_i \,,
\end{equation}
where $k_i$ is the number of hyperparameters for that prior.
This approximation is valid in the infinite-statistics limit when the true function can be exactly reproduced by the model~\cite{Jay:2020jkz}. All three criteria converge in the infinite statistics limit. 

\section{Numerical Implementation}\label{sec:data_tests}

In this section, we apply Gaussian process regression to synthetic data generated from phenomenological PDFs to study a closure test of the methodology. The analysis requires choosing a kernel, a mean function (with associated hyperparameters and priors), and a discretization of the $x$-domain with an appropriate interpolation scheme. Many possible combinations exist, and we restrict ourselves to representative cases.

Two possible choices of the mean function given by
\begin{equation}
    g_{\rm flat} (x;\theta=\{N\}) = N \quad ; \qquad g_{\rm PDF} (x;\theta=\{N,\alpha,\beta\}) = N x^\alpha (1-x)^\beta \,
\end{equation}
will be used. As was done in Refs.~\cite{Dutrieux:2024rem, Dutrieux:2025jed}, we will start with fixed values for these models. An advantage of the fixed models or even the unfixed flat model is that a limited number of parameters makes for a simpler, lower-dimensional $\theta$ integral to perform. A disadvantage is that the $x\to0$ behavior, which is unguided by the data, will always ultimately converge to what this model gives. Even so, all cases will generally give statistically consistent values when in the regimes guided by the data.  

Finally, every kernel suggested has an overall scale hyperparameter $\sigma$, which sets the overall variance of the prior. Furthermore, in a region of $x$ unconstrained by data, $\sigma$ sets the variance $q$  has about the prior's mean $g$.  For example, in Ref.~\cite{Dutrieux:2024rem}, when the mean function can well represent the data, the evidence favors a small value of $\sigma\sim O(0.1)$ which suppresses the non-trivial effects of the GP in the case of the log-RBF kernel. However, if the kernel exhibits divergent behavior in the low $x$ region, as is the case with the CDGP kernel, the $\sigma\sim O(10^{-2})$ removes unphysical oscillations of the PDF reconstruction in this region. The limit $\sigma\to0$ forces the GP prior to be a $\delta$ function, such as in Eq.~\eqref{eq:prior_for_model_fits}, forcing the unknown $q$  to the prior's mean $q=g$. This behavior has been observed in this particular GPR application. Optimizing the hyperparameters is reduced to simply fitting the prior's mean. 

The tendency to small $\sigma$ occurs most frequently when the hyperparameters of the mean function are optimized and the prior mean describes the data well at this optimal hyperparameter point. As suggested in Ref.~\cite{Dutrieux:2024rem}, the prior on $\sigma$ could be a $\delta$ function. The dependence on the choice of fixed $\sigma$ can be studied and is an example of a bias/variance trade-off. This trade-off is between doing a simple fit as $\sigma\to0$ and having an overwhelmingly noisy prior as $\sigma\to \infty$, which fails to regulate the inverse problem. The rest of the hyperparameters will be given less trivializing priors, though some may still be restrictive, but informed by several explorations of the parameter space.

In the remainder of this section, we will study the following list of choices, beyond varying the prior kernel and mean,
\begin{itemize}
    \item Fixing the hyperparameters, i.e., the model's hyperparameters' priors are all $\delta$ functions
    \item Allowing certain hyperparameters to be integrated, i.e., the model's prior has non-trivial hyperparameter dependence
    \item Model averaging the results, i.e., the model's prior is a sum of priors for individual models
\end{itemize}
These represent three levels of inference or levels of complexity of the prior. Within each level, we test the impact of changing the prior. The first level begins with a single Gaussian Process with a $\delta$ function prior for the hyperparameters, followed by a single Gaussian process with a hyperparameter prior which is less constraining, and finally, in the final level, a prior which is a sum of Gaussian processes with hyperparameter priors of either type. The different information criteria are different prescriptions for the bias correction when adding the Gaussian processes. It is from the most general last case that we draw the final results.

\subsection{Synthetic Data Generation}
\begin{figure}
    \centering
    \includegraphics[width=0.95\linewidth]{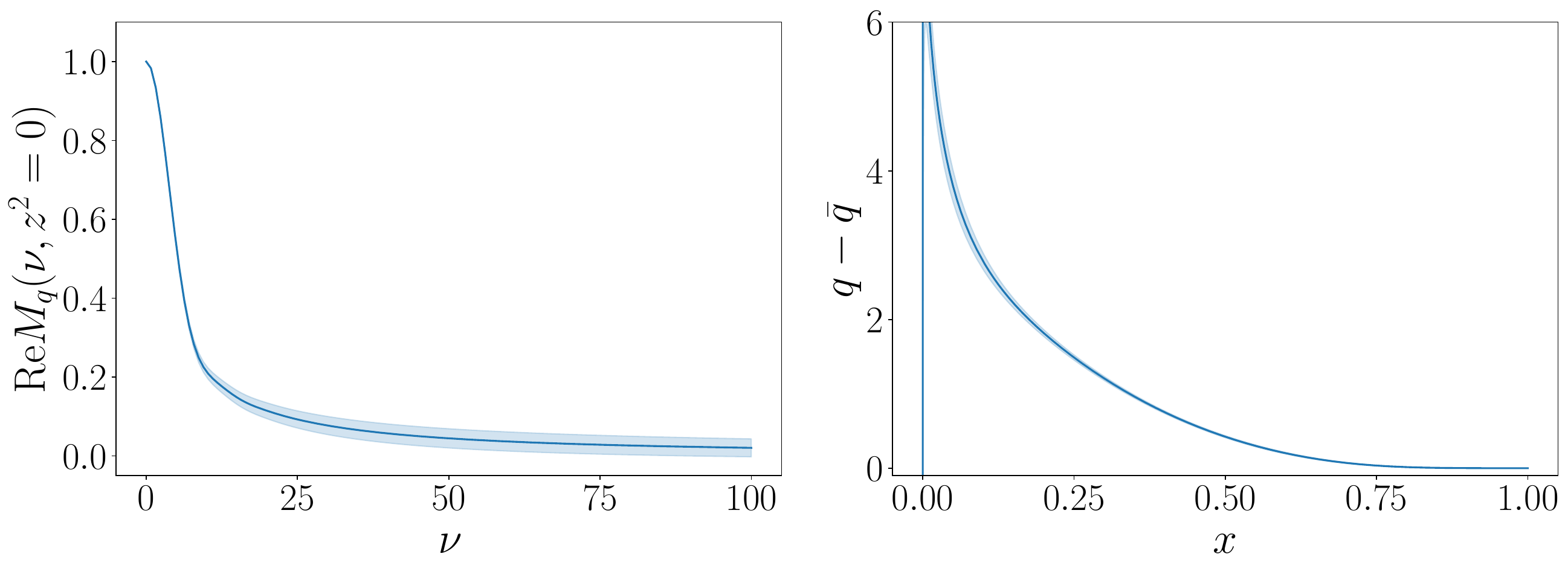}
    \includegraphics[width=0.95\linewidth]{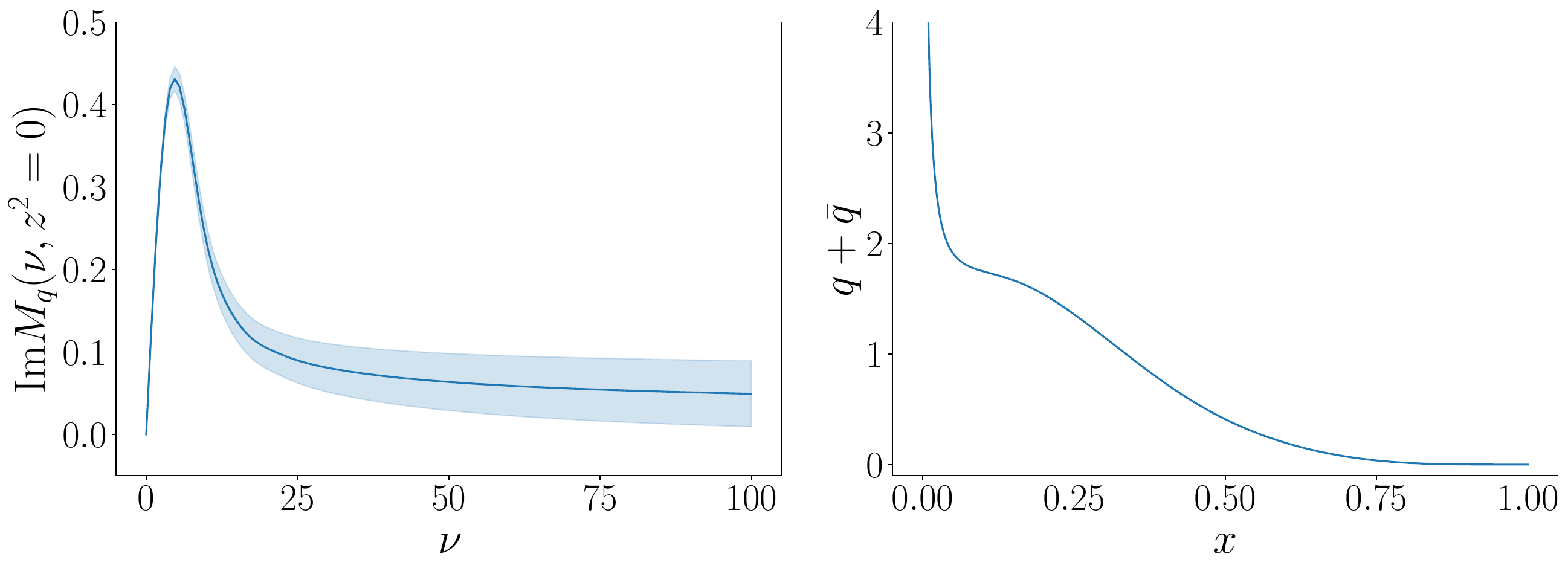}
    \caption{The NNPDF4.0~\cite{NNPDF:2021njg} isovector $u-d$ quark Ioffe time distributions (Left), Parton Distribution Functions (Right), split into CP even $q-\bar{q}$ (Upper) and CP odd $q+\bar{q}$ (Lower) contributions.}
    \label{fig:NNPDF40}
\end{figure}

We generate synthetic data from NNPDF4.0~\cite{NNPDF:2021njg} to test the fidelity of reconstructed central values and error estimates.
The synthetic data are generated for the isovector flavor quark PDFs analogous to the most precise lattice QCD data available to us. Specifically 
\begin{equation}
    q^{NS}_\pm(x) = (u(x)-d(x)) \pm (\bar{u}(x)-\bar{d}(x))
\end{equation} 
at renormalization scale $\mu=2$ GeV. The Ioffe time distribution (ITD) is evaluated on a grid with unit separation in the range $\nu\in [0,\nu_{\rm max}]$ with $\nu_{\rm max} =$ 4, 10, and 25. The Ioffe time and momentum fraction space distributions are shown in Fig.~\ref{fig:NNPDF40}. The all of the methodologies explored in this section are applied to actual lattice datasets in Sec.~\ref{sec:realdata}. 

\subsection{Varying the mean function}

The missing information in the inverse problem comes from regions of $\nu$ beyond the range accessible in lattice QCD data.  How the result extrapolates into this region will change the resulting PDF based on analysis-dependent choices. Ultimately, far from lattice QCD data, the ITD will approach the Fourier transform of the prior's mean. This extrapolation may depend on the resolution of the grid in $x$ in the corresponding region. As stated before, the second-order finite elements method is used to calculate the Fourier transform $B(\nu,x)\circ$. We compare the linear grid $x\in[10^{-3},1]$ to a ``split grid'' with linear spacing for $x\in[0.1,1]$ and logarithmically spacing for $x\in [10^{-8},0.1]$. Both grids consist of 257 points, and we have added a point at $5 \cdot 10^{-9}$ in the case of the split grid for stability purposes.

Following Ref.~\cite{Dutrieux:2024rem}, we begin the numerical studies with a $log$-RBF kernel with $l = \ln2$ (see Eq.~\eqref{eq:exp-rbf}). The value of $\sigma$ will be selected by the smallest value to produce a result satisfying a specific criterion: the maximum size of the variance in the extrapolation of the ITD is compared to either the deviation of the central value from 0 or the size of the error of the data point with the largest $\nu$. Specifically, the criterion is given by
\begin{equation}
   {\rm max}\bigg( \Delta M^{th}(\nu) \bigg) \geq {\rm max}\bigg( (1+p)\Delta M_{\nu_{\rm max}},\,\, p M_{\nu_{\rm max}} \bigg)\label{eq:criteriamax}
\end{equation}
where $p=0.3$ is selected. It was observed in Ref.~\cite{Dutrieux:2024rem} that this criteria would consistently provide error estimates which did not underestimate the true underlying variance.

\begin{figure}
    \centering
    \includegraphics[width=0.95\linewidth]{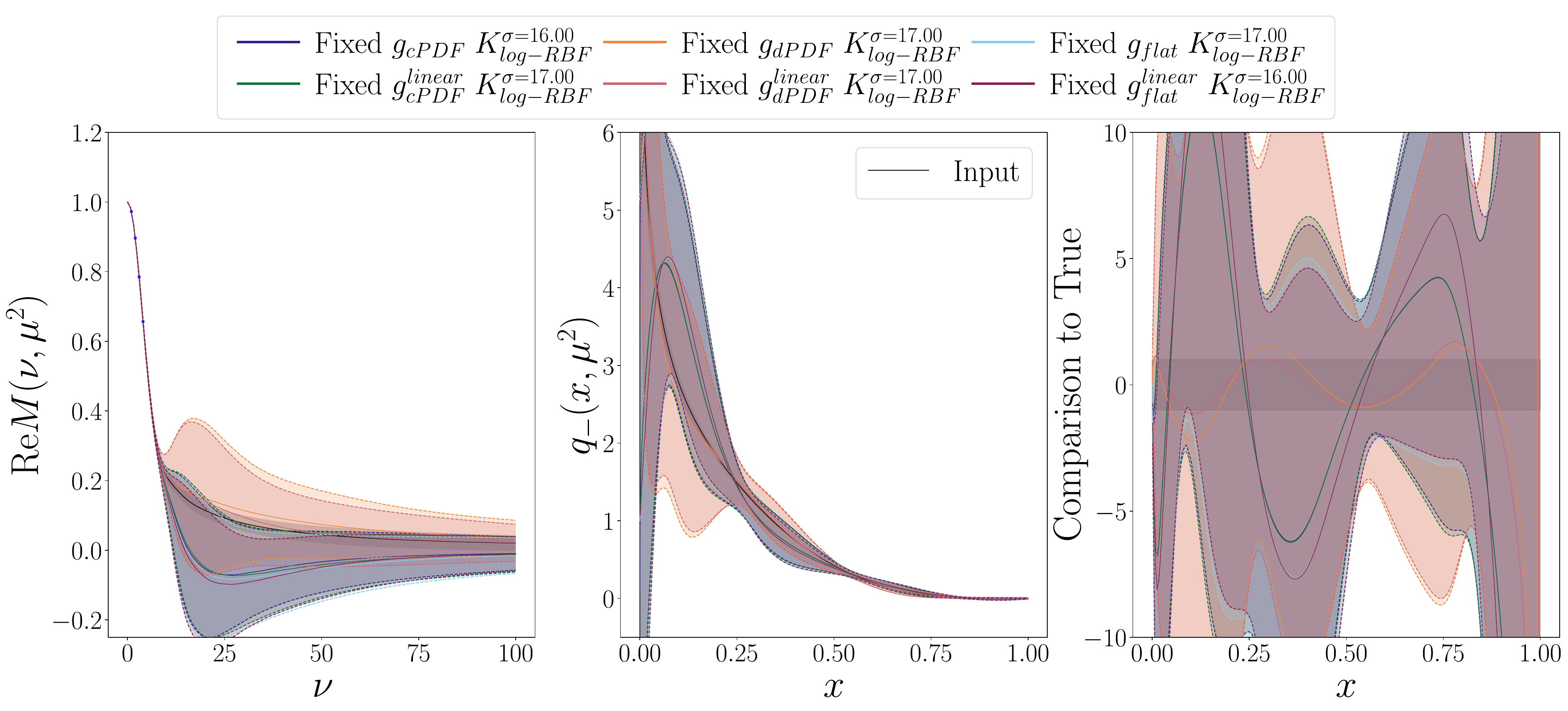}
    \includegraphics[width=0.95\linewidth]{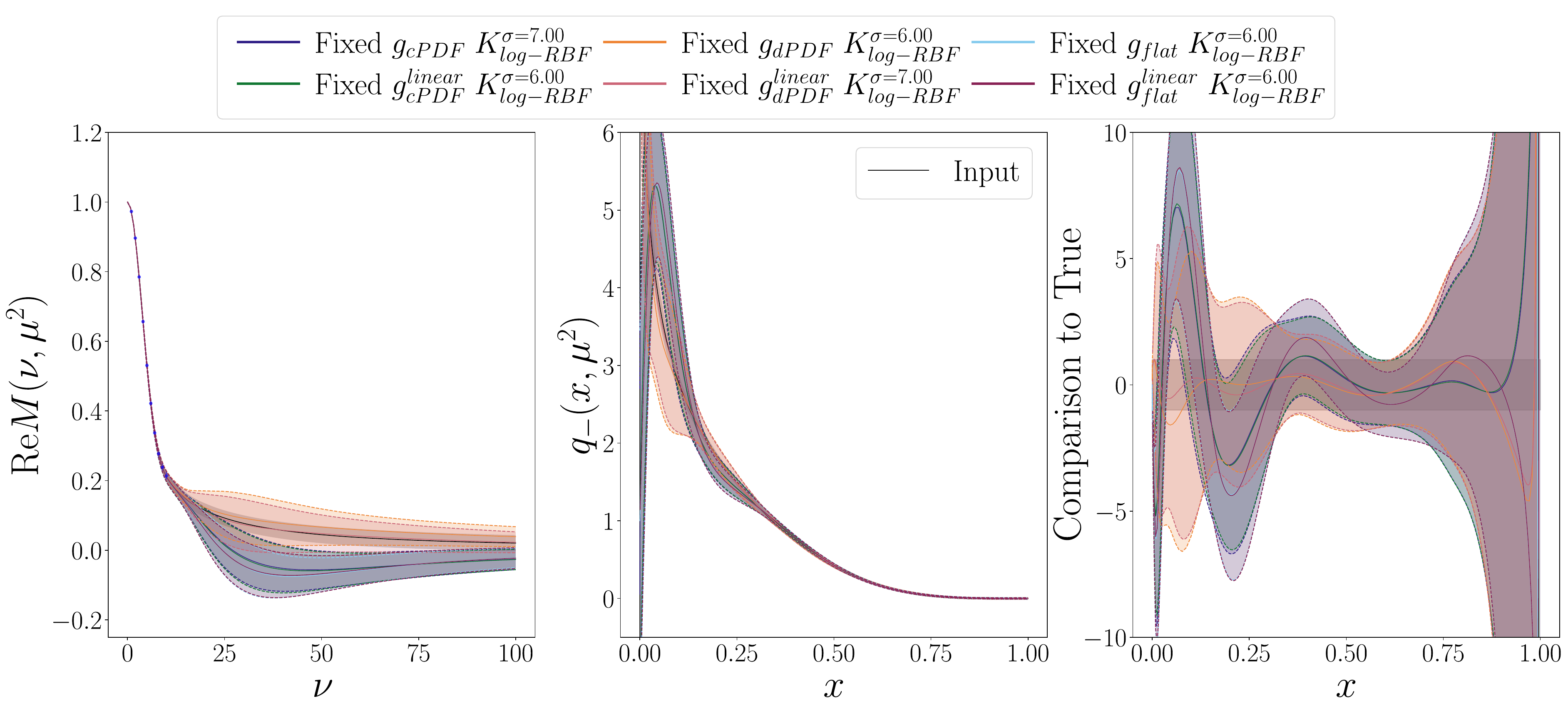}
    \includegraphics[width=0.95\linewidth]{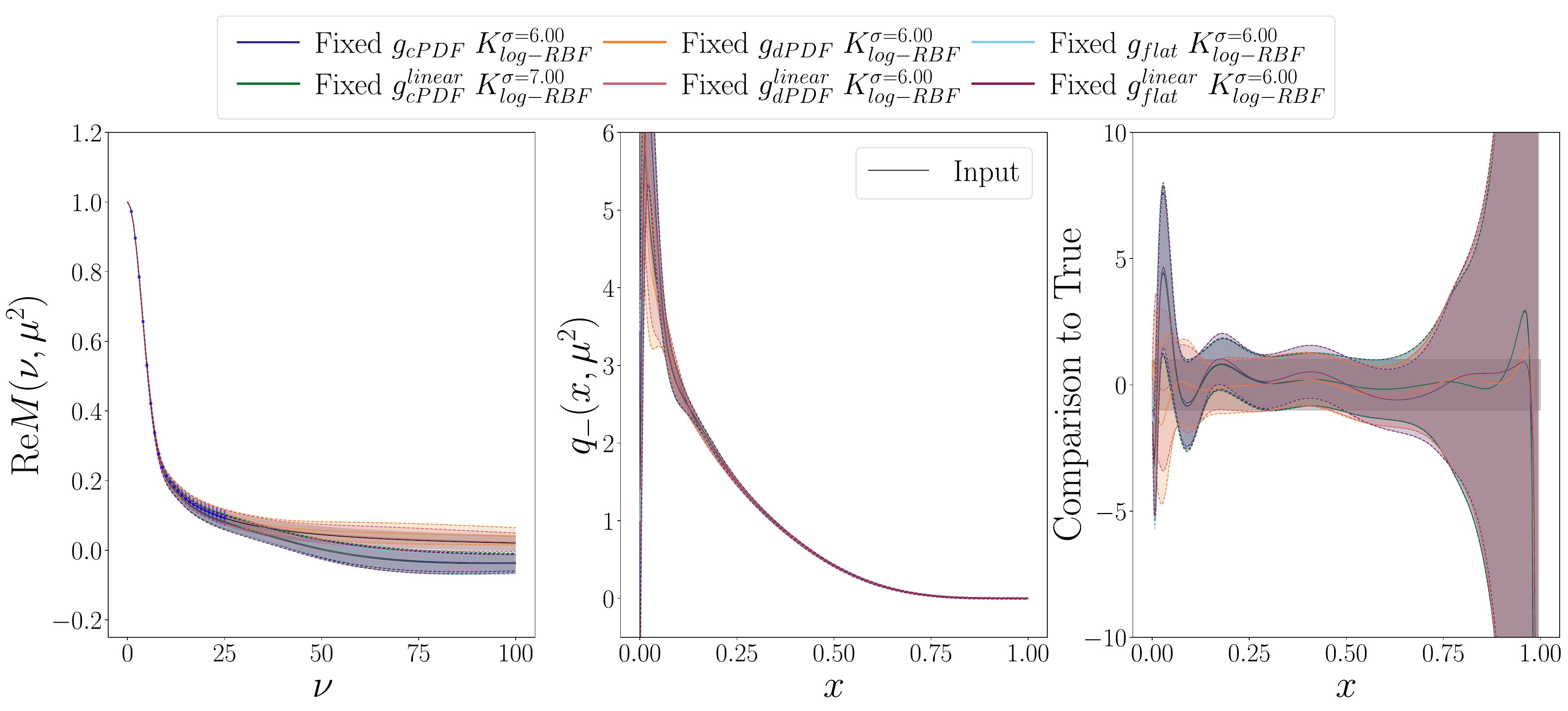}
    \caption{Impact of different prior mean functions for a $log-RBF$ kernel on the reconstruction of the NNPDF4.0 CP even isovector PDF (gray) from synthetic data. The maximum Ioffe time used was 4, 10, and 25, increasing from top to bottom. (Left) The ITD in $\nu$ space. (Middle) The PDF in $x$ space. (Right) Ratio of reconstruction error to true PDF uncertainty. We also highlight the small difference produced by using either a linear or split grid.} 
    \label{fig:convdiv_Re}
\end{figure}

\begin{figure}
    \centering
    \includegraphics[width=0.95\linewidth]{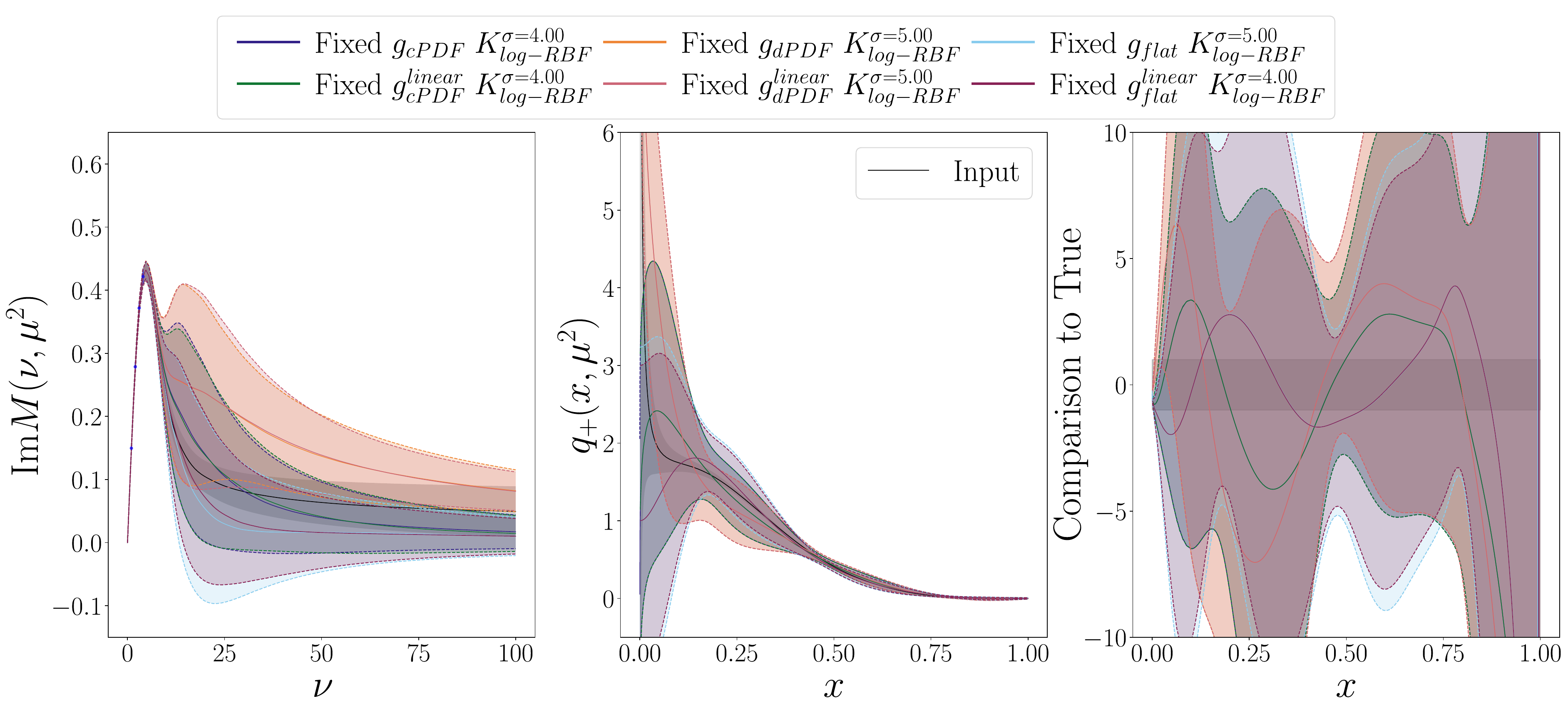}
    \includegraphics[width=0.95\linewidth]{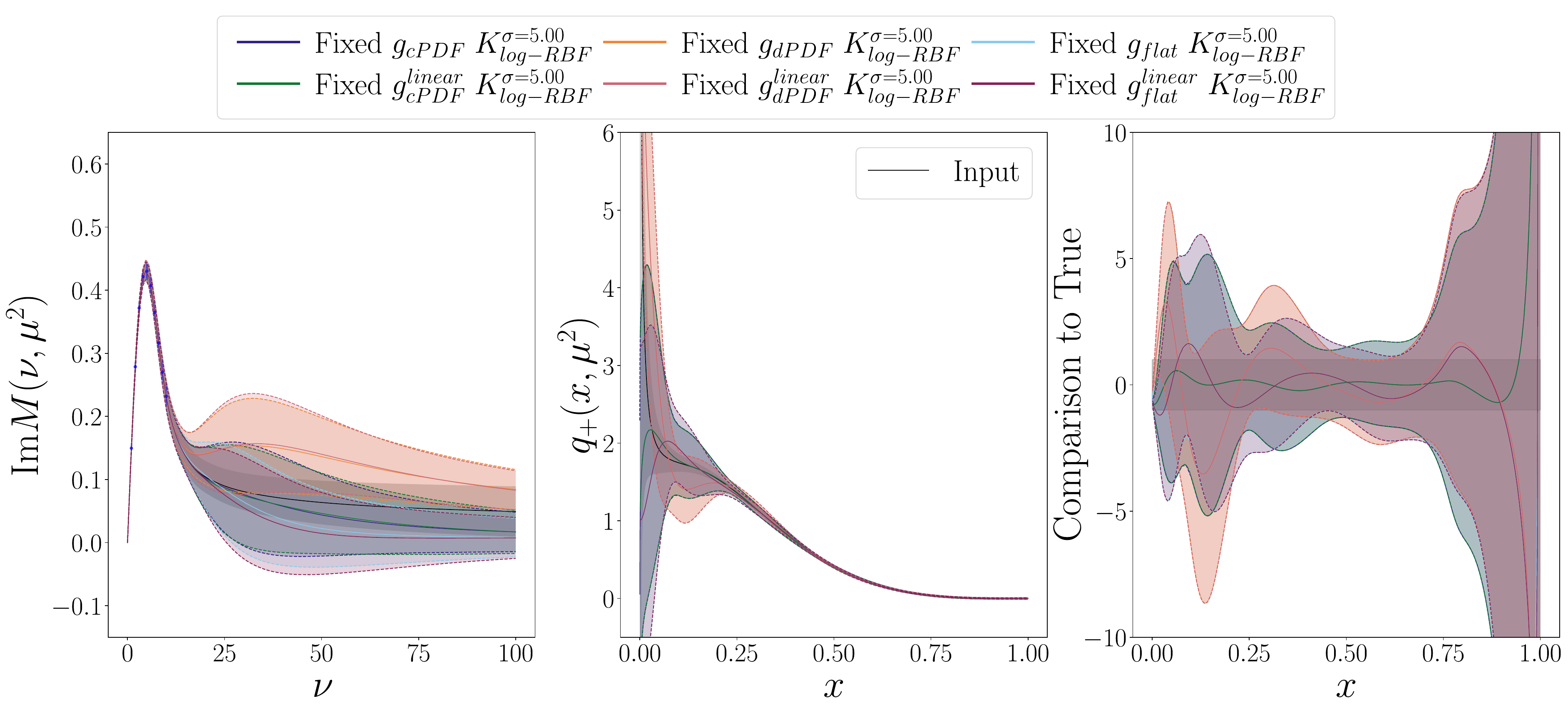}
    \includegraphics[width=0.95\linewidth]{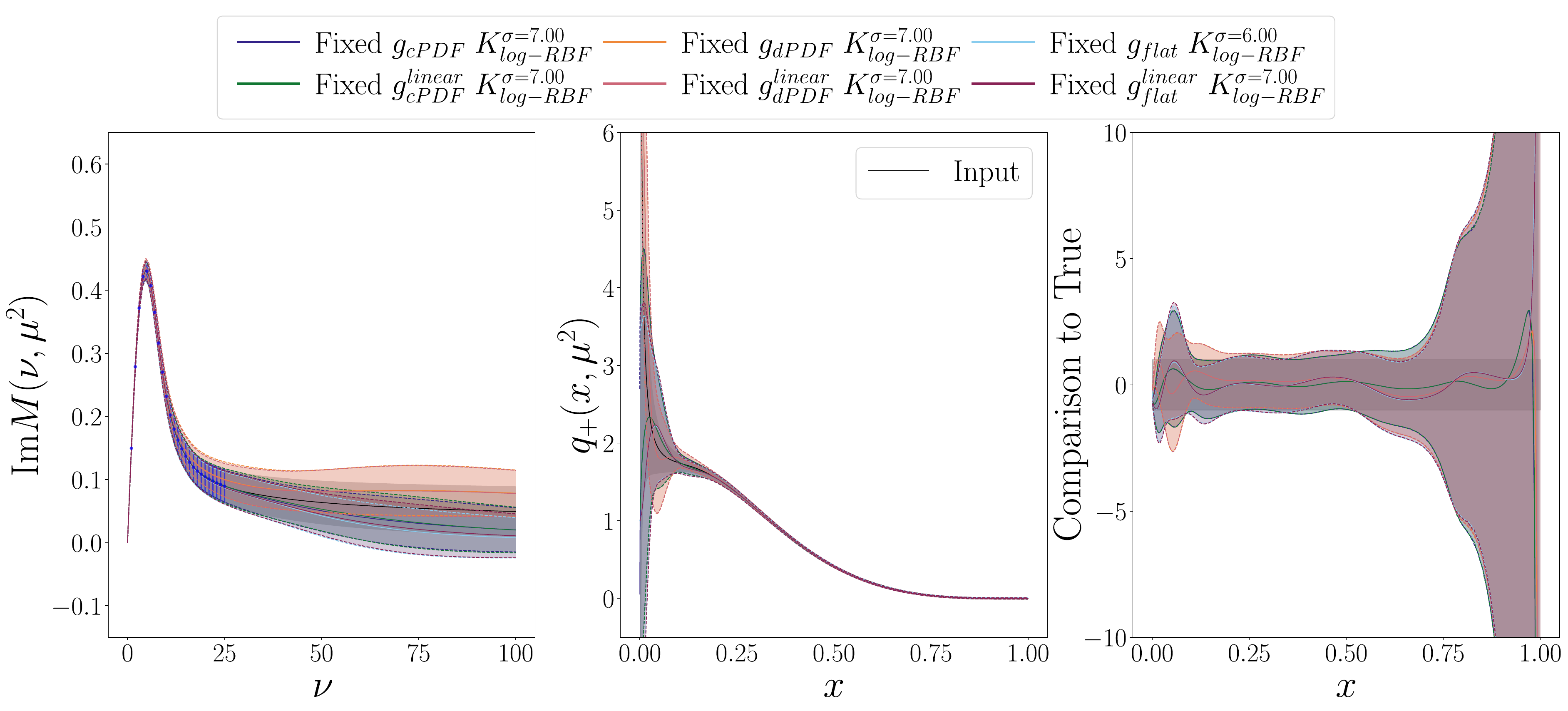}
    \caption{Impact of different prior mean functions for a $log-RBF$ kernel on the NNPDF4.0 CP odd isovector PDF. See caption of Fig.~\ref{fig:convdiv_Re} for more details.
    }
    \label{fig:convdiv_Im}
\end{figure}

Besides $g_{\rm flat}$ with $N=\sigma$ to bias for positive functions through the prior, we also use $g_{\rm PDF}$ with a choice of $\alpha=-0.25$, which diverges as $x\to0$ and another with $\alpha=0.25$ which converges. We label them as $g_{dPDF}$ and $g_{cPDF}$ respectively. In both $g_{\rm PDF}$, the model is normalized by using $N=1/B(\alpha+1,\beta+1)$ with $\beta=3$. The results for the real and components are shown in Fig.~\ref{fig:convdiv_Re} and~\ref{fig:convdiv_Im}.

Indeed, the diverging $g_{\rm PDF}$ for the prior's mean function has different extrapolation behavior than the flat model or converging $g_{\rm PDF}$. The ITD of the divergent PDF decays as $\nu$ increases much more slowly than the converging ones, which is imprinted onto the reconstruction results. Still, the PDF in the majority of the region of $x$ are largely consistent, with the quality of agreement improving at smaller $x$ values as $\nu_{max}$ increases. By using a variety of means, the region of extrapolation, which we identify as low $x$ or large $\nu$,  can be studied to obtain larger and more realistic uncertainty.

\subsection{How much was actually learned?}\label{subsec:KL}

\begin{figure}
    \centering
    \includegraphics[width=0.95\linewidth]{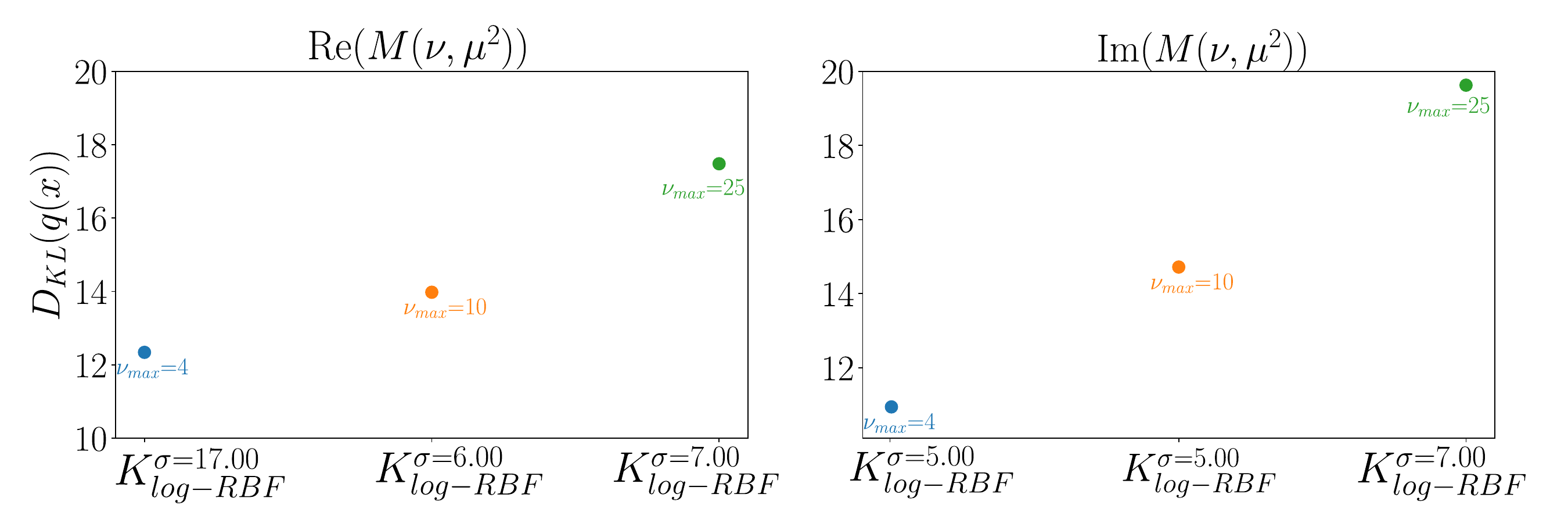}
    \caption{KL divergence in the $x$ and $\nu$ spaces for the real (Right) and imaginary (Left) components for the Fixed $g_{dPDF}$ model.}
    \label{fig:KL_all}
\end{figure}

\begin{figure}
    \centering
    \includegraphics[width=0.95\linewidth]{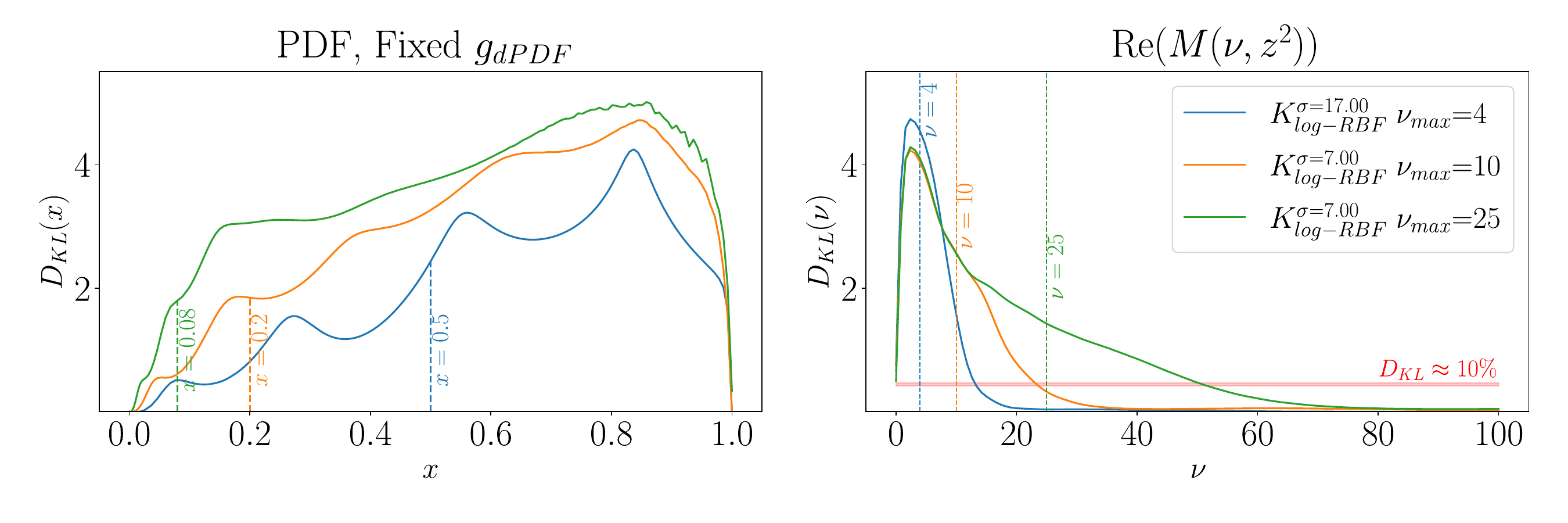}
    \includegraphics[width=0.95\linewidth]{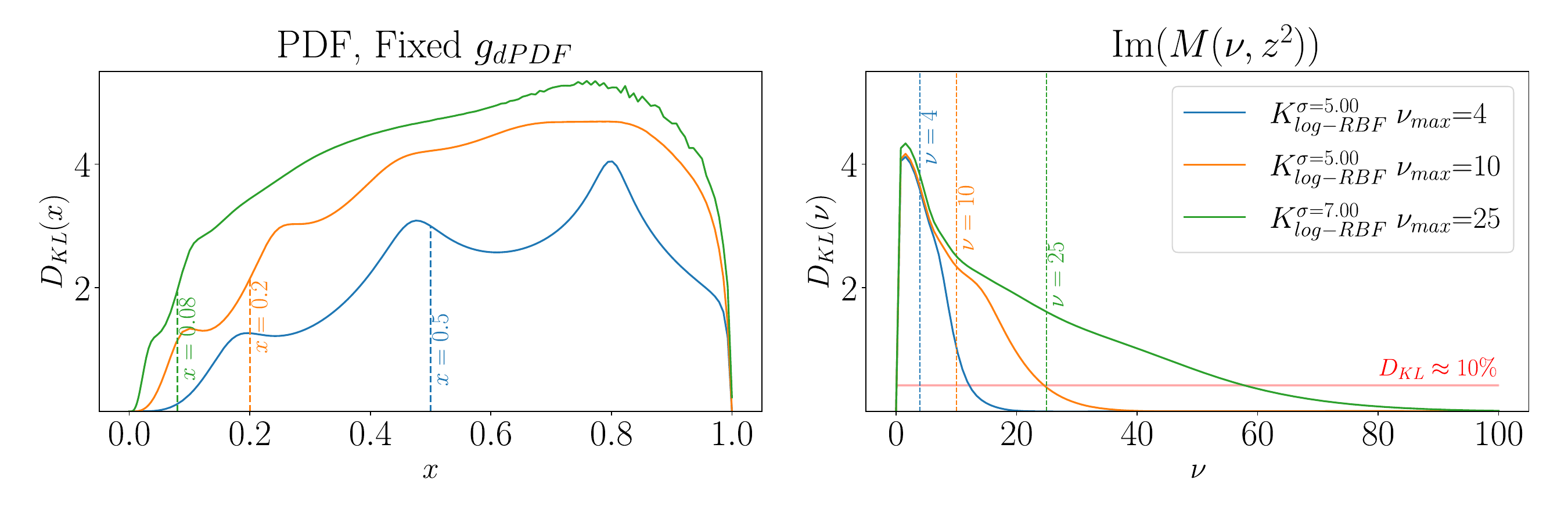}
    \caption{KL divergence in the $x$ (Left) and $\nu$ (Right) spaces for the real (Upper) and imaginary (lower) components. (Left) The vertical lines represent $x=2/\nu_{\rm max}$ and point out approximately where a $50\%$ increase in uncertainty was found in Ref~\cite{Dutrieux:2024rem}. (Right) The vertical lines from below show $\nu_{\rm max}$ and the horizontal line is approximately $10\%$ of the maximum of the KL divergence of all $\nu_{max}$. }
    \label{fig:KL_point}
\end{figure}

We evaluate the KL divergence, Eq.~\eqref{eq:full_KL}, for a representative model, shown in Fig. ~\ref{fig:KL_all}.
The covariance matrices used for the calculation are regulated in accordance with the following sections. The upward trend with $\nu_{\rm max}$ shows how much information is being gained by adding the data likelihood on top of the prior distribution. 
Interestingly, a substantial fraction of the information available at $\nu_{\rm max}=25$   is already captured at $\nu_{\rm max}=4$. In Fig.~\ref{fig:KL_point}, the KL divergence is calculated for individual points in $x$ from Eq.~\eqref{eq:pointKLx} and~\eqref{eq:pointKLnu}. As expected, the information gain is largest in the upper $x$ region and in the low $\nu$ region.

Because of the constraint $q(1)=0$, no additional information is gained exactly at  $x=1$
and information gain peaks just below this point. Similarly, the normalization of $q$ and the vanishing of the sine transform lead to no information gain at $\nu=0$ and a sharp rise shortly above. Interestingly, with this highly correlated data, information was gained at $\nu$ substantially past the $\nu_{\rm max}$ of the data. For actual lattice QCD data in the next section, information was gained noticeably past the last data point, but it decays more rapidly.


\subsection{Which hyperparameters to fix?}
\begin{figure}
    \centering
    \includegraphics[width=0.95\linewidth]{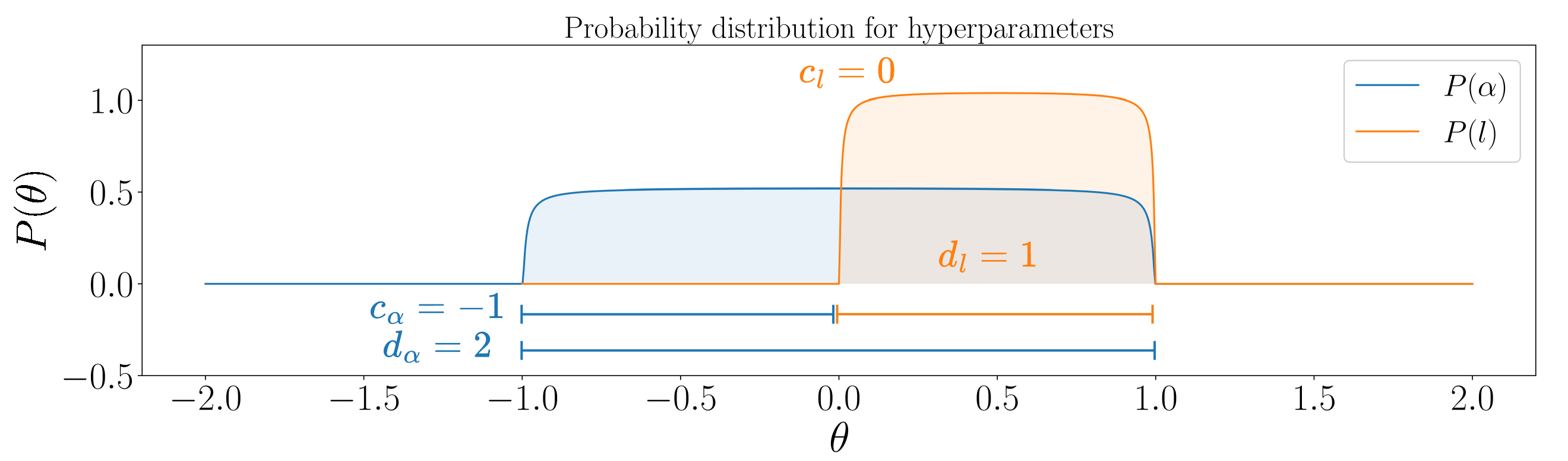}
    \caption{ Examples of prior probability distribution for $\alpha$ and $l$ in the model: Free $g_{PDF}$ Free $K_{log-RBF}^{\sigma=1}$ in the split grid. The upper plot demonstrates the roles of $c$ and $d$ in setting the window of the prior. }
    \label{fig:probtheta}
\end{figure}

As in Ref.~\cite{Dutrieux:2024rem}, all of the hyperparameters were fixed in the previous section in order to have an efficient and simple solution to the inverse problem. This section studies the impact of fixing or varying different combinations of hyperparameters as was done in Ref.~\cite{Candido:2024hjt}. For this section, $\sigma$ will remain fixed but either $l^2$, the parameters of $g$, or both are given non-trivial prior distributions. We propose a smooth and mostly flat prior within a physically meaningful range for the hyperparameters given by 

\begin{equation}\label{eq:expbeta}
    P[\theta|I]=ne^{-\frac{\hat{\theta}^a(1-\hat{\theta})^b}{2\cdot B(a+1,b+1)}}, \quad\text{with} \quad \hat{\theta} = \frac{\theta- c}{d}.
\end{equation}
By setting $a=b=-0.99$, this approximates a uniform prior with smooth boundaries. Parameters $c$ and $d$ allow us to set the starting point and the extension of the distribution, respectively. An example of this probability distribution function for two hyperparameters used in some of our models is presented in Fig.~\ref{fig:probtheta}, where we notice that $P(\theta=c)=0$ and $P(\theta=c+d)=0$. We call these two points the bounds of our distribution because they limit the exploration of $\theta$ to the region $[c,c+d]$ when sampling or minimizing. To find the constants $c$ and $d$ in \eqref{eq:expbeta}, we should specify the region of interest of each hyperparameter. 
The normalization constant $n$ is evaluated numerically to machine precision.
The integrals for $\langle q(x) \rangle$ and $\langle \textrm{Cov}[q(x),q(x')]\rangle$ given in Eq.~\eqref{eq:evidence_integral} and~\eqref{eq:weighted_errors} are performed with Gaussian importance sampling Monte Carlo Integration, described in App.~\ref{app:important_sampling}.

\begin{figure}
    \centering
    \includegraphics[width=0.95\linewidth]{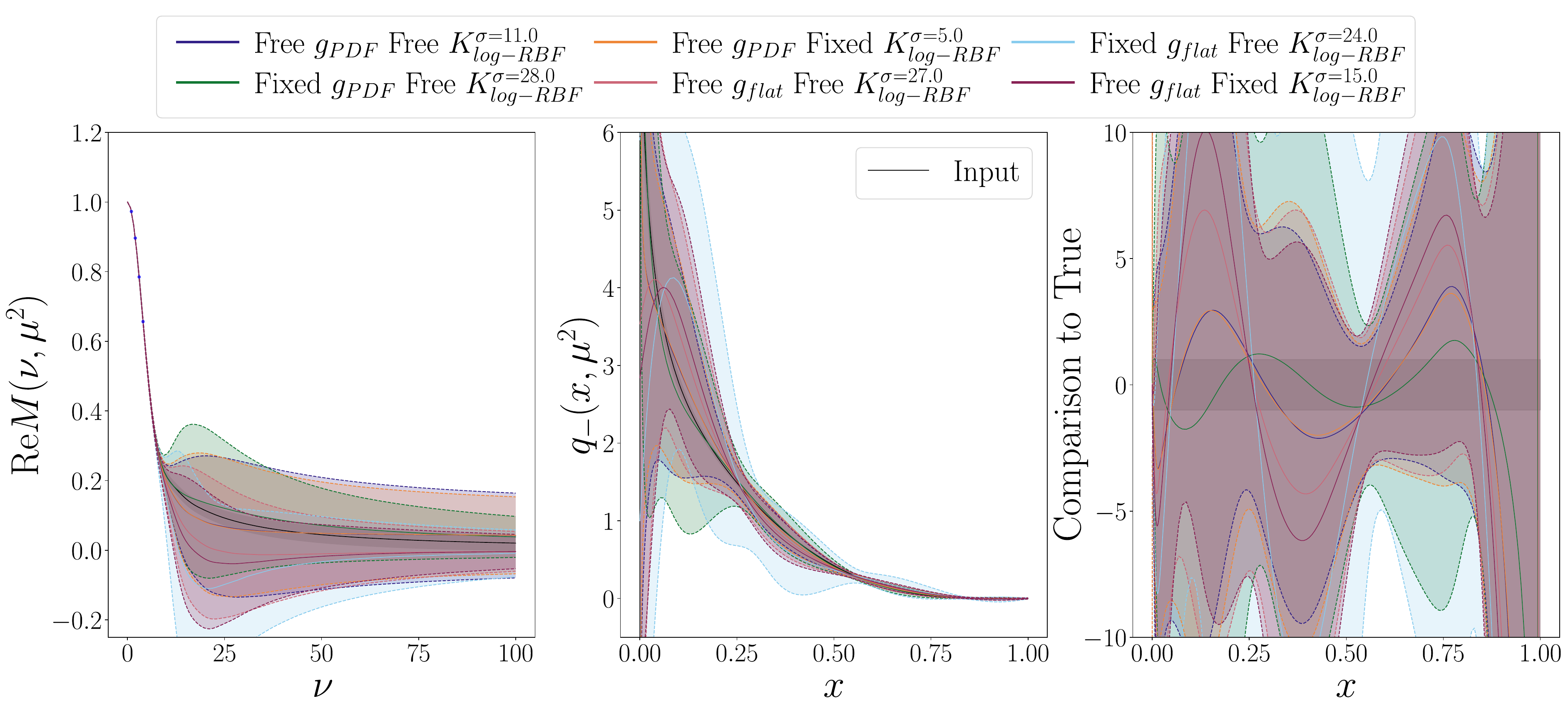}
    \includegraphics[width=0.95\linewidth]{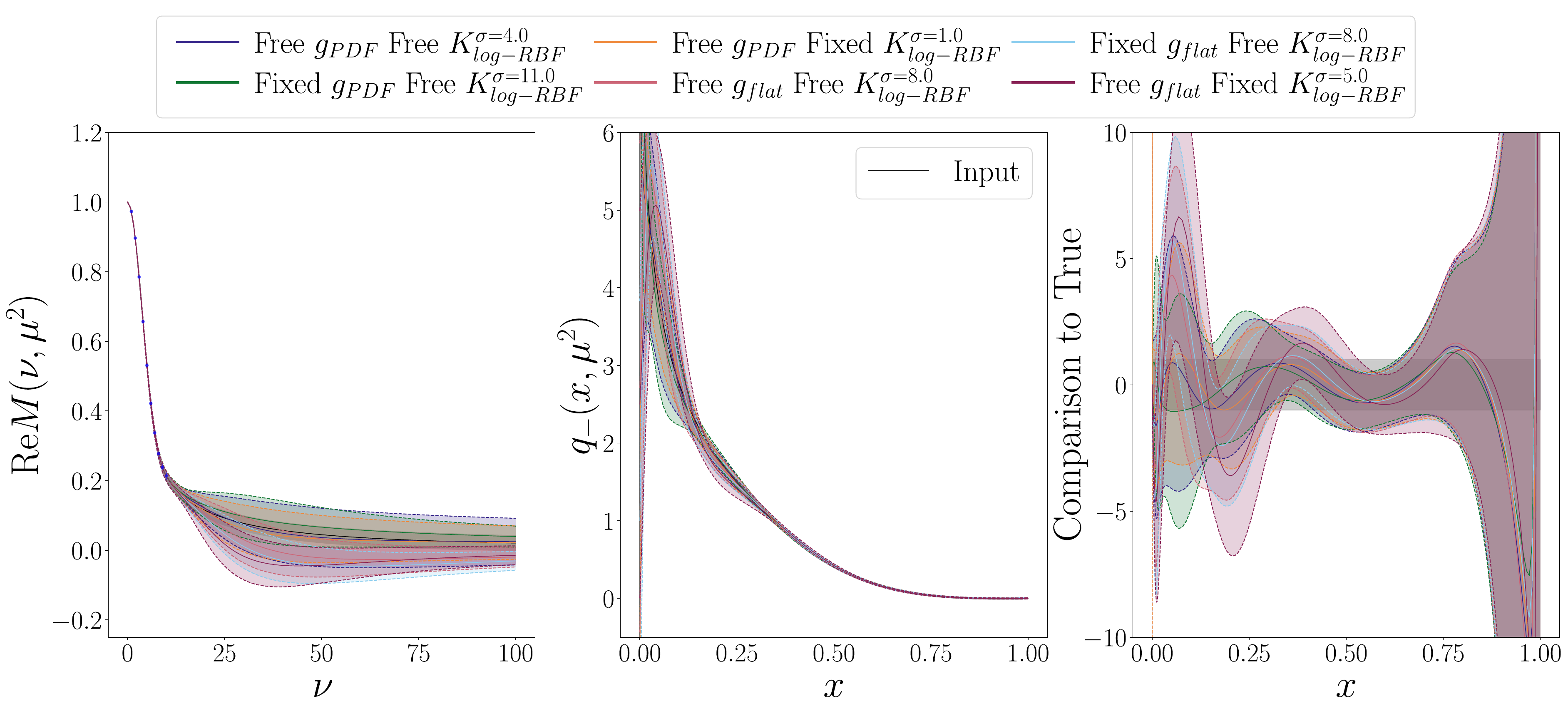}
    \includegraphics[width=0.95\linewidth]{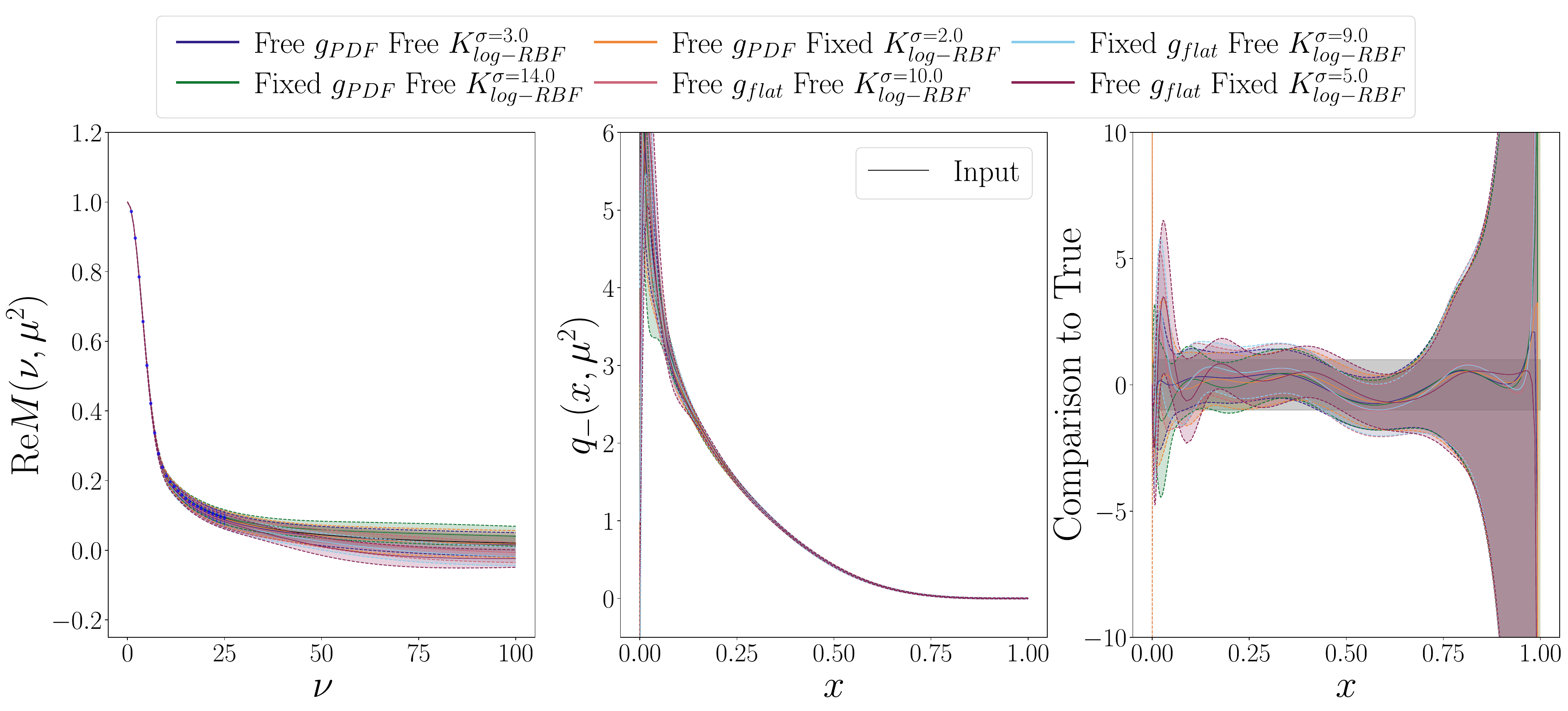}
    \caption{Impact of sampling over the hyperparameters of both the prior mean function and prior kernel, using a $log-\textrm{RBF}$ kernel for the CP even PDF. See caption of Fig.~\ref{fig:convdiv_Re} for more details.}
    \label{fig:logrbf_no_s_Re}
\end{figure}

\begin{figure}
    \centering
    \includegraphics[width=0.95\linewidth]{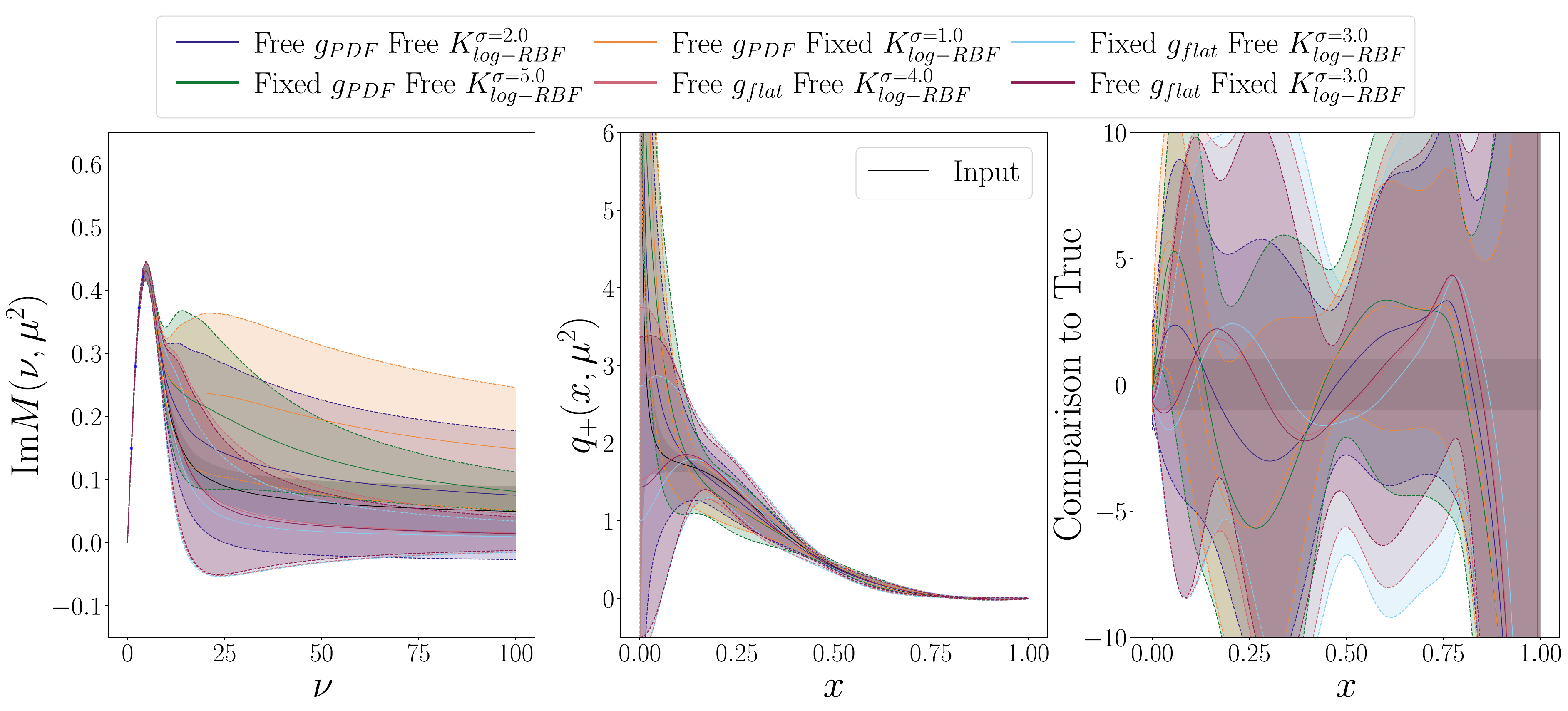}
    \includegraphics[width=0.95\linewidth]{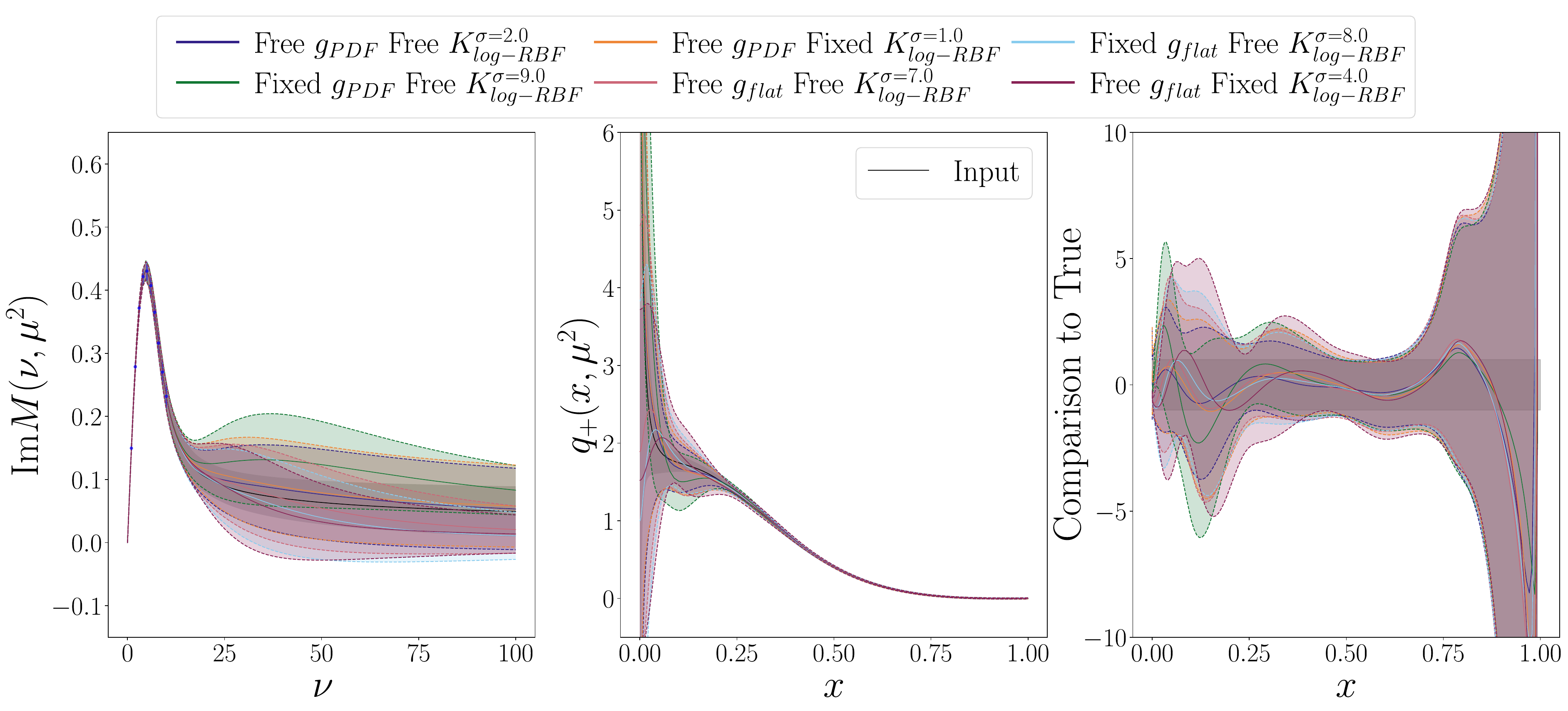}
    \includegraphics[width=0.95\linewidth]{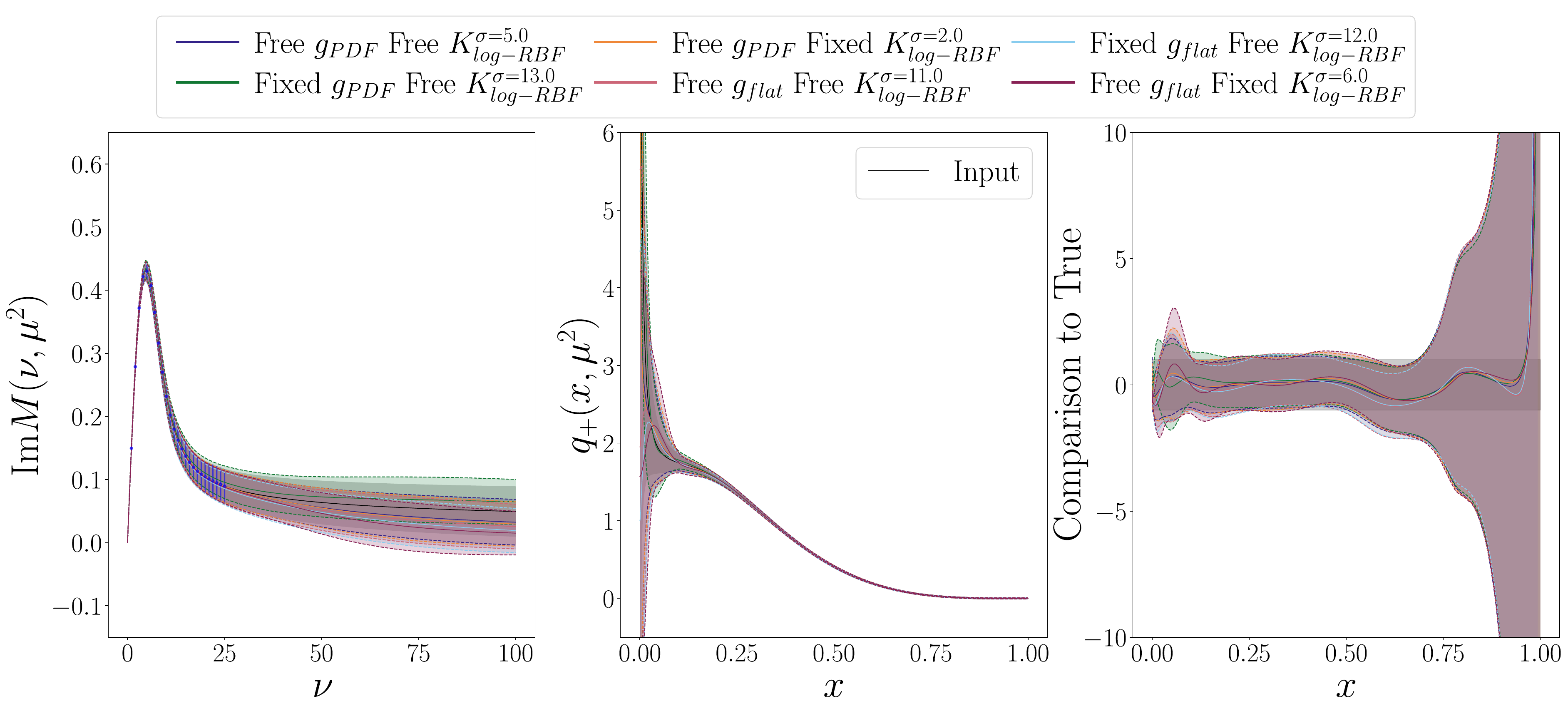}
    \caption{Impact of sampling over the hyperparameters of both the prior mean function and a $log-RBF$ prior kernel for the CP odd PDF. See caption of Fig.~\ref{fig:logrbf_no_s_Re} for more details.}
    \label{fig:logrbf_no_s_Im}
\end{figure}

In Figs. \ref{fig:logrbf_no_s_Re} and~\ref{fig:logrbf_no_s_Im}, the overall scale of the prior distribution was held fixed, based upon the previously mentioned criterion of the extrapolation error \eqref{eq:criteriamax}. Instead, one could give $\sigma$ a non-trivial prior distribution and allow it to vary. However, this approach was explored, and the majority of the predictions underestimated the uncertainty.
Regarding other hyperparameters, we first allowed those that represent a correlation length, labeled $l$, to take any positive value. However, numerical instabilities at $x$ larger than 1 appeared, so we limit ourselves to exploring the interval $[0,1]$. On the other hand in the $g_{PDF}$ mean functions, $\alpha \in [-1,1]$ and $N$, $\beta \in [0,15]$. Furthermore, the parameter $\alpha$ introduced in $K_{CDGP}$ is considered to lie on $\alpha\in[-0.8,0.0]$ to avoid unphysical extrapolations at low $x$. A Tikhonov regulator $r =O(10^{-5}-10^{-6})$ was added to the posterior covariance, $H\rightarrow H+r\cdot I$, to control these instabilities. 
In the case of fixed prior mean, the values of $\sigma$ at the minimum of $E^2$ were $O(1)$ similar to those selected in the previous analyses, while when the prior mean was $g_{\rm PDF}$ which can well reproduce the data, the minimum $\sigma$ was an order of magnitude lower. In this case, the variance of the final PDF was dominated not by the width of the Gaussian Process $H(\theta)$, as it would be with all hyperparameters fixed, but by the variance across hyperparameters $(\bar{q}(x;\theta) - \langle q(x) \rangle)^2$.

\subsection{Comparisons of different kernels}
 In this section, we evaluate the extrapolation behavior at low $x$ of the kernels proposed in Sec.~\ref{sec:examples_of_kernels} and we explore different ways of sampling the hyperparameters. In addition, we only consider the split grid because of its integration properties for divergent PDFs, to which our synthetic data belong. 
 The mean is held fixed with the same parameters as in $g_{dPDF}$ for that reason.

\begin{figure}
    \centering
    \includegraphics[width=0.95\linewidth]{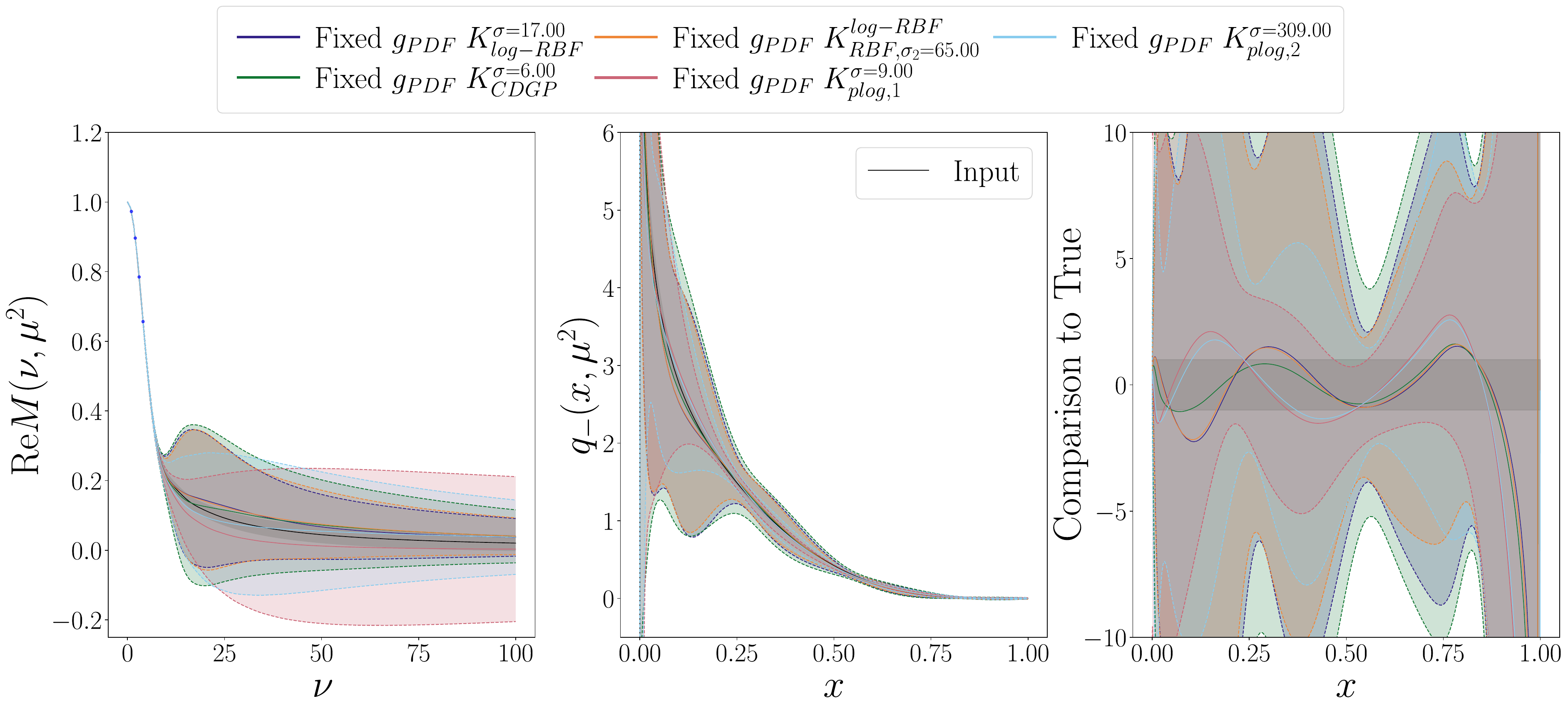}
    \includegraphics[width=0.95\linewidth]{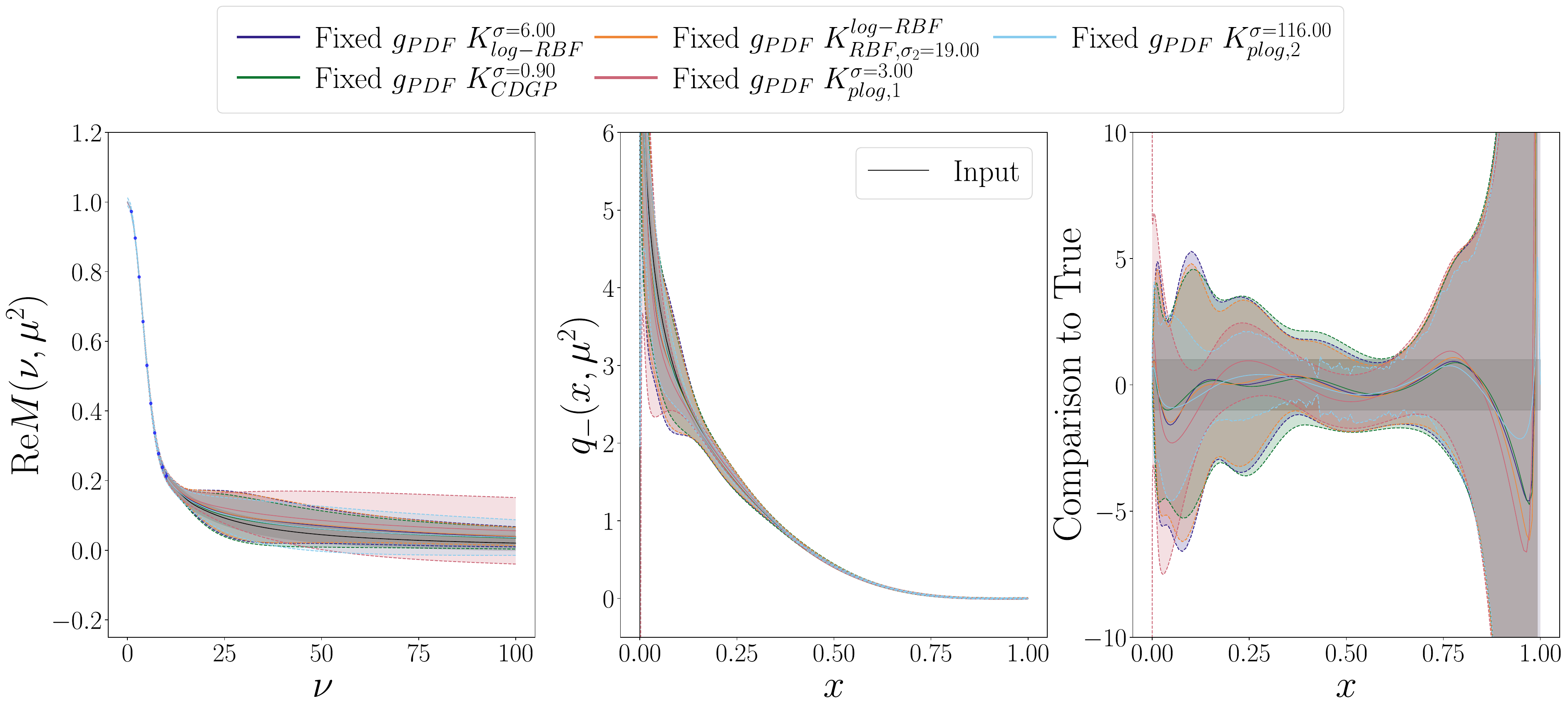}
    \includegraphics[width=0.95\linewidth]{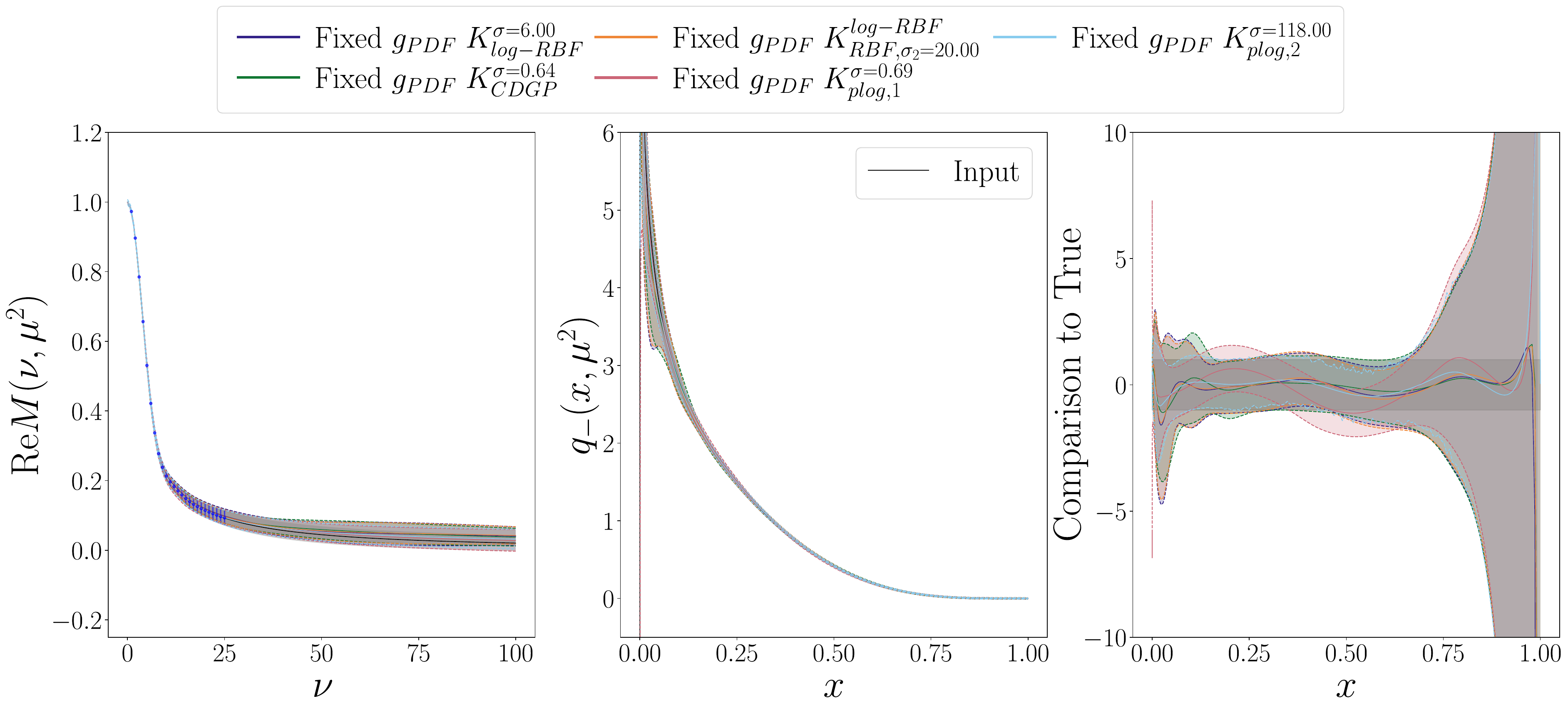}
    \caption{Impact of varying the prior kernel with fixed hyperparameters and a fixed prior mean function for the CP even PDF. See caption of Fig.~\ref{fig:convdiv_Re} for more details.
    }
    \label{fig:Good_models_Re}
\end{figure}

\begin{figure}
    \centering
    \includegraphics[width=0.95\linewidth]{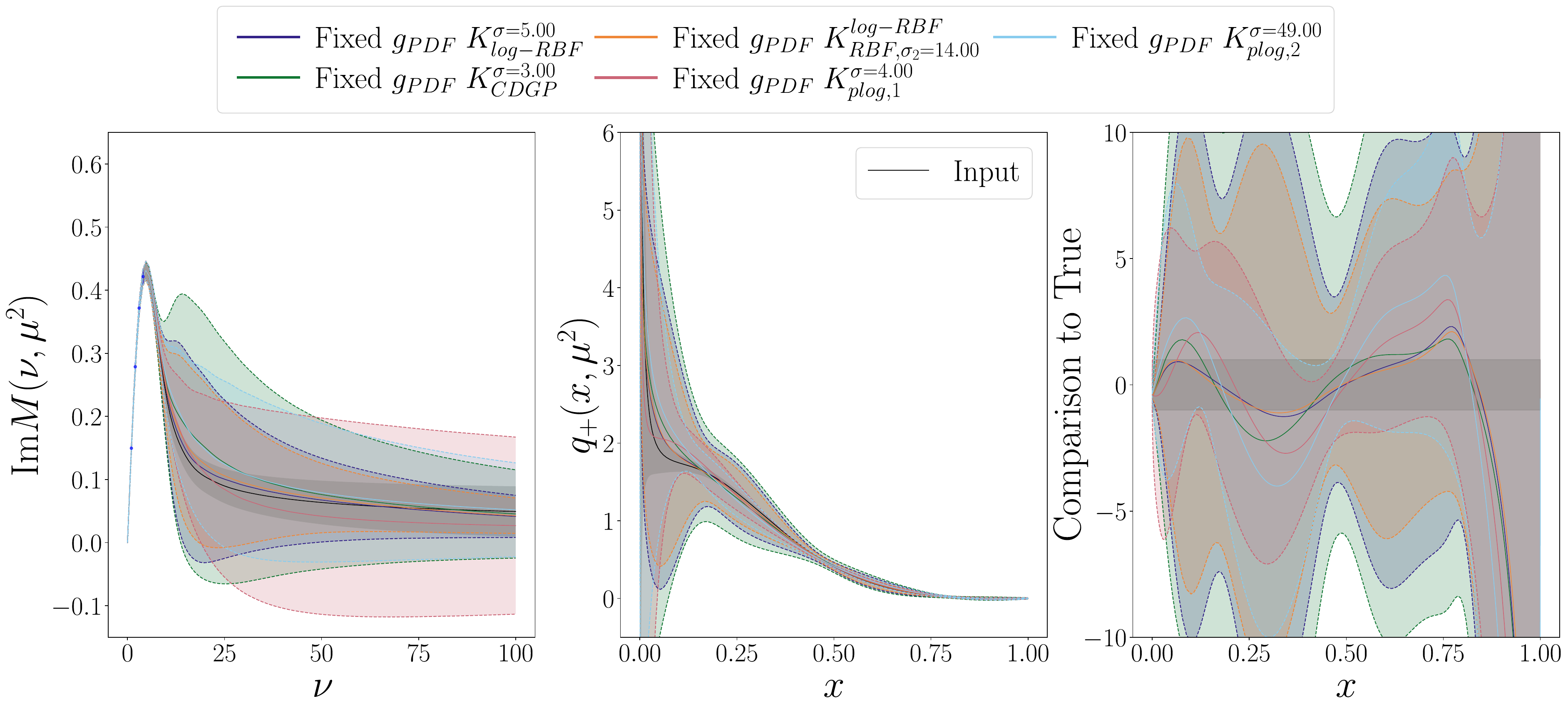}
    \includegraphics[width=0.95\linewidth]{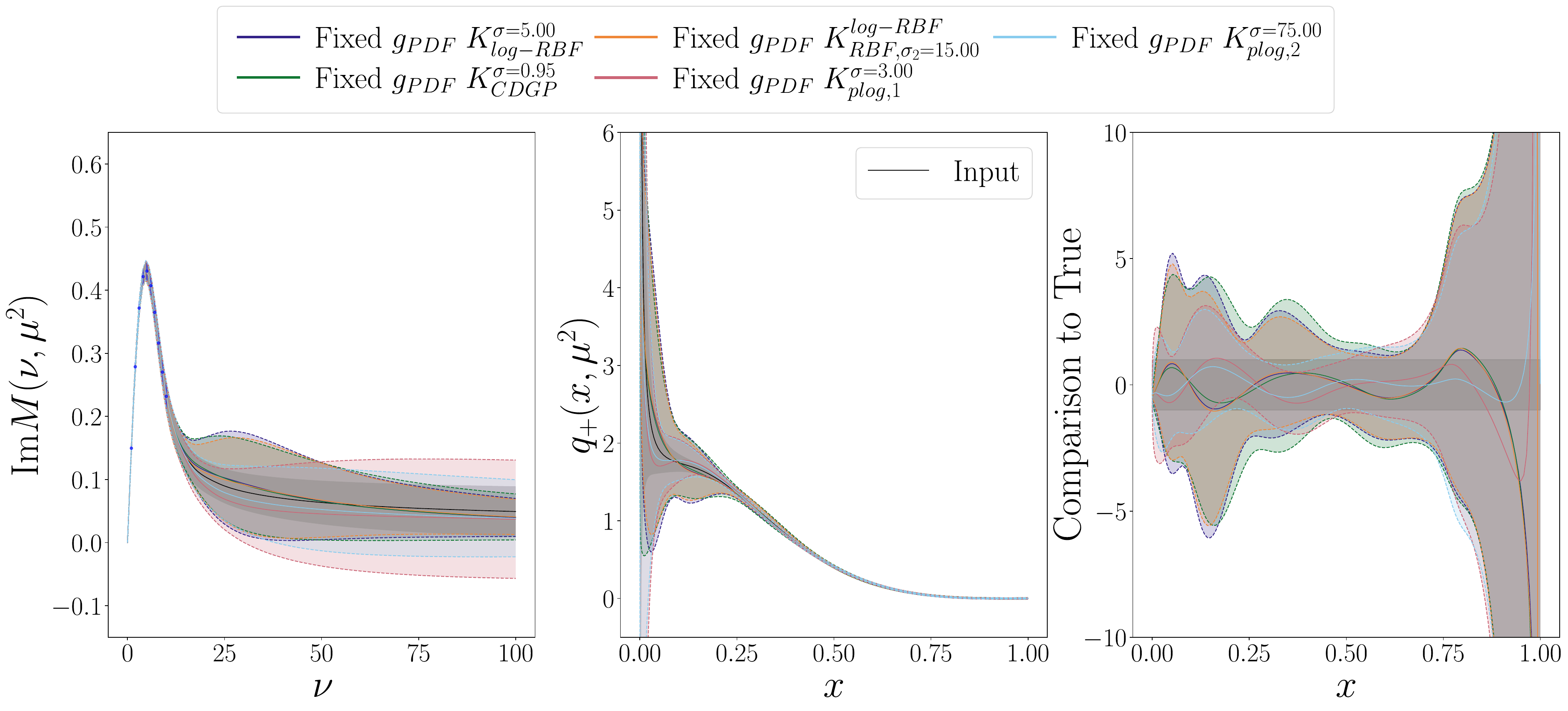}
    \includegraphics[width=0.95\linewidth]{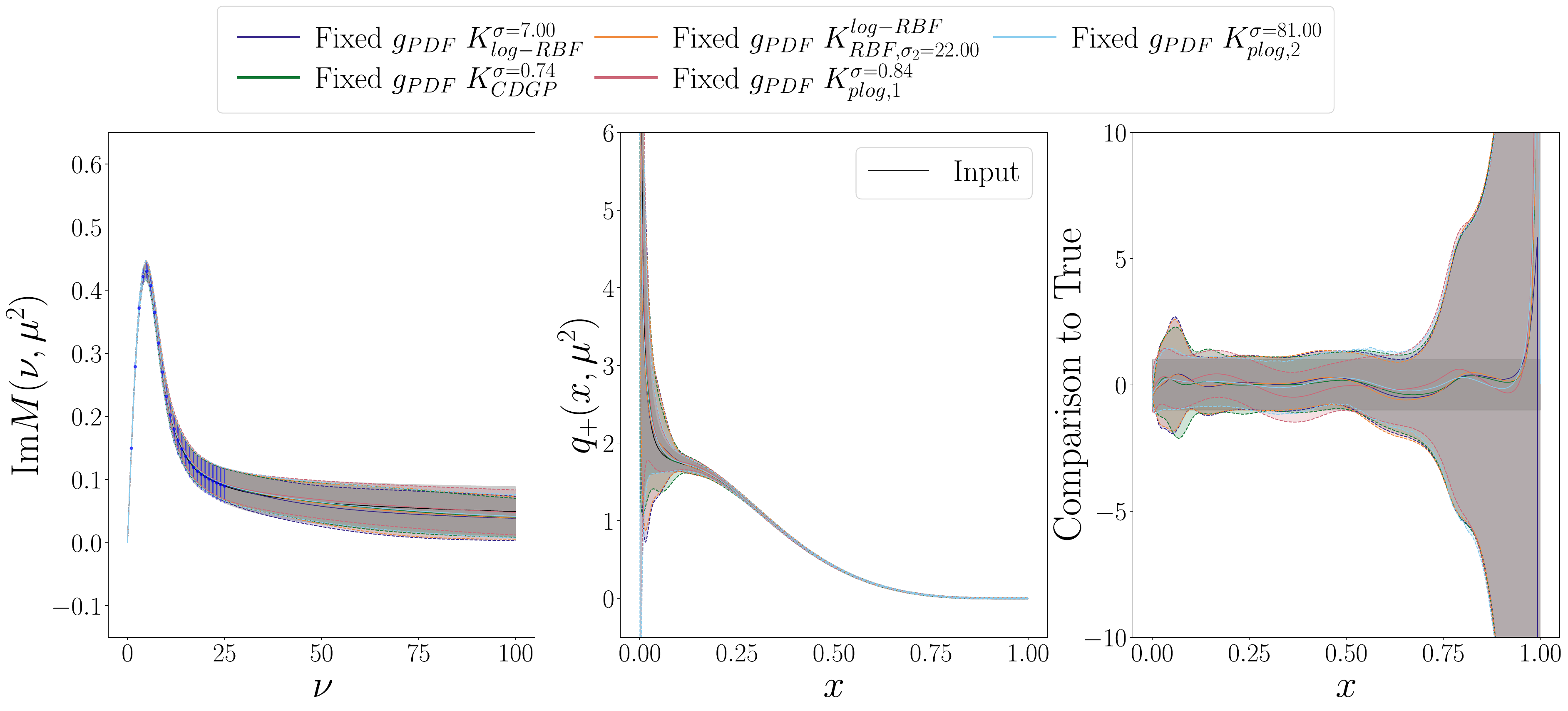}
    \caption{Impact of varying the prior kernel with fixed hyperparameters and a fixed prior mean function for the CP odd PDF. See caption of Fig.~\ref{fig:convdiv_Re} for more details. 
    }
    \label{fig:Good_models_Im}
\end{figure}

All the correlation length parameters are fixed at $l=\log(2)$ when needed, and we use the extrapolation criterion to fix any $\sigma$. Considering the mean and covariance of the posterior expressions in Eqs. \eqref{eq:mean_woodbury} and \eqref{eq:woodbury}, we notice that $K$ and $C$ play a pivotal role in its construction. Particularly, $\Gamma=C+B\circ K\circ B^T$ must satisfy a positive definite condition, but finite statistics and numerical precision can lead to instabilities. Positive definiteness can be achieved by most of the kernels if we add a regulator, $\Gamma\rightarrow\Gamma+r\cdot \mathbf{1}$. In our investigation, $r$ is of the order $O(10^{-13}-10^{-9})$ for $K_{log-RBF}$ and $K_{RBF}^{log-RBF}$ and of the order $O(10^{-9}-10^{-7})$ for the other cases presented. Additional kernels suffering from worse pathologies are presented in Appendix \ref{app:poorkernels}.
Kernels in section \ref{sec:examples_of_kernels} with all parameters fixed are presented in Figs. \ref{fig:Good_models_Re}-\ref{fig:Good_models_Im}. The extrapolation at high $\nu$ is similar in the case of RBF-like covariances, as expected in \cite{Dutrieux:2024rem}. Furthermore, we can notice that  $K_{plog,2}$ has equally optimal performance compared with the previously mentioned kernels, which differ from $K_{plog,2}$, where a more conservative uncertainty is associated with the reconstruction. 

\begin{figure}
    \centering
    \includegraphics[width=0.95\linewidth]{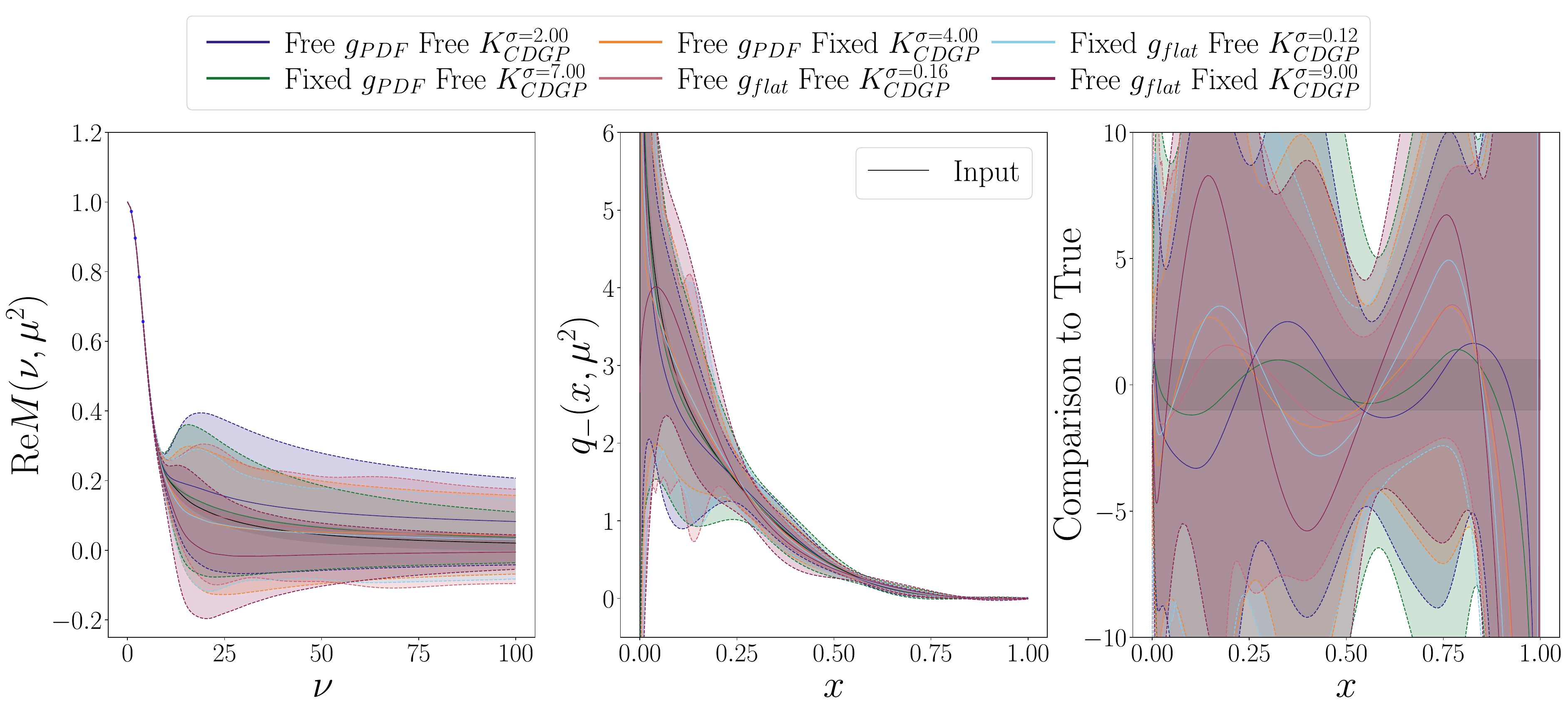}
    \includegraphics[width=0.95\linewidth]{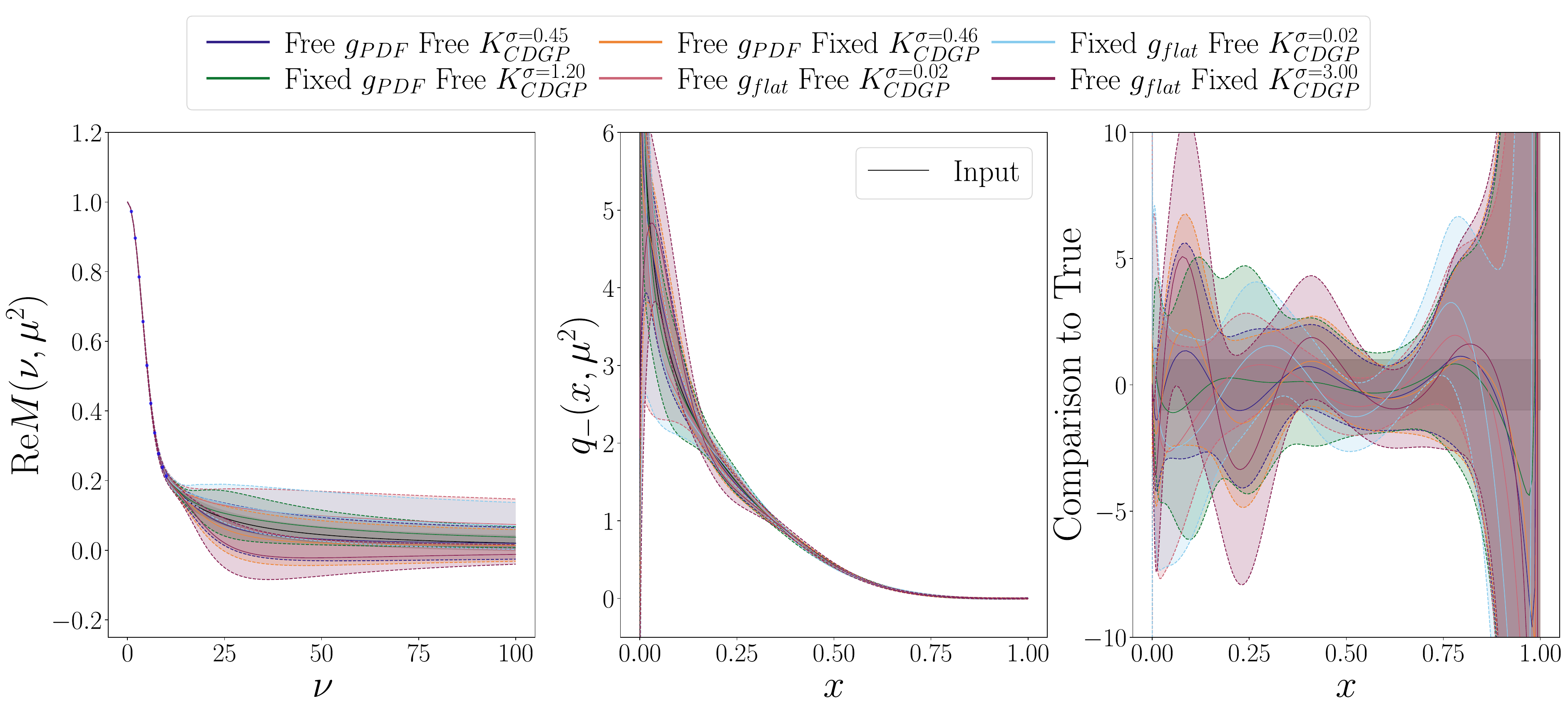}
    \includegraphics[width=0.95\linewidth]{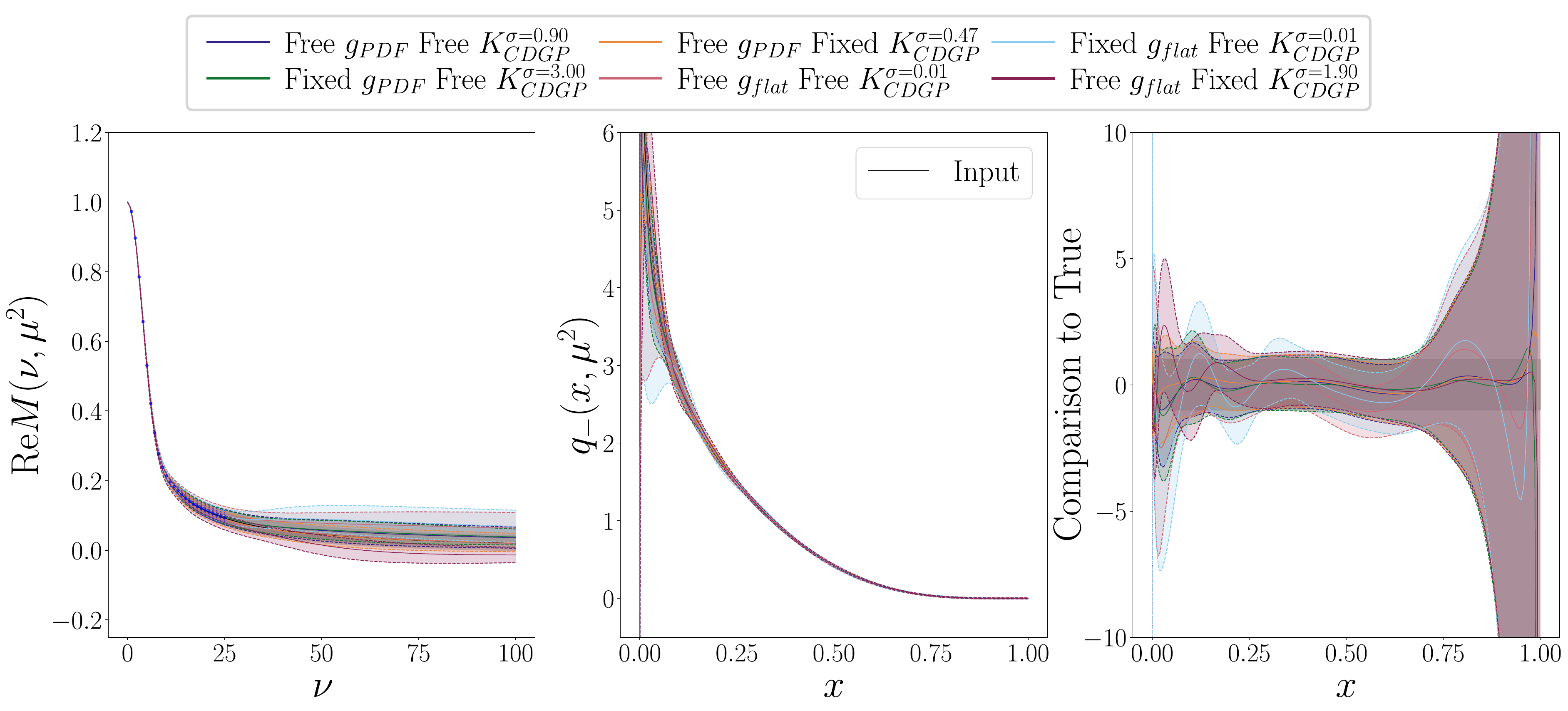}
    \caption{Impact of sampling over the hyperparameters of both the prior mean function and prior kernel, using a $CDGP$ kernel for the CP even PDF. See caption of Fig.~\ref{fig:convdiv_Re} for more details.
    }
    \label{fig:deb_Re}
\end{figure}

\begin{figure}
    \centering
    \includegraphics[width=0.95\linewidth]{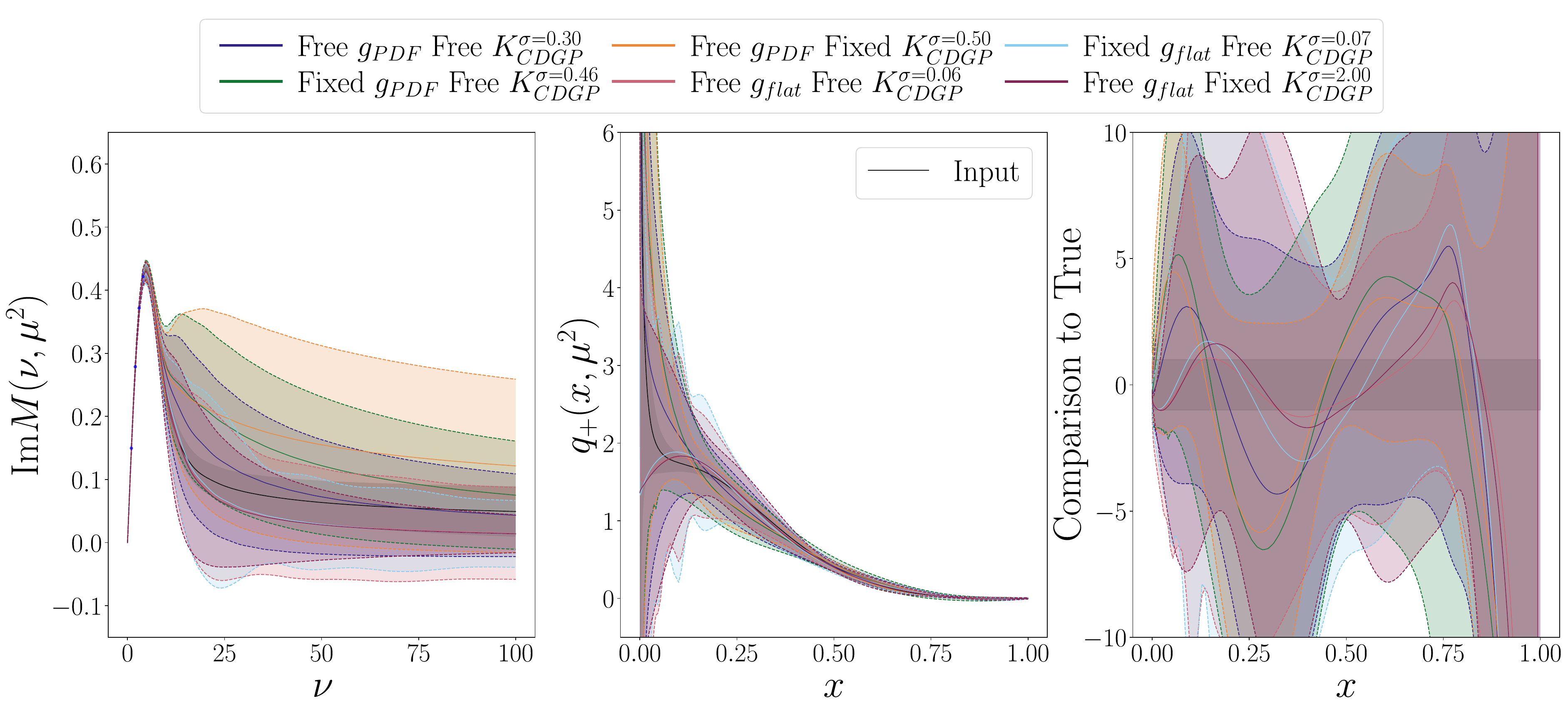}
    \includegraphics[width=0.95\linewidth]{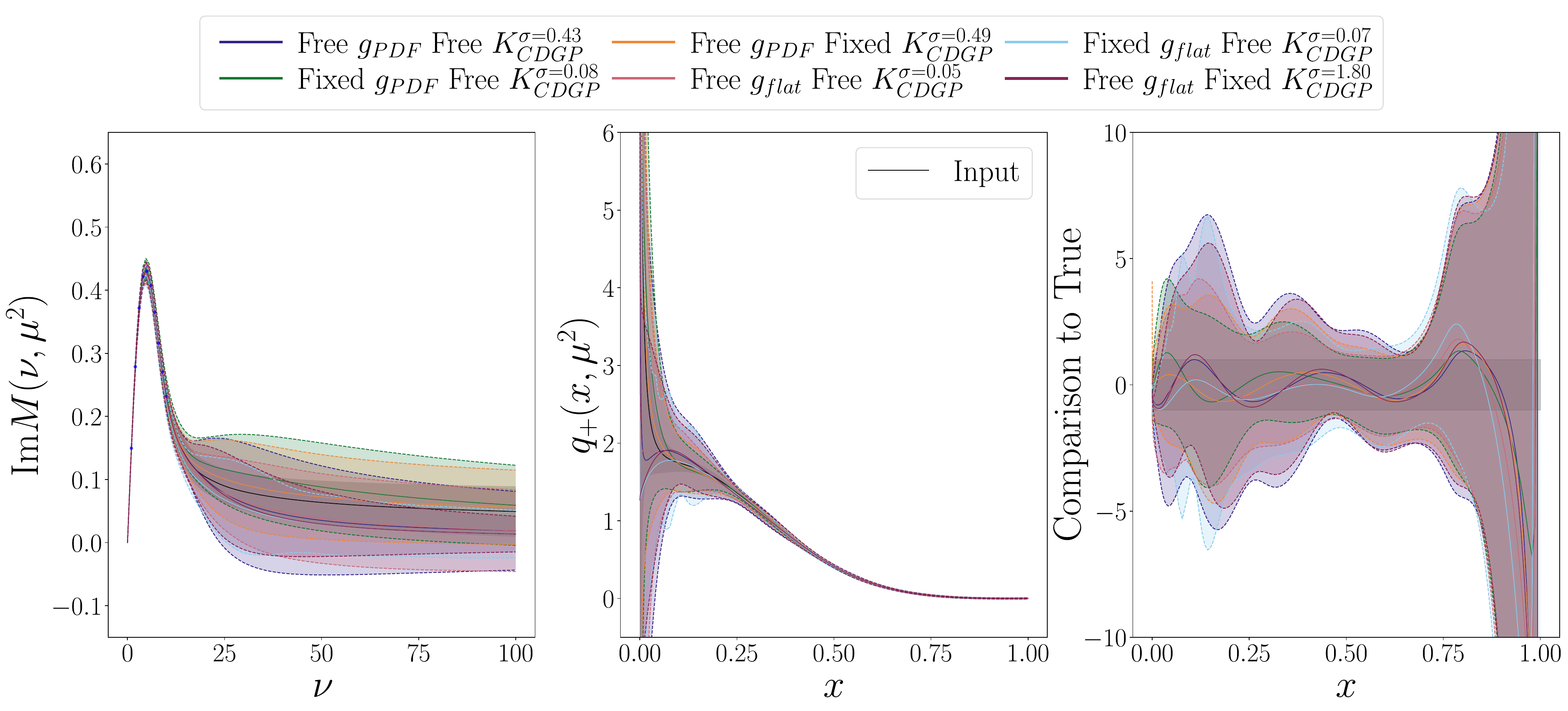}
    \includegraphics[width=0.95\linewidth]{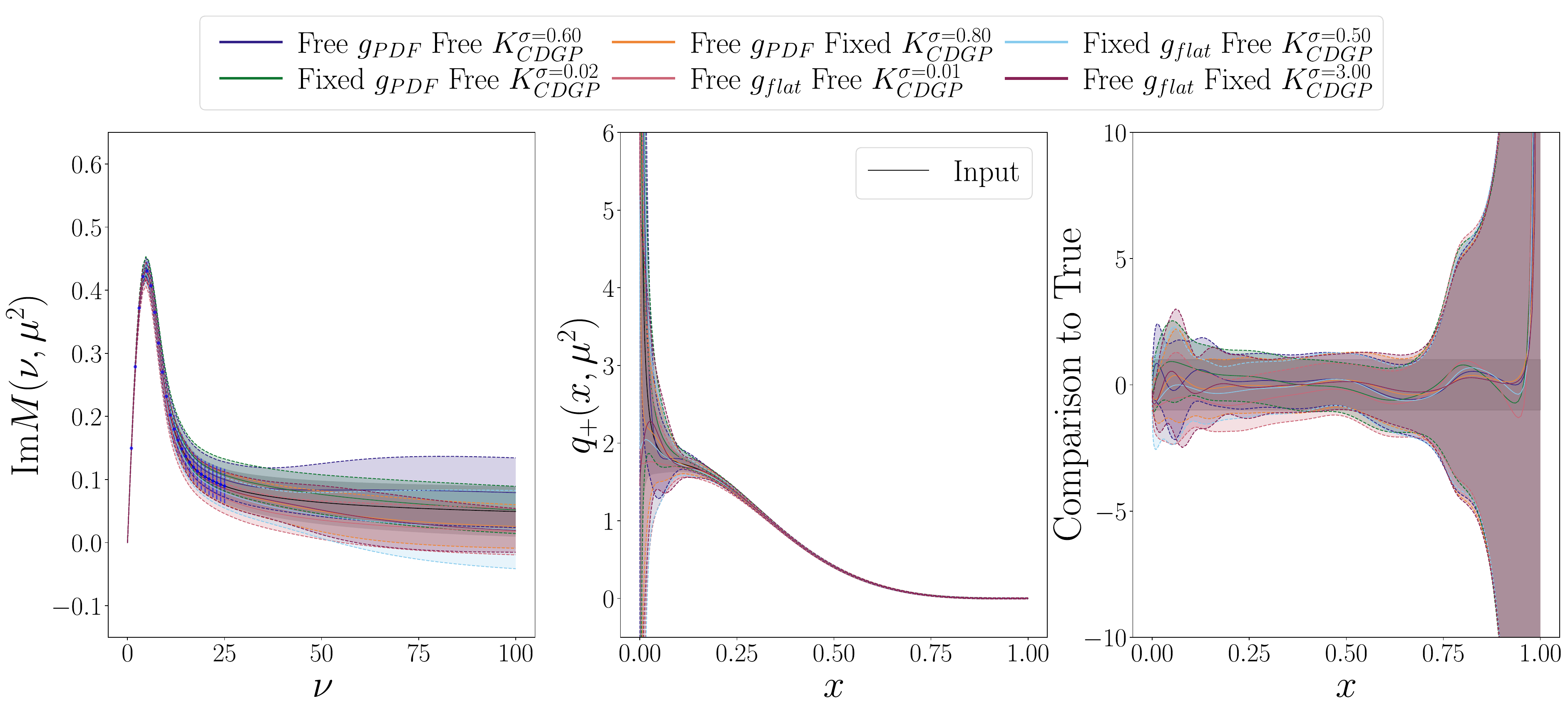}
    \caption{Impact of sampling over the hyperparameters of both the prior mean function and prior kernel, using a $CDGP$ kernel for the CP odd PDF. See caption of Fig.~\ref{fig:convdiv_Re} for more details.
    }
    \label{fig:deb_Im}
\end{figure}

\begin{figure}
    \centering
    \includegraphics[width=0.95\linewidth]{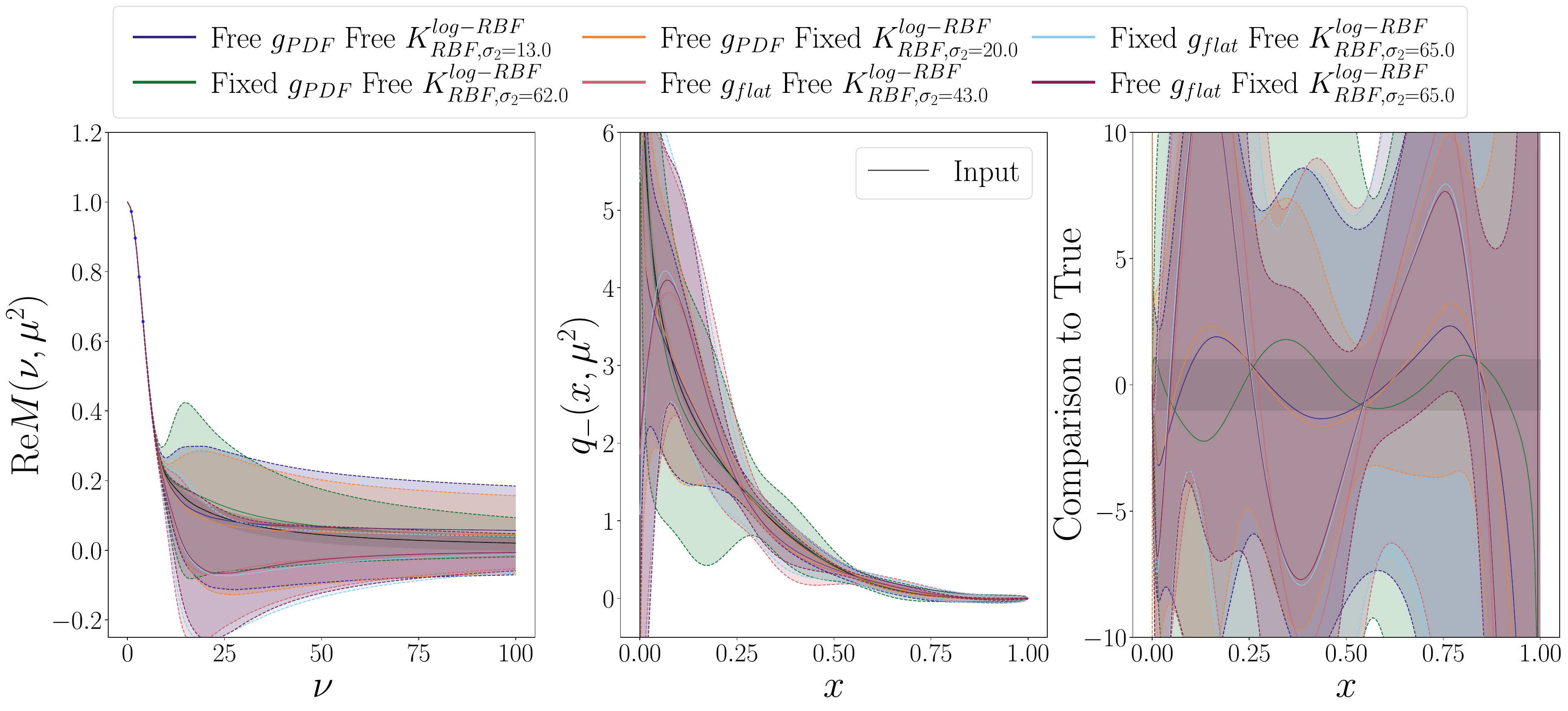}
    \includegraphics[width=0.95\linewidth]{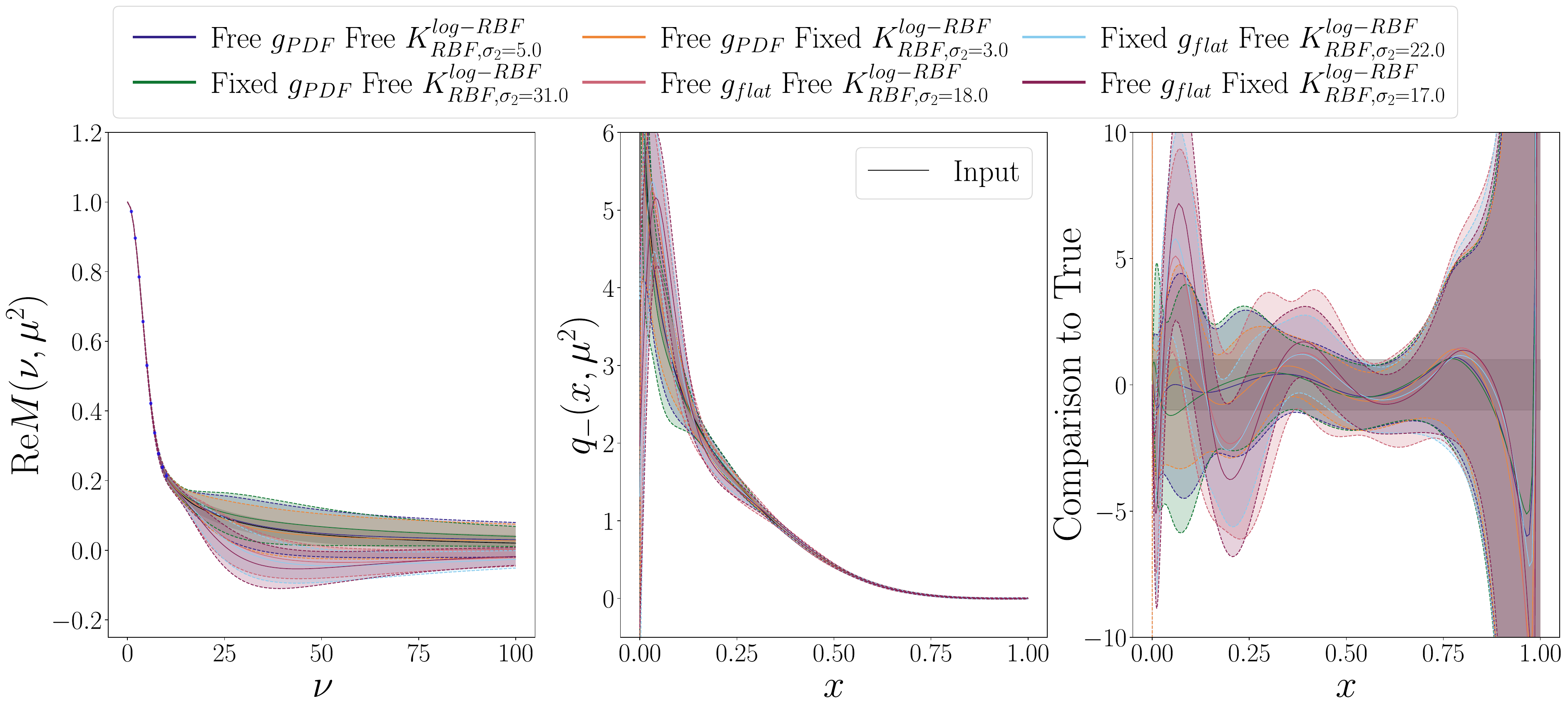}
    \includegraphics[width=0.95\linewidth]{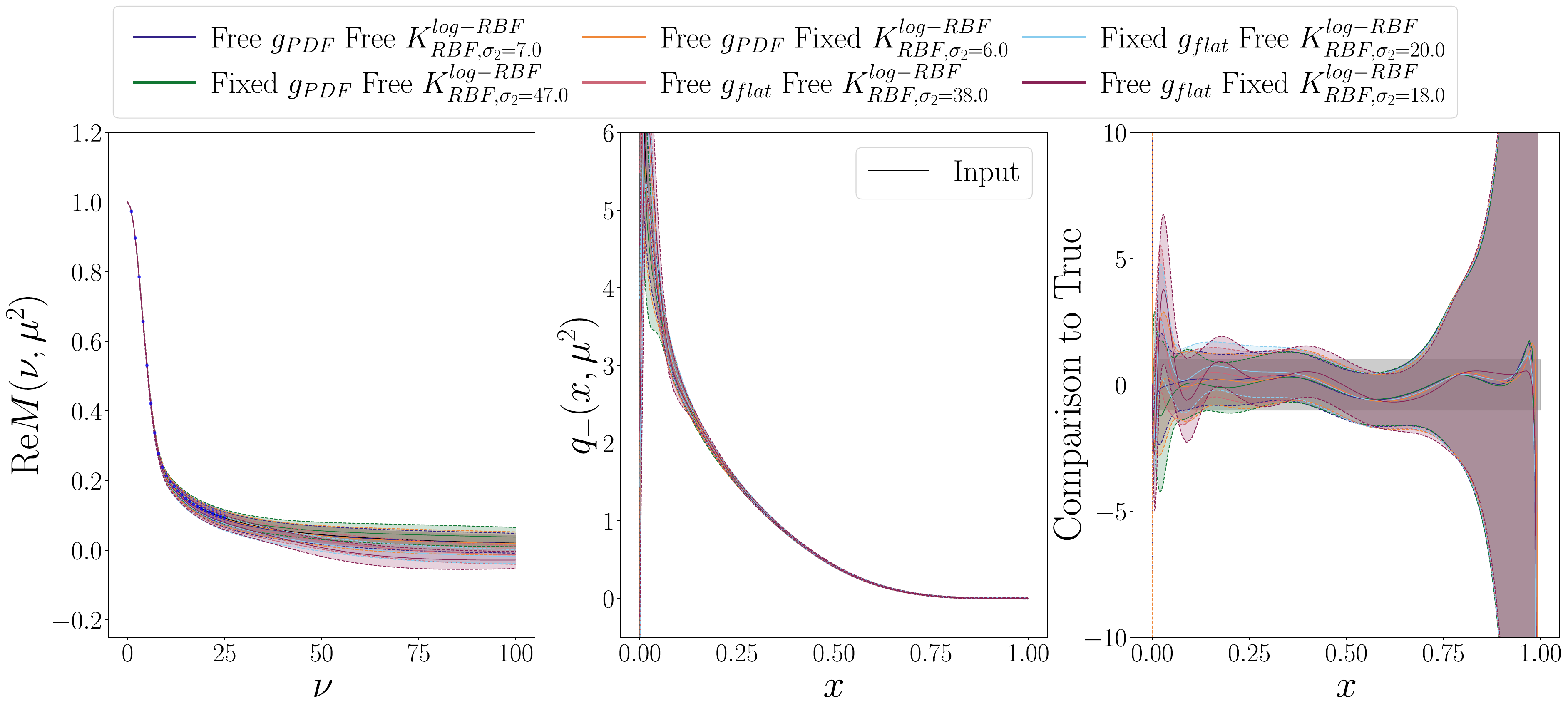}
    \caption{Impact of sampling over the hyperparameters of both the prior mean function and prior kernel, using a combined $log-RBF$ / $RBF$ kernel for the CP even PDF. See caption of Fig.~\ref{fig:convdiv_Re} for more details.}
    \label{fig:rbf_logrbf_Re}
\end{figure}

\begin{figure}
    \centering
    \includegraphics[width=0.95\linewidth]{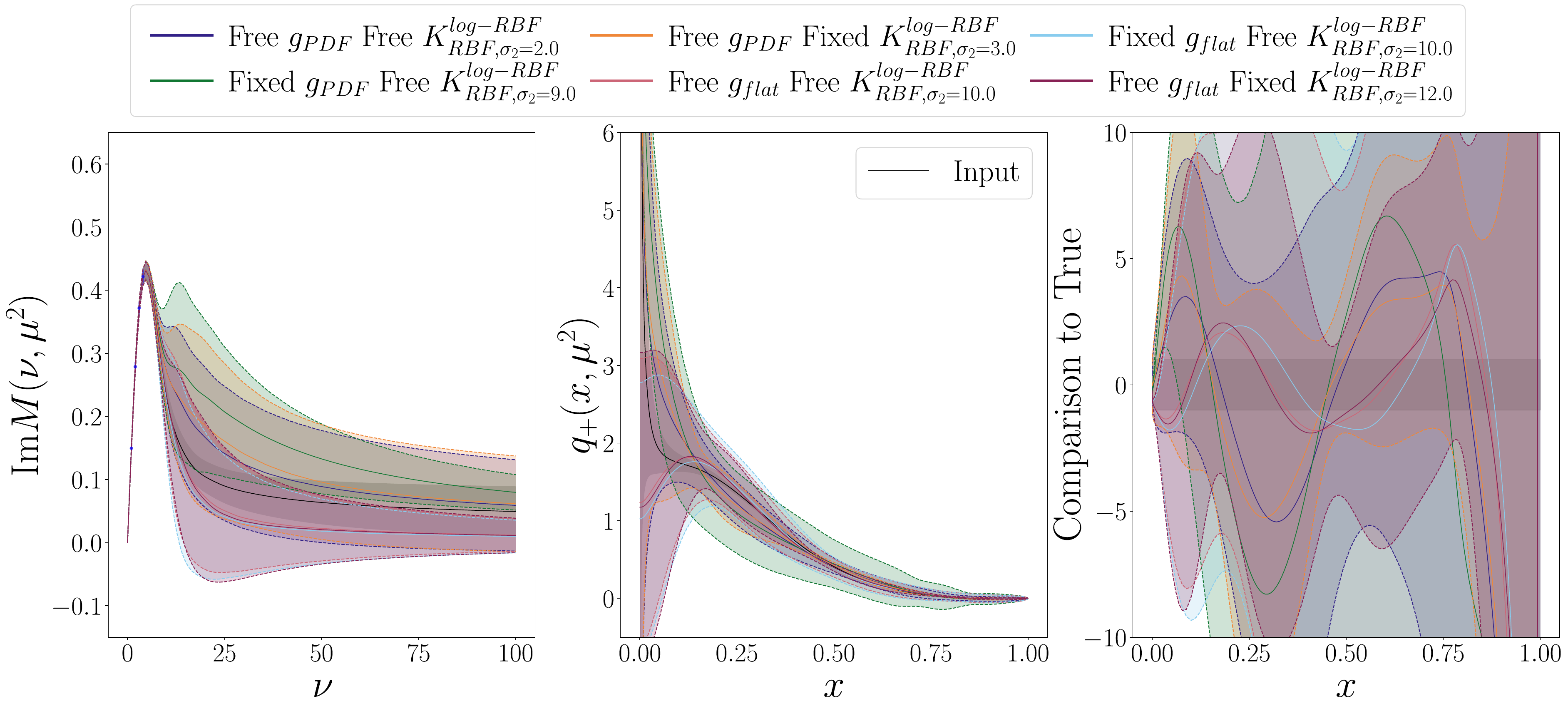}
    \includegraphics[width=0.95\linewidth]{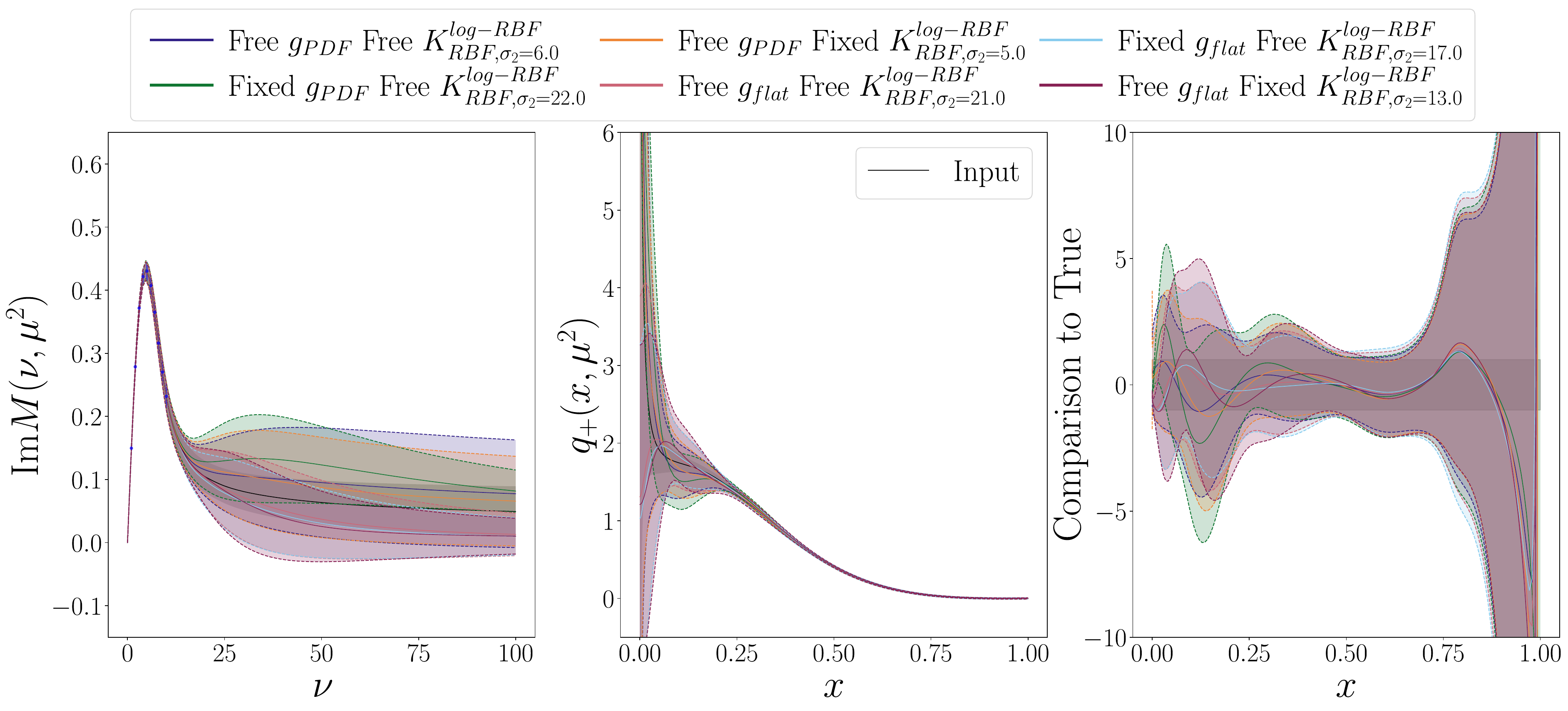}
    \includegraphics[width=0.95\linewidth]{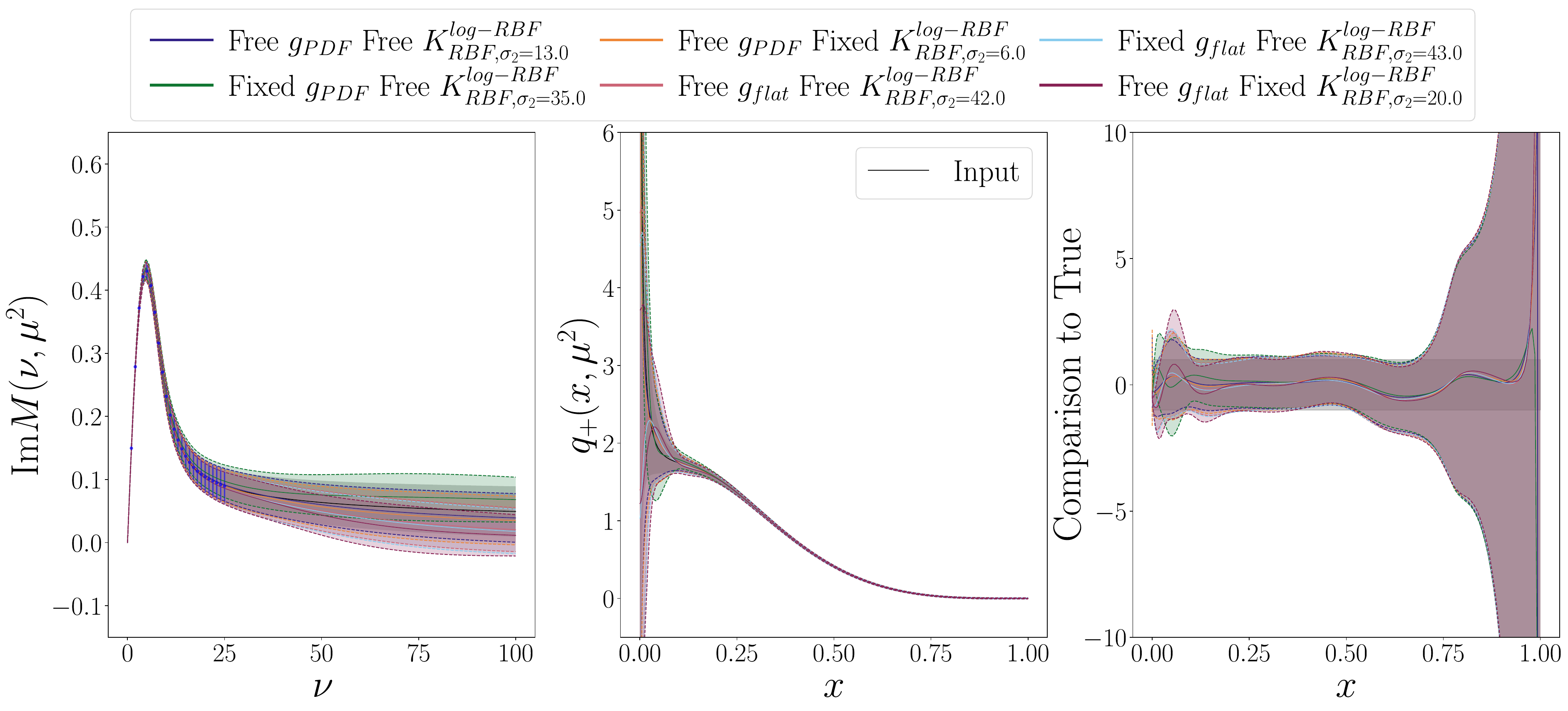}
    \caption{Impact of sampling over the hyperparameters of both the prior mean function and prior kernel, using a combined $log-RBF$ / $RBF$ kernel for the CP odd PDF. See caption of Fig.~\ref{fig:convdiv_Re} for more details. }
    \label{fig:rbf_logrbf_Im}
\end{figure}

To explore the question of which parameters we should sample, we use the models with better performance from our previous analysis. The available options in which effective evidence can be sampled with $N$ parameters are $2^N-1$. However, in this case, we constrain ourselves to explore only three possibilities: sampling all hyperparameters in the kernel, all hyperparameters in the mean, or all hyperparameters in both. Because $\sigma$ influences the reconstruction, in all cases, we fix it with our extrapolation criterion. The alternative of sampling over all the $\sigma$-like parameters in our kernels was explored, and in comparison with this approach, we conclude that the error extrapolation was better estimated with the fixing procedure. For the case of the combined kernel $K_{RBF}^{log-RBF}$, we fix the large $x$ region's variance with $\sigma_1=1.0$ and apply our criteria to determine $\sigma_2$. Results for sampling the combination of kernels and $K_{CDGP}$ are presented in Figs.~\ref{fig:deb_Re}-\ref{fig:rbf_logrbf_Im}.

Our findings about the kernels used for sampling are summarized. We observe two sources of the uncertainty at low $x$, one from setting $\sigma$ and the other from a divergence behavior introduced as an $x^{2\alpha}$ factor in the kernel. We ran test to allow $\alpha\in[0,1]$ which translate in small values of sigma O($10^{-4}$). Thus, by limiting the value of $\alpha \in [-0.8,0]$ we regulate highly oscillatory solutions at a moderate $\sigma$, O($10^{-2}$). This is an improvement in efficiency, given that the sigma can be determined in a grid with sensitivity on the first significant figure. On the other hand, by fixing $\sigma_1=1$ in $K_{RBF}^{log-RBF}$, even if $l_1$ takes values near zero, or using $K_{log-RBF}$, an expected and self-consistent uncertainty was found for these two kernels.
Lastly, in all our results, we have noticed a systematic increase in the relative error's uncertainty as $x\rightarrow1$.
A combination of the regulator in the posterior covariance and the synthetic dataset’s high-$x$ uncertainty (approximately $10^{-6}$–$10^{-3}$ for $x>0.85$) likely explains this.
 More generally, the relative error from comparing sets of shrinking numbers, such as parton distributions at large $x$, can grow large in such a manner simply because the central values shrink faster than the size of the error or from deviations which are small in absolute error but are exacerbated in relative error. Such behavior is frequently seen in comparisons of phenomenological PDFs, such as shown in Refs.~\cite{Bailey:2020ooq,NNPDF:2021njg} and other works.

\subsection{Bayesian Model Averaged results} 
The consistency across diverse models demonstrates that accurate reproduction of the true function is possible despite the variety of researcher-imposed modeling choices.
Given this set of results, we now wish to create a Bayesian model-averaged result weighted by the information criteria discussed in Sec.~\ref{sec:model_average}.

\begin{figure}
    \centering
    \includegraphics[width=0.95\linewidth]{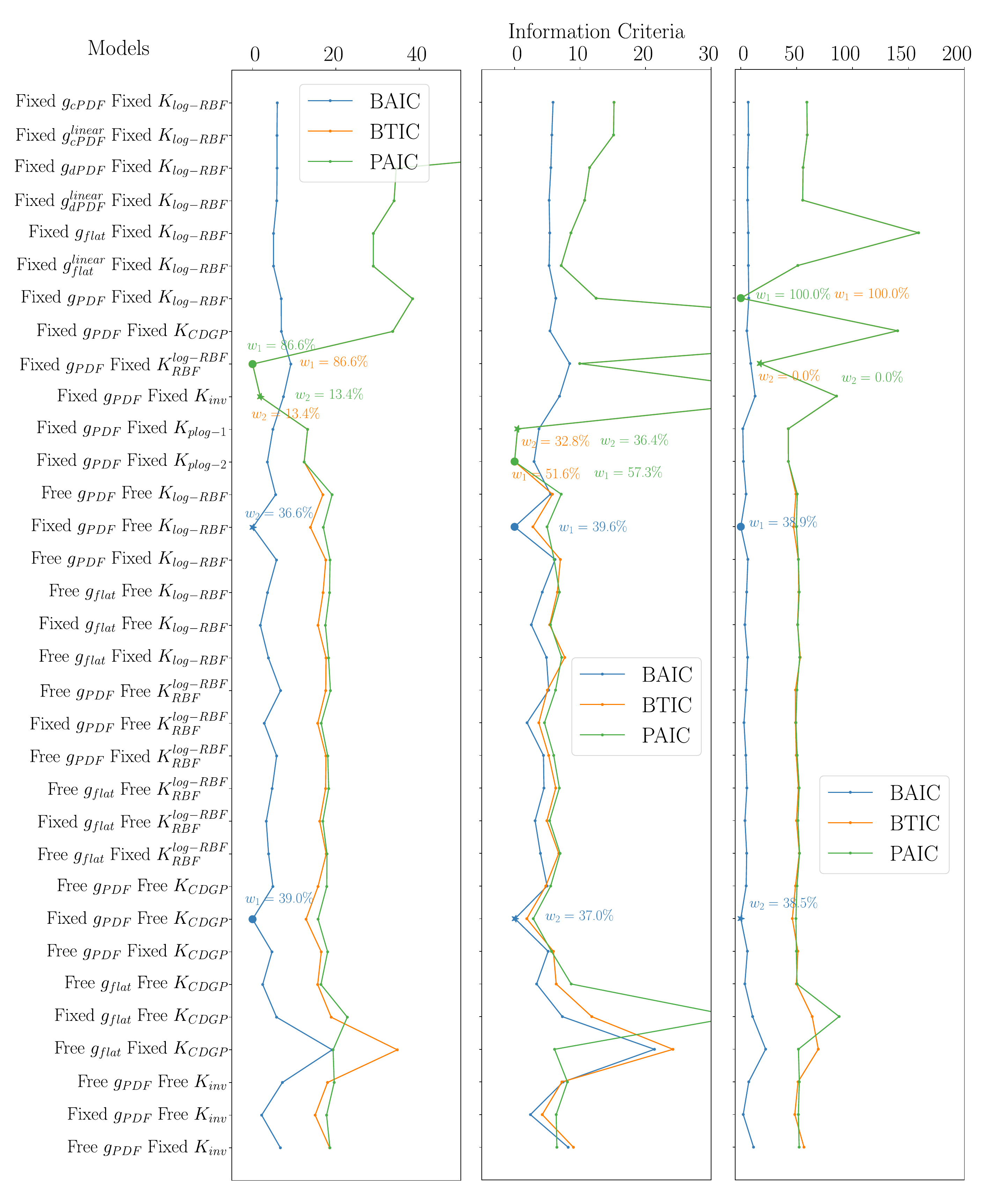}
    \caption{The information criteria for models fit to the real component, where the maximum Ioffe time used was 4, 10, and 25, increasing from top to bottom. $w_1$ and $w_2$ represent the two highest weights for each IC. The lowest IC is marked with a dot and set to zero.}
    \label{fig:IC_all_Re}
\end{figure}

\begin{figure}
    \centering
    \includegraphics[width=0.95\linewidth]{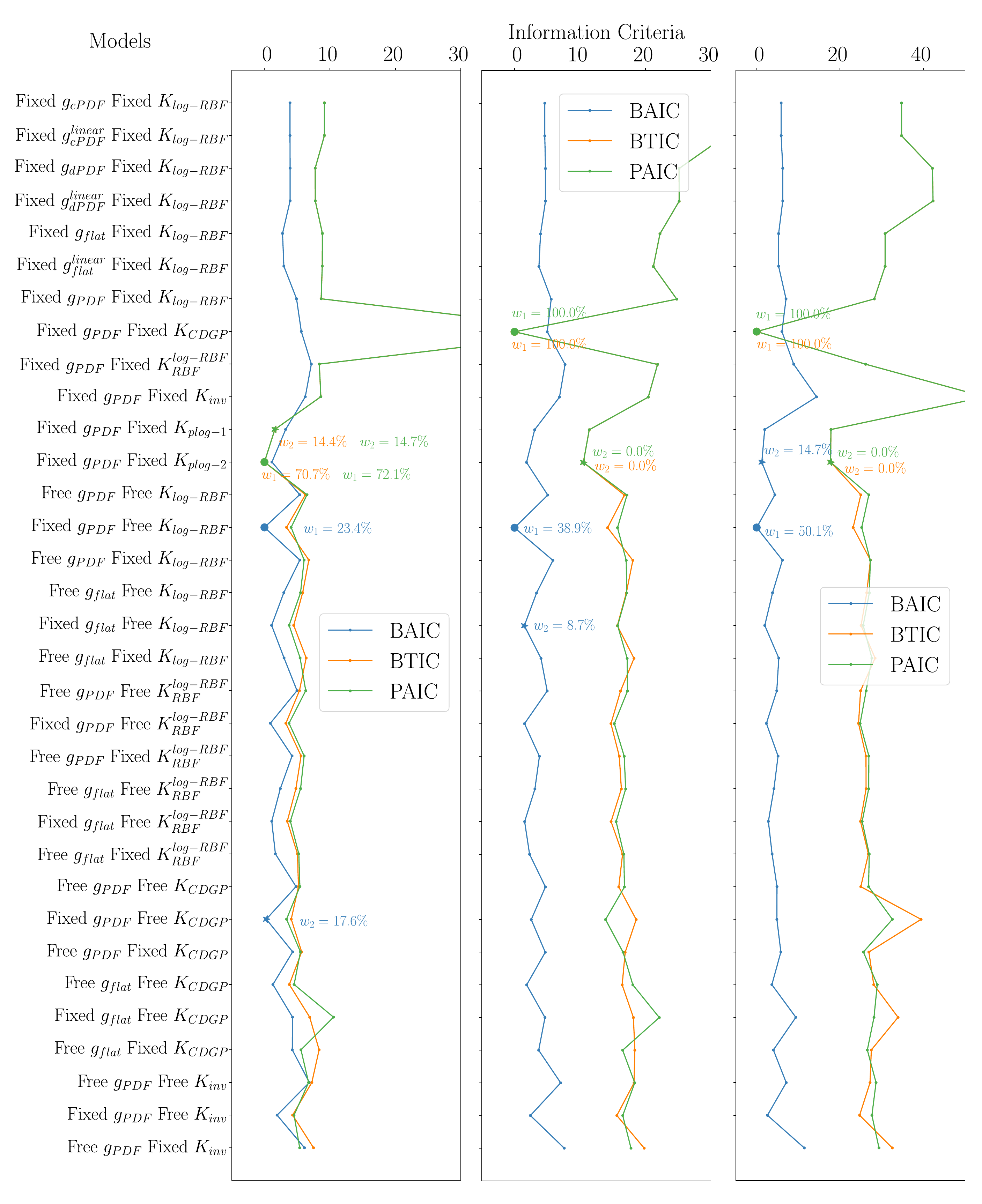}
    \caption{The information criteria for models fit to the imaginary component, where the maximum Ioffe time used was 4, 10, and 25, increasing from top to bottom. $w_1$ and $w_2$ represent the two highest weights for each IC. The lowest IC is marked with a dot and set to zero.}
    \label{fig:IC_all_Im}
\end{figure}

\begin{figure}
    \centering
    \includegraphics[width=1.0\linewidth]{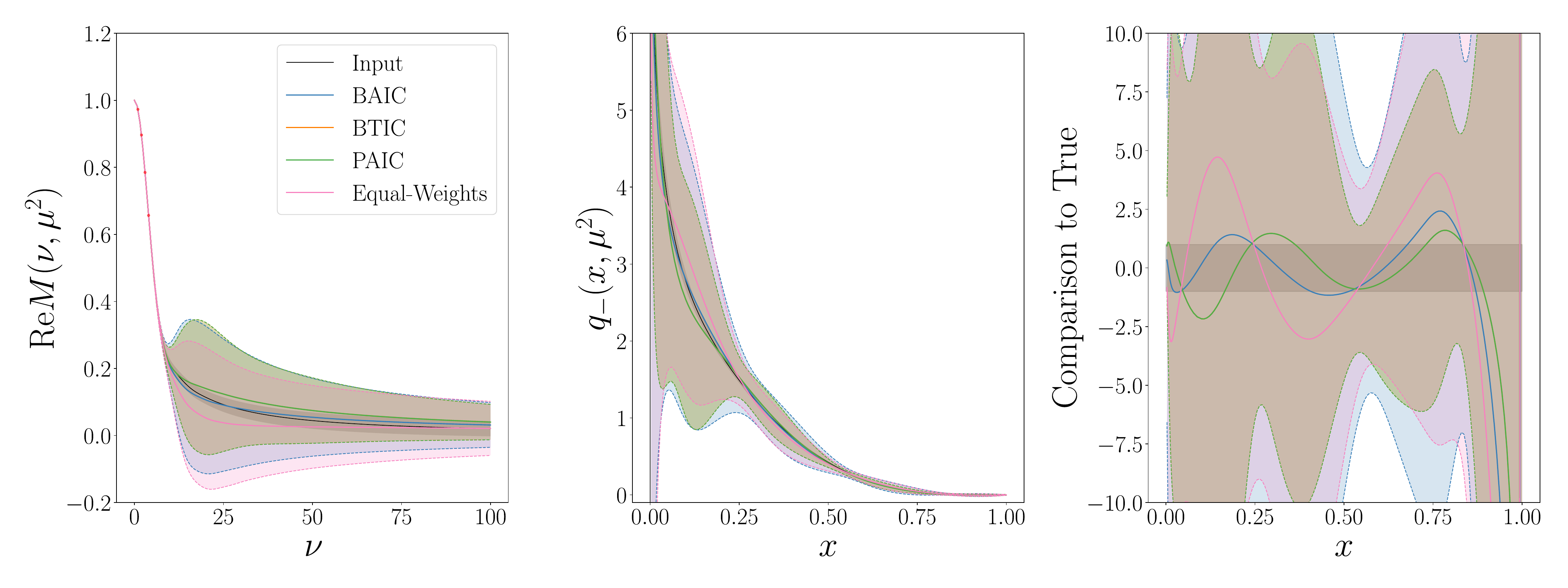}
    \includegraphics[width=1.0\linewidth]{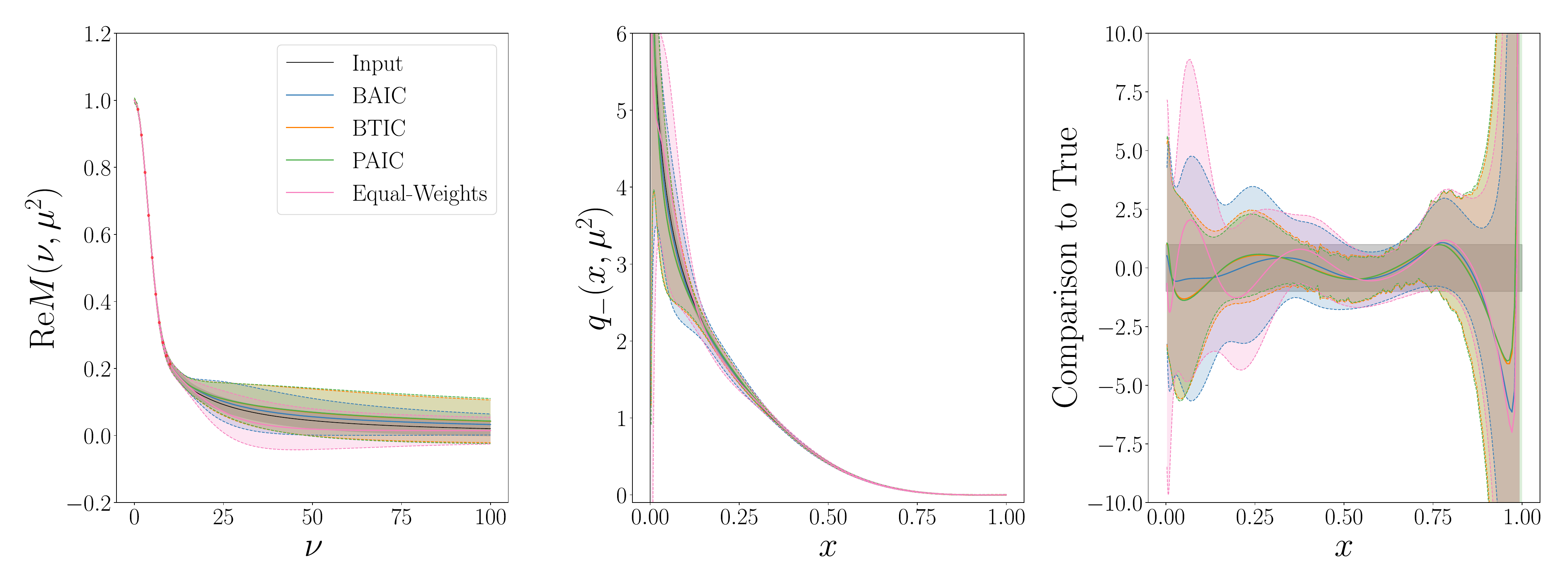}
    \includegraphics[width=1.0\linewidth]{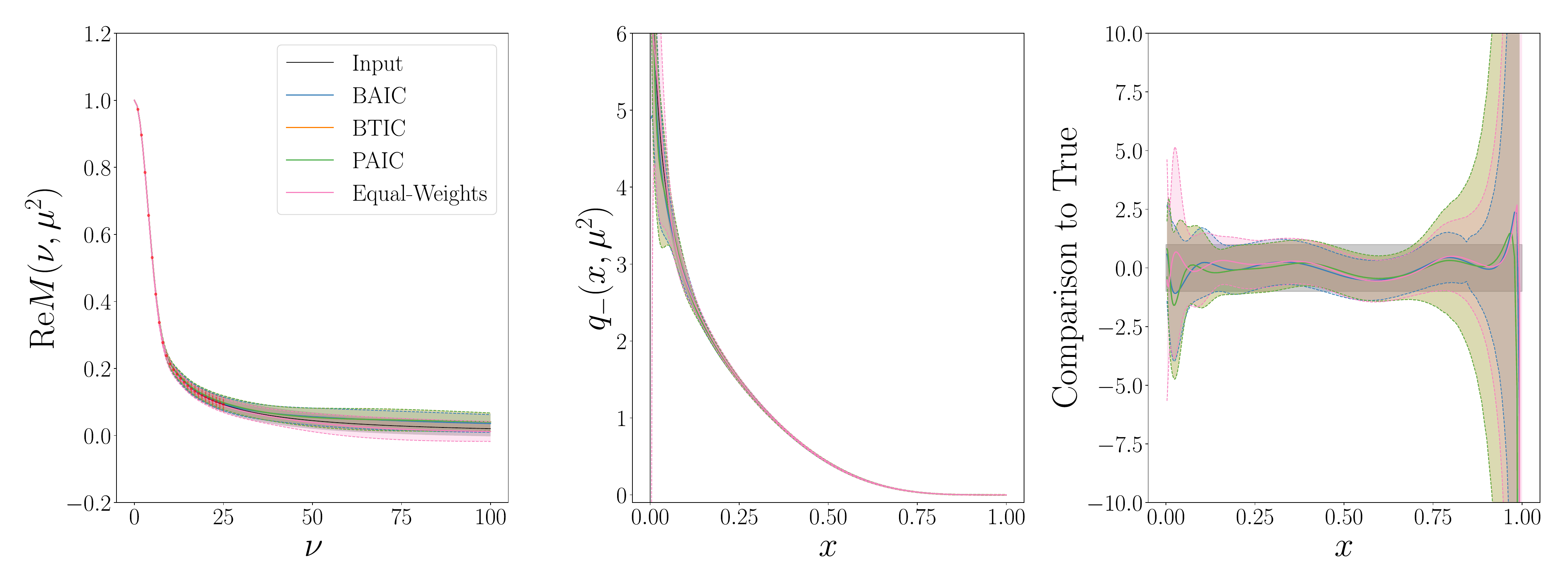}
    \caption{The reconstructions of the NNPDF4.0 CP even isovector PDF (gray) from synthetic data. The maximum Ioffe time used was 4, 10, and 25, increasing from top to bottom. (Right) The ITD in $\nu$ space. (Middle) The PDF in $x$ space. (Left) The difference between the reconstruction and true PDF divided by the error of the true PDF.}
    \label{fig:MA_Recon_Re}
\end{figure}

\begin{figure}
    \centering
    \includegraphics[width=1.0\linewidth]{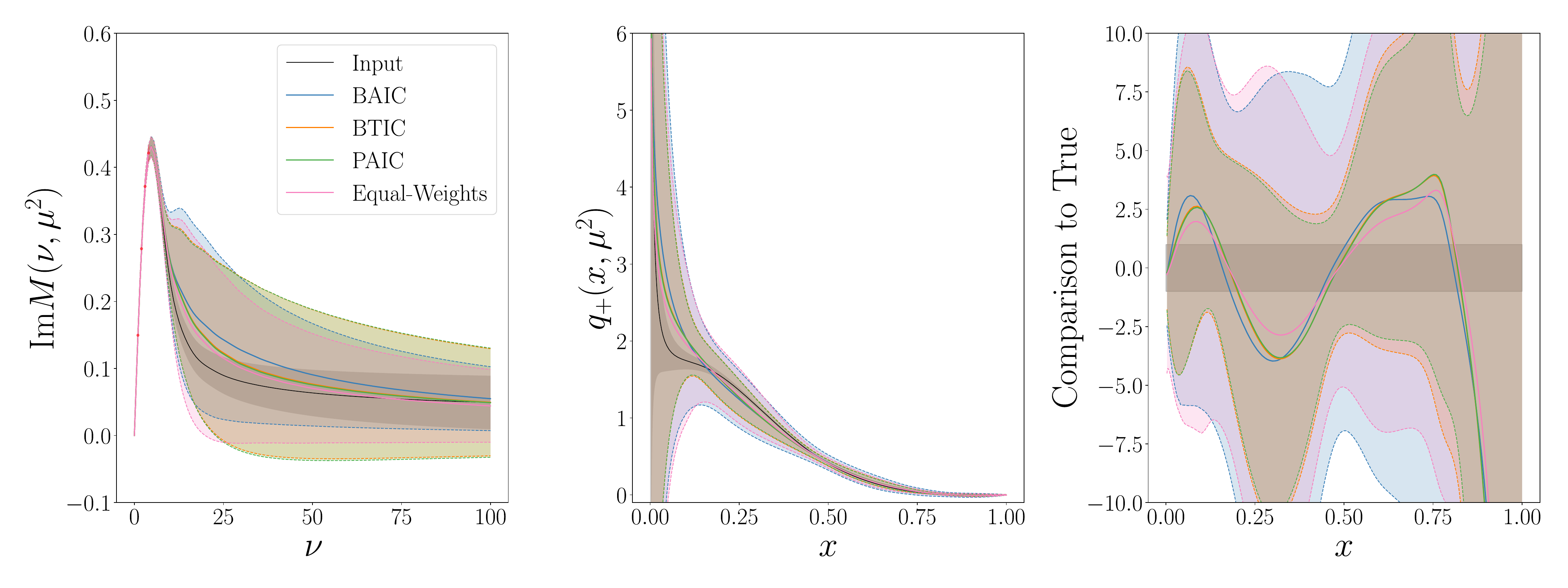}
    \includegraphics[width=1.0\linewidth]{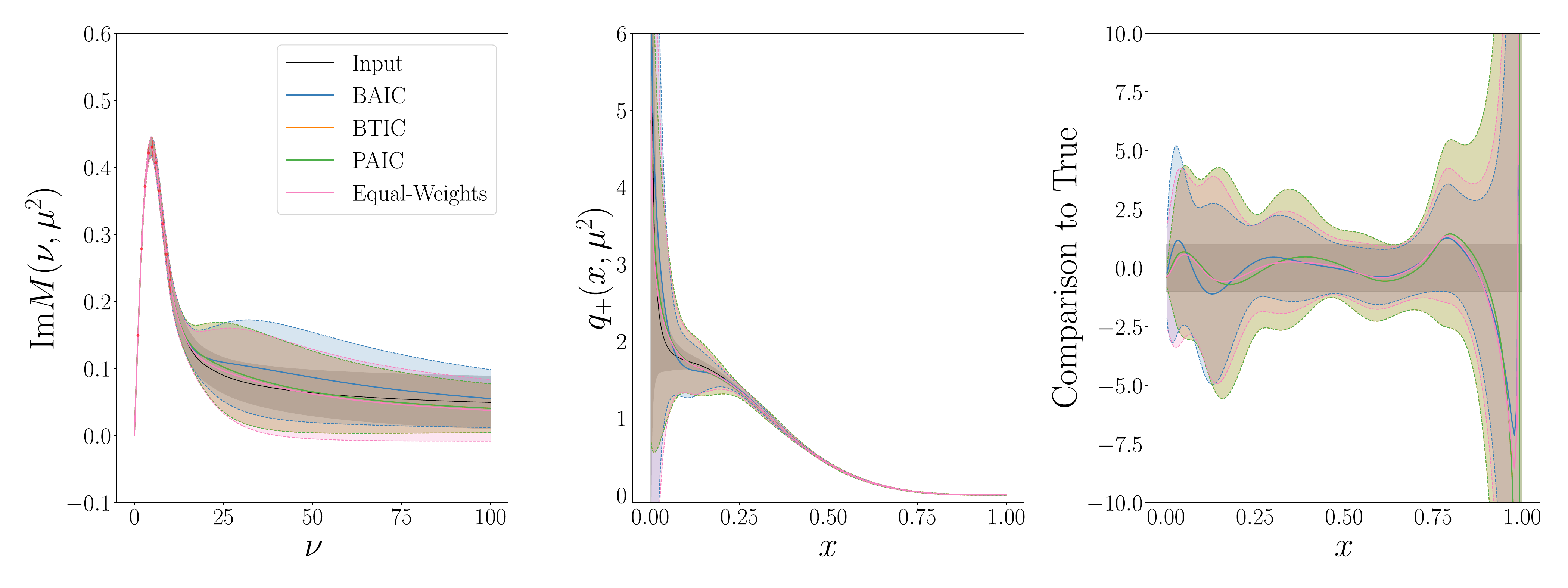}
    \includegraphics[width=1.0\linewidth]{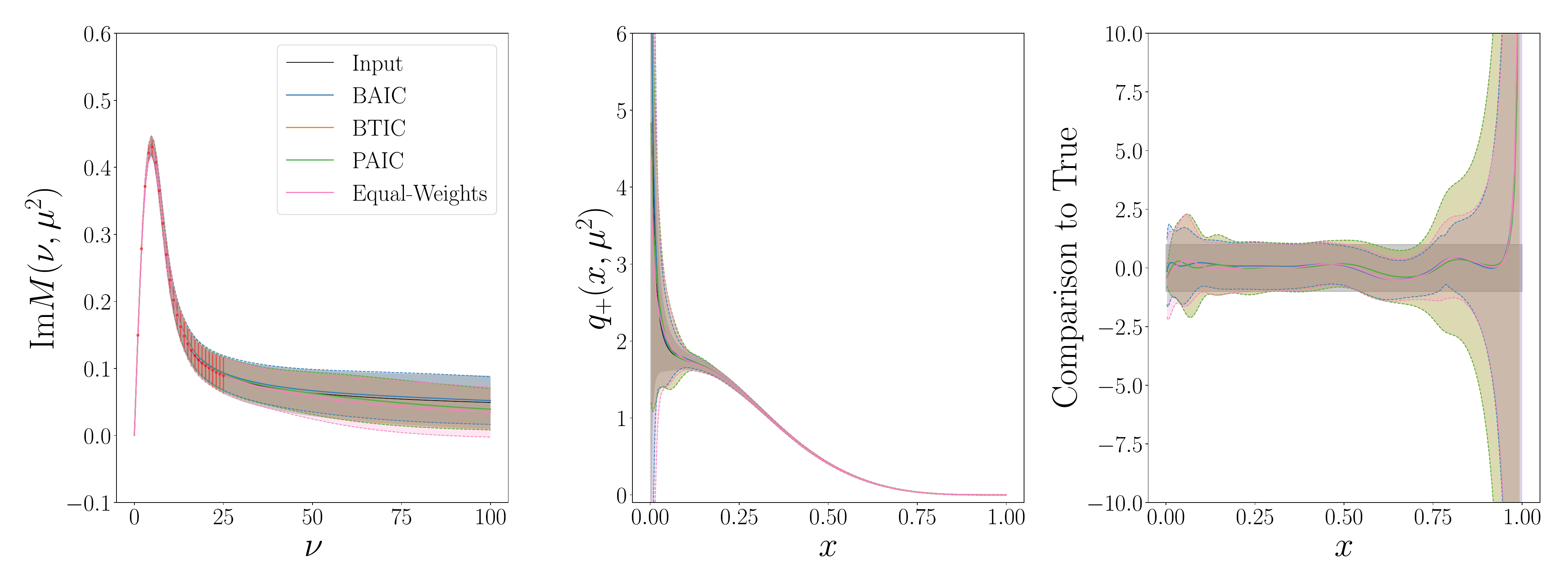}
    \caption{The reconstructions of the NNPDF4.0 CP odd isovector PDF (gray) from synthetic data. The maximum Ioffe times used were 4, 10, and 25 increasing from top to bottom. (Right) The ITD in $\nu$ space. (Middle) The PDF in $x$ space. (Left) The difference between the reconstruction and the true PDF divided by the error of the true PDF. }
    \label{fig:MA_Recon_Im}
\end{figure}

Figs.~\ref{fig:IC_all_Re}-~\ref{fig:IC_all_Im} show the different information criteria for each of the models discussed above. Remarkably, the different information criteria yield broadly consistent weighted posteriors, ie central value and error in Figs.~\ref{fig:MA_Recon_Re}-~\ref{fig:MA_Recon_Im}, over wide ranges of $x$. Constant shifts in the IC value are irrelevant to the definition of the weights, and only relative differences between the models are practically meaningful. 
The lowest IC will be preferred within the weighted average. A unit increase above this minimum will be weighted down by a factor of $1/\sqrt{e}= 0.60\dots$ while an increase of 10 will be weighted by a factor of $e^{-5}=0.0067\dots$, dramatically lowering that model's impact. In each of the datasets, only a few models will have a substantial enough weight to shift the results away from the most preferred.  Using the BTIC or PAIC on the imaginary component with $\nu_{\rm max}=25$, only the model with all hyperparameters free and the mean function $g_{\rm PDF}$ will contribute since the second lowest IC is dozens higher than the minimum IC, while the BAIC actually disfavors this model compared to some others. Despite this difference in contributing models, the final weighted posteriors tend to produce similar means and variances in the relevant regions of $x$.

By construction, equation \eqref{eq:criteriamax} is satisfied by all our models, which are summarized in Figs.~\ref{fig:IC_all_Re} and ~\ref{fig:IC_all_Im} 
 are therefore included in the model-averaged results in Figs.~\ref{fig:MA_Recon_Re} and ~\ref{fig:MA_Recon_Im}. 
 As a baseline, we also consider an unweighted average (Equal-Weights, E–W) in which all models are given equal weight.
 The agreement in most of the scenarios between E-W and any of the different information criteria reconstructions of the PDF is a signal of general consistency between model selection and model averaging procedures in the Bayesian framework. 
 
 The primary example where the information criteria give different-sized errors is the real component with the least extent in $\nu$. The BTIC and PAIC seemed to specifically prefer one model, something which occurred in other cases. On the other hand, the BAIC and Equal Weight are giving an average of different models in all cases. As mentioned in Sec.~\ref{sec:model_average}, the BTIC and PAIC have fewer approximations, and the PAIC proved a more accurate estimator in synthetic data studies, so perhaps they are a preferable choice when discrepancies between IC occur. On the other hand, in this particular setting, with limited Ioffe time extent, one perhaps wishes to take the more conservative choice instead. This scientist made decision is the unfortunate hallmark of ill-posed inverse problems. It is reassuring that the different ICs give similar results in the more stable, larger $\nu_{max}$ cases, including when BTIC and PAIC similarly select a model or average a few. 



\section{Application to Lattice data}\label{sec:realdata}

We perform a re-analysis of real lattice QCD data with the same methodology used in the closure tests. The lattice QCD data were generated in Ref.~\cite{Egerer:2021ymv} for studying the unpolarized isovector quark parton distribution of the nucleon. The lattice ensemble used 2+1 flavors of clover improved Wilson quarks with a lattice spacing $a=0.094(1)$ fm and a pion mass of $m_\pi=358(3)$ MeV. More details of action and generation are given in Ref.~\cite{Egerer:2021ymv} and enclosed references. The matrix element data was generated on a set of integer values for $z/a\in[0,12]$ and $Lp/2\pi\in[0,6]$. For studying the pseudo-PDF determination, we choose $z/a=3,6,9$ as our testbeds to show the impact of Ioffe time extent. Said extent is limited by the largest value of $p$ that is meaningfully precise. Finally, it is worth noting that lattice QCD data present, in most cases, less correlation in the covariance of the likelihood, which translates into a less singular problem compared with the one we regularized for synthetic data.

\begin{figure}
    \centering
    \includegraphics[width=0.95\linewidth]{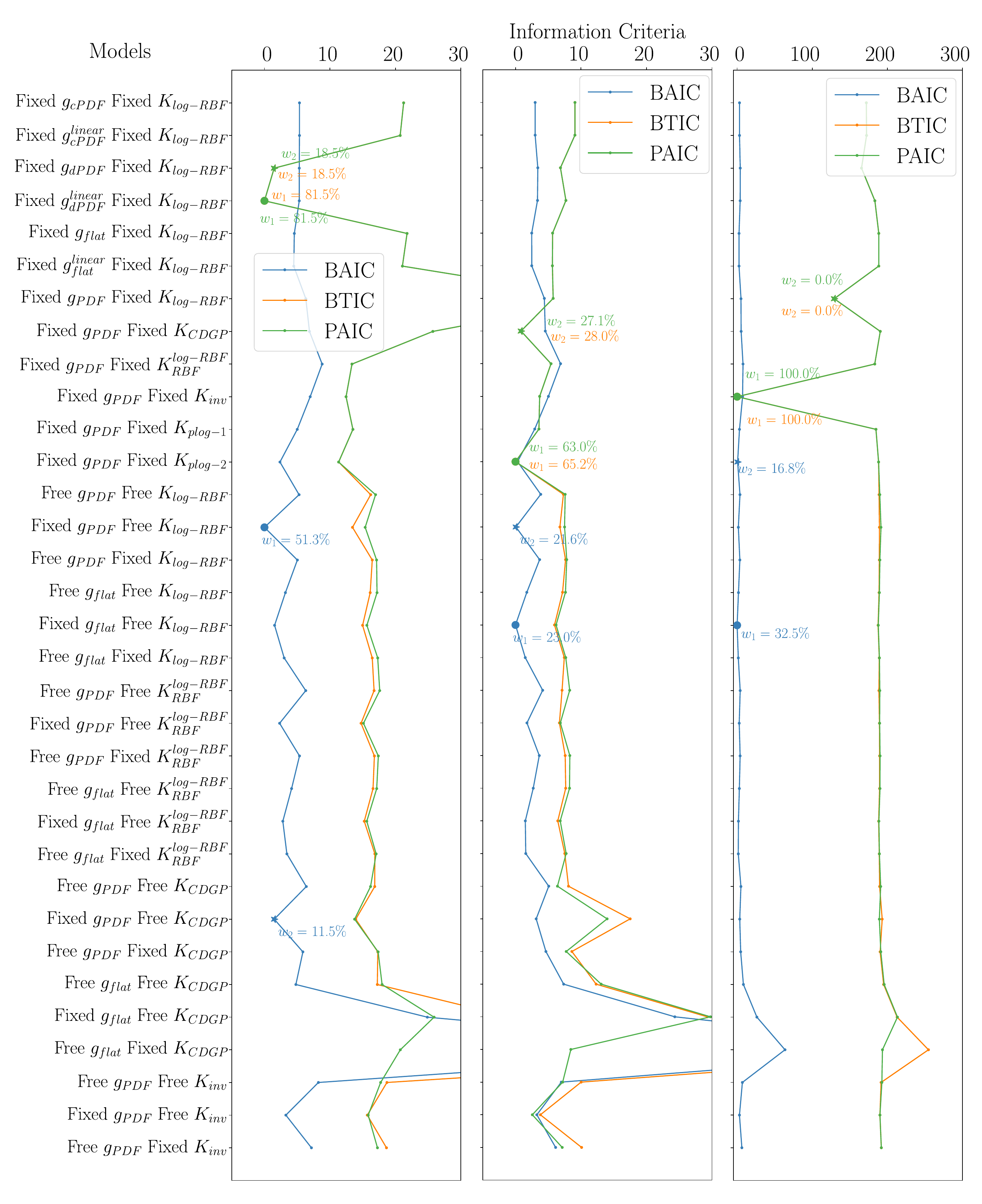}
    \caption{The information criteria for models fit to the real component,  with $z=3a,6a,9a$, increasing from left to right. $w_1$ and $w_2$ represent the two highest weights for each IC. The lowest IC is marked with a dot and set to zero.}
    \label{fig:IC_all_Re_Col}
\end{figure}

\begin{figure}
    \centering
    \includegraphics[width=0.95\linewidth]{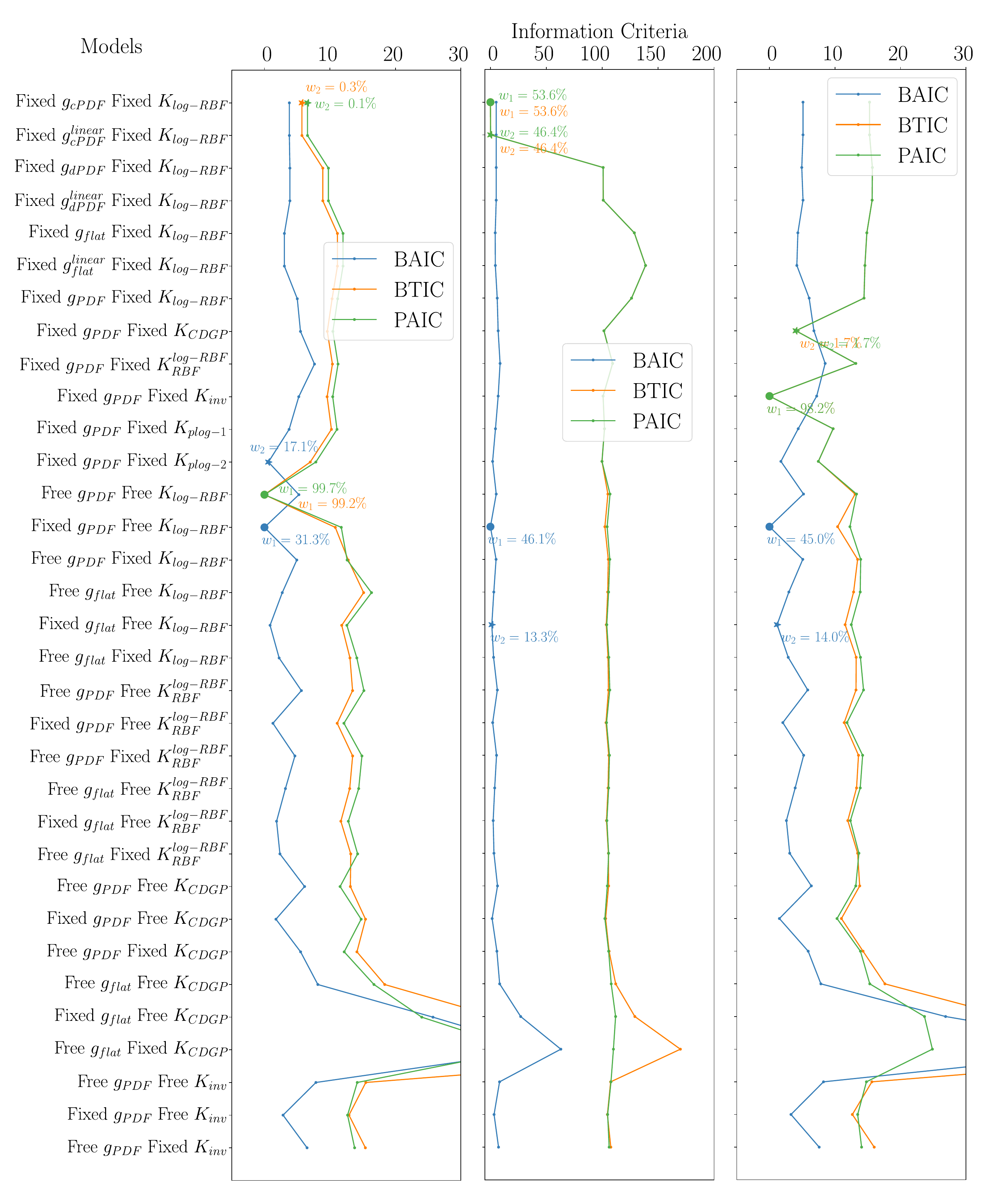}
    \caption{The information criteria for models fit to the imaginary component, with $z=3a,6a,9a$, increasing from left to right. $w_1$ and $w_2$ represent the two highest weights for each IC. The lowest IC is marked with a dot and set to zero.}
    \label{fig:IC_all_Im_Col}
\end{figure}

\begin{figure}
    \centering
    \includegraphics[width=1.0\linewidth]{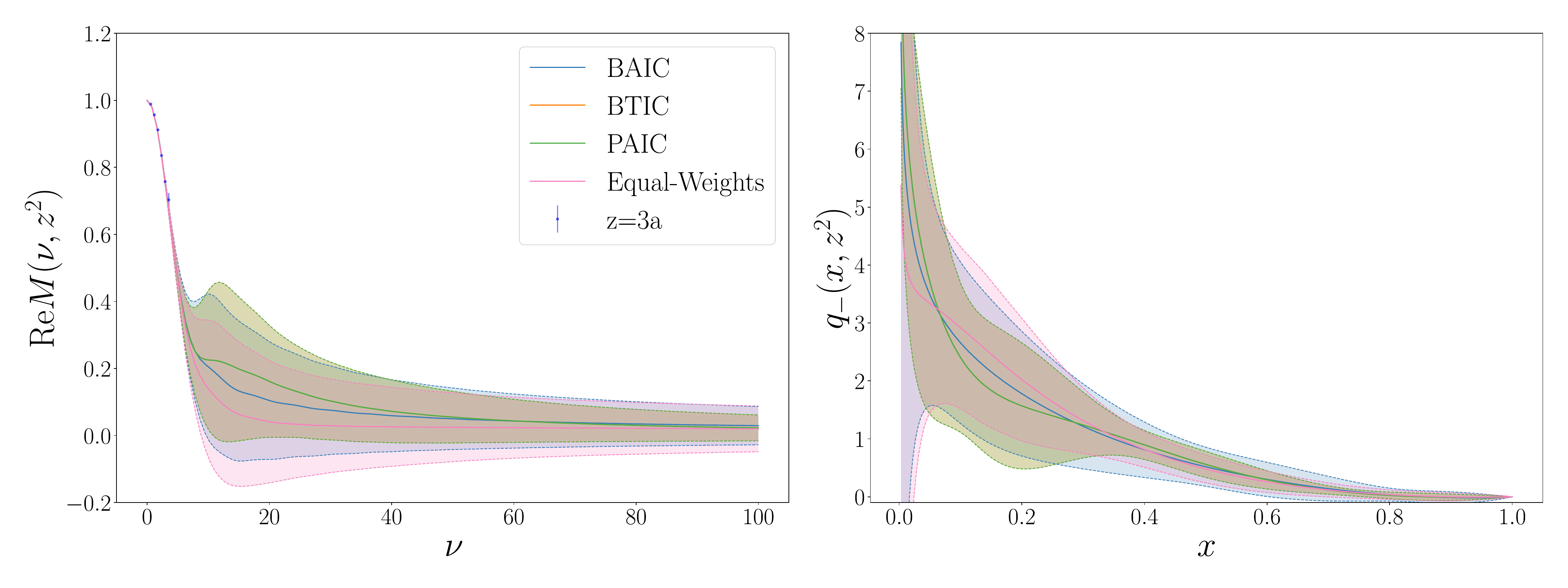}
    \includegraphics[width=1.0\linewidth]{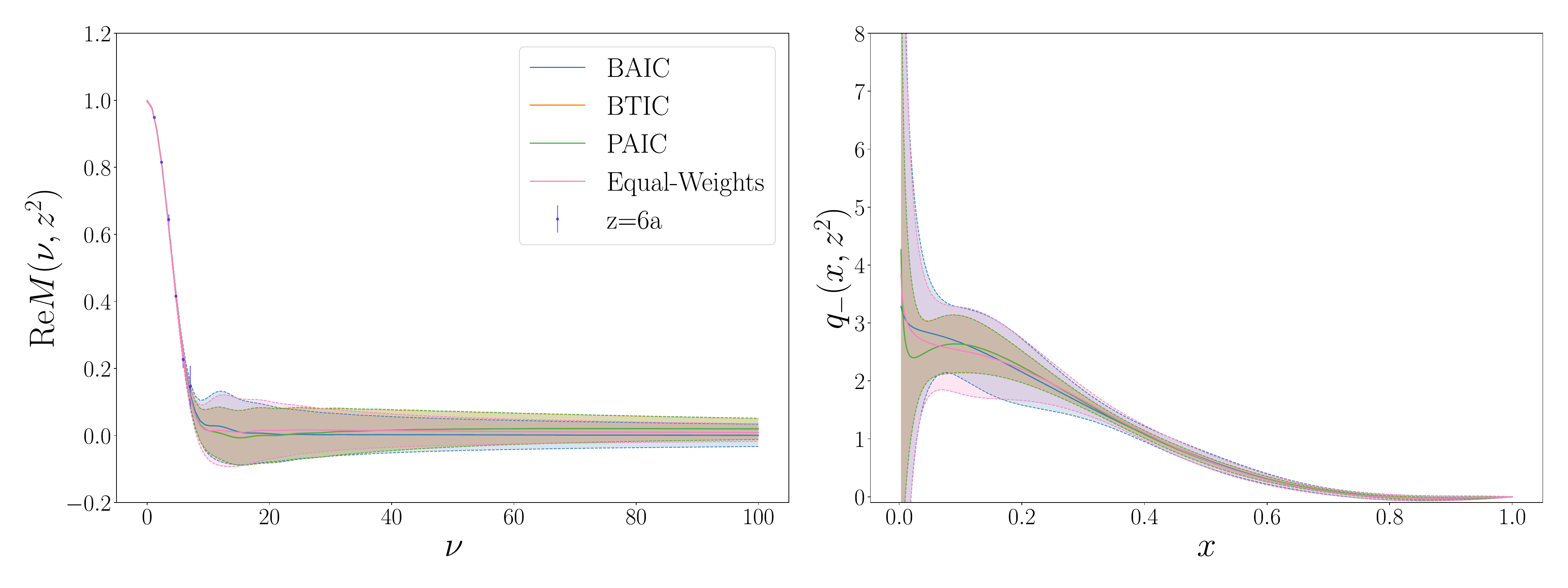}
    \includegraphics[width=1.0\linewidth]{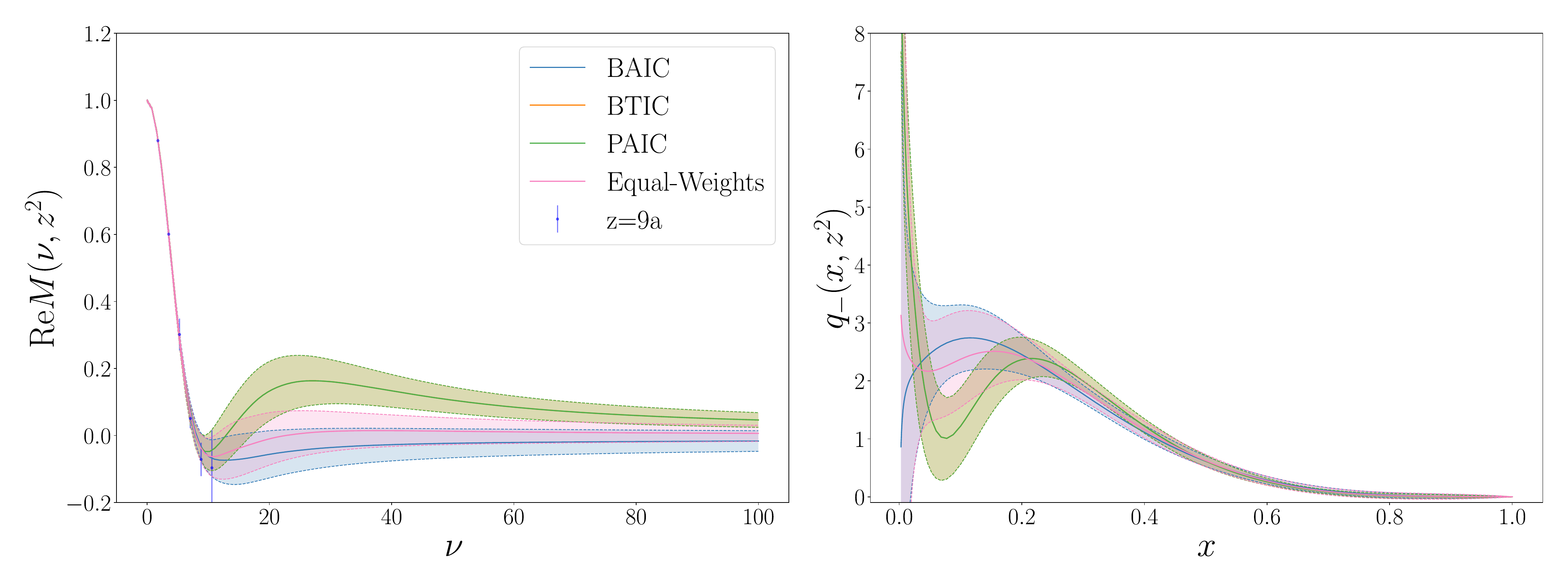}
    \caption{The reconstructions of the CP even isovector PDF (gray) from lattice QCD data. Real ITD components with $z=3a,6a,9a$ from top to bottom. (Right) The ITD in $\nu$ space. (Left) The PDF in $x$ space.}
    \label{fig:MA_Recon_Re_Col}
\end{figure}

\begin{figure}
    \centering
    \includegraphics[width=1.0\linewidth]{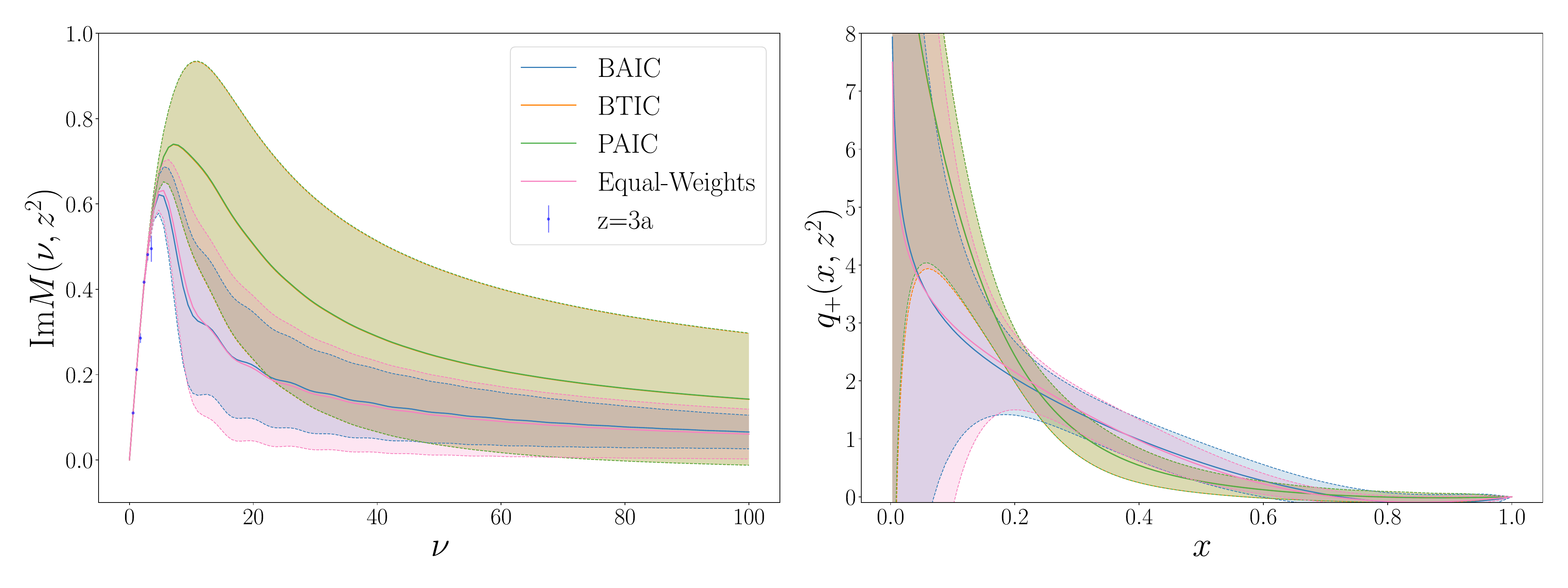}
    \includegraphics[width=1.0\linewidth]{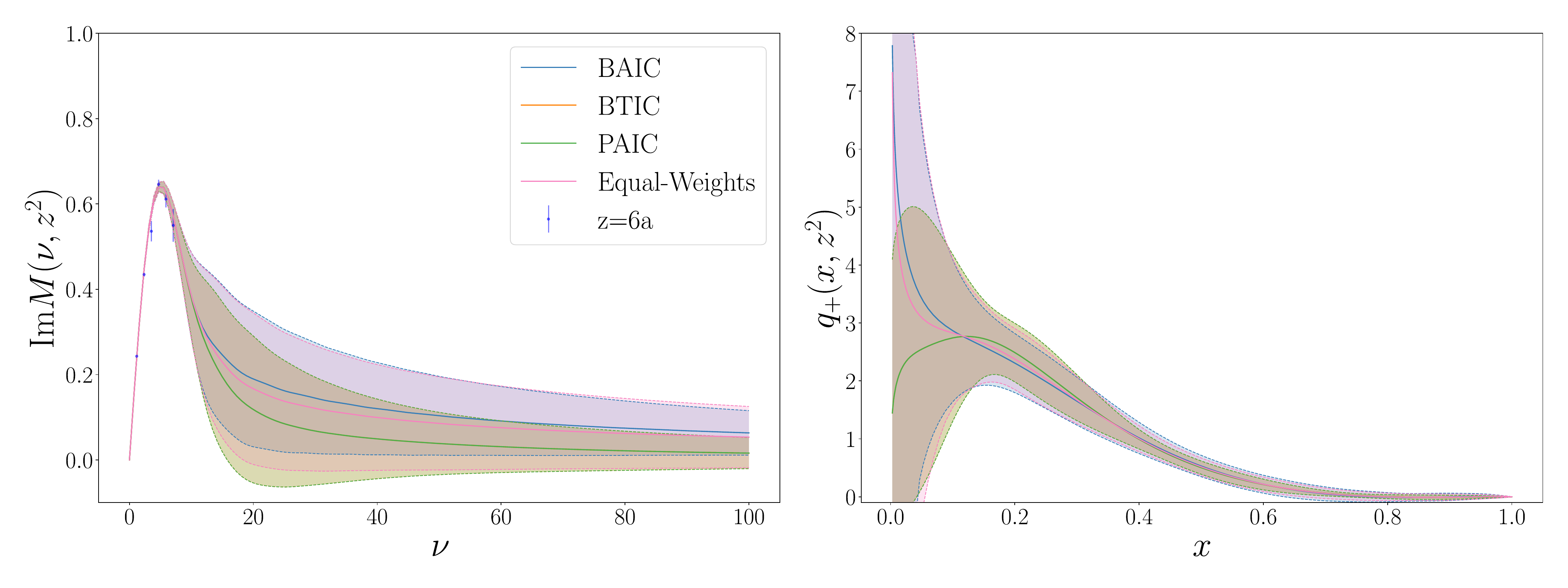}
    \includegraphics[width=1.0\linewidth]{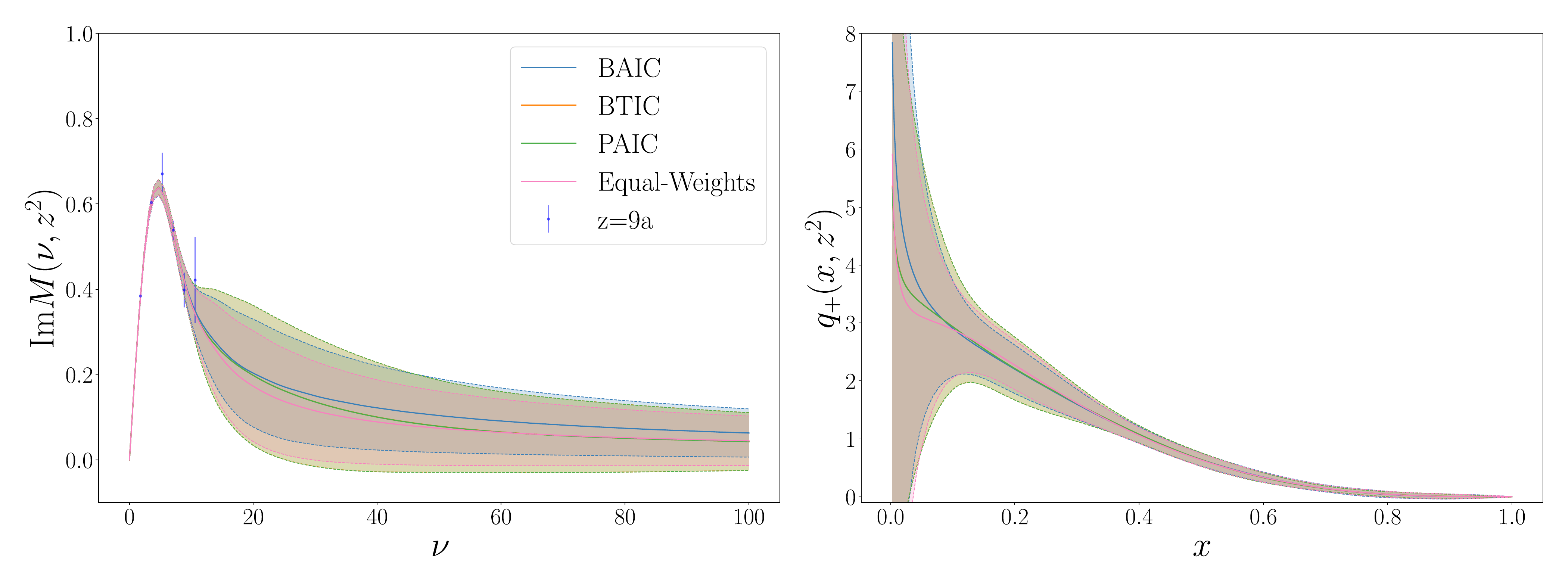}
    \caption{The reconstructions of the CP even isovector PDF (gray) from lattice QCD data. Imaginary ITD components with $z=3a,6a,9a$ from top to bottom. (Right) The ITD in $\nu$ space. (Left) The PDF in $x$ space. }
    \label{fig:MA_Recon_Im_Col}
\end{figure}

Figs.~\ref{fig:IC_all_Re_Col}-\ref{fig:IC_all_Im_Col} show the information criteria for the different $z$ values. In most cases, the PAIC and BTIC tend to select one or two models for averaging. On the other hand, the BAIC allows more to be included with noticeably more models with weights larger than 10\%. Figs.~\ref{fig:MA_Recon_Re_Col}-\ref{fig:MA_Recon_Im_Col} show the results of Bayesian Model Averaging. As with the synthetic data, all the information criteria give generally similar central values. For the realistic data, there tend to be more scenarios where the different information criteria give different estimates of the size of the variance. As before, the PAIC and BTIC tend to give a smaller variance, while the BAIC, averaging over more models, tends to have a larger variance.


\begin{figure}
    \centering
    \includegraphics[width=0.95\linewidth]{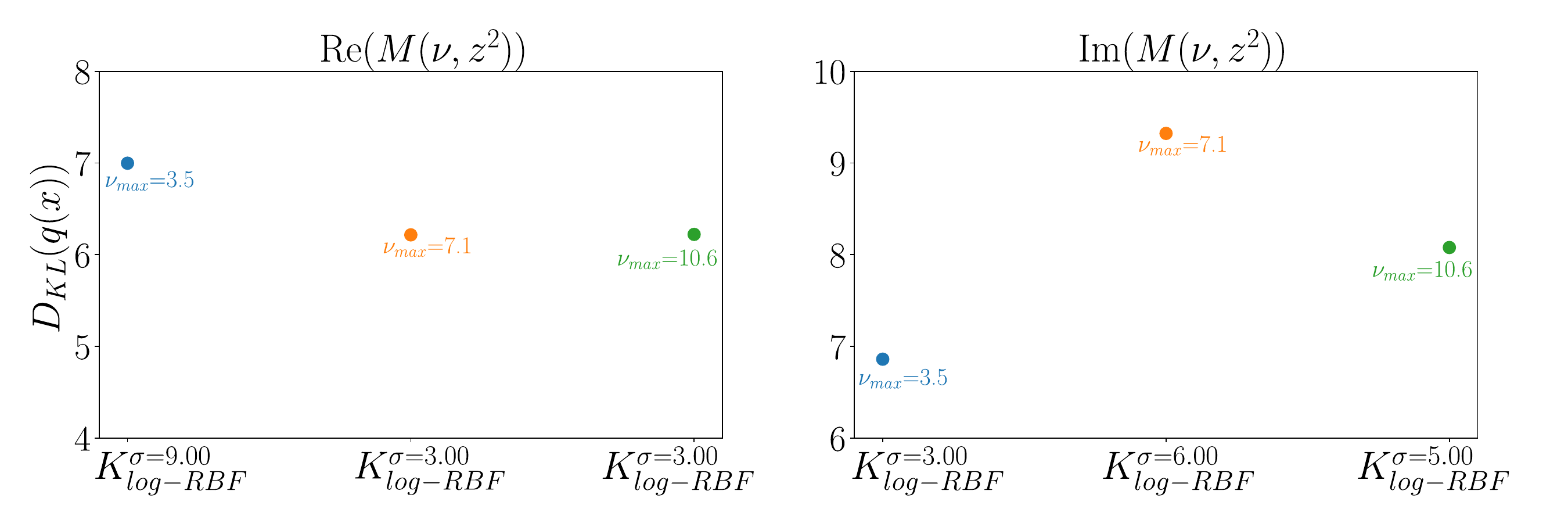}
    \caption{KL divergence in the $x$ and $\nu$ spaces for the real (Right) and imaginary (Left) components for the Fixed $g_{dPDF}$ model.}
    \label{fig:KL_all_C}
\end{figure}

We show the KL divergence for the case of real lattice data. Since there are the same amount of data in each set, the information gained from each data set $z= 3a,6a,9a$ are expected to be more similar in Fig.~\ref{fig:KL_all_C} compared to the upward trend in the synthetic analysis. The global information gained from this data set, compared to the synthetic data analysis, is reduced in size; this can be associated with the smaller correlations between data points in the covariance matrix of the likelihood. 

\begin{figure}
    \centering
    \includegraphics[width=0.95\linewidth]{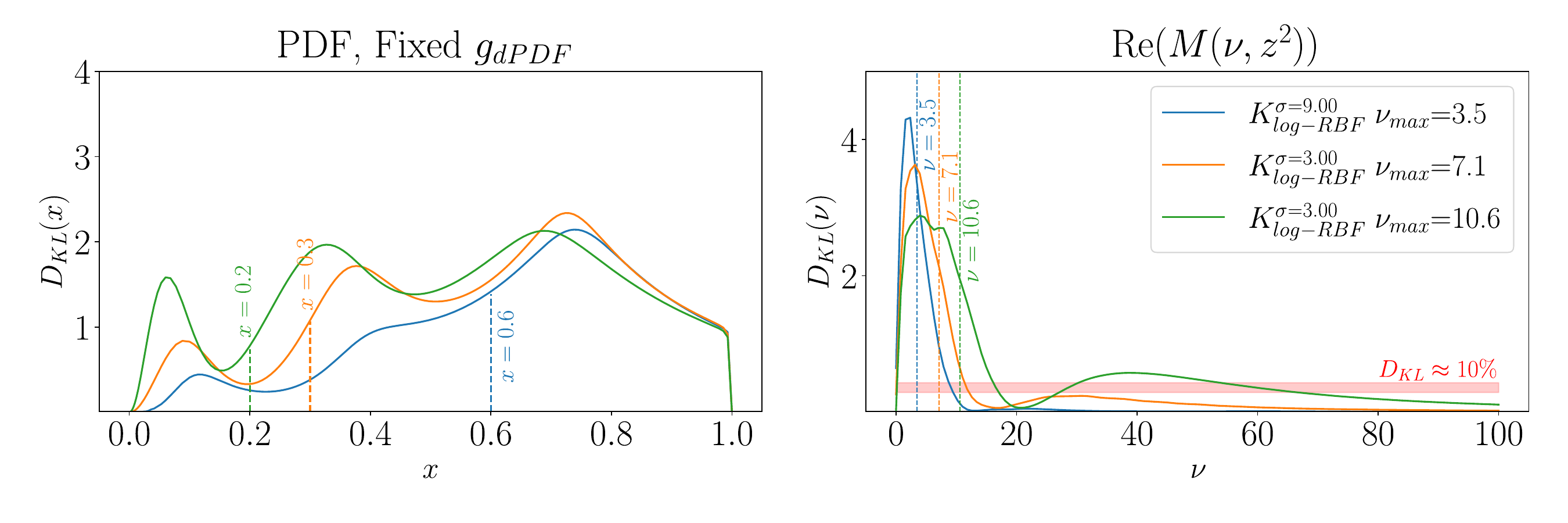}
    \includegraphics[width=0.95\linewidth]{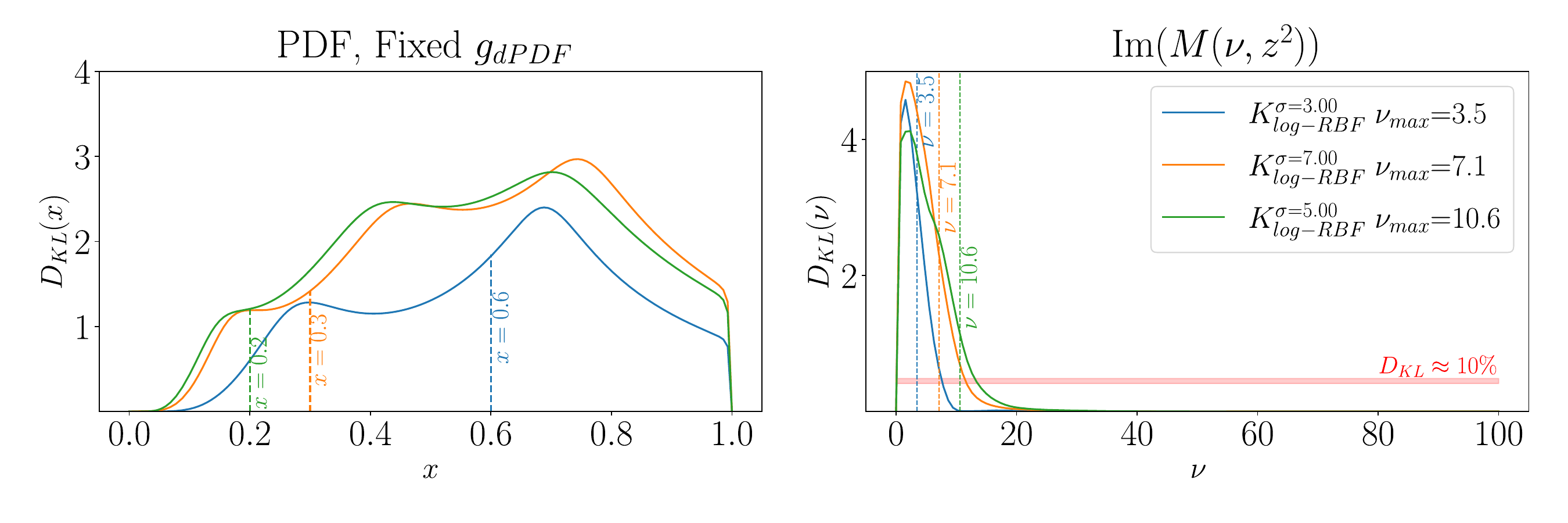}
    \caption{KL divergence in the $x$ (Left) and $\nu$ (Right) spaces for the real (Upper) and imaginary (lower) components. (Left) The vertical lines represent $x=2/\nu_{\rm max}$ and point out approximately where a $50\%$ increase in uncertainty was found in Ref~\cite{Dutrieux:2024rem}. (Right) The vertical lines from below show $\nu_{\rm max}$ and the horizontal line is approximately $10\%$ of the maximum of the KL divergence.}
    \label{fig:KL_point_C}
\end{figure}

Compared with the point-by-point KL divergence for the synthetic data set in $\nu$, the decay is noticeably stronger in Fig.~\ref{fig:KL_point_C} than with the synthetic data analysis, which can be explained in this case with the use of fewer number of less correlated datapoints. The information gain in $x$ is also fairly similar for each $z$ or $\nu_{max}$ value, unlike the synthetic data. There is a trade-off between the higher precision but limited information from the limited Ioffe time range in the smaller $z$ analysis and the noisier information extended further in Ioffe time for the large $z$ analysis. The specifics come out to give a total KL divergence, which is not monotonically increasing like the synthetic dataset.  


\section{Conclusions}\label{sec:conclusion}

Reconstructing parton distributions from lattice QCD requires addressing an ill-posed inverse problem, where prior assumptions strongly influence the outcome. In this work, extending the work in Ref.~\cite{Candido:2024hjt,Dutrieux:2024rem}, we have shown that Gaussian Process Regression provides a systematic framework to regulate this problem while retaining flexibility and offering transparent control over uncertainties.

Our tests with synthetic data demonstrate that GPR reconstructions reliably reproduce the underlying PDFs, with error bands that do not underestimate the true variance. The KL divergence analysis highlights which regions of $x$ are genuinely informed by the data, identifying the limits of what lattice results can constrain. In addition, by varying kernels, mean functions, and hyperparameters, and then applying Bayesian model averaging, we obtain results that are both consistent and robust against modeling choices.

The central message of this study is that GPR delivers a non-parametric and systematically improvable approach to PDF reconstruction, with reliable uncertainty quantification and reduced model bias compared to more restrictive methods. In particular, two aspects are especially informative: (i) the point-by-point KL divergence, which quantifies where the data contributes information across $x$, and (ii) the model-averaged reconstructions, which demonstrate that stable PDFs can be obtained despite large variations in prior assumptions.

Looking ahead, this framework can be extended to incorporate additional physics, such as constraints relevant for quasi-PDFs, alternative priors on hyperparameters, and higher-dimensional distributions like GPDs. Together, these developments position GPR as a powerful and flexible tool for lattice QCD studies of hadron structure.

\FloatBarrier
\section{Acknowledgements}

This project has benefited from the generous support of multiple funding sources.

JK acknowledges support from the U.S. Department of Energy (DOE), Office of Science, under Contract No. DE-AC05-06OR23177, under which Jefferson Science Associates, LLC operates Jefferson Lab.

YC acknowledges support in part from the Southeastern Universities Research Association (SURA) through Grant C2024-FEMT-011-03, awarded by the Center for Nuclear Femtography (CNF) and in part by the Fulbright-Garcia Robles scholarship. 

KO and HD were supported in part by U.S. DOE Grant No. DE-FG02-04ER41302. Their work was also conducted in part under the Laboratory Directed Research and Development (LDRD) Program (Projects LD2412 and LD2510) at Thomas Jefferson National Accelerator Facility for the U.S. DOE.

SZ and HD acknowledge support in part from the Agence Nationale de la Recherche (ANR) under Project No. ANR-23-CE31-0019. For open access, the author has applied a CC-BY public copyright licence to any Author Accepted Manuscript (AAM) version arising from this submission.

 This work has benefited from the collaboration enabled by the Quark-Gluon Tomography (QGT) Topical Collaboration, U.S.~DOE Award \mbox{\#DE-SC0023646}. Computations for this work were carried out in part on facilities of the USQCD Collaboration, which are funded by the Office of Science of the U.S.~Department of Energy. This work was performed in part using computing facilities at William \& Mary, which were provided by contributions from the National Science Foundation (MRI grant PHY-1626177), and the Commonwealth of Virginia Equipment Trust Fund. In addition, this work used resources at NERSC, a DOE Office of Science User Facility supported by the Office of Science of the U.S. Department of Energy under Contract \#DE-AC02-05CH11231. The software codes {\tt Chroma} \cite{Edwards:2004sx}, {\tt QUDA} \cite{Clark:2009wm, Babich:2010mu}, and {\tt Redstar} \cite{Chen:2023zyy} were used to generate the lattice QCD data. The authors acknowledge support from the U.S. Department of Energy, Office of Science, Office of Advanced Scientific Computing Research, and Office of Nuclear Physics, Scientific Discovery through Advanced Computing (SciDAC) program, and of the U.S. Department of Energy Exascale Computing Project (ECP). The authors also acknowledge the Texas Advanced Computing Center (TACC) at The University of Texas at Austin for providing HPC resources, like the Frontera computing system~\cite{10.1145/3311790.3396656} that has contributed to the research results reported within this paper. The authors acknowledge William \& Mary Research Computing for providing computational resources and/or technical support that have contributed to the results reported within this paper.
 We acknowledge the EuroHPC Joint Undertaking for awarding this project access to the EuroHPC supercomputer LUMI, hosted by CSC (Finland) and the LUMI consortium through a EuroHPC Extreme Scale access call. This work also benefited from access to the Jean Zay supercomputer at the Institute for Development and Resources in Intensive Scientific Computing (IDRIS) in Orsay, France, under project (2020-2024)-A0080511504.

\FloatBarrier

\appendix
\section{Importance Sampling Integrals}\label{app:important_sampling}
 
The Bayesian inferences in this study require evaluating integrals of the form
\begin{equation}
    \langle O \rangle = \int d \theta \, O(\theta) \exp[-E^2(\theta)/2]  / Z
\end{equation}
where $Z= \int d \theta \exp[-E^2(\theta)/2]$. Monte Carlo integration begins by choosing a probability distribution and performing a weighted average of $O(\theta)$ with weights based upon the distribution and $\exp[-E^2(\theta)/2]/Z$. The simplest Monte Carlo integration algorithm is to uniformly sample over the domain of $\theta$ and evaluate
\begin{equation}
    \langle O \rangle \approx \frac{\sum_i w_i O(\theta_i)} {\sum_i w_i}
\end{equation}
where $w_i = \exp[-E^2(\theta_i)]$. In importance sampling Monte Carlo integration, the set of sample $\theta_i$ values are drawn from a probability distribution $f(\theta)$, with the goal of not having a significant number of weights be near 0. In this case, the weights will be
\begin{equation}
w_i = \frac{\exp[-E^2(\theta_i)]}{ f(\theta_i)} \,.
\end{equation}

Using Hybrid Monte Carlo (HMC) or its variant used in Ref.~\cite{Candido:2024hjt}, the posterior of $\theta$ is explicitly sampled, giving $f(\theta) = \exp[-E^2(\theta)/2]$. In this case, the expectation value is still given by the weighted sum, except that the weights are trivial $w_i = \exp[-E^2(\theta_i)]/\exp[-E^2(\theta_i)] = 1$ and the expectation value becomes the standard unweighted mean of $O(\theta_i)$ given $\sum_i w_i=N$.

Finally, the Gaussian importance sampling used in this work employs a normal distribution
\begin{equation}f(\theta) = \exp[-(\theta-\theta_0)^T\Sigma^{-1}(\theta-\theta_0)/2]  / \sqrt {|2\pi \Sigma|}\,.
\end{equation} 
The weights are 
\begin{equation}
    w_i = \exp[-\frac12 \bigg(E^2(\theta_i)-(\theta-\theta_0)^T\Sigma^{-1}(\theta-\theta_0)  - \log |2\pi\Sigma|  \bigg)]\,.
\end{equation}
Any Gaussian distribution could have been used, but the principle of importance sampling is to design the sampler to mimic the features of the dominant region of the integrand in order to maximize the weights. The Gaussian approximation to the posterior about the minimum of $E^2$ is a natural choice to obtain the majority of the integration region. The minimum of $E^2$ is found using the ADAM optimizer, and the second derivative at that minimum is calculated with PyTorch’s automatic differentiation. These are used to set the mean and covariance, respectively, for a Gaussian distribution with which $\theta$ values are sampled for the Monte Carlo. In the current applications with small number of hyperparameters being sampled, Gaussian MC can be significantly more efficient that HMC. Eventually when the number of parameters increases the topology of the log-posterior could become more complex and the Gaussian sampling would waste significant amounts of time sampling unimportant regions and HMC will become more useful at fixed cost.

To ensure that the important regions of the integrand are sampled, the sampling covariance is tripled at the cost of potentially wasted samples. In the limit of infinite samples, it was not necessary to find the true minimum or Hessian for the Gaussian distribution as long as the important regions are sufficiently covered. There are cases of a relatively flat $E^2$ or some other condition that the true minimum may not be found quickly, and the Hessian estimated at the result may not be positive definite. When this occurs, the minimum is taken to be wherever the minimizer stops, and the eigenvalues of the estimate are made positive prior to sampling.

 If the posterior is wider than the Gaussian approximation, this will require more samples for the wider parameter space to be reached, while if it is narrower, some samples will be wasted, giving little contribution to the integral. Since Gaussian sampling and the evaluation of $\bar{q}$ and $H$ are quite efficient, the inaccuracies from not sufficiently sampling the full space are a more worrying problem than computations wasted from sampling too much. As such, to ensure appropriate sampling, $\Sigma$ is inflated by a factor of three compared to the Gaussian approximation of the posterior to ensure that the relevant information is captured.

\section{Derivation of the point by point KL divergence} \label{app:kl}

Consider having a Gaussian probability distribution for a vector $q$ given by
\begin{equation}
    P[q] =  \frac{1}{\sqrt{\det{2\pi H}}} \exp\left[ -(q-\bar{q})^T H^{-1}(q-\bar{q}) \right] \,.
\end{equation}
The probability of a single element of $q$ is given by integrating the other elements
\begin{equation}
    P[q_i] = \int \prod_{j\neq i} dq_j P[q].
\end{equation}
Changing variables to $f=q-\bar{q}$ for brevity,
\begin{equation}
    P[q_i] = \frac{1}{\sqrt{\det{2\pi H}}} \int \prod_{j\neq i} df_j \exp [ -\frac12f_j^T H^{-1}_{jj'} f_{j'} - f_j^T H^{-1}_{ji}f_i - \frac12 f_iH^{-1}_{ii}f_i]
\end{equation}
where $j,j'\neq i$ are always summed over while $i$ is fixed. Integrating gives
\begin{equation}
    P[q_i] = \sqrt{\frac{\det \hat{H}^{-1}}{2\pi \det H}}\exp[-\frac 12  (q_i-\bar{q_i})(H^{-1}_{ii} - H^{-1}_{ij}[\hat{H}^{-1}]_{jj'}H^{-1}_{j'i}) (q_i-\bar{q_i}))] 
\end{equation}
where $\hat{H}$ represents $H^{-1}$ with the $i^{\rm th}$ row and column deleted. Note that it is a smaller dimension matrix, so all but a single factor of $2\pi$ remains when pulling them out of the determinants. For brevity, we label the Schur complement as $\tilde{H}_{ii}=H^{-1}/\hat{H} = H^{-1}_{ii} - H^{-1}_{ij}[\hat{H}^{-1}]_{jj'}H^{-1}_{j'i}$ which in this case is a single number.  

Similarly the prior distribution at a single point has a similar form with $H\to K$, $\tilde{H}\to\tilde{K}$, $\bar{q}\to g$, and $\hat{H}\to \hat{K}$. The $KL$ divergence of the posterior and prior distributions at a single point in $x$ is given by
\begin{eqnarray}
    D_{KL}(Post[q_i] || Prior[q_i]) = \sqrt{\frac{\det \hat{H}^{-1}}{2\pi \det H}}  \int dq_i e^{-\frac 12 (q_i-\bar{q}_i)\tilde{H}_{ii}(q_i-\bar{q}_i) } \nonumber\\\times \bigg[ \log\frac{\det \hat{H}^{-1}}{2\pi\det H} - (q_i-\bar{q_i})\tilde{H} (q_i-\bar{q_i}) -  \log\frac{\det \hat{K}^{-1}}{2\pi\det K} + (q_i-g_i)\tilde{K} (q_i-g_i) \bigg]
\end{eqnarray}
which integrates to
\begin{eqnarray}
    D_{KL}(Post[q_i] || Prior[q_i]) =\frac 12 \sqrt{\frac{\det \hat{H}^{-1}}{\tilde{H}_{ii}\det H}}  \Bigg(\frac{\tilde{K}_{ii}}{\tilde{H}_{ii}}-\log \frac{\det H \det  \hat{K}^{-1}}{\det \hat{H}^{-1} \det K}  -1  \nonumber\\ + (\bar{q}_i - g_i)\tilde{K}_{ii}(\bar{q}_i- g_i)\Bigg) \,.
\end{eqnarray}
In the special case that the prior covariance $K$, the data covariance $C$, and thereby the posterior covariance $H$ were diagonal matrices, then this formula would reduce to Eq 23 of Ref.~\cite{ValentineSambridgeGP}. 
\section{Logarithmic plots}

\begin{figure}
    \centering
    \includegraphics[width=0.95\linewidth]{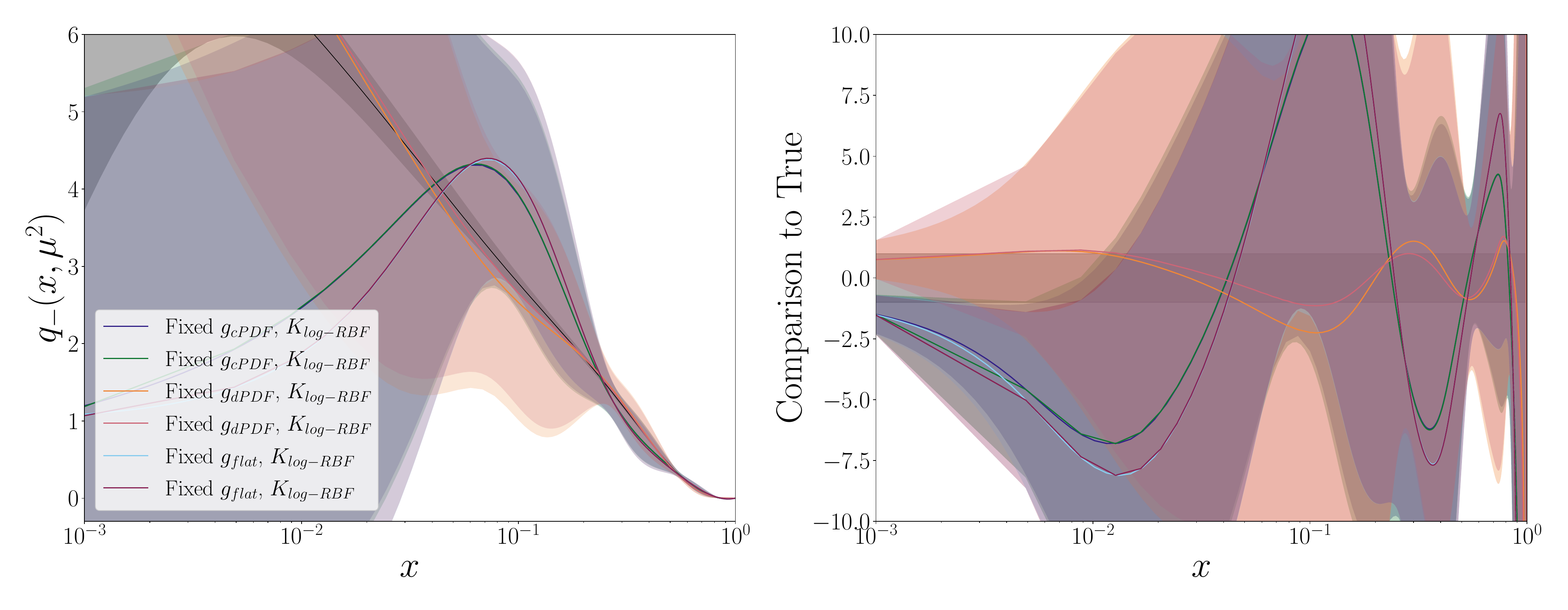}
    \includegraphics[width=0.95\linewidth]{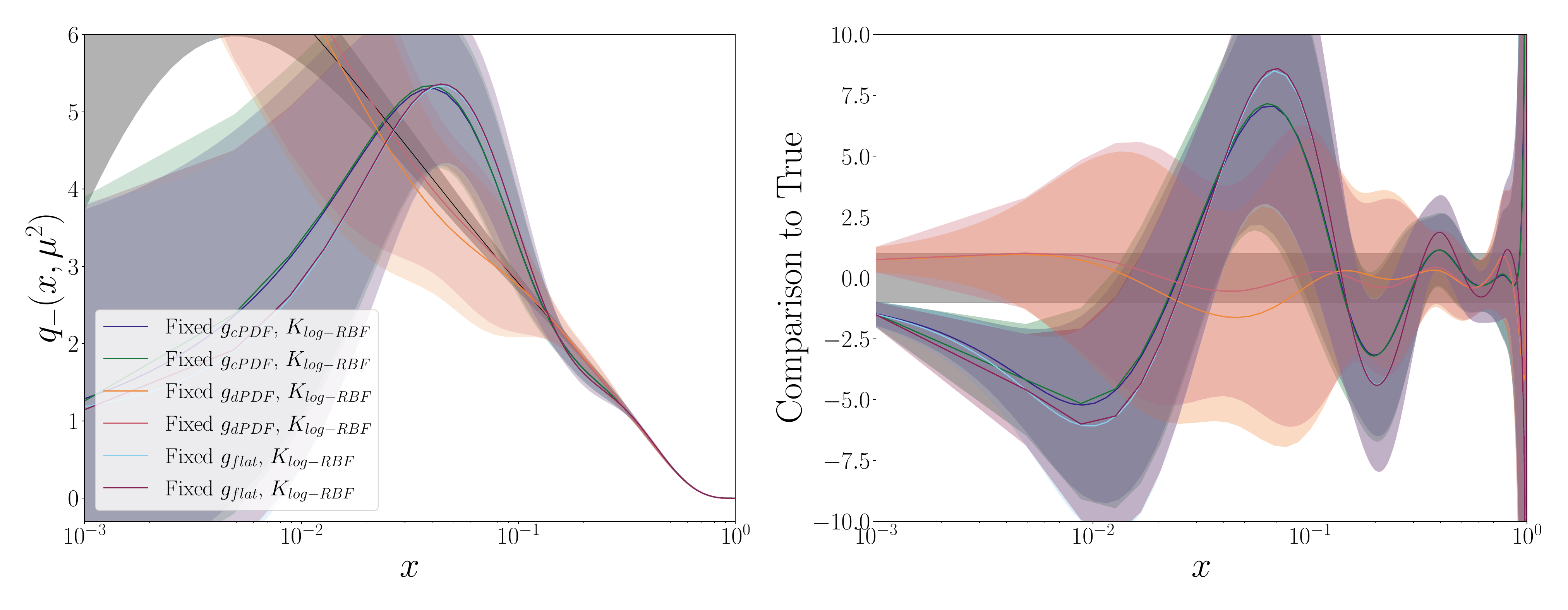}
    \includegraphics[width=0.95\linewidth]{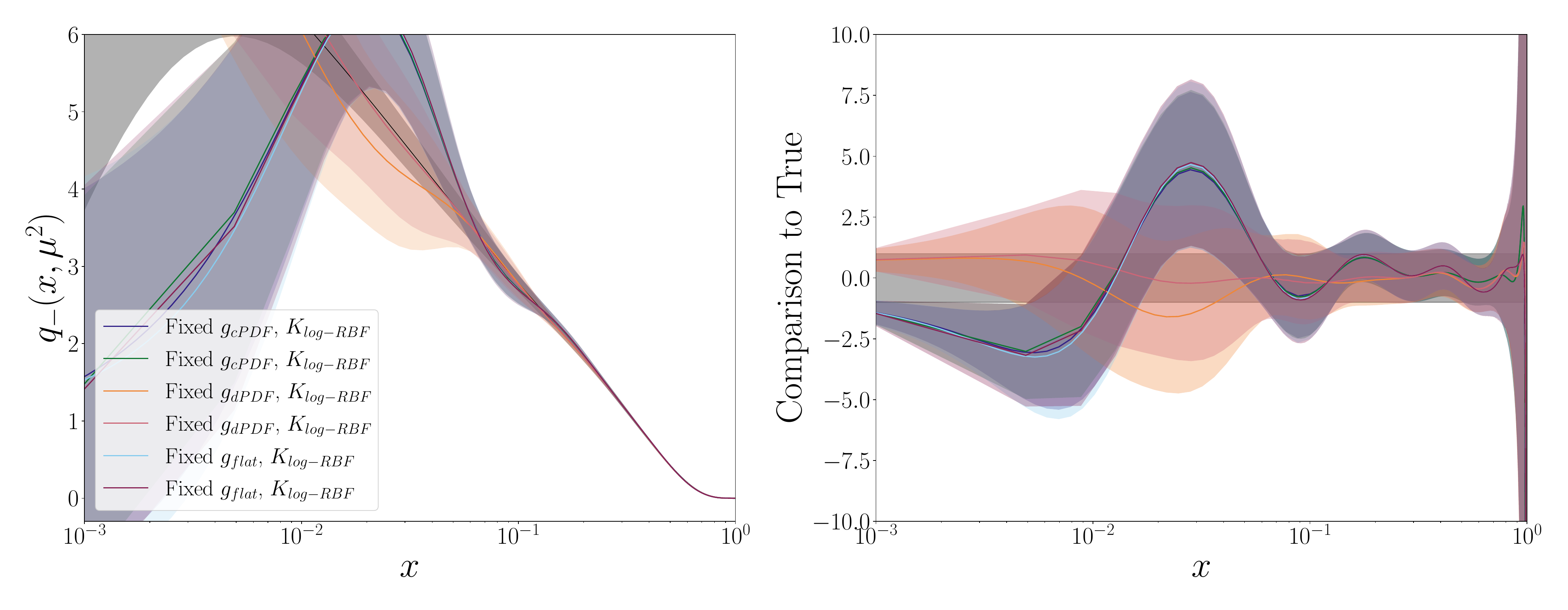}
    \caption{Impact of different prior mean functions for a $log-RBF$ kernel on the reconstruction of the NNPDF4.0 CP even isovector PDF in logarithmic scale. See caption of Fig.~\ref{fig:convdiv_Re} for more details.} 
    \label{fig:convdiv_Re_log}
\end{figure}

\begin{figure}
    \centering
    \includegraphics[width=0.95\linewidth]{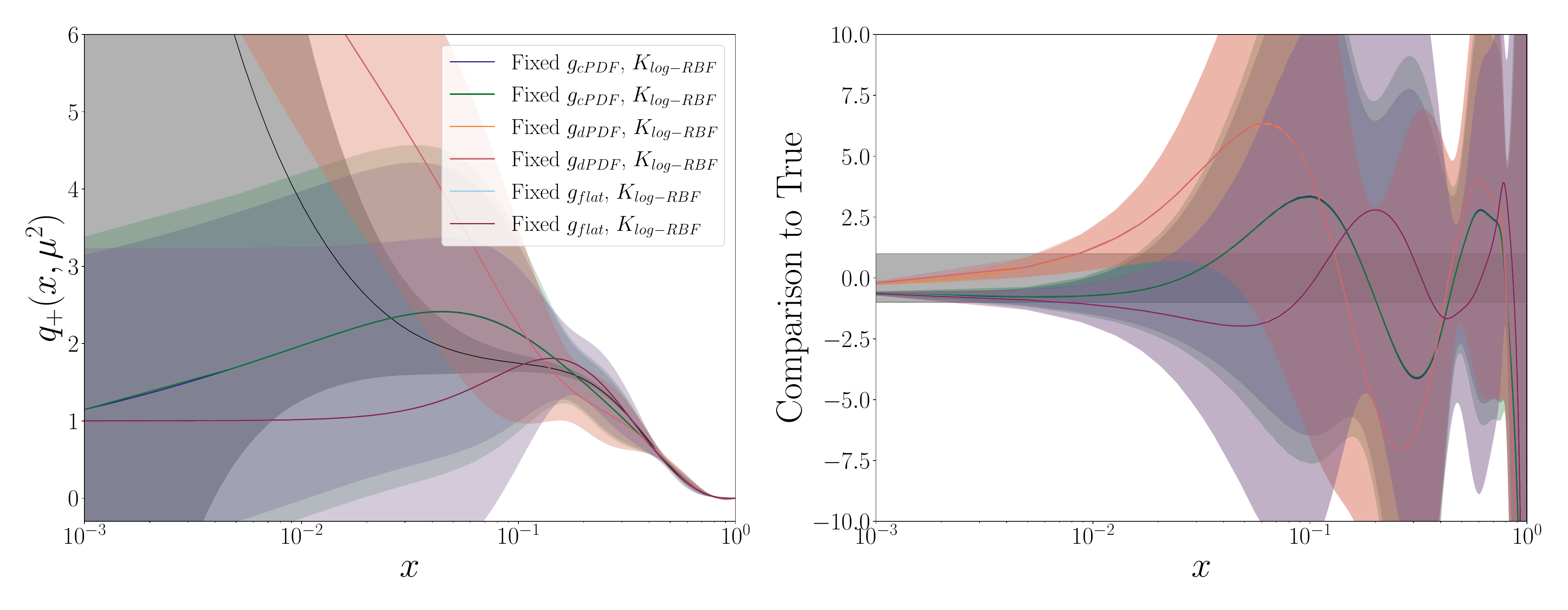}
    \includegraphics[width=0.95\linewidth]{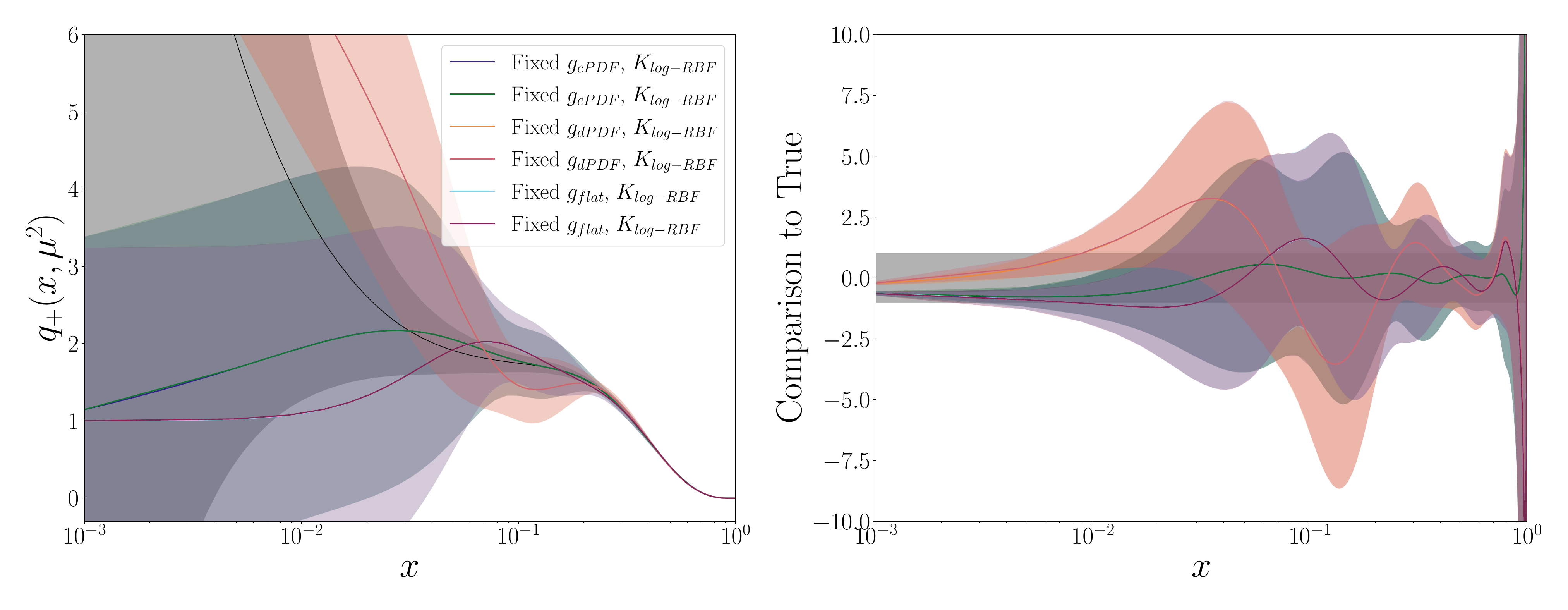}
    \includegraphics[width=0.95\linewidth]{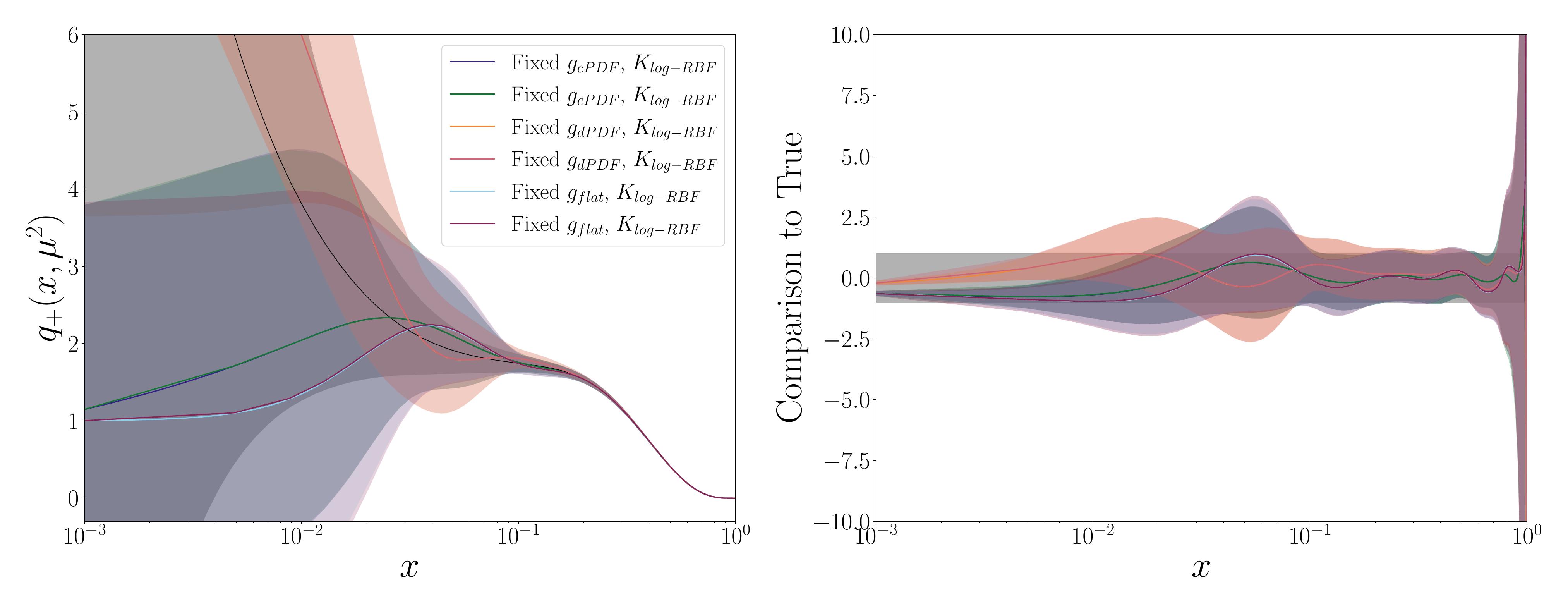}
    \caption{Impact of different prior mean functions for a $log-RBF$ kernel on the NNPDF4.0 CP odd isovector PDF in logarithmic scale. See caption of Fig.~\ref{fig:convdiv_Re} for more details.
    }
    \label{fig:convdiv_Im_log}
\end{figure}

Rescaling the $x$-axis to a logarithmic scale allows us to visualize the region where $x<0.1$ alongside the uncertainty reconstruction, as shown in Figs. \ref{fig:convdiv_Re_log} and \ref{fig:convdiv_Im_log}. However, as demonstrated in Section \ref{subsec:KL}, the information gain in this particular region is negligible. Consequently, the posterior behavior is primarily governed by the prior choice of kernel and mean functions, with sensitivity in this region depending on the number of data points available at large $\nu$. Therefore, throughout this work, we emphasize the analysis of the $x>0.1$ region, where the information gain is significantly greater.

\section{Pathologies on different kernels}
\label{app:poorkernels}

As we have evoked in Sec.\ref{sec:ktrick}, some kernels behave poorly if they do not implement properly the physics constraints. We give here a brief discussion and illustration of some of those kernels. 

A first natural idea could be to use a kernel trick with the family of mean functions $\phi_n(x;\alpha,\beta) = x^{\alpha+n} (1-x)^\beta$, which is a model of $x^\alpha(1-x)^\beta$ multiplying a polynomial. If we use the model parameter's prior covariance given by $\Sigma_{nn'}= \sigma t^{n+n'}$, using the identity $\sum_{n=0}^\infty (tx)^n = (1-tx)^{-1}$, the kernel 
\begin{equation}
    K_{\rm Poly}(x,x'; \theta=\{\sigma, t, \alpha,\beta\}) = \frac{\sigma (xx')^\alpha \big((1-x)(1-x'))^\beta }{ (1-tx)(1-tx')} \, , \label{eq:poly_kernelapp}
\end{equation}
can be obtained. Instead, one could have chosen a prior covariance on the parameters $c_n$, which is diagonal $\Sigma'_{nn'} = \sigma \delta^{nn'}$. This would generate the kernel
\begin{equation}
    K_{\rm Poly'}(x,x'; \theta=\{\sigma, \alpha, \beta\}) =  \frac{\sigma (xx')^\alpha \big((1-x)(1-x')\big)^\beta }{ (1-xx')} \,. 
\end{equation}
These two priors, and the associated kernels, may have quite different behavior. 
In fact, $\Sigma$, the covariance used in Eq.~\eqref{eq:poly_kernelapp}, has a single non-zero eigenvalue that depends on $t$. This implies that the prior on the infinite set of $c_n$ contains a single degree of freedom whose variance depends on $t$. $\Sigma'$, on the other hand, is of full rank and each of the infinite $c_n$ will be an independent degree of freedom in the prior with a fixed prior variance of $\sigma'$. It will be shown that $K_{\rm Poly}$ is extremely constrained, giving results in disagreement with the data. While less dramatically, $K_{\rm Poly'}$ will be over-constrained, leading to an underestimation of the error at low $x$. Both cases demonstrate the importance of closure tests to ensure that not only central values can be reproduced but also that errors are not underestimated.

Another kernel of this type can be constructed from Jacobi polynomials $J_n$ reweighted as
\begin{equation}
    Q(x;c_n, \alpha, \beta) = \sum_{n=0}^\infty c_n \phi_n(x;\alpha,\beta) \quad;\qquad  \phi_n(x;\alpha,\beta) = x^\alpha (1-x)^\beta J_n^{(\alpha,\beta)}(2x-1)
\end{equation} whose truncated series was used in Ref.~\cite{Parisi:1978jv,Barker:1980wu,Barker:1982rv,Chyla:1986eb,Krivokhizhin:1987rz,Krivokhizhin:1990ct,AtashbarTehrani:2007odq,Khorramian:2009asi,Taghavi-Shahri:2016idw,Karpie:2021pap} to define a parameterized model of the PDF. With Jacobi polynomials, for convenience, we modify the normalization of the covariance as $\Sigma_{nn'} = \frac{\sigma}{4^{\alpha + \beta}} t^{n+n'}$. The generating function of Jacobi polynomials, given by the relation
\begin{equation}
    \sum_{n=0}^\infty t^n J_n^{(\alpha,\beta)}(z)  = 2^{\alpha+\beta} \left[R(z,t) \big(1-t+R(z,t)\big)^{\alpha} \big(1+t+R(z,t)\big)^{\beta} \right]^{-1} = 2^{\alpha+\beta} F(z,t;\alpha,\beta)
\end{equation}
with $R(z,t)= (1-2zt+t^2)^{1/2}$, allows us to derive the kernel for $z=2x-1$,
\begin{eqnarray}
    K_{\rm Jacobi}&(x,x'; \theta=\{\sigma, t, \alpha,\beta\})= \frac{\sigma   (xx')^\alpha \big((1-x)(1-x')\big)^\beta  }{4^{\alpha + \beta}}\sum_{n,n'=0}^\infty J_n^{(\alpha,\beta)}(2x-1) t^{n+n'} J_{n'}^{(\alpha,\beta)}(2x'-1) \nonumber \\
    =&  \sigma (xx')^\alpha \big((1-x)(1-x'))^\beta F(2x-1,t;\alpha,\beta)F(2x'-1,t;\alpha,\beta)\,.
\end{eqnarray}
Using this kernel, the analysis is equivalent to fitting the model from the complete set of Jacobi polynomials. Once again, however, it suffers from excessive correlations related to the rank 1 of the covariance matrix.

\begin{figure}
    \centering
    \includegraphics[width=0.95\linewidth]{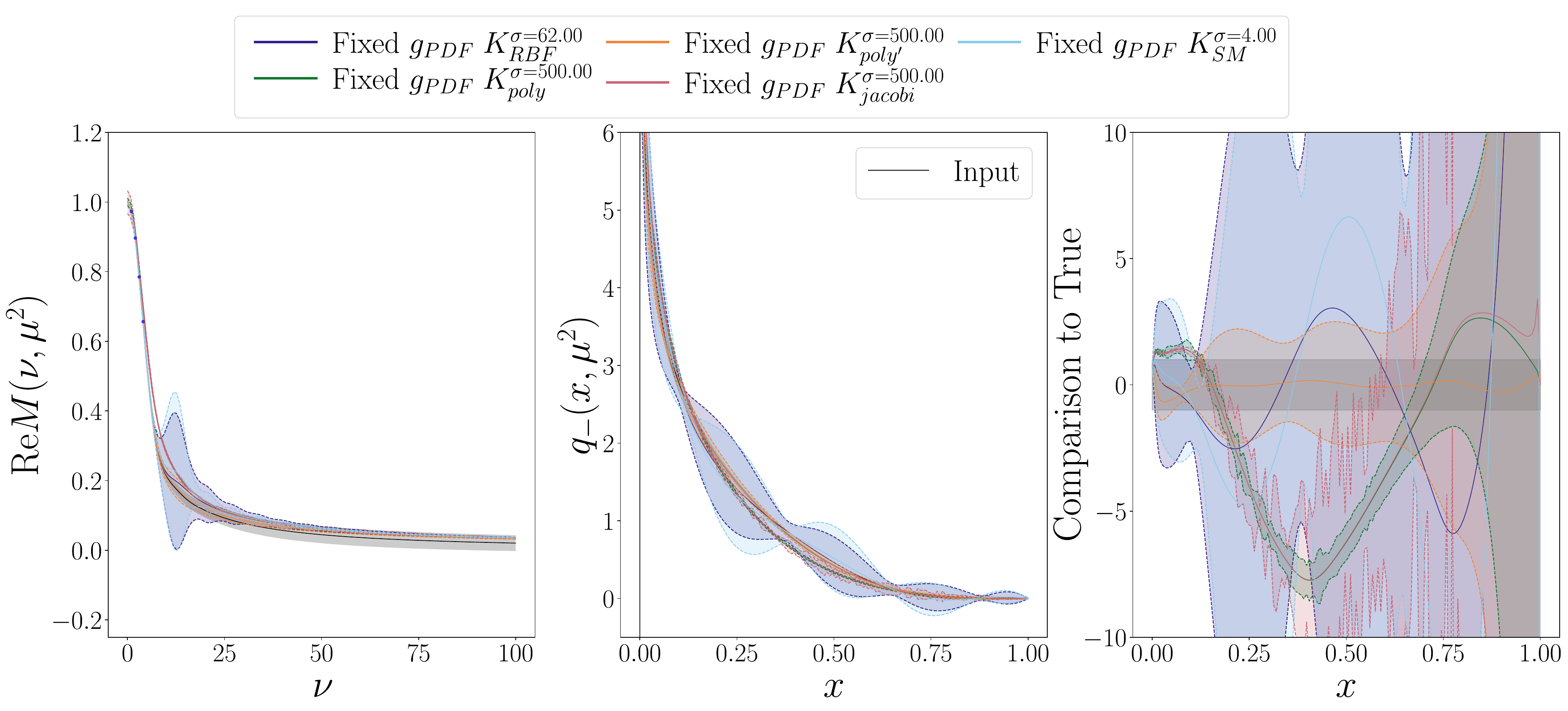}
    \includegraphics[width=0.95\linewidth]{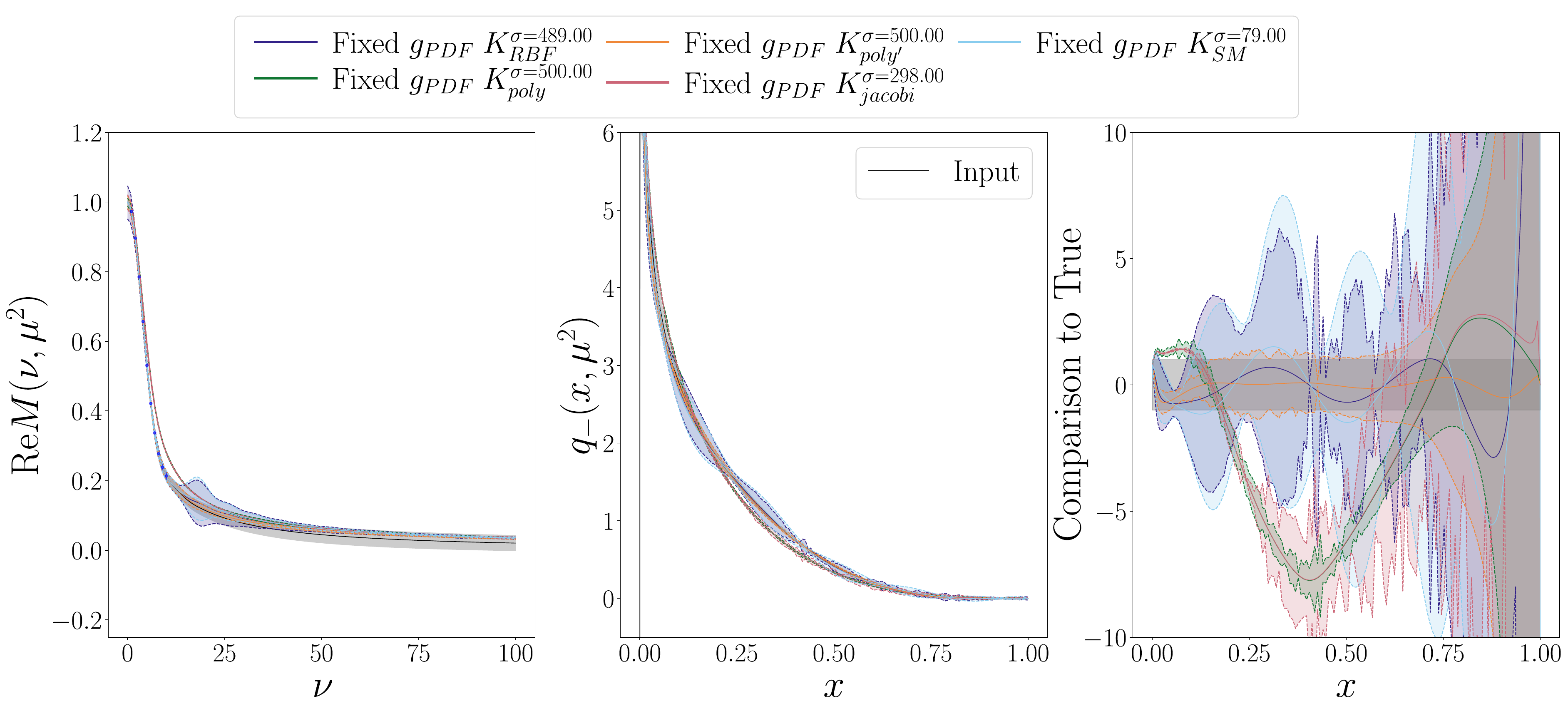}
    \includegraphics[width=0.95\linewidth]{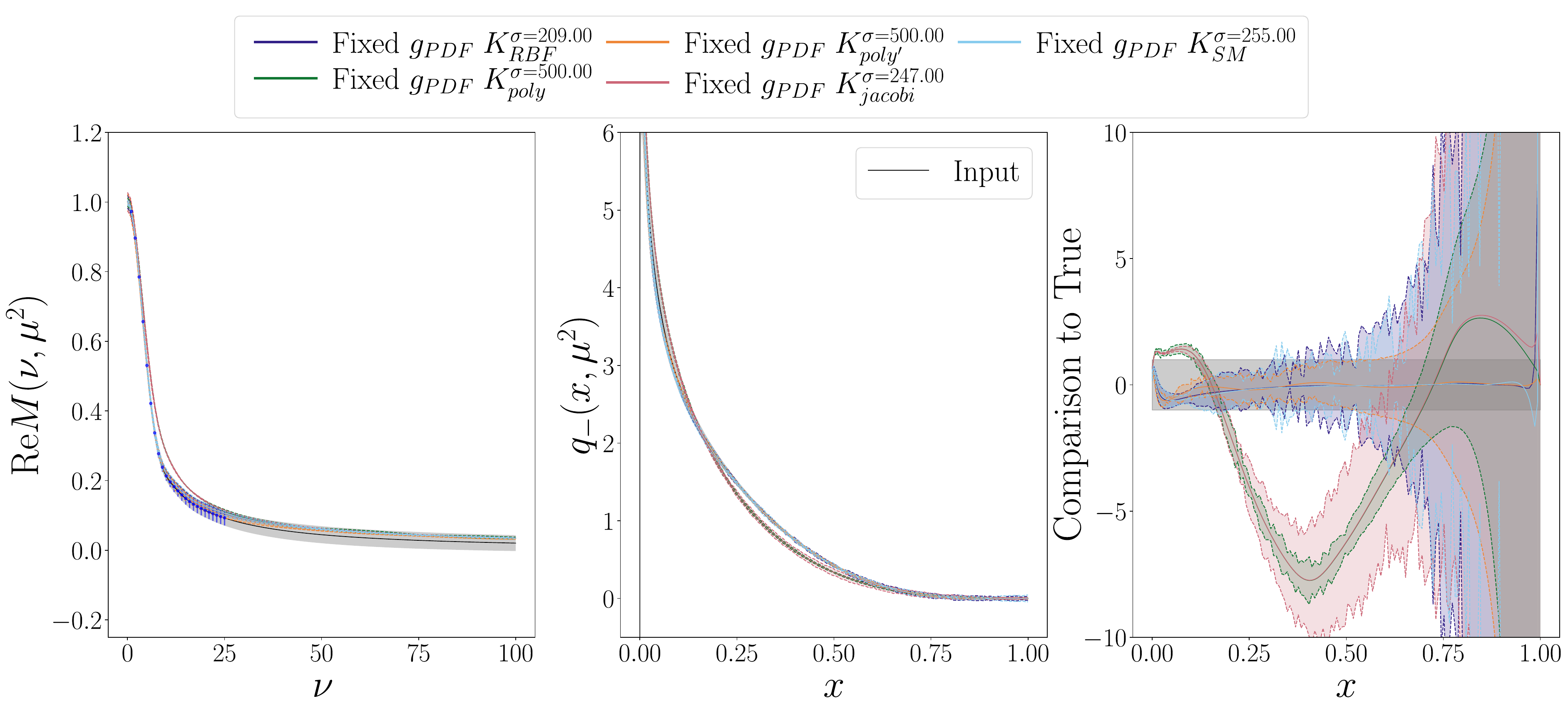}
    \caption{The reconstructions of the NNPDF4.0 CP even isovector PDF (gray) from synthetic data. The maximum Ioffe time used was 4, 10, and 25, increasing from top to bottom. (Right) The ITD in $\nu$ space. (Middle) The PDF in $x$ space. (Left) The difference between the reconstruction and the true PDF divided by the error of the true PDF.}
    \label{fig:Bad_models_Re}
\end{figure}

\begin{figure}
    \centering
    \includegraphics[width=0.95\linewidth]{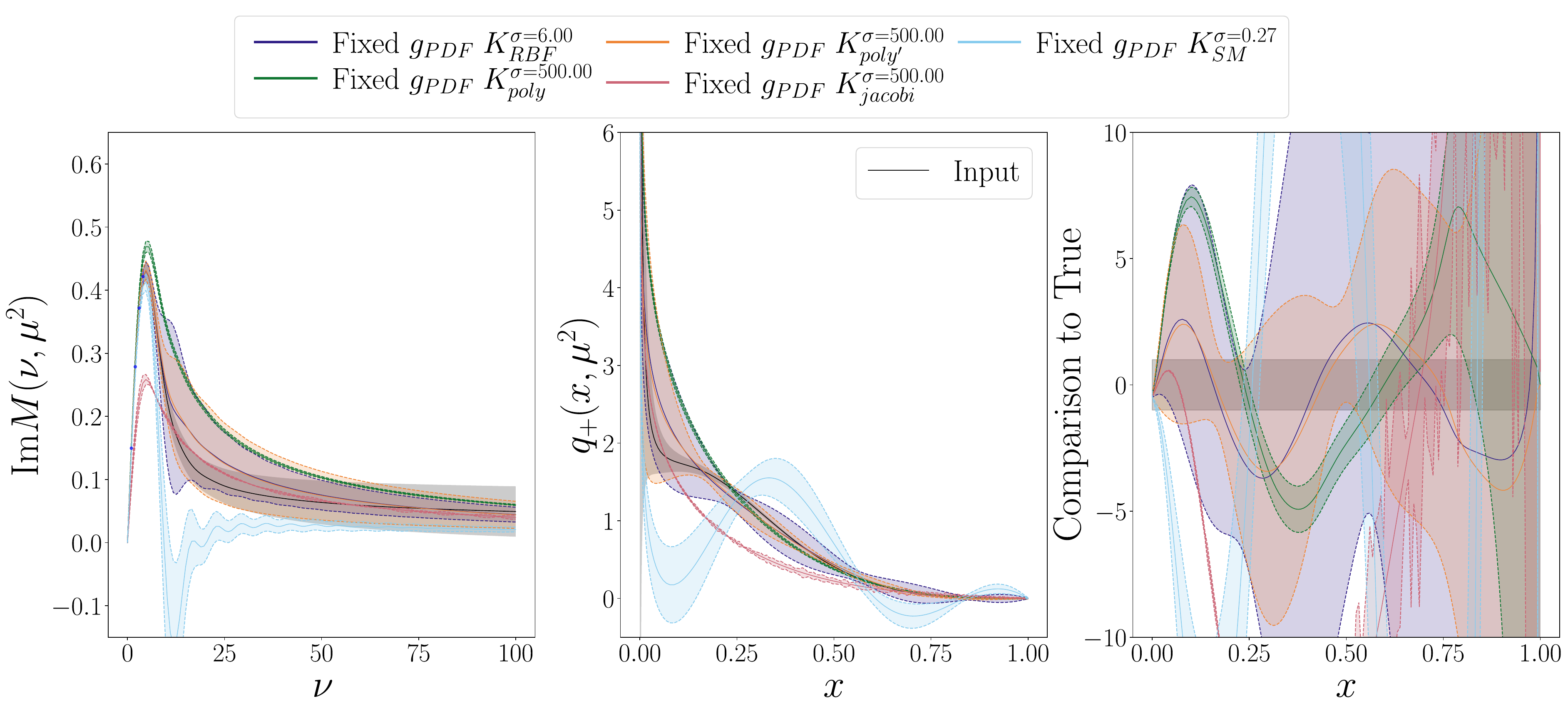}
    \includegraphics[width=0.95\linewidth]{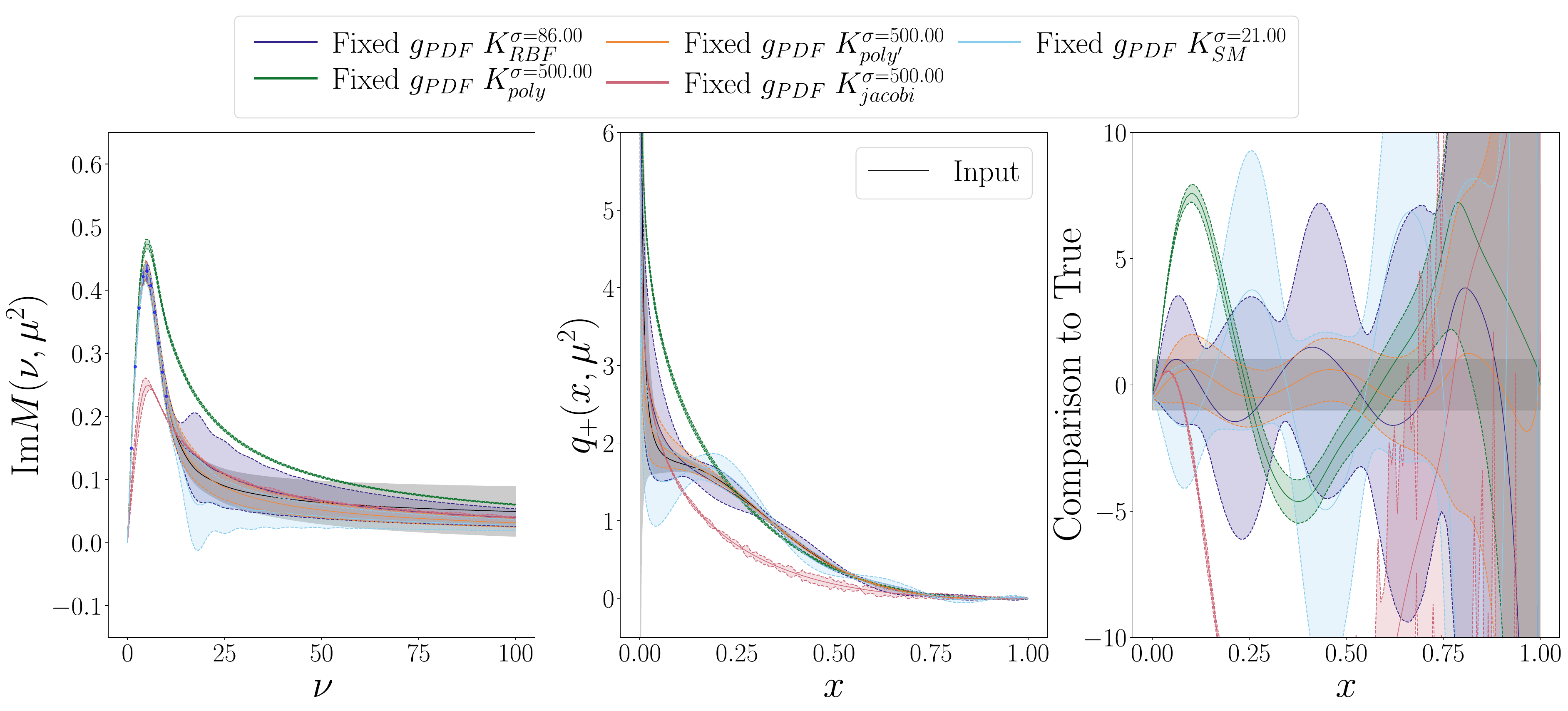}
    \includegraphics[width=0.95\linewidth]{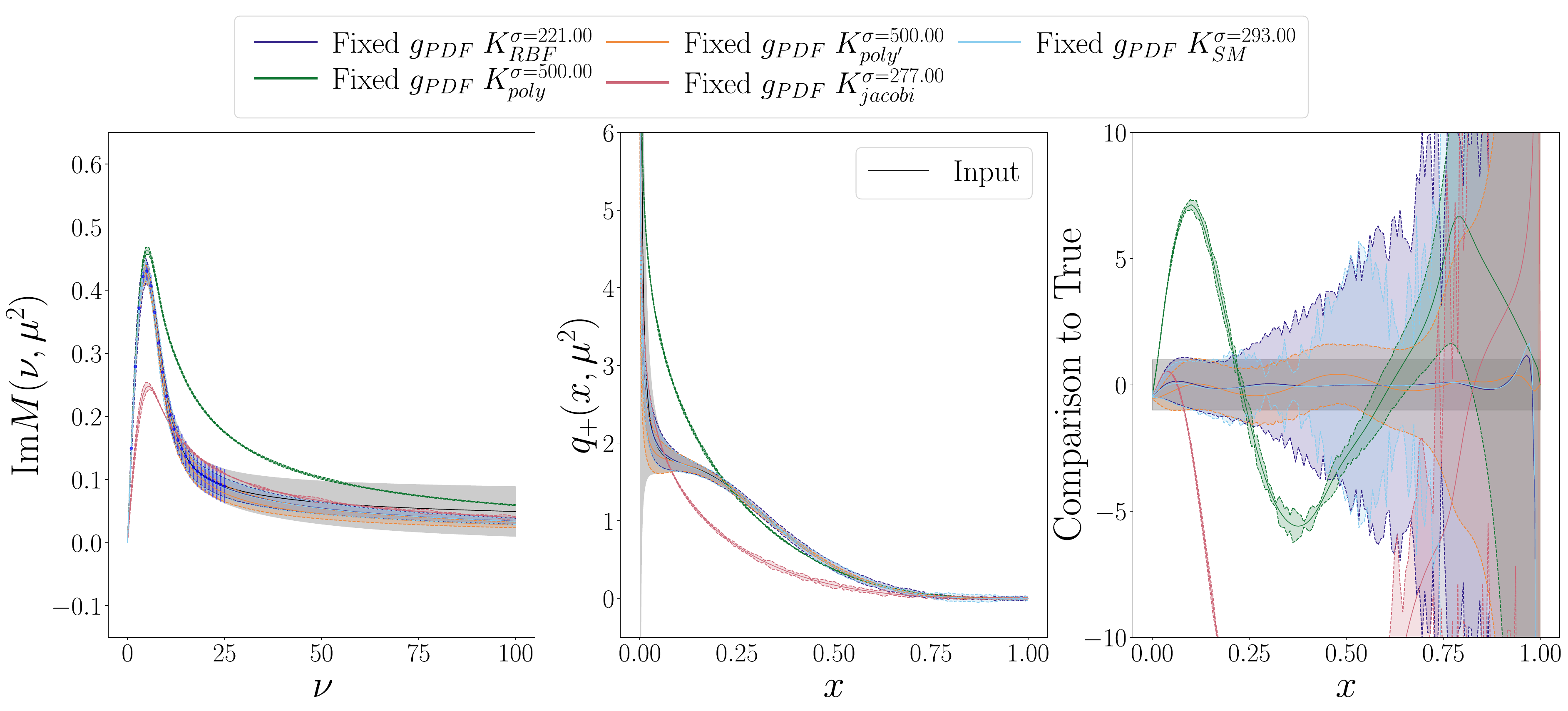}
    \caption{The reconstructions of the NNPDF4.0 CP odd isovector PDF (gray) from synthetic data. The maximum Ioffe time used was 4, 10, and 25, increasing from top to bottom. (Right) The ITD in $\nu$ space. (Middle) The PDF in $x$ space. (Left) The difference between the reconstruction and the true PDF divided by the error of the true PDF. }
    \label{fig:Bad_models_Im}
\end{figure}

The result of those kernels, as well as some we have introduced earlier in the document with poor performance, are shown in Figs. \ref{fig:Bad_models_Re}-\ref{fig:Bad_models_Im}. 
While adding a regulator to $\Gamma = C + B \circ K \circ B^T$ helps with some particular parameters, if we want to perform sampling or minimization over some parameters, this regulator should alleviate numerical instabilities for any $\theta$ in the allowed parameter space. Furthermore, terms such as the evidence $E(\theta)$ or its derivatives $\nabla_\theta log(E(\theta))$ include the calculation of $\log\det(\Gamma)$, which, if not regulated, runs into problems from erroneously small negative eigenvalues. $K_{\rm SM}$, $K_{\rm Jacobi}$, and $K_{\rm Poly}$ are kernels that present such problems, even after introducing a regulator, which become more severe as the dataset grows. A possible reason for this behavior is that we are working with highly correlated data, which has the effect of making $C$ singular. 

Finally, a phenomenological criterion of the kernel comes from Eq. \eqref{eq:criteriamax}, which evaluates the extrapolation power of the kernel at low $x$ or equivalently $\nu>\nu_{max}$. In general,  $K_{\rm Poly}$, $K_{\rm Poly'}$, and $K_{\rm Jacobi}$ fail this test due to the way these kernels correlate large portions of the points in $x$, which reduces the ability to increase the extrapolation variance by merely increasing $\sigma$. 
The resulting PDFs and ITDs in Figs.~\ref{fig:Bad_models_Re} and~\ref{fig:Bad_models_Im} 
show pathological behavior, such as oscillatory artifacts or underestimated uncertainties.


    \FloatBarrier
\bibliography{bib}

\begin{thebibliography}{96}%
\makeatletter
\providecommand \@ifxundefined [1]{%
 \@ifx{#1\undefined}
}%
\providecommand \@ifnum [1]{%
 \ifnum #1\expandafter \@firstoftwo
 \else \expandafter \@secondoftwo
 \fi
}%
\providecommand \@ifx [1]{%
 \ifx #1\expandafter \@firstoftwo
 \else \expandafter \@secondoftwo
 \fi
}%
\providecommand \natexlab [1]{#1}%
\providecommand \enquote  [1]{``#1''}%
\providecommand \bibnamefont  [1]{#1}%
\providecommand \bibfnamefont [1]{#1}%
\providecommand \citenamefont [1]{#1}%
\providecommand \href@noop [0]{\@secondoftwo}%
\providecommand \href [0]{\begingroup \@sanitize@url \@href}%
\providecommand \@href[1]{\@@startlink{#1}\@@href}%
\providecommand \@@href[1]{\endgroup#1\@@endlink}%
\providecommand \@sanitize@url [0]{\catcode `\\12\catcode `\$12\catcode `\&12\catcode `\#12\catcode `\^12\catcode `\_12\catcode `\%12\relax}%
\providecommand \@@startlink[1]{}%
\providecommand \@@endlink[0]{}%
\providecommand \url  [0]{\begingroup\@sanitize@url \@url }%
\providecommand \@url [1]{\endgroup\@href {#1}{\urlprefix }}%
\providecommand \urlprefix  [0]{URL }%
\providecommand \Eprint [0]{\href }%
\providecommand \doibase [0]{http://dx.doi.org/}%
\providecommand \selectlanguage [0]{\@gobble}%
\providecommand \bibinfo  [0]{\@secondoftwo}%
\providecommand \bibfield  [0]{\@secondoftwo}%
\providecommand \translation [1]{[#1]}%
\providecommand \BibitemOpen [0]{}%
\providecommand \bibitemStop [0]{}%
\providecommand \bibitemNoStop [0]{.\EOS\space}%
\providecommand \EOS [0]{\spacefactor3000\relax}%
\providecommand \BibitemShut  [1]{\csname bibitem#1\endcsname}%
\let\auto@bib@innerbib\@empty
\bibitem [{\citenamefont {Meyer}(2007)}]{Meyer:2007ic}%
  \BibitemOpen
  \bibfield  {author} {\bibinfo {author} {\bibfnamefont {H.~B.}\ \bibnamefont {Meyer}},\ }\href {\doibase 10.1103/PhysRevD.76.101701} {\bibfield  {journal} {\bibinfo  {journal} {Phys. Rev. D}\ }\textbf {\bibinfo {volume} {76}},\ \bibinfo {pages} {101701} (\bibinfo {year} {2007})},\ \Eprint {http://arxiv.org/abs/0704.1801} {arXiv:0704.1801 [hep-lat]} \BibitemShut {NoStop}%
\bibitem [{\citenamefont {Alexandrou}\ \emph {et~al.}(2024)\citenamefont {Alexandrou} \emph {et~al.}}]{ExtendedTwistedMass:2024myu}%
  \BibitemOpen
  \bibfield  {author} {\bibinfo {author} {\bibfnamefont {C.}~\bibnamefont {Alexandrou}} \emph {et~al.} (\bibinfo {collaboration} {Extended Twisted Mass}),\ }\href {\doibase 10.1103/PhysRevLett.132.261901} {\bibfield  {journal} {\bibinfo  {journal} {Phys. Rev. Lett.}\ }\textbf {\bibinfo {volume} {132}},\ \bibinfo {pages} {261901} (\bibinfo {year} {2024})},\ \Eprint {http://arxiv.org/abs/2403.05404} {arXiv:2403.05404 [hep-lat]} \BibitemShut {NoStop}%
\bibitem [{\citenamefont {Alexandrou}\ \emph {et~al.}(2023)\citenamefont {Alexandrou} \emph {et~al.}}]{ExtendedTwistedMassCollaborationETMC:2022sta}%
  \BibitemOpen
  \bibfield  {author} {\bibinfo {author} {\bibfnamefont {C.}~\bibnamefont {Alexandrou}} \emph {et~al.} (\bibinfo {collaboration} {Extended Twisted Mass Collaboration (ETMC)}),\ }\href {\doibase 10.1103/PhysRevLett.130.241901} {\bibfield  {journal} {\bibinfo  {journal} {Phys. Rev. Lett.}\ }\textbf {\bibinfo {volume} {130}},\ \bibinfo {pages} {241901} (\bibinfo {year} {2023})},\ \Eprint {http://arxiv.org/abs/2212.08467} {arXiv:2212.08467 [hep-lat]} \BibitemShut {NoStop}%
\bibitem [{\citenamefont {Fowlie}\ and\ \citenamefont {Li}(2023)}]{Fowlie:2023cta}%
  \BibitemOpen
  \bibfield  {author} {\bibinfo {author} {\bibfnamefont {A.}~\bibnamefont {Fowlie}}\ and\ \bibinfo {author} {\bibfnamefont {Q.}~\bibnamefont {Li}},\ }\href {\doibase 10.1140/epjc/s10052-023-12110-9} {\bibfield  {journal} {\bibinfo  {journal} {Eur. Phys. J. C}\ }\textbf {\bibinfo {volume} {83}},\ \bibinfo {pages} {943} (\bibinfo {year} {2023})},\ \Eprint {http://arxiv.org/abs/2306.17385} {arXiv:2306.17385 [hep-ph]} \BibitemShut {NoStop}%
\bibitem [{\citenamefont {Salg}\ \emph {et~al.}(2025)\citenamefont {Salg}, \citenamefont {Romero-L{\'o}pez},\ and\ \citenamefont {Jay}}]{Salg:2025now}%
  \BibitemOpen
  \bibfield  {author} {\bibinfo {author} {\bibfnamefont {M.}~\bibnamefont {Salg}}, \bibinfo {author} {\bibfnamefont {F.}~\bibnamefont {Romero-L{\'o}pez}}, \ and\ \bibinfo {author} {\bibfnamefont {W.~I.}\ \bibnamefont {Jay}},\ }\href@noop {} {\  (\bibinfo {year} {2025})},\ \Eprint {http://arxiv.org/abs/2506.16161} {arXiv:2506.16161 [hep-lat]} \BibitemShut {NoStop}%
\bibitem [{\citenamefont {Liu}\ and\ \citenamefont {Dong}(1994)}]{Liu:1993cv}%
  \BibitemOpen
  \bibfield  {author} {\bibinfo {author} {\bibfnamefont {K.-F.}\ \bibnamefont {Liu}}\ and\ \bibinfo {author} {\bibfnamefont {S.-J.}\ \bibnamefont {Dong}},\ }\href {\doibase 10.1103/PhysRevLett.72.1790} {\bibfield  {journal} {\bibinfo  {journal} {Phys. Rev. Lett.}\ }\textbf {\bibinfo {volume} {72}},\ \bibinfo {pages} {1790} (\bibinfo {year} {1994})},\ \Eprint {http://arxiv.org/abs/hep-ph/9306299} {arXiv:hep-ph/9306299} \BibitemShut {NoStop}%
\bibitem [{\citenamefont {Aglietti}\ \emph {et~al.}(1998)\citenamefont {Aglietti}, \citenamefont {Ciuchini}, \citenamefont {Corbo}, \citenamefont {Franco}, \citenamefont {Martinelli},\ and\ \citenamefont {Silvestrini}}]{Aglietti:1998ur}%
  \BibitemOpen
  \bibfield  {author} {\bibinfo {author} {\bibfnamefont {U.}~\bibnamefont {Aglietti}}, \bibinfo {author} {\bibfnamefont {M.}~\bibnamefont {Ciuchini}}, \bibinfo {author} {\bibfnamefont {G.}~\bibnamefont {Corbo}}, \bibinfo {author} {\bibfnamefont {E.}~\bibnamefont {Franco}}, \bibinfo {author} {\bibfnamefont {G.}~\bibnamefont {Martinelli}}, \ and\ \bibinfo {author} {\bibfnamefont {L.}~\bibnamefont {Silvestrini}},\ }\href {\doibase 10.1016/S0370-2693(98)01138-1} {\bibfield  {journal} {\bibinfo  {journal} {Phys. Lett. B}\ }\textbf {\bibinfo {volume} {441}},\ \bibinfo {pages} {371} (\bibinfo {year} {1998})},\ \Eprint {http://arxiv.org/abs/hep-ph/9806277} {arXiv:hep-ph/9806277} \BibitemShut {NoStop}%
\bibitem [{\citenamefont {Detmold}\ and\ \citenamefont {Lin}(2006)}]{Detmold:2005gg}%
  \BibitemOpen
  \bibfield  {author} {\bibinfo {author} {\bibfnamefont {W.}~\bibnamefont {Detmold}}\ and\ \bibinfo {author} {\bibfnamefont {C.~J.~D.}\ \bibnamefont {Lin}},\ }\href {\doibase 10.1103/PhysRevD.73.014501} {\bibfield  {journal} {\bibinfo  {journal} {Phys. Rev.}\ }\textbf {\bibinfo {volume} {D73}},\ \bibinfo {pages} {014501} (\bibinfo {year} {2006})},\ \Eprint {http://arxiv.org/abs/hep-lat/0507007} {arXiv:hep-lat/0507007 [hep-lat]} \BibitemShut {NoStop}%
\bibitem [{\citenamefont {Braun}\ and\ \citenamefont {M{\"u}ller}(2008)}]{Braun:2007wv}%
  \BibitemOpen
  \bibfield  {author} {\bibinfo {author} {\bibfnamefont {V.}~\bibnamefont {Braun}}\ and\ \bibinfo {author} {\bibfnamefont {D.}~\bibnamefont {M{\"u}ller}},\ }\href {\doibase 10.1140/epjc/s10052-008-0608-4} {\bibfield  {journal} {\bibinfo  {journal} {Eur. Phys. J.}\ }\textbf {\bibinfo {volume} {C55}},\ \bibinfo {pages} {349} (\bibinfo {year} {2008})},\ \Eprint {http://arxiv.org/abs/0709.1348} {arXiv:0709.1348 [hep-ph]} \BibitemShut {NoStop}%
\bibitem [{\citenamefont {Ji}(2013)}]{Ji:2013dva}%
  \BibitemOpen
  \bibfield  {author} {\bibinfo {author} {\bibfnamefont {X.}~\bibnamefont {Ji}},\ }\href {\doibase 10.1103/PhysRevLett.110.262002} {\bibfield  {journal} {\bibinfo  {journal} {Phys. Rev. Lett.}\ }\textbf {\bibinfo {volume} {110}},\ \bibinfo {pages} {262002} (\bibinfo {year} {2013})},\ \Eprint {http://arxiv.org/abs/1305.1539} {arXiv:1305.1539 [hep-ph]} \BibitemShut {NoStop}%
\bibitem [{\citenamefont {Monahan}\ and\ \citenamefont {Orginos}(2015)}]{Monahan:2015lha}%
  \BibitemOpen
  \bibfield  {author} {\bibinfo {author} {\bibfnamefont {C.}~\bibnamefont {Monahan}}\ and\ \bibinfo {author} {\bibfnamefont {K.}~\bibnamefont {Orginos}},\ }\href {\doibase 10.1103/PhysRevD.91.074513} {\bibfield  {journal} {\bibinfo  {journal} {Phys. Rev. D}\ }\textbf {\bibinfo {volume} {91}},\ \bibinfo {pages} {074513} (\bibinfo {year} {2015})},\ \Eprint {http://arxiv.org/abs/1501.05348} {arXiv:1501.05348 [hep-lat]} \BibitemShut {NoStop}%
\bibitem [{\citenamefont {Monahan}\ and\ \citenamefont {Orginos}(2017)}]{Monahan:2016bvm}%
  \BibitemOpen
  \bibfield  {author} {\bibinfo {author} {\bibfnamefont {C.}~\bibnamefont {Monahan}}\ and\ \bibinfo {author} {\bibfnamefont {K.}~\bibnamefont {Orginos}},\ }\href {\doibase 10.1007/JHEP03(2017)116} {\bibfield  {journal} {\bibinfo  {journal} {JHEP}\ }\textbf {\bibinfo {volume} {03}},\ \bibinfo {pages} {116} (\bibinfo {year} {2017})},\ \Eprint {http://arxiv.org/abs/1612.01584} {arXiv:1612.01584 [hep-lat]} \BibitemShut {NoStop}%
\bibitem [{\citenamefont {Radyushkin}(2017)}]{Radyushkin:2017cyf}%
  \BibitemOpen
  \bibfield  {author} {\bibinfo {author} {\bibfnamefont {A.~V.}\ \bibnamefont {Radyushkin}},\ }\href {\doibase 10.1103/PhysRevD.96.034025} {\bibfield  {journal} {\bibinfo  {journal} {Phys. Rev.}\ }\textbf {\bibinfo {volume} {D96}},\ \bibinfo {pages} {034025} (\bibinfo {year} {2017})},\ \Eprint {http://arxiv.org/abs/1705.01488} {arXiv:1705.01488 [hep-ph]} \BibitemShut {NoStop}%
\bibitem [{\citenamefont {Chambers}\ \emph {et~al.}(2017)\citenamefont {Chambers}, \citenamefont {Horsley}, \citenamefont {Nakamura}, \citenamefont {Perlt}, \citenamefont {Rakow}, \citenamefont {Schierholz}, \citenamefont {Schiller}, \citenamefont {Somfleth}, \citenamefont {Young},\ and\ \citenamefont {Zanotti}}]{Chambers:2017dov}%
  \BibitemOpen
  \bibfield  {author} {\bibinfo {author} {\bibfnamefont {A.~J.}\ \bibnamefont {Chambers}}, \bibinfo {author} {\bibfnamefont {R.}~\bibnamefont {Horsley}}, \bibinfo {author} {\bibfnamefont {Y.}~\bibnamefont {Nakamura}}, \bibinfo {author} {\bibfnamefont {H.}~\bibnamefont {Perlt}}, \bibinfo {author} {\bibfnamefont {P.~E.~L.}\ \bibnamefont {Rakow}}, \bibinfo {author} {\bibfnamefont {G.}~\bibnamefont {Schierholz}}, \bibinfo {author} {\bibfnamefont {A.}~\bibnamefont {Schiller}}, \bibinfo {author} {\bibfnamefont {K.}~\bibnamefont {Somfleth}}, \bibinfo {author} {\bibfnamefont {R.~D.}\ \bibnamefont {Young}}, \ and\ \bibinfo {author} {\bibfnamefont {J.~M.}\ \bibnamefont {Zanotti}},\ }\href {\doibase 10.1103/PhysRevLett.118.242001} {\bibfield  {journal} {\bibinfo  {journal} {Phys. Rev. Lett.}\ }\textbf {\bibinfo {volume} {118}},\ \bibinfo {pages} {242001} (\bibinfo {year} {2017})},\ \Eprint {http://arxiv.org/abs/1703.01153} {arXiv:1703.01153 [hep-lat]} \BibitemShut {NoStop}%
\bibitem [{\citenamefont {Karpie}\ \emph {et~al.}(2018)\citenamefont {Karpie}, \citenamefont {Orginos},\ and\ \citenamefont {Zafeiropoulos}}]{Karpie:2018zaz}%
  \BibitemOpen
  \bibfield  {author} {\bibinfo {author} {\bibfnamefont {J.}~\bibnamefont {Karpie}}, \bibinfo {author} {\bibfnamefont {K.}~\bibnamefont {Orginos}}, \ and\ \bibinfo {author} {\bibfnamefont {S.}~\bibnamefont {Zafeiropoulos}},\ }\href {\doibase 10.1007/JHEP11(2018)178} {\bibfield  {journal} {\bibinfo  {journal} {JHEP}\ }\textbf {\bibinfo {volume} {11}},\ \bibinfo {pages} {178} (\bibinfo {year} {2018})},\ \Eprint {http://arxiv.org/abs/1807.10933} {arXiv:1807.10933 [hep-lat]} \BibitemShut {NoStop}%
\bibitem [{\citenamefont {Ma}\ and\ \citenamefont {Qiu}(2018)}]{Ma:2017pxb}%
  \BibitemOpen
  \bibfield  {author} {\bibinfo {author} {\bibfnamefont {Y.-Q.}\ \bibnamefont {Ma}}\ and\ \bibinfo {author} {\bibfnamefont {J.-W.}\ \bibnamefont {Qiu}},\ }\href {\doibase 10.1103/PhysRevLett.120.022003} {\bibfield  {journal} {\bibinfo  {journal} {Phys. Rev. Lett.}\ }\textbf {\bibinfo {volume} {120}},\ \bibinfo {pages} {022003} (\bibinfo {year} {2018})},\ \Eprint {http://arxiv.org/abs/1709.03018} {arXiv:1709.03018 [hep-ph]} \BibitemShut {NoStop}%
\bibitem [{\citenamefont {Gao}\ \emph {et~al.}(2024)\citenamefont {Gao}, \citenamefont {Liu},\ and\ \citenamefont {Zhao}}]{Gao:2023lny}%
  \BibitemOpen
  \bibfield  {author} {\bibinfo {author} {\bibfnamefont {X.}~\bibnamefont {Gao}}, \bibinfo {author} {\bibfnamefont {W.-Y.}\ \bibnamefont {Liu}}, \ and\ \bibinfo {author} {\bibfnamefont {Y.}~\bibnamefont {Zhao}},\ }\href {\doibase 10.1103/PhysRevD.109.094506} {\bibfield  {journal} {\bibinfo  {journal} {Phys. Rev. D}\ }\textbf {\bibinfo {volume} {109}},\ \bibinfo {pages} {094506} (\bibinfo {year} {2024})},\ \Eprint {http://arxiv.org/abs/2306.14960} {arXiv:2306.14960 [hep-ph]} \BibitemShut {NoStop}%
\bibitem [{\citenamefont {Shindler}(2024)}]{Shindler:2023xpd}%
  \BibitemOpen
  \bibfield  {author} {\bibinfo {author} {\bibfnamefont {A.}~\bibnamefont {Shindler}},\ }\href {\doibase 10.1103/PhysRevD.110.L051503} {\bibfield  {journal} {\bibinfo  {journal} {Phys. Rev. D}\ }\textbf {\bibinfo {volume} {110}},\ \bibinfo {pages} {L051503} (\bibinfo {year} {2024})},\ \Eprint {http://arxiv.org/abs/2311.18704} {arXiv:2311.18704 [hep-lat]} \BibitemShut {NoStop}%
\bibitem [{\citenamefont {Dutrieux}\ \emph {et~al.}(2024)\citenamefont {Dutrieux}, \citenamefont {Karpie}, \citenamefont {Monahan}, \citenamefont {Orginos},\ and\ \citenamefont {Zafeiropoulos}}]{Dutrieux:2023zpy}%
  \BibitemOpen
  \bibfield  {author} {\bibinfo {author} {\bibfnamefont {H.}~\bibnamefont {Dutrieux}}, \bibinfo {author} {\bibfnamefont {J.}~\bibnamefont {Karpie}}, \bibinfo {author} {\bibfnamefont {C.}~\bibnamefont {Monahan}}, \bibinfo {author} {\bibfnamefont {K.}~\bibnamefont {Orginos}}, \ and\ \bibinfo {author} {\bibfnamefont {S.}~\bibnamefont {Zafeiropoulos}} (\bibinfo {collaboration} {HadStruc}),\ }\href {\doibase 10.1007/JHEP04(2024)061} {\bibfield  {journal} {\bibinfo  {journal} {JHEP}\ }\textbf {\bibinfo {volume} {04}},\ \bibinfo {pages} {061} (\bibinfo {year} {2024})},\ \Eprint {http://arxiv.org/abs/2310.19926} {arXiv:2310.19926 [hep-lat]} \BibitemShut {NoStop}%
\bibitem [{\citenamefont {Karpie}\ \emph {et~al.}(2019)\citenamefont {Karpie}, \citenamefont {Orginos}, \citenamefont {Rothkopf},\ and\ \citenamefont {Zafeiropoulos}}]{Karpie:2019eiq}%
  \BibitemOpen
  \bibfield  {author} {\bibinfo {author} {\bibfnamefont {J.}~\bibnamefont {Karpie}}, \bibinfo {author} {\bibfnamefont {K.}~\bibnamefont {Orginos}}, \bibinfo {author} {\bibfnamefont {A.}~\bibnamefont {Rothkopf}}, \ and\ \bibinfo {author} {\bibfnamefont {S.}~\bibnamefont {Zafeiropoulos}},\ }\href {\doibase 10.1007/JHEP04(2019)057} {\bibfield  {journal} {\bibinfo  {journal} {JHEP}\ }\textbf {\bibinfo {volume} {04}},\ \bibinfo {pages} {057} (\bibinfo {year} {2019})},\ \Eprint {http://arxiv.org/abs/1901.05408} {arXiv:1901.05408 [hep-lat]} \BibitemShut {NoStop}%
\bibitem [{\citenamefont {Xiong}\ \emph {et~al.}(2025)\citenamefont {Xiong}, \citenamefont {Hua}, \citenamefont {Wei}, \citenamefont {Yu}, \citenamefont {Zhang},\ and\ \citenamefont {Zheng}}]{Xiong:2025obq}%
  \BibitemOpen
  \bibfield  {author} {\bibinfo {author} {\bibfnamefont {A.-S.}\ \bibnamefont {Xiong}}, \bibinfo {author} {\bibfnamefont {J.}~\bibnamefont {Hua}}, \bibinfo {author} {\bibfnamefont {T.}~\bibnamefont {Wei}}, \bibinfo {author} {\bibfnamefont {F.-S.}\ \bibnamefont {Yu}}, \bibinfo {author} {\bibfnamefont {Q.-A.}\ \bibnamefont {Zhang}}, \ and\ \bibinfo {author} {\bibfnamefont {Y.}~\bibnamefont {Zheng}},\ }\href@noop {} {\  (\bibinfo {year} {2025})},\ \Eprint {http://arxiv.org/abs/2506.16689} {arXiv:2506.16689 [hep-lat]} \BibitemShut {NoStop}%
\bibitem [{\citenamefont {Bailey}\ \emph {et~al.}(2021)\citenamefont {Bailey}, \citenamefont {Cridge}, \citenamefont {Harland-Lang}, \citenamefont {Martin},\ and\ \citenamefont {Thorne}}]{Bailey:2020ooq}%
  \BibitemOpen
  \bibfield  {author} {\bibinfo {author} {\bibfnamefont {S.}~\bibnamefont {Bailey}}, \bibinfo {author} {\bibfnamefont {T.}~\bibnamefont {Cridge}}, \bibinfo {author} {\bibfnamefont {L.~A.}\ \bibnamefont {Harland-Lang}}, \bibinfo {author} {\bibfnamefont {A.~D.}\ \bibnamefont {Martin}}, \ and\ \bibinfo {author} {\bibfnamefont {R.~S.}\ \bibnamefont {Thorne}},\ }\href {\doibase 10.1140/epjc/s10052-021-09057-0} {\bibfield  {journal} {\bibinfo  {journal} {Eur. Phys. J. C}\ }\textbf {\bibinfo {volume} {81}},\ \bibinfo {pages} {341} (\bibinfo {year} {2021})},\ \Eprint {http://arxiv.org/abs/2012.04684} {arXiv:2012.04684 [hep-ph]} \BibitemShut {NoStop}%
\bibitem [{\citenamefont {Moffat}\ \emph {et~al.}(2021)\citenamefont {Moffat}, \citenamefont {Melnitchouk}, \citenamefont {Rogers},\ and\ \citenamefont {Sato}}]{PhysRevD.104.016015}%
  \BibitemOpen
  \bibfield  {author} {\bibinfo {author} {\bibfnamefont {E.}~\bibnamefont {Moffat}}, \bibinfo {author} {\bibfnamefont {W.}~\bibnamefont {Melnitchouk}}, \bibinfo {author} {\bibfnamefont {T.~C.}\ \bibnamefont {Rogers}}, \ and\ \bibinfo {author} {\bibfnamefont {N.}~\bibnamefont {Sato}} (\bibinfo {collaboration} {Jefferson Lab Angular Momentum (JAM) Collaboration}),\ }\href {\doibase 10.1103/PhysRevD.104.016015} {\bibfield  {journal} {\bibinfo  {journal} {Phys. Rev. D}\ }\textbf {\bibinfo {volume} {104}},\ \bibinfo {pages} {016015} (\bibinfo {year} {2021})}\BibitemShut {NoStop}%
\bibitem [{\citenamefont {{Ball, Richard D.}}\ \emph {et~al.}(2022)\citenamefont {{Ball, Richard D.}}, \citenamefont {{Carrazza, Stefano}}, \citenamefont {{Cruz-Martinez, Juan}}, \citenamefont {{Del Debbio, Luigi}}, \citenamefont {{Forte, Stefano}}, \citenamefont {{Giani, Tommaso}}, \citenamefont {{Iranipour, Shayan}}, \citenamefont {{Kassabov, Zahari}}, \citenamefont {{Latorre, Jose I.}}, \citenamefont {{Nocera, Emanuele R.}}, \citenamefont {{Pearson, Rosalyn L.}}, \citenamefont {{Rojo, Juan}}, \citenamefont {{Stegeman, Roy}}, \citenamefont {{Schwan, Christopher}}, \citenamefont {{Ubiali, Maria}}, \citenamefont {{Voisey, Cameron}},\ and\ \citenamefont {{Wilson, Michael}}}]{refId0}%
  \BibitemOpen
  \bibfield  {author} {\bibinfo {author} {\bibnamefont {{Ball, Richard D.}}}, \bibinfo {author} {\bibnamefont {{Carrazza, Stefano}}}, \bibinfo {author} {\bibnamefont {{Cruz-Martinez, Juan}}}, \bibinfo {author} {\bibnamefont {{Del Debbio, Luigi}}}, \bibinfo {author} {\bibnamefont {{Forte, Stefano}}}, \bibinfo {author} {\bibnamefont {{Giani, Tommaso}}}, \bibinfo {author} {\bibnamefont {{Iranipour, Shayan}}}, \bibinfo {author} {\bibnamefont {{Kassabov, Zahari}}}, \bibinfo {author} {\bibnamefont {{Latorre, Jose I.}}}, \bibinfo {author} {\bibnamefont {{Nocera, Emanuele R.}}}, \bibinfo {author} {\bibnamefont {{Pearson, Rosalyn L.}}}, \bibinfo {author} {\bibnamefont {{Rojo, Juan}}}, \bibinfo {author} {\bibnamefont {{Stegeman, Roy}}}, \bibinfo {author} {\bibnamefont {{Schwan, Christopher}}}, \bibinfo {author} {\bibnamefont {{Ubiali, Maria}}}, \bibinfo {author} {\bibnamefont {{Voisey, Cameron}}}, \ and\ \bibinfo {author} {\bibnamefont {{Wilson, Michael}}},\ }\href {\doibase 10.1140/epjc/s10052-022-10328-7} {\bibfield  {journal} {\bibinfo  {journal} {Eur. Phys. J. C}\ }\textbf {\bibinfo {volume} {82}},\ \bibinfo {pages} {428} (\bibinfo {year} {2022})}\BibitemShut {NoStop}%
\bibitem [{\citenamefont {Chiefa}\ \emph {et~al.}(2025)\citenamefont {Chiefa}, \citenamefont {Costantini}, \citenamefont {Cruz-Martinez}, \citenamefont {Nocera}, \citenamefont {Rabemananjara}, \citenamefont {Rojo}, \citenamefont {Sharma}, \citenamefont {Stegeman},\ and\ \citenamefont {Ubiali}}]{Chiefa:2025loi}%
  \BibitemOpen
  \bibfield  {author} {\bibinfo {author} {\bibfnamefont {A.}~\bibnamefont {Chiefa}}, \bibinfo {author} {\bibfnamefont {M.~N.}\ \bibnamefont {Costantini}}, \bibinfo {author} {\bibfnamefont {J.}~\bibnamefont {Cruz-Martinez}}, \bibinfo {author} {\bibfnamefont {E.~R.}\ \bibnamefont {Nocera}}, \bibinfo {author} {\bibfnamefont {T.~R.}\ \bibnamefont {Rabemananjara}}, \bibinfo {author} {\bibfnamefont {J.}~\bibnamefont {Rojo}}, \bibinfo {author} {\bibfnamefont {T.}~\bibnamefont {Sharma}}, \bibinfo {author} {\bibfnamefont {R.}~\bibnamefont {Stegeman}}, \ and\ \bibinfo {author} {\bibfnamefont {M.}~\bibnamefont {Ubiali}},\ }\href {\doibase 10.1007/JHEP07(2025)067} {\bibfield  {journal} {\bibinfo  {journal} {JHEP}\ }\textbf {\bibinfo {volume} {07}},\ \bibinfo {pages} {067} (\bibinfo {year} {2025})},\ \Eprint {http://arxiv.org/abs/2501.10359} {arXiv:2501.10359 [hep-ph]} \BibitemShut {NoStop}%
\bibitem [{\citenamefont {Lin}\ \emph {et~al.}(2018)\citenamefont {Lin} \emph {et~al.}}]{Lin:2017snn}%
  \BibitemOpen
  \bibfield  {author} {\bibinfo {author} {\bibfnamefont {H.-W.}\ \bibnamefont {Lin}} \emph {et~al.},\ }\href {\doibase 10.1016/j.ppnp.2018.01.007} {\bibfield  {journal} {\bibinfo  {journal} {Prog. Part. Nucl. Phys.}\ }\textbf {\bibinfo {volume} {100}},\ \bibinfo {pages} {107} (\bibinfo {year} {2018})},\ \Eprint {http://arxiv.org/abs/1711.07916} {arXiv:1711.07916 [hep-ph]} \BibitemShut {NoStop}%
\bibitem [{\citenamefont {Constantinou}\ \emph {et~al.}(2021)\citenamefont {Constantinou} \emph {et~al.}}]{Constantinou:2020hdm}%
  \BibitemOpen
  \bibfield  {author} {\bibinfo {author} {\bibfnamefont {M.}~\bibnamefont {Constantinou}} \emph {et~al.},\ }\href {\doibase 10.1016/j.ppnp.2021.103908} {\bibfield  {journal} {\bibinfo  {journal} {Prog. Part. Nucl. Phys.}\ }\textbf {\bibinfo {volume} {121}},\ \bibinfo {pages} {103908} (\bibinfo {year} {2021})},\ \Eprint {http://arxiv.org/abs/2006.08636} {arXiv:2006.08636 [hep-ph]} \BibitemShut {NoStop}%
\bibitem [{\citenamefont {Constantinou}(2021)}]{Constantinou:2020pek}%
  \BibitemOpen
  \bibfield  {author} {\bibinfo {author} {\bibfnamefont {M.}~\bibnamefont {Constantinou}},\ }\href {\doibase 10.1140/epja/s10050-021-00353-7} {\bibfield  {journal} {\bibinfo  {journal} {Eur. Phys. J. A}\ }\textbf {\bibinfo {volume} {57}},\ \bibinfo {pages} {77} (\bibinfo {year} {2021})},\ \Eprint {http://arxiv.org/abs/2010.02445} {arXiv:2010.02445 [hep-lat]} \BibitemShut {NoStop}%
\bibitem [{\citenamefont {Lin}(2023)}]{Lin:2023kxn}%
  \BibitemOpen
  \bibfield  {author} {\bibinfo {author} {\bibfnamefont {H.-W.}\ \bibnamefont {Lin}},\ }\href {\doibase 10.1007/s00601-023-01842-9} {\bibfield  {journal} {\bibinfo  {journal} {Few Body Syst.}\ }\textbf {\bibinfo {volume} {64}},\ \bibinfo {pages} {58} (\bibinfo {year} {2023})}\BibitemShut {NoStop}%
\bibitem [{\citenamefont {Rasmussen}\ and\ \citenamefont {Williams}(2006{\natexlab{a}})}]{RasmussenW06}%
  \BibitemOpen
  \bibfield  {author} {\bibinfo {author} {\bibfnamefont {C.~E.}\ \bibnamefont {Rasmussen}}\ and\ \bibinfo {author} {\bibfnamefont {C.~K.~I.}\ \bibnamefont {Williams}},\ }\href@noop {} {\emph {\bibinfo {title} {Gaussian processes for machine learning.}}},\ Adaptive computation and machine learning\ (\bibinfo  {publisher} {MIT Press},\ \bibinfo {year} {2006})\ pp.\ \bibinfo {pages} {I--XVIII, 1--248}\BibitemShut {NoStop}%
\bibitem [{\citenamefont {Gramacy}(2020)}]{gramacy2020surrogates}%
  \BibitemOpen
  \bibfield  {author} {\bibinfo {author} {\bibfnamefont {R.~B.}\ \bibnamefont {Gramacy}},\ }\href@noop {} {\emph {\bibinfo {title} {Surrogates: {G}aussian Process Modeling, Design and Optimization for the Applied Sciences}}}\ (\bibinfo  {publisher} {Chapman Hall/CRC},\ \bibinfo {address} {Boca Raton, Florida},\ \bibinfo {year} {2020})\ \bibinfo {note} {\url{http://bobby.gramacy.com/surrogates/}}\BibitemShut {NoStop}%
\bibitem [{\citenamefont {Alexandrou}\ \emph {et~al.}(2020)\citenamefont {Alexandrou}, \citenamefont {Iannelli}, \citenamefont {Jansen},\ and\ \citenamefont {Manigrasso}}]{Alexandrou:2020tqq}%
  \BibitemOpen
  \bibfield  {author} {\bibinfo {author} {\bibfnamefont {C.}~\bibnamefont {Alexandrou}}, \bibinfo {author} {\bibfnamefont {G.}~\bibnamefont {Iannelli}}, \bibinfo {author} {\bibfnamefont {K.}~\bibnamefont {Jansen}}, \ and\ \bibinfo {author} {\bibfnamefont {F.}~\bibnamefont {Manigrasso}} (\bibinfo {collaboration} {Extended Twisted Mass}),\ }\href {\doibase 10.1103/PhysRevD.102.094508} {\bibfield  {journal} {\bibinfo  {journal} {Phys. Rev. D}\ }\textbf {\bibinfo {volume} {102}},\ \bibinfo {pages} {094508} (\bibinfo {year} {2020})},\ \Eprint {http://arxiv.org/abs/2007.13800} {arXiv:2007.13800 [hep-lat]} \BibitemShut {NoStop}%
\bibitem [{\citenamefont {Candido}\ \emph {et~al.}(2024)\citenamefont {Candido}, \citenamefont {Del~Debbio}, \citenamefont {Giani},\ and\ \citenamefont {Petrillo}}]{Candido:2024hjt}%
  \BibitemOpen
  \bibfield  {author} {\bibinfo {author} {\bibfnamefont {A.}~\bibnamefont {Candido}}, \bibinfo {author} {\bibfnamefont {L.}~\bibnamefont {Del~Debbio}}, \bibinfo {author} {\bibfnamefont {T.}~\bibnamefont {Giani}}, \ and\ \bibinfo {author} {\bibfnamefont {G.}~\bibnamefont {Petrillo}},\ }\href {\doibase 10.1140/epjc/s10052-024-13100-1} {\bibfield  {journal} {\bibinfo  {journal} {Eur. Phys. J. C}\ }\textbf {\bibinfo {volume} {84}},\ \bibinfo {pages} {716} (\bibinfo {year} {2024})},\ \Eprint {http://arxiv.org/abs/2404.07573} {arXiv:2404.07573 [hep-ph]} \BibitemShut {NoStop}%
\bibitem [{\citenamefont {Dutrieux}\ \emph {et~al.}(2025{\natexlab{a}})\citenamefont {Dutrieux}, \citenamefont {Karpie}, \citenamefont {Orginos},\ and\ \citenamefont {Zafeiropoulos}}]{Dutrieux:2024rem}%
  \BibitemOpen
  \bibfield  {author} {\bibinfo {author} {\bibfnamefont {H.}~\bibnamefont {Dutrieux}}, \bibinfo {author} {\bibfnamefont {J.}~\bibnamefont {Karpie}}, \bibinfo {author} {\bibfnamefont {K.}~\bibnamefont {Orginos}}, \ and\ \bibinfo {author} {\bibfnamefont {S.}~\bibnamefont {Zafeiropoulos}},\ }\href {\doibase 10.1103/PhysRevD.111.034515} {\bibfield  {journal} {\bibinfo  {journal} {Phys. Rev. D}\ }\textbf {\bibinfo {volume} {111}},\ \bibinfo {pages} {034515} (\bibinfo {year} {2025}{\natexlab{a}})},\ \Eprint {http://arxiv.org/abs/2412.05227} {arXiv:2412.05227 [hep-lat]} \BibitemShut {NoStop}%
\bibitem [{\citenamefont {Dutrieux}\ \emph {et~al.}(2025{\natexlab{b}})\citenamefont {Dutrieux}, \citenamefont {Karpie}, \citenamefont {Monahan}, \citenamefont {Orginos}, \citenamefont {Radyushkin}, \citenamefont {Richards},\ and\ \citenamefont {Zafeiropoulos}}]{Dutrieux:2025jed}%
  \BibitemOpen
  \bibfield  {author} {\bibinfo {author} {\bibfnamefont {H.}~\bibnamefont {Dutrieux}}, \bibinfo {author} {\bibfnamefont {J.}~\bibnamefont {Karpie}}, \bibinfo {author} {\bibfnamefont {C.~J.}\ \bibnamefont {Monahan}}, \bibinfo {author} {\bibfnamefont {K.}~\bibnamefont {Orginos}}, \bibinfo {author} {\bibfnamefont {A.}~\bibnamefont {Radyushkin}}, \bibinfo {author} {\bibfnamefont {D.}~\bibnamefont {Richards}}, \ and\ \bibinfo {author} {\bibfnamefont {S.}~\bibnamefont {Zafeiropoulos}},\ }\href@noop {} {\  (\bibinfo {year} {2025}{\natexlab{b}})},\ \Eprint {http://arxiv.org/abs/2504.17706} {arXiv:2504.17706 [hep-lat]} \BibitemShut {NoStop}%
\bibitem [{\citenamefont {Horak}\ \emph {et~al.}(2022)\citenamefont {Horak}, \citenamefont {Pawlowski}, \citenamefont {Rodr\'\i{}guez-Quintero}, \citenamefont {Turnwald}, \citenamefont {Urban}, \citenamefont {Wink},\ and\ \citenamefont {Zafeiropoulos}}]{Horak:2021syv}%
  \BibitemOpen
  \bibfield  {author} {\bibinfo {author} {\bibfnamefont {J.}~\bibnamefont {Horak}}, \bibinfo {author} {\bibfnamefont {J.~M.}\ \bibnamefont {Pawlowski}}, \bibinfo {author} {\bibfnamefont {J.}~\bibnamefont {Rodr\'\i{}guez-Quintero}}, \bibinfo {author} {\bibfnamefont {J.}~\bibnamefont {Turnwald}}, \bibinfo {author} {\bibfnamefont {J.~M.}\ \bibnamefont {Urban}}, \bibinfo {author} {\bibfnamefont {N.}~\bibnamefont {Wink}}, \ and\ \bibinfo {author} {\bibfnamefont {S.}~\bibnamefont {Zafeiropoulos}},\ }\href {\doibase 10.1103/PhysRevD.105.036014} {\bibfield  {journal} {\bibinfo  {journal} {Phys. Rev. D}\ }\textbf {\bibinfo {volume} {105}},\ \bibinfo {pages} {036014} (\bibinfo {year} {2022})},\ \Eprint {http://arxiv.org/abs/2107.13464} {arXiv:2107.13464 [hep-ph]} \BibitemShut {NoStop}%
\bibitem [{\citenamefont {Pawlowski}\ \emph {et~al.}(2023)\citenamefont {Pawlowski}, \citenamefont {Schneider}, \citenamefont {Turnwald}, \citenamefont {Urban},\ and\ \citenamefont {Wink}}]{Pawlowski:2022zhh}%
  \BibitemOpen
  \bibfield  {author} {\bibinfo {author} {\bibfnamefont {J.~M.}\ \bibnamefont {Pawlowski}}, \bibinfo {author} {\bibfnamefont {C.~S.}\ \bibnamefont {Schneider}}, \bibinfo {author} {\bibfnamefont {J.}~\bibnamefont {Turnwald}}, \bibinfo {author} {\bibfnamefont {J.~M.}\ \bibnamefont {Urban}}, \ and\ \bibinfo {author} {\bibfnamefont {N.}~\bibnamefont {Wink}},\ }\href {\doibase 10.1103/PhysRevD.108.076018} {\bibfield  {journal} {\bibinfo  {journal} {Phys. Rev. D}\ }\textbf {\bibinfo {volume} {108}},\ \bibinfo {pages} {076018} (\bibinfo {year} {2023})},\ \Eprint {http://arxiv.org/abs/2212.01113} {arXiv:2212.01113 [hep-ph]} \BibitemShut {NoStop}%
\bibitem [{\citenamefont {Horak}\ \emph {et~al.}(2023)\citenamefont {Horak}, \citenamefont {Pawlowski}, \citenamefont {Turnwald}, \citenamefont {Urban}, \citenamefont {Wink},\ and\ \citenamefont {Zafeiropoulos}}]{Horak:2023xfb}%
  \BibitemOpen
  \bibfield  {author} {\bibinfo {author} {\bibfnamefont {J.}~\bibnamefont {Horak}}, \bibinfo {author} {\bibfnamefont {J.~M.}\ \bibnamefont {Pawlowski}}, \bibinfo {author} {\bibfnamefont {J.}~\bibnamefont {Turnwald}}, \bibinfo {author} {\bibfnamefont {J.~M.}\ \bibnamefont {Urban}}, \bibinfo {author} {\bibfnamefont {N.}~\bibnamefont {Wink}}, \ and\ \bibinfo {author} {\bibfnamefont {S.}~\bibnamefont {Zafeiropoulos}},\ }\href {\doibase 10.1103/PhysRevD.107.076019} {\bibfield  {journal} {\bibinfo  {journal} {Phys. Rev. D}\ }\textbf {\bibinfo {volume} {107}},\ \bibinfo {pages} {076019} (\bibinfo {year} {2023})},\ \Eprint {http://arxiv.org/abs/2301.07785} {arXiv:2301.07785 [hep-ph]} \BibitemShut {NoStop}%
\bibitem [{\citenamefont {Del~Debbio}\ \emph {et~al.}(2025)\citenamefont {Del~Debbio}, \citenamefont {Lupo}, \citenamefont {Panero},\ and\ \citenamefont {Tantalo}}]{DelDebbio:2024lwm}%
  \BibitemOpen
  \bibfield  {author} {\bibinfo {author} {\bibfnamefont {L.}~\bibnamefont {Del~Debbio}}, \bibinfo {author} {\bibfnamefont {A.}~\bibnamefont {Lupo}}, \bibinfo {author} {\bibfnamefont {M.}~\bibnamefont {Panero}}, \ and\ \bibinfo {author} {\bibfnamefont {N.}~\bibnamefont {Tantalo}},\ }\href {\doibase 10.1140/epjc/s10052-025-13885-9} {\bibfield  {journal} {\bibinfo  {journal} {Eur. Phys. J. C}\ }\textbf {\bibinfo {volume} {85}},\ \bibinfo {pages} {185} (\bibinfo {year} {2025})},\ \Eprint {http://arxiv.org/abs/2409.04413} {arXiv:2409.04413 [hep-lat]} \BibitemShut {NoStop}%
\bibitem [{\citenamefont {Jay}(2025)}]{Jay:2025dzl}%
  \BibitemOpen
  \bibfield  {author} {\bibinfo {author} {\bibfnamefont {W.}~\bibnamefont {Jay}},\ }in\ \href@noop {} {\emph {\bibinfo {booktitle} {{41st International Symposium on Lattice Field Theory}}}}\ (\bibinfo {year} {2025})\ \Eprint {http://arxiv.org/abs/2501.12259} {arXiv:2501.12259 [hep-lat]} \BibitemShut {NoStop}%
\bibitem [{\citenamefont {Braun}\ \emph {et~al.}(1995)\citenamefont {Braun}, \citenamefont {Gornicki},\ and\ \citenamefont {Mankiewicz}}]{Braun:1994jq}%
  \BibitemOpen
  \bibfield  {author} {\bibinfo {author} {\bibfnamefont {V.}~\bibnamefont {Braun}}, \bibinfo {author} {\bibfnamefont {P.}~\bibnamefont {Gornicki}}, \ and\ \bibinfo {author} {\bibfnamefont {L.}~\bibnamefont {Mankiewicz}},\ }\href {\doibase 10.1103/PhysRevD.51.6036} {\bibfield  {journal} {\bibinfo  {journal} {Phys. Rev.}\ }\textbf {\bibinfo {volume} {D51}},\ \bibinfo {pages} {6036} (\bibinfo {year} {1995})},\ \Eprint {http://arxiv.org/abs/hep-ph/9410318} {arXiv:hep-ph/9410318 [hep-ph]} \BibitemShut {NoStop}%
\bibitem [{\citenamefont {Radyushkin}(2020)}]{Radyushkin:2019mye}%
  \BibitemOpen
  \bibfield  {author} {\bibinfo {author} {\bibfnamefont {A.}~\bibnamefont {Radyushkin}},\ }\href {\doibase 10.1142/S0217751X20300021} {\bibfield  {journal} {\bibinfo  {journal} {Int. J. Mod. Phys. A}\ }\textbf {\bibinfo {volume} {35}},\ \bibinfo {pages} {2030002} (\bibinfo {year} {2020})},\ \Eprint {http://arxiv.org/abs/1912.04244} {arXiv:1912.04244 [hep-ph]} \BibitemShut {NoStop}%
\bibitem [{\citenamefont {Sato}\ \emph {et~al.}(2016)\citenamefont {Sato}, \citenamefont {Melnitchouk}, \citenamefont {Kuhn}, \citenamefont {Ethier},\ and\ \citenamefont {Accardi}}]{Sato:2016tuz}%
  \BibitemOpen
  \bibfield  {author} {\bibinfo {author} {\bibfnamefont {N.}~\bibnamefont {Sato}}, \bibinfo {author} {\bibfnamefont {W.}~\bibnamefont {Melnitchouk}}, \bibinfo {author} {\bibfnamefont {S.~E.}\ \bibnamefont {Kuhn}}, \bibinfo {author} {\bibfnamefont {J.~J.}\ \bibnamefont {Ethier}}, \ and\ \bibinfo {author} {\bibfnamefont {A.}~\bibnamefont {Accardi}} (\bibinfo {collaboration} {Jefferson Lab Angular Momentum}),\ }\href {\doibase 10.1103/PhysRevD.93.074005} {\bibfield  {journal} {\bibinfo  {journal} {Phys. Rev. D}\ }\textbf {\bibinfo {volume} {93}},\ \bibinfo {pages} {074005} (\bibinfo {year} {2016})},\ \Eprint {http://arxiv.org/abs/1601.07782} {arXiv:1601.07782 [hep-ph]} \BibitemShut {NoStop}%
\bibitem [{\citenamefont {Martin}\ \emph {et~al.}(2009)\citenamefont {Martin}, \citenamefont {Stirling}, \citenamefont {Thorne},\ and\ \citenamefont {Watt}}]{Martin:2009iq}%
  \BibitemOpen
  \bibfield  {author} {\bibinfo {author} {\bibfnamefont {A.~D.}\ \bibnamefont {Martin}}, \bibinfo {author} {\bibfnamefont {W.~J.}\ \bibnamefont {Stirling}}, \bibinfo {author} {\bibfnamefont {R.~S.}\ \bibnamefont {Thorne}}, \ and\ \bibinfo {author} {\bibfnamefont {G.}~\bibnamefont {Watt}},\ }\href {\doibase 10.1140/epjc/s10052-009-1072-5} {\bibfield  {journal} {\bibinfo  {journal} {Eur. Phys. J. C}\ }\textbf {\bibinfo {volume} {63}},\ \bibinfo {pages} {189} (\bibinfo {year} {2009})},\ \Eprint {http://arxiv.org/abs/0901.0002} {arXiv:0901.0002 [hep-ph]} \BibitemShut {NoStop}%
\bibitem [{\citenamefont {Dulat}\ \emph {et~al.}(2016)\citenamefont {Dulat}, \citenamefont {Hou}, \citenamefont {Gao}, \citenamefont {Guzzi}, \citenamefont {Huston}, \citenamefont {Nadolsky}, \citenamefont {Pumplin}, \citenamefont {Schmidt}, \citenamefont {Stump},\ and\ \citenamefont {Yuan}}]{Dulat:2015mca}%
  \BibitemOpen
  \bibfield  {author} {\bibinfo {author} {\bibfnamefont {S.}~\bibnamefont {Dulat}}, \bibinfo {author} {\bibfnamefont {T.-J.}\ \bibnamefont {Hou}}, \bibinfo {author} {\bibfnamefont {J.}~\bibnamefont {Gao}}, \bibinfo {author} {\bibfnamefont {M.}~\bibnamefont {Guzzi}}, \bibinfo {author} {\bibfnamefont {J.}~\bibnamefont {Huston}}, \bibinfo {author} {\bibfnamefont {P.}~\bibnamefont {Nadolsky}}, \bibinfo {author} {\bibfnamefont {J.}~\bibnamefont {Pumplin}}, \bibinfo {author} {\bibfnamefont {C.}~\bibnamefont {Schmidt}}, \bibinfo {author} {\bibfnamefont {D.}~\bibnamefont {Stump}}, \ and\ \bibinfo {author} {\bibfnamefont {C.~P.}\ \bibnamefont {Yuan}},\ }\href {\doibase 10.1103/PhysRevD.93.033006} {\bibfield  {journal} {\bibinfo  {journal} {Phys. Rev. D}\ }\textbf {\bibinfo {volume} {93}},\ \bibinfo {pages} {033006} (\bibinfo {year} {2016})},\ \Eprint {http://arxiv.org/abs/1506.07443} {arXiv:1506.07443 [hep-ph]} \BibitemShut {NoStop}%
\bibitem [{\citenamefont {Forte}\ \emph {et~al.}(2002)\citenamefont {Forte}, \citenamefont {Garrido}, \citenamefont {Latorre},\ and\ \citenamefont {Piccione}}]{Forte:2002fg}%
  \BibitemOpen
  \bibfield  {author} {\bibinfo {author} {\bibfnamefont {S.}~\bibnamefont {Forte}}, \bibinfo {author} {\bibfnamefont {L.}~\bibnamefont {Garrido}}, \bibinfo {author} {\bibfnamefont {J.~I.}\ \bibnamefont {Latorre}}, \ and\ \bibinfo {author} {\bibfnamefont {A.}~\bibnamefont {Piccione}},\ }\href {\doibase 10.1088/1126-6708/2002/05/062} {\bibfield  {journal} {\bibinfo  {journal} {JHEP}\ }\textbf {\bibinfo {volume} {05}},\ \bibinfo {pages} {062} (\bibinfo {year} {2002})},\ \Eprint {http://arxiv.org/abs/hep-ph/0204232} {arXiv:hep-ph/0204232 [hep-ph]} \BibitemShut {NoStop}%
\bibitem [{\citenamefont {Kades}\ \emph {et~al.}(2020)\citenamefont {Kades}, \citenamefont {Pawlowski}, \citenamefont {Rothkopf}, \citenamefont {Scherzer}, \citenamefont {Urban}, \citenamefont {Wetzel}, \citenamefont {Wink},\ and\ \citenamefont {Ziegler}}]{Kades:2019wtd}%
  \BibitemOpen
  \bibfield  {author} {\bibinfo {author} {\bibfnamefont {L.}~\bibnamefont {Kades}}, \bibinfo {author} {\bibfnamefont {J.~M.}\ \bibnamefont {Pawlowski}}, \bibinfo {author} {\bibfnamefont {A.}~\bibnamefont {Rothkopf}}, \bibinfo {author} {\bibfnamefont {M.}~\bibnamefont {Scherzer}}, \bibinfo {author} {\bibfnamefont {J.~M.}\ \bibnamefont {Urban}}, \bibinfo {author} {\bibfnamefont {S.~J.}\ \bibnamefont {Wetzel}}, \bibinfo {author} {\bibfnamefont {N.}~\bibnamefont {Wink}}, \ and\ \bibinfo {author} {\bibfnamefont {F.~P.~G.}\ \bibnamefont {Ziegler}},\ }\href {\doibase 10.1103/PhysRevD.102.096001} {\bibfield  {journal} {\bibinfo  {journal} {Phys. Rev. D}\ }\textbf {\bibinfo {volume} {102}},\ \bibinfo {pages} {096001} (\bibinfo {year} {2020})},\ \Eprint {http://arxiv.org/abs/1905.04305} {arXiv:1905.04305 [physics.comp-ph]} \BibitemShut {NoStop}%
\bibitem [{\citenamefont {Cichy}\ \emph {et~al.}(2019)\citenamefont {Cichy}, \citenamefont {Del~Debbio},\ and\ \citenamefont {Giani}}]{Cichy:2019ebf}%
  \BibitemOpen
  \bibfield  {author} {\bibinfo {author} {\bibfnamefont {K.}~\bibnamefont {Cichy}}, \bibinfo {author} {\bibfnamefont {L.}~\bibnamefont {Del~Debbio}}, \ and\ \bibinfo {author} {\bibfnamefont {T.}~\bibnamefont {Giani}},\ }\href {\doibase 10.1007/JHEP10(2019)137} {\bibfield  {journal} {\bibinfo  {journal} {JHEP}\ }\textbf {\bibinfo {volume} {10}},\ \bibinfo {pages} {137} (\bibinfo {year} {2019})},\ \Eprint {http://arxiv.org/abs/1907.06037} {arXiv:1907.06037 [hep-ph]} \BibitemShut {NoStop}%
\bibitem [{\citenamefont {Del~Debbio}\ \emph {et~al.}(2021)\citenamefont {Del~Debbio}, \citenamefont {Giani}, \citenamefont {Karpie}, \citenamefont {Orginos}, \citenamefont {Radyushkin},\ and\ \citenamefont {Zafeiropoulos}}]{DelDebbio:2020rgv}%
  \BibitemOpen
  \bibfield  {author} {\bibinfo {author} {\bibfnamefont {L.}~\bibnamefont {Del~Debbio}}, \bibinfo {author} {\bibfnamefont {T.}~\bibnamefont {Giani}}, \bibinfo {author} {\bibfnamefont {J.}~\bibnamefont {Karpie}}, \bibinfo {author} {\bibfnamefont {K.}~\bibnamefont {Orginos}}, \bibinfo {author} {\bibfnamefont {A.}~\bibnamefont {Radyushkin}}, \ and\ \bibinfo {author} {\bibfnamefont {S.}~\bibnamefont {Zafeiropoulos}},\ }\href {\doibase 10.1007/JHEP02(2021)138} {\bibfield  {journal} {\bibinfo  {journal} {JHEP}\ }\textbf {\bibinfo {volume} {02}},\ \bibinfo {pages} {138} (\bibinfo {year} {2021})},\ \Eprint {http://arxiv.org/abs/2010.03996} {arXiv:2010.03996 [hep-ph]} \BibitemShut {NoStop}%
\bibitem [{\citenamefont {Dutrieux}\ \emph {et~al.}(2022)\citenamefont {Dutrieux}, \citenamefont {Grocholski}, \citenamefont {Moutarde},\ and\ \citenamefont {Sznajder}}]{Dutrieux:2021wll}%
  \BibitemOpen
  \bibfield  {author} {\bibinfo {author} {\bibfnamefont {H.}~\bibnamefont {Dutrieux}}, \bibinfo {author} {\bibfnamefont {O.}~\bibnamefont {Grocholski}}, \bibinfo {author} {\bibfnamefont {H.}~\bibnamefont {Moutarde}}, \ and\ \bibinfo {author} {\bibfnamefont {P.}~\bibnamefont {Sznajder}},\ }\href {\doibase 10.1140/epjc/s10052-022-10211-5} {\bibfield  {journal} {\bibinfo  {journal} {Eur. Phys. J. C}\ }\textbf {\bibinfo {volume} {82}},\ \bibinfo {pages} {252} (\bibinfo {year} {2022})},\ \bibinfo {note} {[Erratum: Eur.Phys.J.C 82, 389 (2022)]},\ \Eprint {http://arxiv.org/abs/2112.10528} {arXiv:2112.10528 [hep-ph]} \BibitemShut {NoStop}%
\bibitem [{\citenamefont {Khan}\ \emph {et~al.}(2023)\citenamefont {Khan}, \citenamefont {Liu},\ and\ \citenamefont {Sufian}}]{Khan:2022vot}%
  \BibitemOpen
  \bibfield  {author} {\bibinfo {author} {\bibfnamefont {T.}~\bibnamefont {Khan}}, \bibinfo {author} {\bibfnamefont {T.}~\bibnamefont {Liu}}, \ and\ \bibinfo {author} {\bibfnamefont {R.~S.}\ \bibnamefont {Sufian}},\ }\href {\doibase 10.1103/PhysRevD.108.074502} {\bibfield  {journal} {\bibinfo  {journal} {Phys. Rev. D}\ }\textbf {\bibinfo {volume} {108}},\ \bibinfo {pages} {074502} (\bibinfo {year} {2023})},\ \Eprint {http://arxiv.org/abs/2211.15587} {arXiv:2211.15587 [hep-lat]} \BibitemShut {NoStop}%
\bibitem [{\citenamefont {Chowdhury}\ \emph {et~al.}(2025)\citenamefont {Chowdhury}, \citenamefont {Izubuchi}, \citenamefont {Kamruzzaman}, \citenamefont {Karthik}, \citenamefont {Khan}, \citenamefont {Liu}, \citenamefont {Paul}, \citenamefont {Schoenleber},\ and\ \citenamefont {Sufian}}]{Chowdhury:2024ymm}%
  \BibitemOpen
  \bibfield  {author} {\bibinfo {author} {\bibfnamefont {T.~A.}\ \bibnamefont {Chowdhury}}, \bibinfo {author} {\bibfnamefont {T.}~\bibnamefont {Izubuchi}}, \bibinfo {author} {\bibfnamefont {M.}~\bibnamefont {Kamruzzaman}}, \bibinfo {author} {\bibfnamefont {N.}~\bibnamefont {Karthik}}, \bibinfo {author} {\bibfnamefont {T.}~\bibnamefont {Khan}}, \bibinfo {author} {\bibfnamefont {T.}~\bibnamefont {Liu}}, \bibinfo {author} {\bibfnamefont {A.}~\bibnamefont {Paul}}, \bibinfo {author} {\bibfnamefont {J.}~\bibnamefont {Schoenleber}}, \ and\ \bibinfo {author} {\bibfnamefont {R.~S.}\ \bibnamefont {Sufian}},\ }\href {\doibase 10.1103/PhysRevD.111.074509} {\bibfield  {journal} {\bibinfo  {journal} {Phys. Rev. D}\ }\textbf {\bibinfo {volume} {111}},\ \bibinfo {pages} {074509} (\bibinfo {year} {2025})},\ \Eprint {http://arxiv.org/abs/2409.17234} {arXiv:2409.17234 [hep-lat]} \BibitemShut {NoStop}%
\bibitem [{\citenamefont {Chu}\ \emph {et~al.}(2025)\citenamefont {Chu}, \citenamefont {Cichy}, \citenamefont {Constantinou}, \citenamefont {Sznajder},\ and\ \citenamefont {Wagner}}]{Chu:2025jsi}%
  \BibitemOpen
  \bibfield  {author} {\bibinfo {author} {\bibfnamefont {M.-H.}\ \bibnamefont {Chu}}, \bibinfo {author} {\bibfnamefont {K.}~\bibnamefont {Cichy}}, \bibinfo {author} {\bibfnamefont {M.}~\bibnamefont {Constantinou}}, \bibinfo {author} {\bibfnamefont {P.}~\bibnamefont {Sznajder}}, \ and\ \bibinfo {author} {\bibfnamefont {J.}~\bibnamefont {Wagner}},\ }\href@noop {} {\  (\bibinfo {year} {2025})},\ \Eprint {http://arxiv.org/abs/2509.15931} {arXiv:2509.15931 [hep-lat]} \BibitemShut {NoStop}%
\bibitem [{\citenamefont {Parisi}\ and\ \citenamefont {Sourlas}(1979)}]{Parisi:1978jv}%
  \BibitemOpen
  \bibfield  {author} {\bibinfo {author} {\bibfnamefont {G.}~\bibnamefont {Parisi}}\ and\ \bibinfo {author} {\bibfnamefont {N.}~\bibnamefont {Sourlas}},\ }\href {\doibase 10.1016/0550-3213(79)90448-6} {\bibfield  {journal} {\bibinfo  {journal} {Nucl. Phys. B}\ }\textbf {\bibinfo {volume} {151}},\ \bibinfo {pages} {421} (\bibinfo {year} {1979})}\BibitemShut {NoStop}%
\bibitem [{\citenamefont {Barker}\ \emph {et~al.}(1981)\citenamefont {Barker}, \citenamefont {Langensiepen},\ and\ \citenamefont {Shaw}}]{Barker:1980wu}%
  \BibitemOpen
  \bibfield  {author} {\bibinfo {author} {\bibfnamefont {I.~S.}\ \bibnamefont {Barker}}, \bibinfo {author} {\bibfnamefont {C.~S.}\ \bibnamefont {Langensiepen}}, \ and\ \bibinfo {author} {\bibfnamefont {G.}~\bibnamefont {Shaw}},\ }\href {\doibase 10.1016/0550-3213(81)90093-6} {\bibfield  {journal} {\bibinfo  {journal} {Nucl. Phys. B}\ }\textbf {\bibinfo {volume} {186}},\ \bibinfo {pages} {61} (\bibinfo {year} {1981})}\BibitemShut {NoStop}%
\bibitem [{\citenamefont {Barker}\ \emph {et~al.}(1983)\citenamefont {Barker}, \citenamefont {Martin},\ and\ \citenamefont {Shaw}}]{Barker:1982rv}%
  \BibitemOpen
  \bibfield  {author} {\bibinfo {author} {\bibfnamefont {I.~S.}\ \bibnamefont {Barker}}, \bibinfo {author} {\bibfnamefont {B.~R.}\ \bibnamefont {Martin}}, \ and\ \bibinfo {author} {\bibfnamefont {G.}~\bibnamefont {Shaw}},\ }\href {\doibase 10.1007/BF01571777} {\bibfield  {journal} {\bibinfo  {journal} {Z. Phys. C}\ }\textbf {\bibinfo {volume} {19}},\ \bibinfo {pages} {147} (\bibinfo {year} {1983})}\BibitemShut {NoStop}%
\bibitem [{\citenamefont {Chyla}\ and\ \citenamefont {Rames}(1986)}]{Chyla:1986eb}%
  \BibitemOpen
  \bibfield  {author} {\bibinfo {author} {\bibfnamefont {J.}~\bibnamefont {Chyla}}\ and\ \bibinfo {author} {\bibfnamefont {J.}~\bibnamefont {Rames}},\ }\href {\doibase 10.1007/BF01559606} {\bibfield  {journal} {\bibinfo  {journal} {Z. Phys. C}\ }\textbf {\bibinfo {volume} {31}},\ \bibinfo {pages} {151} (\bibinfo {year} {1986})}\BibitemShut {NoStop}%
\bibitem [{\citenamefont {Krivokhizhin}\ \emph {et~al.}(1987)\citenamefont {Krivokhizhin}, \citenamefont {Kurlovich}, \citenamefont {Sanadze}, \citenamefont {Savin}, \citenamefont {Sidorov},\ and\ \citenamefont {Skachkov}}]{Krivokhizhin:1987rz}%
  \BibitemOpen
  \bibfield  {author} {\bibinfo {author} {\bibfnamefont {V.~G.}\ \bibnamefont {Krivokhizhin}}, \bibinfo {author} {\bibfnamefont {S.~P.}\ \bibnamefont {Kurlovich}}, \bibinfo {author} {\bibfnamefont {V.~V.}\ \bibnamefont {Sanadze}}, \bibinfo {author} {\bibfnamefont {I.~A.}\ \bibnamefont {Savin}}, \bibinfo {author} {\bibfnamefont {A.~V.}\ \bibnamefont {Sidorov}}, \ and\ \bibinfo {author} {\bibfnamefont {N.~B.}\ \bibnamefont {Skachkov}},\ }\href {\doibase 10.1007/BF01556164} {\bibfield  {journal} {\bibinfo  {journal} {Z. Phys. C}\ }\textbf {\bibinfo {volume} {36}},\ \bibinfo {pages} {51} (\bibinfo {year} {1987})}\BibitemShut {NoStop}%
\bibitem [{\citenamefont {Krivokhizhin}\ \emph {et~al.}(1990)\citenamefont {Krivokhizhin}, \citenamefont {Kurlovich}, \citenamefont {Lednicky}, \citenamefont {Nemecek}, \citenamefont {Sanadze}, \citenamefont {Savin}, \citenamefont {Sidorov},\ and\ \citenamefont {Skachkov}}]{Krivokhizhin:1990ct}%
  \BibitemOpen
  \bibfield  {author} {\bibinfo {author} {\bibfnamefont {V.~G.}\ \bibnamefont {Krivokhizhin}}, \bibinfo {author} {\bibfnamefont {S.~P.}\ \bibnamefont {Kurlovich}}, \bibinfo {author} {\bibfnamefont {R.}~\bibnamefont {Lednicky}}, \bibinfo {author} {\bibfnamefont {S.}~\bibnamefont {Nemecek}}, \bibinfo {author} {\bibfnamefont {V.~V.}\ \bibnamefont {Sanadze}}, \bibinfo {author} {\bibfnamefont {I.~A.}\ \bibnamefont {Savin}}, \bibinfo {author} {\bibfnamefont {A.~V.}\ \bibnamefont {Sidorov}}, \ and\ \bibinfo {author} {\bibfnamefont {N.~B.}\ \bibnamefont {Skachkov}},\ }\href {\doibase 10.1007/BF01554485} {\bibfield  {journal} {\bibinfo  {journal} {Z. Phys. C}\ }\textbf {\bibinfo {volume} {48}},\ \bibinfo {pages} {347} (\bibinfo {year} {1990})}\BibitemShut {NoStop}%
\bibitem [{\citenamefont {Atashbar~Tehrani}\ and\ \citenamefont {Khorramian}(2007)}]{AtashbarTehrani:2007odq}%
  \BibitemOpen
  \bibfield  {author} {\bibinfo {author} {\bibfnamefont {S.}~\bibnamefont {Atashbar~Tehrani}}\ and\ \bibinfo {author} {\bibfnamefont {A.~N.}\ \bibnamefont {Khorramian}},\ }\href {\doibase 10.1088/1126-6708/2007/07/048} {\bibfield  {journal} {\bibinfo  {journal} {JHEP}\ }\textbf {\bibinfo {volume} {07}},\ \bibinfo {pages} {048} (\bibinfo {year} {2007})},\ \Eprint {http://arxiv.org/abs/0705.2647} {arXiv:0705.2647 [hep-ph]} \BibitemShut {NoStop}%
\bibitem [{\citenamefont {Khorramian}\ \emph {et~al.}(2009)\citenamefont {Khorramian}, \citenamefont {Atashbar~Tehrani}, \citenamefont {Khanpour},\ and\ \citenamefont {Monfared}}]{Khorramian:2009asi}%
  \BibitemOpen
  \bibfield  {author} {\bibinfo {author} {\bibfnamefont {A.~N.}\ \bibnamefont {Khorramian}}, \bibinfo {author} {\bibfnamefont {S.}~\bibnamefont {Atashbar~Tehrani}}, \bibinfo {author} {\bibfnamefont {H.}~\bibnamefont {Khanpour}}, \ and\ \bibinfo {author} {\bibfnamefont {S.~T.}\ \bibnamefont {Monfared}},\ }\href {\doibase 10.1007/s10751-009-0090-x} {\bibfield  {journal} {\bibinfo  {journal} {Hyperfine Interact.}\ }\textbf {\bibinfo {volume} {194}},\ \bibinfo {pages} {337} (\bibinfo {year} {2009})}\BibitemShut {NoStop}%
\bibitem [{\citenamefont {Taghavi-Shahri}\ \emph {et~al.}(2016)\citenamefont {Taghavi-Shahri}, \citenamefont {Khanpour}, \citenamefont {Atashbar~Tehrani},\ and\ \citenamefont {Alizadeh~Yazdi}}]{Taghavi-Shahri:2016idw}%
  \BibitemOpen
  \bibfield  {author} {\bibinfo {author} {\bibfnamefont {F.}~\bibnamefont {Taghavi-Shahri}}, \bibinfo {author} {\bibfnamefont {H.}~\bibnamefont {Khanpour}}, \bibinfo {author} {\bibfnamefont {S.}~\bibnamefont {Atashbar~Tehrani}}, \ and\ \bibinfo {author} {\bibfnamefont {Z.}~\bibnamefont {Alizadeh~Yazdi}},\ }\href {\doibase 10.1103/PhysRevD.93.114024} {\bibfield  {journal} {\bibinfo  {journal} {Phys. Rev. D}\ }\textbf {\bibinfo {volume} {93}},\ \bibinfo {pages} {114024} (\bibinfo {year} {2016})},\ \Eprint {http://arxiv.org/abs/1603.03157} {arXiv:1603.03157 [hep-ph]} \BibitemShut {NoStop}%
\bibitem [{\citenamefont {Karpie}\ \emph {et~al.}(2021)\citenamefont {Karpie}, \citenamefont {Orginos}, \citenamefont {Radyushkin},\ and\ \citenamefont {Zafeiropoulos}}]{Karpie:2021pap}%
  \BibitemOpen
  \bibfield  {author} {\bibinfo {author} {\bibfnamefont {J.}~\bibnamefont {Karpie}}, \bibinfo {author} {\bibfnamefont {K.}~\bibnamefont {Orginos}}, \bibinfo {author} {\bibfnamefont {A.}~\bibnamefont {Radyushkin}}, \ and\ \bibinfo {author} {\bibfnamefont {S.}~\bibnamefont {Zafeiropoulos}} (\bibinfo {collaboration} {HadStruc}),\ }\href {\doibase 10.1007/JHEP11(2021)024} {\bibfield  {journal} {\bibinfo  {journal} {JHEP}\ }\textbf {\bibinfo {volume} {11}},\ \bibinfo {pages} {024} (\bibinfo {year} {2021})},\ \Eprint {http://arxiv.org/abs/2105.13313} {arXiv:2105.13313 [hep-lat]} \BibitemShut {NoStop}%
\bibitem [{\citenamefont {Kotz}\ \emph {et~al.}(2024)\citenamefont {Kotz}, \citenamefont {Courtoy}, \citenamefont {Nadolsky}, \citenamefont {Olness},\ and\ \citenamefont {Ponce-Chavez}}]{Kotz:2023pbu}%
  \BibitemOpen
  \bibfield  {author} {\bibinfo {author} {\bibfnamefont {L.}~\bibnamefont {Kotz}}, \bibinfo {author} {\bibfnamefont {A.}~\bibnamefont {Courtoy}}, \bibinfo {author} {\bibfnamefont {P.}~\bibnamefont {Nadolsky}}, \bibinfo {author} {\bibfnamefont {F.}~\bibnamefont {Olness}}, \ and\ \bibinfo {author} {\bibfnamefont {M.}~\bibnamefont {Ponce-Chavez}},\ }\href {\doibase 10.1103/PhysRevD.109.074027} {\bibfield  {journal} {\bibinfo  {journal} {Phys. Rev. D}\ }\textbf {\bibinfo {volume} {109}},\ \bibinfo {pages} {074027} (\bibinfo {year} {2024})},\ \Eprint {http://arxiv.org/abs/2311.08447} {arXiv:2311.08447 [hep-ph]} \BibitemShut {NoStop}%
\bibitem [{\citenamefont {Liang}\ \emph {et~al.}(2020)\citenamefont {Liang}, \citenamefont {Draper}, \citenamefont {Liu}, \citenamefont {Rothkopf},\ and\ \citenamefont {Yang}}]{Liang:2019frk}%
  \BibitemOpen
  \bibfield  {author} {\bibinfo {author} {\bibfnamefont {J.}~\bibnamefont {Liang}}, \bibinfo {author} {\bibfnamefont {T.}~\bibnamefont {Draper}}, \bibinfo {author} {\bibfnamefont {K.-F.}\ \bibnamefont {Liu}}, \bibinfo {author} {\bibfnamefont {A.}~\bibnamefont {Rothkopf}}, \ and\ \bibinfo {author} {\bibfnamefont {Y.-B.}\ \bibnamefont {Yang}} (\bibinfo {collaboration} {XQCD}),\ }\href {\doibase 10.1103/PhysRevD.101.114503} {\bibfield  {journal} {\bibinfo  {journal} {Phys. Rev. D}\ }\textbf {\bibinfo {volume} {101}},\ \bibinfo {pages} {114503} (\bibinfo {year} {2020})},\ \Eprint {http://arxiv.org/abs/1906.05312} {arXiv:1906.05312 [hep-ph]} \BibitemShut {NoStop}%
\bibitem [{\citenamefont {Egerer}\ \emph {et~al.}(2021)\citenamefont {Egerer}, \citenamefont {Edwards}, \citenamefont {Kallidonis}, \citenamefont {Orginos}, \citenamefont {Radyushkin}, \citenamefont {Richards}, \citenamefont {Romero},\ and\ \citenamefont {Zafeiropoulos}}]{Egerer:2021ymv}%
  \BibitemOpen
  \bibfield  {author} {\bibinfo {author} {\bibfnamefont {C.}~\bibnamefont {Egerer}}, \bibinfo {author} {\bibfnamefont {R.~G.}\ \bibnamefont {Edwards}}, \bibinfo {author} {\bibfnamefont {C.}~\bibnamefont {Kallidonis}}, \bibinfo {author} {\bibfnamefont {K.}~\bibnamefont {Orginos}}, \bibinfo {author} {\bibfnamefont {A.~V.}\ \bibnamefont {Radyushkin}}, \bibinfo {author} {\bibfnamefont {D.~G.}\ \bibnamefont {Richards}}, \bibinfo {author} {\bibfnamefont {E.}~\bibnamefont {Romero}}, \ and\ \bibinfo {author} {\bibfnamefont {S.}~\bibnamefont {Zafeiropoulos}} (\bibinfo {collaboration} {HadStruc}),\ }\href {\doibase 10.1007/JHEP11(2021)148} {\bibfield  {journal} {\bibinfo  {journal} {JHEP}\ }\textbf {\bibinfo {volume} {11}},\ \bibinfo {pages} {148} (\bibinfo {year} {2021})},\ \Eprint {http://arxiv.org/abs/2107.05199} {arXiv:2107.05199 [hep-lat]} \BibitemShut {NoStop}%
\bibitem [{\citenamefont {Dutrieux}\ \emph {et~al.}(2025{\natexlab{c}})\citenamefont {Dutrieux}, \citenamefont {Karpie}, \citenamefont {Monahan}, \citenamefont {Orginos}, \citenamefont {Radyushkin}, \citenamefont {Richards},\ and\ \citenamefont {Zafeiropoulos}}]{Dutrieux:2025axb}%
  \BibitemOpen
  \bibfield  {author} {\bibinfo {author} {\bibfnamefont {H.}~\bibnamefont {Dutrieux}}, \bibinfo {author} {\bibfnamefont {J.}~\bibnamefont {Karpie}}, \bibinfo {author} {\bibfnamefont {C.~J.}\ \bibnamefont {Monahan}}, \bibinfo {author} {\bibfnamefont {K.}~\bibnamefont {Orginos}}, \bibinfo {author} {\bibfnamefont {A.}~\bibnamefont {Radyushkin}}, \bibinfo {author} {\bibfnamefont {D.}~\bibnamefont {Richards}}, \ and\ \bibinfo {author} {\bibfnamefont {S.}~\bibnamefont {Zafeiropoulos}},\ }\href@noop {} {\  (\bibinfo {year} {2025}{\natexlab{c}})},\ \Eprint {http://arxiv.org/abs/2506.24037} {arXiv:2506.24037 [hep-lat]} \BibitemShut {NoStop}%
\bibitem [{\citenamefont {Gao}\ \emph {et~al.}(2022)\citenamefont {Gao}, \citenamefont {Hanlon}, \citenamefont {Mukherjee}, \citenamefont {Petreczky}, \citenamefont {Scior}, \citenamefont {Syritsyn},\ and\ \citenamefont {Zhao}}]{Gao:2021dbh}%
  \BibitemOpen
  \bibfield  {author} {\bibinfo {author} {\bibfnamefont {X.}~\bibnamefont {Gao}}, \bibinfo {author} {\bibfnamefont {A.~D.}\ \bibnamefont {Hanlon}}, \bibinfo {author} {\bibfnamefont {S.}~\bibnamefont {Mukherjee}}, \bibinfo {author} {\bibfnamefont {P.}~\bibnamefont {Petreczky}}, \bibinfo {author} {\bibfnamefont {P.}~\bibnamefont {Scior}}, \bibinfo {author} {\bibfnamefont {S.}~\bibnamefont {Syritsyn}}, \ and\ \bibinfo {author} {\bibfnamefont {Y.}~\bibnamefont {Zhao}},\ }\href {\doibase 10.1103/PhysRevLett.128.142003} {\bibfield  {journal} {\bibinfo  {journal} {Phys. Rev. Lett.}\ }\textbf {\bibinfo {volume} {128}},\ \bibinfo {pages} {142003} (\bibinfo {year} {2022})},\ \Eprint {http://arxiv.org/abs/2112.02208} {arXiv:2112.02208 [hep-lat]} \BibitemShut {NoStop}%
\bibitem [{\citenamefont {Chen}\ \emph {et~al.}(2025)\citenamefont {Chen} \emph {et~al.}}]{Chen:2025cxr}%
  \BibitemOpen
  \bibfield  {author} {\bibinfo {author} {\bibfnamefont {J.-W.}\ \bibnamefont {Chen}} \emph {et~al.},\ }\href@noop {} {\  (\bibinfo {year} {2025})},\ \Eprint {http://arxiv.org/abs/2505.14619} {arXiv:2505.14619 [hep-lat]} \BibitemShut {NoStop}%
\bibitem [{\citenamefont {Skilling}\ and\ \citenamefont {Gull}(1991)}]{MEM}%
  \BibitemOpen
  \bibfield  {author} {\bibinfo {author} {\bibfnamefont {J.}~\bibnamefont {Skilling}}\ and\ \bibinfo {author} {\bibfnamefont {S.~F.}\ \bibnamefont {Gull}},\ }\href {http://www.jstor.org/stable/4355715} {\bibfield  {journal} {\bibinfo  {journal} {Lecture Notes-Monograph Series}\ }\textbf {\bibinfo {volume} {20}},\ \bibinfo {pages} {341} (\bibinfo {year} {1991})}\BibitemShut {NoStop}%
\bibitem [{\citenamefont {Jarrell}\ and\ \citenamefont {Gubernatis}(1996)}]{MEM2}%
  \BibitemOpen
  \bibfield  {author} {\bibinfo {author} {\bibfnamefont {M.}~\bibnamefont {Jarrell}}\ and\ \bibinfo {author} {\bibfnamefont {J.}~\bibnamefont {Gubernatis}},\ }\href {\doibase https://doi.org/10.1016/0370-1573(95)00074-7} {\bibfield  {journal} {\bibinfo  {journal} {Physics Reports}\ }\textbf {\bibinfo {volume} {269}},\ \bibinfo {pages} {133} (\bibinfo {year} {1996})}\BibitemShut {NoStop}%
\bibitem [{\citenamefont {Burnier}\ and\ \citenamefont {Rothkopf}(2013)}]{Burnier:2013nla}%
  \BibitemOpen
  \bibfield  {author} {\bibinfo {author} {\bibfnamefont {Y.}~\bibnamefont {Burnier}}\ and\ \bibinfo {author} {\bibfnamefont {A.}~\bibnamefont {Rothkopf}},\ }\href {\doibase 10.1103/PhysRevLett.111.182003} {\bibfield  {journal} {\bibinfo  {journal} {Phys. Rev. Lett.}\ }\textbf {\bibinfo {volume} {111}},\ \bibinfo {pages} {182003} (\bibinfo {year} {2013})},\ \Eprint {http://arxiv.org/abs/1307.6106} {arXiv:1307.6106 [hep-lat]} \BibitemShut {NoStop}%
\bibitem [{\citenamefont {Gibbs}(1998)}]{gibbs1998bayesian}%
  \BibitemOpen
  \bibfield  {author} {\bibinfo {author} {\bibfnamefont {M.~N.}\ \bibnamefont {Gibbs}},\ }\emph {\bibinfo {title} {Bayesian Gaussian processes for regression and classification}},\ \href@noop {} {Ph.D. thesis},\ \bibinfo  {school} {Citeseer} (\bibinfo {year} {1998})\BibitemShut {NoStop}%
\bibitem [{\citenamefont {Bochner}\ \emph {et~al.}(1959)\citenamefont {Bochner}, \citenamefont {Functions}, \citenamefont {Integrals}, \citenamefont {Analysis}, \citenamefont {Tenenbaum},\ and\ \citenamefont {Pollard}}]{bochner}%
  \BibitemOpen
  \bibfield  {author} {\bibinfo {author} {\bibfnamefont {S.}~\bibnamefont {Bochner}}, \bibinfo {author} {\bibfnamefont {M.}~\bibnamefont {Functions}}, \bibinfo {author} {\bibfnamefont {S.}~\bibnamefont {Integrals}}, \bibinfo {author} {\bibfnamefont {H.}~\bibnamefont {Analysis}}, \bibinfo {author} {\bibfnamefont {M.}~\bibnamefont {Tenenbaum}}, \ and\ \bibinfo {author} {\bibfnamefont {H.}~\bibnamefont {Pollard}},\ }\href {http://www.jstor.org/stable/j.ctt1b9s09r} {\emph {\bibinfo {title} {Lectures on Fourier Integrals. (AM-42)}}}\ (\bibinfo  {publisher} {Princeton University Press},\ \bibinfo {year} {1959})\BibitemShut {NoStop}%
\bibitem [{\citenamefont {Stein}(1999)}]{stein1999interpolation}%
  \BibitemOpen
  \bibfield  {author} {\bibinfo {author} {\bibfnamefont {M.}~\bibnamefont {Stein}},\ }\href@noop {} {\emph {\bibinfo {title} {Interpolation of Spatial Data: Some Theory for Kriging}}},\ Springer Series in Statistics\ (\bibinfo  {publisher} {Springer New York},\ \bibinfo {year} {1999})\BibitemShut {NoStop}%
\bibitem [{\citenamefont {Gihman}\ and\ \citenamefont {Skorokhod}(2004)}]{gihmanskorohod1974}%
  \BibitemOpen
  \bibfield  {author} {\bibinfo {author} {\bibfnamefont {I.}~\bibnamefont {Gihman}}\ and\ \bibinfo {author} {\bibfnamefont {A.}~\bibnamefont {Skorokhod}},\ }\href@noop {} {\emph {\bibinfo {title} {The Theory of Stochastic Processes I}}},\ Springer Series in Classics in Mathematics; Reprint of 1974 edition\ (\bibinfo  {publisher} {Springer Berlin},\ \bibinfo {year} {2004})\BibitemShut {NoStop}%
\bibitem [{\citenamefont {Wilson}\ and\ \citenamefont {Adams}(2013)}]{pmlr-v28-wilson13}%
  \BibitemOpen
  \bibfield  {author} {\bibinfo {author} {\bibfnamefont {A.}~\bibnamefont {Wilson}}\ and\ \bibinfo {author} {\bibfnamefont {R.}~\bibnamefont {Adams}},\ }in\ \href {https://proceedings.mlr.press/v28/wilson13.html} {\emph {\bibinfo {booktitle} {Proceedings of the 30th International Conference on Machine Learning}}},\ \bibinfo {series} {Proceedings of Machine Learning Research}, Vol.~\bibinfo {volume} {28},\ \bibinfo {editor} {edited by\ \bibinfo {editor} {\bibfnamefont {S.}~\bibnamefont {Dasgupta}}\ and\ \bibinfo {editor} {\bibfnamefont {D.}~\bibnamefont {McAllester}}}\ (\bibinfo  {publisher} {PMLR},\ \bibinfo {address} {Atlanta, Georgia, USA},\ \bibinfo {year} {2013})\ pp.\ \bibinfo {pages} {1067--1075}\BibitemShut {NoStop}%
\bibitem [{\citenamefont {Rasmussen}\ and\ \citenamefont {Williams}(2006{\natexlab{b}})}]{Rasmussen2006Gaussian}%
  \BibitemOpen
  \bibfield  {author} {\bibinfo {author} {\bibfnamefont {C.~E.}\ \bibnamefont {Rasmussen}}\ and\ \bibinfo {author} {\bibfnamefont {C.~K.~I.}\ \bibnamefont {Williams}},\ }\href@noop {} {\emph {\bibinfo {title} {Gaussian Processes for Machine Learning}}}\ (\bibinfo  {publisher} {The MIT Press},\ \bibinfo {year} {2006})\BibitemShut {NoStop}%
\bibitem [{\citenamefont {Zhu}\ \emph {et~al.}(1997)\citenamefont {Zhu}, \citenamefont {Williams}, \citenamefont {Rohwer}, \citenamefont {Morciniec},\ and\ \citenamefont {Hammel}}]{Zhu1998}%
  \BibitemOpen
  \bibfield  {author} {\bibinfo {author} {\bibfnamefont {H.}~\bibnamefont {Zhu}}, \bibinfo {author} {\bibfnamefont {C.}~\bibnamefont {Williams}}, \bibinfo {author} {\bibfnamefont {R.}~\bibnamefont {Rohwer}}, \bibinfo {author} {\bibfnamefont {M.}~\bibnamefont {Morciniec}}, \ and\ \bibinfo {author} {\bibfnamefont {M.}~\bibnamefont {Hammel}},\ }\href {https://www.researchgate.net/publication/2770035_Gaussian_Regression_and_Optimal_Finite_Dimensional_Linear_Models} {\emph {\bibinfo {title} {Gaussian Regression and Optimal Finite Dimensional Linear Models}}},\ \bibinfo {type} {WorkingPaper}\ (\bibinfo {year} {1997})\BibitemShut {NoStop}%
\bibitem [{\citenamefont {Duvenaud}(2014)}]{duvenaud_2014}%
  \BibitemOpen
  \bibfield  {author} {\bibinfo {author} {\bibfnamefont {D.}~\bibnamefont {Duvenaud}},\ }\emph {\bibinfo {title} {Automatic model construction with Gaussian processes}},\ \href {\doibase 10.17863/CAM.14087} {Ph.D. thesis},\ \bibinfo  {school} {Apollo - University of Cambridge Repository} (\bibinfo {year} {2014})\BibitemShut {NoStop}%
\bibitem [{\citenamefont {Ji}\ \emph {et~al.}(2021)\citenamefont {Ji}, \citenamefont {Liu}, \citenamefont {Sch{\"a}fer}, \citenamefont {Wang}, \citenamefont {Yang}, \citenamefont {Zhang},\ and\ \citenamefont {Zhao}}]{Ji:2020brr}%
  \BibitemOpen
  \bibfield  {author} {\bibinfo {author} {\bibfnamefont {X.}~\bibnamefont {Ji}}, \bibinfo {author} {\bibfnamefont {Y.}~\bibnamefont {Liu}}, \bibinfo {author} {\bibfnamefont {A.}~\bibnamefont {Sch{\"a}fer}}, \bibinfo {author} {\bibfnamefont {W.}~\bibnamefont {Wang}}, \bibinfo {author} {\bibfnamefont {Y.-B.}\ \bibnamefont {Yang}}, \bibinfo {author} {\bibfnamefont {J.-H.}\ \bibnamefont {Zhang}}, \ and\ \bibinfo {author} {\bibfnamefont {Y.}~\bibnamefont {Zhao}},\ }\href {\doibase 10.1016/j.nuclphysb.2021.115311} {\bibfield  {journal} {\bibinfo  {journal} {Nucl. Phys. B}\ }\textbf {\bibinfo {volume} {964}},\ \bibinfo {pages} {115311} (\bibinfo {year} {2021})},\ \Eprint {http://arxiv.org/abs/2008.03886} {arXiv:2008.03886 [hep-ph]} \BibitemShut {NoStop}%
\bibitem [{\citenamefont {Max}(1950)}]{woodbury}%
  \BibitemOpen
  \bibfield  {author} {\bibinfo {author} {\bibfnamefont {A.~W.}\ \bibnamefont {Max}},\ }in\ \href@noop {} {\emph {\bibinfo {booktitle} {Memorandum Rept. 42, Statistical Research Group}}}\ (\bibinfo  {publisher} {Princeton Univ.},\ \bibinfo {year} {1950})\ p.~\bibinfo {pages} {4}\BibitemShut {NoStop}%
\bibitem [{\citenamefont {Valentine}\ and\ \citenamefont {Sambridge}(2019)}]{ValentineSambridgeGP}%
  \BibitemOpen
  \bibfield  {author} {\bibinfo {author} {\bibfnamefont {A.~P.}\ \bibnamefont {Valentine}}\ and\ \bibinfo {author} {\bibfnamefont {M.}~\bibnamefont {Sambridge}},\ }\href {\doibase 10.1093/gji/ggz520} {\bibfield  {journal} {\bibinfo  {journal} {Geophysical Journal International}\ }\textbf {\bibinfo {volume} {220}},\ \bibinfo {pages} {1632} (\bibinfo {year} {2019})},\ \Eprint {http://arxiv.org/abs/https://academic.oup.com/gji/article-pdf/220/3/1632/31578341/ggz520.pdf} {https://academic.oup.com/gji/article-pdf/220/3/1632/31578341/ggz520.pdf} \BibitemShut {NoStop}%
\bibitem [{\citenamefont {Leamer}(1978)}]{leamer1978specification}%
  \BibitemOpen
  \bibfield  {author} {\bibinfo {author} {\bibfnamefont {E.}~\bibnamefont {Leamer}},\ }\href {https://books.google.com/books?id=sYVYAAAAMAAJ} {\emph {\bibinfo {title} {Specification Searches: Ad Hoc Inference with Nonexperimental Data}}},\ Wiley Series in Probability and Statistics - Applied Probability and Statistics Section\ (\bibinfo  {publisher} {Wiley},\ \bibinfo {year} {1978})\BibitemShut {NoStop}%
\bibitem [{\citenamefont {Kass}\ and\ \citenamefont {Raftery}(1995)}]{kass1995}%
  \BibitemOpen
  \bibfield  {author} {\bibinfo {author} {\bibfnamefont {R.~E.}\ \bibnamefont {Kass}}\ and\ \bibinfo {author} {\bibfnamefont {A.~E.}\ \bibnamefont {Raftery}},\ }\href {\doibase 10.1080/01621459.1995.10476572} {\bibfield  {journal} {\bibinfo  {journal} {Journal of the American Statistical Association}\ }\textbf {\bibinfo {volume} {90}},\ \bibinfo {pages} {773} (\bibinfo {year} {1995})},\ \Eprint {http://arxiv.org/abs/https://www.tandfonline.com/doi/pdf/10.1080/01621459.1995.10476572} {https://www.tandfonline.com/doi/pdf/10.1080/01621459.1995.10476572} \BibitemShut {NoStop}%
\bibitem [{\citenamefont {Jay}\ and\ \citenamefont {Neil}(2021)}]{Jay:2020jkz}%
  \BibitemOpen
  \bibfield  {author} {\bibinfo {author} {\bibfnamefont {W.~I.}\ \bibnamefont {Jay}}\ and\ \bibinfo {author} {\bibfnamefont {E.~T.}\ \bibnamefont {Neil}},\ }\href {\doibase 10.1103/PhysRevD.103.114502} {\bibfield  {journal} {\bibinfo  {journal} {Phys. Rev. D}\ }\textbf {\bibinfo {volume} {103}},\ \bibinfo {pages} {114502} (\bibinfo {year} {2021})},\ \Eprint {http://arxiv.org/abs/2008.01069} {arXiv:2008.01069 [stat.ME]} \BibitemShut {NoStop}%
\bibitem [{\citenamefont {Zhou}(2023)}]{zhou2009}%
  \BibitemOpen
  \bibfield  {author} {\bibinfo {author} {\bibfnamefont {S.}~\bibnamefont {Zhou}},\ }\href {\doibase 10.3390/e25030468} {\bibfield  {journal} {\bibinfo  {journal} {Entropy}\ }\textbf {\bibinfo {volume} {25}} (\bibinfo {year} {2023}),\ 10.3390/e25030468}\BibitemShut {NoStop}%
\bibitem [{\citenamefont {Takeuchi}(1976)}]{takeuchiinfo}%
  \BibitemOpen
  \bibfield  {author} {\bibinfo {author} {\bibfnamefont {K.}~\bibnamefont {Takeuchi}},\ }\href@noop {} {\bibfield  {journal} {\bibinfo  {journal} {Suri-Kagaku (Mathematical Sciences)}\ }\textbf {\bibinfo {volume} {153}} (\bibinfo {year} {1976})}\BibitemShut {NoStop}%
\bibitem [{\citenamefont {Neil}\ and\ \citenamefont {Sitison}(2024)}]{Neil:2022joj}%
  \BibitemOpen
  \bibfield  {author} {\bibinfo {author} {\bibfnamefont {E.~T.}\ \bibnamefont {Neil}}\ and\ \bibinfo {author} {\bibfnamefont {J.~W.}\ \bibnamefont {Sitison}},\ }\href {\doibase 10.1103/PhysRevD.109.014510} {\bibfield  {journal} {\bibinfo  {journal} {Phys. Rev. D}\ }\textbf {\bibinfo {volume} {109}},\ \bibinfo {pages} {014510} (\bibinfo {year} {2024})},\ \Eprint {http://arxiv.org/abs/2208.14983} {arXiv:2208.14983 [stat.ME]} \BibitemShut {NoStop}%
\bibitem [{\citenamefont {Akaike}(1974)}]{akaikeinfo}%
  \BibitemOpen
  \bibfield  {author} {\bibinfo {author} {\bibfnamefont {H.}~\bibnamefont {Akaike}},\ }\href {\doibase 10.1109/TAC.1974.1100705} {\bibfield  {journal} {\bibinfo  {journal} {IEEE Transactions on Automatic Control}\ }\textbf {\bibinfo {volume} {19}},\ \bibinfo {pages} {716} (\bibinfo {year} {1974})}\BibitemShut {NoStop}%
\bibitem [{\citenamefont {Ball}\ \emph {et~al.}(2022)\citenamefont {Ball} \emph {et~al.}}]{NNPDF:2021njg}%
  \BibitemOpen
  \bibfield  {author} {\bibinfo {author} {\bibfnamefont {R.~D.}\ \bibnamefont {Ball}} \emph {et~al.} (\bibinfo {collaboration} {NNPDF}),\ }\href {\doibase 10.1140/epjc/s10052-022-10328-7} {\bibfield  {journal} {\bibinfo  {journal} {Eur. Phys. J. C}\ }\textbf {\bibinfo {volume} {82}},\ \bibinfo {pages} {428} (\bibinfo {year} {2022})},\ \Eprint {http://arxiv.org/abs/2109.02653} {arXiv:2109.02653 [hep-ph]} \BibitemShut {NoStop}%
\bibitem [{\citenamefont {Edwards}\ and\ \citenamefont {Joo}(2005)}]{Edwards:2004sx}%
  \BibitemOpen
  \bibfield  {author} {\bibinfo {author} {\bibfnamefont {R.~G.}\ \bibnamefont {Edwards}}\ and\ \bibinfo {author} {\bibfnamefont {B.}~\bibnamefont {Joo}} (\bibinfo {collaboration} {SciDAC, LHPC, UKQCD}),\ }\href {\doibase 10.1016/j.nuclphysbps.2004.11.254} {\bibfield  {journal} {\bibinfo  {journal} {Nucl. Phys. B Proc. Suppl.}\ }\textbf {\bibinfo {volume} {140}},\ \bibinfo {pages} {832} (\bibinfo {year} {2005})},\ \Eprint {http://arxiv.org/abs/hep-lat/0409003} {arXiv:hep-lat/0409003} \BibitemShut {NoStop}%
\bibitem [{\citenamefont {Clark}\ \emph {et~al.}(2010)\citenamefont {Clark}, \citenamefont {Babich}, \citenamefont {Barros}, \citenamefont {Brower},\ and\ \citenamefont {Rebbi}}]{Clark:2009wm}%
  \BibitemOpen
  \bibfield  {author} {\bibinfo {author} {\bibfnamefont {M.}~\bibnamefont {Clark}}, \bibinfo {author} {\bibfnamefont {R.}~\bibnamefont {Babich}}, \bibinfo {author} {\bibfnamefont {K.}~\bibnamefont {Barros}}, \bibinfo {author} {\bibfnamefont {R.}~\bibnamefont {Brower}}, \ and\ \bibinfo {author} {\bibfnamefont {C.}~\bibnamefont {Rebbi}},\ }\href {\doibase 10.1016/j.cpc.2010.05.002} {\bibfield  {journal} {\bibinfo  {journal} {Comput. Phys. Commun.}\ }\textbf {\bibinfo {volume} {181}},\ \bibinfo {pages} {1517} (\bibinfo {year} {2010})},\ \Eprint {http://arxiv.org/abs/0911.3191} {arXiv:0911.3191 [hep-lat]} \BibitemShut {NoStop}%
\bibitem [{\citenamefont {Babich}\ \emph {et~al.}(2010)\citenamefont {Babich}, \citenamefont {Clark},\ and\ \citenamefont {Joo}}]{Babich:2010mu}%
  \BibitemOpen
  \bibfield  {author} {\bibinfo {author} {\bibfnamefont {R.}~\bibnamefont {Babich}}, \bibinfo {author} {\bibfnamefont {M.~A.}\ \bibnamefont {Clark}}, \ and\ \bibinfo {author} {\bibfnamefont {B.}~\bibnamefont {Joo}},\ }in\ \href@noop {} {\emph {\bibinfo {booktitle} {{SC 10 (Supercomputing 2010)}}}}\ (\bibinfo {year} {2010})\ \Eprint {http://arxiv.org/abs/1011.0024} {arXiv:1011.0024 [hep-lat]} \BibitemShut {NoStop}%
\bibitem [{\citenamefont {Chen}\ \emph {et~al.}(2023)\citenamefont {Chen}, \citenamefont {Edwards},\ and\ \citenamefont {Mao}}]{Chen:2023zyy}%
  \BibitemOpen
  \bibfield  {author} {\bibinfo {author} {\bibfnamefont {J.}~\bibnamefont {Chen}}, \bibinfo {author} {\bibfnamefont {R.~G.}\ \bibnamefont {Edwards}}, \ and\ \bibinfo {author} {\bibfnamefont {W.}~\bibnamefont {Mao}},\ }in\ \href {\doibase 10.1145/3592979.3593409} {\emph {\bibinfo {booktitle} {{Platform for Advanced Scientific Computing}}}}\ (\bibinfo {year} {2023})\BibitemShut {NoStop}%
\bibitem [{\citenamefont {Stanzione}\ \emph {et~al.}(2020)\citenamefont {Stanzione}, \citenamefont {West}, \citenamefont {Evans}, \citenamefont {Minyard}, \citenamefont {Ghattas},\ and\ \citenamefont {Panda}}]{10.1145/3311790.3396656}%
  \BibitemOpen
  \bibfield  {author} {\bibinfo {author} {\bibfnamefont {D.}~\bibnamefont {Stanzione}}, \bibinfo {author} {\bibfnamefont {J.}~\bibnamefont {West}}, \bibinfo {author} {\bibfnamefont {R.~T.}\ \bibnamefont {Evans}}, \bibinfo {author} {\bibfnamefont {T.}~\bibnamefont {Minyard}}, \bibinfo {author} {\bibfnamefont {O.}~\bibnamefont {Ghattas}}, \ and\ \bibinfo {author} {\bibfnamefont {D.~K.}\ \bibnamefont {Panda}},\ }in\ \href {\doibase 10.1145/3311790.3396656} {\emph {\bibinfo {booktitle} {Practice and Experience in Advanced Research Computing}}},\ \bibinfo {series and number} {PEARC '20}\ (\bibinfo  {publisher} {Association for Computing Machinery},\ \bibinfo {address} {New York, NY, USA},\ \bibinfo {year} {2020})\ p.\ \bibinfo {pages} {106–111}\BibitemShut {NoStop}%
\end{thebibliography}%
\end{document}